\documentclass[a4paper,11pt]{article}
\pdfoutput=1 % if your are submitting a pdflatex (i.e. if you have
\usepackage{jheppub} % for details on the use of the package, please
\usepackage[T1]{fontenc} % if needed

\usepackage{amsmath,amsfonts,amsbsy,amssymb,array,accents,dsfont}
\usepackage{enumerate,latexsym,graphicx}
\usepackage{xcolor,tikz}
\usepackage{tcolorbox}
\usepackage{mathtools}
\usepackage{braket}
\usepackage{subfigure}
\usepackage{stmaryrd}

\def\cV{\mathcal {V}} \def\cW{\mathcal {W}}

\newcommand{\beq}{\begin{equation}}
\newcommand{\eeq}{\end{equation}}
\newcommand{\bea}{\begin{eqnarray}}
\newcommand{\eea}{\end{eqnarray}}

\newcommand{\vep}{\varepsilon}
\newcommand{\ep}{\epsilon}

\newcommand{\der}{\partial}
\newcommand{\derb}{\bar{\partial}}
\newcommand{\nn}{\nonumber}
\newcommand{\<}{\langle}
\renewcommand{\>}{\rangle}
\newcommand{\N}{{\cal{N}}}
\newcommand{\M}{{\cal{M}}}
\newcommand{\im}{\mathrm{Im}\,}
\newcommand{\re}{\mathrm{Re}\,}

\newcommand{\Bw}{\boldsymbol{\w}}

\usepackage{braket}
\usepackage{tikz}
\usetikzlibrary{matrix,calc,positioning,decorations.markings,decorations.pathmorphing,decorations.pathreplacing}%tikzmark,
\usetikzlibrary{arrows,cd}
\usepackage{bbding}
\usetikzlibrary{positioning}
\tikzset{>=stealth}

\newcommand{\psisl}{\psi}
\newcommand{\psisu}{\chi}

\newcommand{\del}{\partial}

\newcommand{\qqquad}{\;, \quad\qquad}  %more space, with a comma in between

\newcommand{\RR}{\mathbb{R}}

\newcommand{\Aa}{\mathcal{A}}
\newcommand{\Bb}{\mathcal{B}}
\newcommand{\Cc}{\mathcal{C}}
\newcommand{\Dd}{\mathcal{D}}

\newcommand{\Ff}{\mathcal{F}}
\newcommand{\Gg}{\mathcal{G}}
\newcommand{\Hh}{\mathcal{H}}
\newcommand{\Kk}{\mathcal{K}}
\newcommand{\Ii}{\mathcal{I}}
\newcommand{\Jj}{\mathcal{J}}
\newcommand{\Ll}{\mathcal{L}}

\newcommand{\Nn}{\mathcal{N}}
\newcommand{\Oo}{\mathcal{O}}

\newcommand{\Qq}{\mathcal{Q}}

\newcommand{\Ss}{\mathcal{S}}
\newcommand{\Tt}{\mathcal{T}}

\newcommand{\Vv}{\mathcal{V}}

\newcommand{\Yy}{\mathcal{Y}}
\newcommand{\Ww}{\mathcal{W}}

\newcommand{\ttt}{\tilde{t}}
\newcommand{\yt}{\tilde{y}}

\DeclareMathOperator{\Tr}{Tr}

\newcommand{\R}{\ensuremath{\mathbb{R}}}

\newcommand{\mb}{\bar{m}}
\newcommand{\zb}{\bar{z}}
\newcommand{\hb}{\bar{h}}
\newcommand{\Nb}{\bar{N}}
\newcommand{\wb}{\bar{w}}
\newcommand{\xb}{\bar{x}}
\newcommand{\Jb}{\bar{J}}

\newcommand{\gammab}{\bar{\gamma}}
\newcommand{\betab}{\bar{\beta}}
\newcommand{\yb}{\bar{y}}
\newcommand{\w}{\omega}

\usepackage{color}

\usepackage[normalem]{ulem}  % \sout{old text} for strikeout

\definecolor{darkred}{rgb}{0.6,0,0}
\definecolor{darkblue}{rgb}{0,0,0.6}

\newcommand\p{\partial}
\renewcommand\L{\mathcal{L}}

\newcommand{\be}{\begin{equation}}
\newcommand{\ee}{\end{equation}}

\usepackage[bbgreekl]{mathbbol}
\usepackage{amsfonts}
\DeclareSymbolFontAlphabet{\mathbb}{AMSb} % This makes you use the "usual" mathbb font for upper case letters
\DeclareSymbolFontAlphabet{\mathbbl}{bbold} % Use this for mathbb lower case letters! You have two commands now and you do not overlap with the usual mathbb

\title{\boldmath 
Lecture notes on strings in AdS$_3$ from the worldsheet and the AdS$_3$/CFT$_2$ duality
}

\author[a]{Nicolas Kovensky}
\affiliation[a]{Instituto de F\'isica de La Plata - CONICET, Diagonal 113 e/ 63 y 64, 1900 - La Plata, Argentina.}
\emailAdd{nicolas.kovensky@iflp.unlp.edu.ar}

\abstract{
These lecture notes, based  on the course given at IPhT in November/December 2023, provide a pedagogical introduction to the study of strings in AdS$_3$ backgrounds supported by NSNS flux from the worldsheet perspective, including a number of updates incorporating recent results. We attempt to give a self-contained overview of the state-of-the-art understanding of this topic, describing key aspects of its 25-year-long rich history alongside some important recent developments,  with an emphasis on the computation of worldsheet correlation functions involving spectrally flowed insertions.    
}

\newcommand{\of}[1]{\left(#1\right)}
\newcommand{\off}[1]{\left[#1\right]}
  
\newcommand{\T}[1]{\Tilde{#1}}
\begin{document} 
\maketitle
\flushbottom

\section{Introduction}
\label{sec: intro}

How can we describe the universe moments after the big bang? What sets the scale of the cosmological constant derived from current astrophysical observations? How can we think of four-dimensional black holes at the microscopic level? What is quantum gravity? Can we unify it with the quantum field theory framework? How can we compute observables in generic quantum field theories at strong coupling? These are some of the burning open questions in high-energy physics. Most readers will probably be familiar with them, and they would also recognize that they are both conceptually and technically extremely hard to answer.  Nevertheless, it is believed that, at least in principle, string theory could potentially provide answers to all of them! 

One way to try to deal with these difficulties is to study some appropriate toy model. We could, for instance, consider larger symmetry structures such as conformal symmetry or supersymmetry, which may allow us to compute some relevant observables without having to resort to perturbative expansions. We could also lower the number of dimensions in order to reduce the number of propagating degrees of freedom, effectively simplifying the dynamics. Conversely, we could increase the number of dimensions, such that we have the possibility of landing on a critical string theory solution. Finally, we could also change the characteristics of the geometry under consideration, studying some of these problems in (asymptotically) Anti-de Sitter (AdS) spacetimes, as opposed to  asymptotically flat or asymptotically de Sitter backgrounds which are directly relevant for particle physics and cosmology. This would allow us to make use of the AdS/CFT correspondence, and more generically of the holographic principle, which states  that certain theories of quantum gravity in $d+1$ dimensions can be described equivalently in terms of their $d$-dimensional field theory duals. However, in doing so we must be careful when deciding what toy model to focus on. It should be simple enough so that it helps us in answering some of the above questions in a more controlled setting. On the other hand, it should not be oversimplified, otherwise it would fail to capture the relevant ingredients of the original system, and the whole exercise would become pointless.  

In these notes we discuss one of the most interesting and most studied such toy models: that of strings in AdS$_3$. More precisely, this 2+1-dimensional background should be thought of as the extended part of a supergravity solution of the form AdS$_3 \times M_7$, where $M_7$ is some appropriate seven-dimensional compact manifold. We will mostly be agnostic about this internal sector, even though some of the results that we will present where the  precise structure of $M_7$ comes into play correspond to the cases with $M_7= S^3 \times M_4$, and where the four-dimensional sector is either $T^4$ or K3. 

The propagation of strings in AdS$_3$ spacetimes combines highly non-trivial dynamics with a number of key ingredients which allow for an unprecedented level of computability, together with a considerable number of interesting applications. Let us briefly comment on some of these aspects that will be relevant for us:  

\begin{itemize}
    \item The AdS$_3$ geometry can arise as part of a superstring solution in different ways. For instance, it describes the near-horizon region of the D1D5 system, which describes configurations sourced by a collection of $n_1$ D1-branes and $n_5$ D5-branes with a common spatial circle. One also takes the remaining four spatial directions of the D5 worldvolume as compactified on $T^4$ or K3. In the decoupling limit, this provides one of the original examples of the gauge-gravity duality envisioned in \cite{Maldacena:1997re}, which can be stated  schematically as  
    \begin{center}
        String theory on AdS$_3$ $\Leftrightarrow$ CFT$_2$ on its asymptotic boundary. 
    \end{center}

    \item The dual conformal field theory (CFT) is known as the D1D5 CFT. As opposed to the well-known case of 4d $\N=4$ U($N$) super Yang-Mills (SYM) describing the low-energy dynamics of $N$ overlapping D3-branes, the gauge theory associated to the D1D5 system is not conformal. It does, however, flow to a fixed point in the infrared (IR). Even though pinpointing the precise theory one lands on is perhaps surprisingly difficult, we can get a lot of mileage just from knowing  that it must be a two-dimensional $\N = (4,4)$ superconformal field theory (SCFT). We will refer to it as the holographic CFT, boundary CFT, or spacetime theory interchangeably. The corresponding central charge must be of the form\footnote{More precisely, the central charge is $c=6n_1n_5+6$, but the extra term comes from a free, decoupled sector, which we will ignore from now on \cite{Giveon:1998ns,Aharony:2024fid}.} 
    \begin{equation}
    \label{total c HCFT}
        c_{\rm st} = 6 n_1 n_5. 
    \end{equation}
    This  was originally derived in \cite{Brown:1986nw} from the gravity perspective by studying asymptotic symmetries, and only later related to the string theory/black hole perspective in \cite{Strominger:1996sh,Strominger:1997eq}. 

    \item Alternatively, the D1-branes can be seen as string-like instantons of the six dimensional gauge theory on the D5-branes. The low-energy dynamics of the system should then describe the zero-modes of the corresponding topologically non-trivial solutions. For a single instanton, this can be thought of as an $\N=(4,4)$ CFT with target space given by the corresponding moduli space $\M_{\rm inst}$. The latter has (bosonic) dimension $4n_5$. Including the fermionic contributions, this gives a central charge $c^{(1)}_{\rm st} = 6 n_5$, hence we denote this moduli space as $\M_{\rm inst}^{(1)} = \M_{6n_5}$. 
    
    \item If one could ignore interactions, the fact that each D1-brane is indistinguishable from each other implies that the holographic CFT (HCFT) would be a symmetric product orbifold theory given by 
    \begin{equation}
    \label{SymOrbifold in n1}
        {\rm Sym}^{n_1} \left(\M_{6n_5}\right).  
    \end{equation}
    By this we simply mean the product of $n_1$ copies of the SCFT with target space $\M_{6 n_5}$, quotiented by the action of the permutation group $S_{n_1}$. The total central charge is then \eqref{total c HCFT}. Of course, this is an uncontrolled approximation. More generally, we can only expect the HCFT to be a deformation of a symmetric orbifold. 
    
    \item One can further use U-duality \cite{Larsen:1999uk} to relate the system under consideration with an analogous one with a single D5-brane and $N= n_1n_5$ D1-branes. Here the single instanton moduli space has central charge $6$, and we can heuristically think of it as parametrizing the location of the corresponding D1-brane on, say,  $T_4$. At this point in moduli space, it is reasonable to expect that the dual theory is closely related to the orbifold model
    \begin{equation}
    \label{D1D5 CFT}
        {\rm Sym}^{N} \left( T^4 \right).  
    \end{equation}
    This is usually referred to as the orbifold point (in the CFT moduli space). Here the holographic CFT is under exact control. One can use it, for instance, to calculate protected quantities such as the three-point functions of chiral primary operators in order to compare them with computations on the gravity side.  

    \item The instanton moduli space $\M_{\rm inst}$ should not be confused with the moduli space of the HCFT, parametrized by all its marginal deformations \cite{deBoer:2008ss,OhlssonSax:2018hgc}. For the latter, the holographic correspondence dictates that it should be identified with the type II supergravity moduli space in the near-horizon regime of the D1D5 configuration compactified on $M_4$. Locally, this takes the form $\M_{\rm sugra}^{\rm IR}=$SO(4,$n$)/SO(4)$\times$SO($n$), with $n=5$ for $M_4 = T^4$ and $n=21$ for $M_4 =$ K3 \cite{Romans:1984an,Seiberg:1988pf}. Note that this is only a subgroup of the of the full $\M_{\rm sugra}^{\rm UV}=$SO(5,$n$)/SO(5)$\times$SO($n$)     
    for the asymptotically flat solution. As we flow from the UV region of the geometry to the IR one, some of the moduli get fixed. This is known as the \textit{attractor mechanism} \cite{Ferrara:1995ih}. 

    \item By means of $S$-duality one can relate the above system with a solution sourced by a NS5-branes and (fundamental) F1-strings. Thus, in sharp contrast with higher-dimensional cases such as that of AdS$_5 \times S^5$, where the presence of D-branes implies that Ramond-Ramond (RR) fluxes must be turned on, one can obtain AdS$_3\times S^3 \times M_4$ backgrounds sourced solely by fluxes of the Neveu-Shwarz-Neveu-Shwarz (NSNS) type. As a consequence, the worldsheet theory can be quantized, and the (perturbative) string dynamics can be studied exactly in $\alpha'$, i.e.~well-beyond  the supergravity regime. In particular, this allows us to study the mechanisms at work behind the holography duality with a far greater precision than in almost all other instances of AdS/CFT. The main ingredient of the worldsheet theory is the SL(2,$\R$) Wess-Zumino-Witten (WZW) model, since the corresponding group manifold (or rather its universal covering) is precisely global AdS$_3$ \cite{Giveon:1998ns,deBoer:1998gyt,Kutasov:1999xu,Maldacena:2000hw,Maldacena:2000kv,Maldacena:2001km}.  

    \item Moreover, the D1D5 system has been a cornerstone for the string-theoretical description black holes. Black holes have a macroscopic entropy dictated by the Bekenstein-Hawking formula $S = A_{\rm h}/4G$, where $A_{\rm h}$ is the area of the horizon and $G$ is the gravitational coupling. 
    It is in the context of the D1D5 system that the first string-theoretical microscopic computation of 
    this entropy was obtained \cite{Strominger:1996sh} by counting BPS excitations of the brane configuration in the weak coupling regime. This has also been the more fruitful arena for the Fuzzball program \cite{Lunin:2001jy,Mathur:2005zp,Bena:2022rna}, which postulates that BHs can be understood as an averaged  description of their microstates, some of which are coherent enough to be described in classical terms as  smooth and horizonless geometries which look like the corresponding BHs for a distant observer.   

    \item All the tools discussed above can be used to describe string propagation in black hole (BH) backgrounds. For BHs in two dimensions, both Lorentzian and Euclidean, this is because they can be obtained by starting with the SL(2,$\R$) model and gauging appropriate currents \cite{Witten:1991yr}. For instance, the 2d Euclidean BH corresponds to the \textit{cigar} geometry SL(2,$\R$)/U(1). On the other hand, three-dimensional asymptotically AdS$_3$ BHs are known as BTZ black holes \cite{Banados:1992wn}. They can be obtained by a set (discrete) orbifoldings of AdS$_3$. More recently, it was shown that, in the context of the Fuzzball program, one can also use related coset models to describe a (particularly symmetric) set of black hole microstates \cite{Martinec:2017ztd,Martinec:2018nco,Martinec:2019wzw,Martinec:2020gkv,Martinec:2022okx,Bufalini:2021ndn,Bufalini:2022wyp,Bufalini:2022wzu}. Hence, understanding string dynamics in AdS$_3$ can be seen as a necessary step for obtaining an exact description of string propagation in black hole and black hole microstate backgrounds.

    \item Finally, one also has an interesting relation 
    with the decoupled theory on the NS5-branes, which is known as little string theory \cite{Kutasov:2001uf}. This is a non-gravitational theory with several stringy properties such as Hagedorn thermodynamics at high energies. It also constitutes one of the building blocks of an instance of holography beyond AdS, where the gravity background includes a dilaton that runs linearly with the radial direction. Here the dual theory is not a CFT, and not even a local quantum field theory. More recently, it was argued that this can also be understood in terms of certain particularly tractable irrelevant deformations of the original HCFT of the $T\bar{T}$-type \cite{Zamolodchikov:2004ce,Smirnov:2016lqw,Giveon:2017nie,Chakraborty:2023wel}.  
\end{itemize}

The D1D5 system and strings in AdS$_3$ in particular therefore have an important number of very interesting aspects, which have been studied at length for over 25 years. In these notes we will work in the NS5-F1 frame and focus on the worldsheet theory, with a particular emphasis on the computation of its correlation functions. Other aspects of the model are discussed in several reviews, see \cite{Giveon:2001up,Kutasov:2001uf,Dolan:2002uy,Schomerus:2002dc,Nunez:2004xw,Schomerus:2005aq,Kraus:2006wn,McElgin:2015eho} for an incomplete list of references.  

As stated above, in the RNS formalism employed here the relevant theory is the SL(2,$\R$) WZW model. Although the latter is believed to be exactly solvable, it presents a number of highly non-trivial features. This is mainly due to the fact that the target space under consideration is both Lorentzian and non-compact. As a result, one finds that the spectrum includes both a discrete sector, which encompasses the so-called short strings, and, above a certain threshold, a continuous one, representing the long strings configurations. Moreover, the spectral flow operation \cite{Schwimmer:1986mf}, which in simpler contexts such as the SU(2) WZW model \cite{Zamolodchikov:1986bd} merely reshuffles the different states, generates new, inequivalent representations in the AdS$_3$ setting. These must be included in the spectrum, otherwise the string theory description is simply inconsistent. The corresponding vertex operators have quite complicated operator product expansions (OPEs) with the symmetry currents of the model, which renders the computation of their correlation functions particularly  involved. We will discuss recent results on how to overcome these difficulties\footnote{For simplicity we will restrict to studying the properties of the worldsheet CFT on the sphere, which gives the leading contribution at large $N$ (which for us effectively means large $n_1$), although many of the results we described have been extended to higher genus, see for instance \cite{Eberhardt:2019ywk, Eberhardt:2020akk,Knighton:2020kuh,Eberhardt:2021jvj} and references therein.}, and also focus on the role of the dual CFT as seen from the worldsheet theory.

The structure of these notes is as follows. Section \ref{sec: bosonic AdS3 spectrum} is devoted to discussing the spectrum of the SL(2,$\R$) WZW model \cite{Maldacena:2000hw,Maldacena:2000kv}. For this, we start by presenting the asymptotically flat NS5-F1 solution in type IIB supergravity compactified on $S^1 \times T^4$ and show how the decoupling limit leads to the near-horizon AdS$_3\times S^3 \times T^4$. In the string frame, the corresponding (squared) AdS$_3$ radius is set by $n_5$, in units of $\alpha'$. We then focus on bosonic strings propagating in AdS$_3$ for simplicity, characterized by the SL(2,$\R$) WZW model at level $k = n_5+2$. The classical limit, sometimes called the mini-superspace limit, then corresponds to large $k \sim n_5$. We discuss the isometries of AdS$_3$, which form an SL$(2,\R)_L \times $SL$(2,\R)_R$ algebra, and present the particle and string geodesics, highlighting the appearance of those associated to long strings states. We then move to the Euclidean geometry. We derive the semi-classical expression for the (unflowed) vertex operators of the model, and discuss their normalization conditions \cite{Teschner:1997fv}. The rest of the section deals with the quantum theory, first using the Wakimoto free field representation, valid in the region near the boundary of AdS$_3$ \cite{Giveon:1998ns,deBoer:1998gyt}, and then moving to the exact theory \cite{Teschner:1997ft,Teschner:1999ug,Kutasov:1999xu,Maldacena:2001km}. Here we derive the spacetime symmetry generators from the worldsheet. We further highlight the role of the spectrally flowed representations, showing how they allow us to overcome an unphysical bound on the spacetime dimension of the states in the model, and sketching how they are taken into account in the torus partition function. This is followed by Section \ref{sec: Exact theory, unflowed sector}, which is devoted to the exact description of the symmetry algebra, the vertex operators and their correlation functions, and the OPE structure of the theory in the unflowed sector. 

Section \ref{sec: bosonic correlators} is perhaps the more involved section of these notes at the technical level. It focuses on the computation of correlators involving spectrally flowed insertions. We discuss in detail the different representations of the corresponding vertex operators, and their interplay \cite{Fateev,Maldacena:2001km,Dei:2021xgh,Iguri:2022eat}. The $m$-basis is that of eigenstates of the zero-modes of the Cartan currents. It allows for a pedagogical discussion of spectral flow and is best suited for applications to certain gauged models, such as the cigar theory. On the other hand, the conjugate $x$-basis resums the action of the rising zero-mode operators, and is best suited for holographic applications, as the complex coordinate $x$ is identified with that of the boundary plane. Finally, an additional complex variable $y$ is used to resum the action of  \textit{flowed} currents, allowing one to rewrite a number of recursion relations satisfied by correlators involving operators with shifted spacetime weights \cite{Eberhardt:2019ywk} in terms of differential equations. We describe in detail how the appearance of (largely unknown) higher order poles in the OPEs between currents and flowed operators greatly complicates the computation of correlation functions. Then, we discuss how one can use a number of holomorphic covering maps to bypass these issues, leading to explicit, integral expressions for correlators with arbitrary spectral flow charges \cite{Eberhardt:2019ywk,Dei:2021xgh,Dei:2021yom,Iguri:2022eat,Bufalini:2022toj}. As it turns out, the way these covering maps are used is reminiscent of their appearance in the computation of $n$-point functions in symmetric orbifold models, see \cite{Lunin:2000yv,Lunin:2001pw,Pakman:2009ab,Pakman:2009zz,Dei:2019iym}. In view of the above discussion on the nature of the dual CFT, this is of course not a coincidence.

%The following sections are a slightly more streamlined, and, in a sense, can be considered as applications of the methods described in Secs. \ref{sec: bosonic AdS3 spectrum} and \ref{sec: bosonic correlators}. They are devoted to the physical interpretation of those results, with a particular emphasis on holographic aspects related to the precise form of the boundary theory. 

In Section \ref{sec:tensionless} we discuss the tensionless limit of the model, corresponding to $n_5=1$, or equivalently $k=3$ in the bosonic language. We first show how massless higher-spin states arise in this context, where the AdS$_3$ space is string-size \cite{Gaberdiel:2017oqg,Gaberdiel:2018rqv,Giribet:2018ada}. We then argue that it is at this point in the moduli space that the dual theory is exactly given by a symmetric orbifold, following \cite{Eberhardt:2018ouy,Eberhardt:2019niq,Eberhardt:2019ywk,Eberhardt:2020bgq}. In particular, for strings in AdS$_3\times S^3\times T^4$ at $n_5=1$, the HCFT is precisely that of Eq.~\eqref{SymOrbifold in n1}, with $\M_{6n_5} = \M_{6} = T^4$. Strictly speaking, for this value of $n_5$ the RNS formalism breaks down, and one should use the hybrid formalism as in \cite{Eberhardt:2018ouy}. Nevertheless, many of the relevant features are visible in the bosonic SL(2,$\R$) model at level $k=3$. We first motivate this by looking at the spectrum, and then describe how correlation functions -- in particular those involving the worldsheet avatars of spacetime twist operators -- localize on the loci defined by the existence of the corresponding covering maps \cite{Eberhardt:2019ywk,Dei:2019osr,Bertle:2020sgd}. Finally, even though the geometry is highly curved in this regime, we discuss the "geometric" interpretation of these features in terms of Wakimoto fields. 

In Section \ref{sec: AdS3xS3xT4} we continue our holographic exploration, now for $n_5\geq 2$. We present the superstring worldsheet model on AdS$_3\times S^3\times T^4$, and construct the vertex operators associated to the chiral primary operators of the boundary theory \cite{Kutasov:1998zh,Argurio:2000tb,Gaberdiel:2007vu,Dabholkar:2007ey,Giribet:2007wp}. We then extend the techniques reviewed in the previous sections in order to compute their three-point functions for arbitrary spectral flow charges \cite{Iguri:2022pbp,Iguri:2023khc}. We show that the worldsheet results precisely reproduce the structure constants and fusion rules of the exact chiral ring as computed from the D1D5 CFT at the orbifold point, see Eq.~\eqref{D1D5 CFT}. This had to be the case, as these observables are protected by supersymmetry \cite{deBoer:2008ss}. 

In Section \ref{sec: holography} we move past protected observables, go back to the bosonic setting, and discuss the structure of the putative holographic CFT for $k>3$. Here we mostly follow the presentation recently given in \cite{Balthazar:2021xeh,Eberhardt:2021vsx,Dei:2022pkr}. We motivate the holographic proposal by studying the theory on the long strings \cite{Seiberg:1999xz}, and discuss the main differences between the cases with $k>3$ and those with $2<k<3$. (As the target space is non-compact, the affine level of the SL(2,$\R$) WZW model is not necessarily integer). This leads to a deformation of a different symmetric orbifold model, where the structure is as in Eq.~\eqref{SymOrbifold in n1}. The seed theory with target space $\M_{6n_5}$ now contains a Liouville-type factor, such that the total seed central charge is $c^{(1)}=6n_5$. The non-compact scalar in question is related to the worldsheet field representing the radial direction in AdS$_3$. We motivate the precise marginal deformation involved, briefly discuss the differences between long and short strings from the holographic perspective, and sketch how the matching between certain residues of correlators at special points in the space of complex SL(2,$\R$) spins was recently obtained \cite{Eberhardt:2021vsx,Dei:2022pkr,Dei:2023ivl,Knighton:2023mhq,Knighton:2024qxd}. 

Finally, Section \ref{sec: 3 roads to LST} is devoted to an application of the worldsheet methods in a slightly more general contexts.  We show how implementing a marginal deformation of the worldsheet CFT modifies the asymptotic structure of the geometry the strings propagate on \cite{Giveon:2017myj}. In the UV region, the spacetime becomes flat, while the dilaton runs linearly along the radial direction. This is interpreted as a \textit{reverse} RG flow from the boundary point of view, which is implemented by a specific, highly tractable, $T\bar{T}$-type irrelevant deformation \cite{Zamolodchikov:2004ce,Smirnov:2016lqw,Giveon:2017nie,Apolo:2019zai,Georgescu:2022iyx,Chakraborty:2023wel}. This constitutes an interesting instance of non-AdS holography, where the dual theory, which lives on the NS5-branes, is nothing but the little string theory mentioned above. We finish by discussing the Hagedorn behavior of the high-energy thermodynamics of these models.   

%Finally, we focus on black hole applications...

Some relevant background material is provided in the appendices.  Appendix \ref{sec: appB WZW models} dwells  on WZW models in general, with a particular focus on the SU(2) example \cite{DiFrancesco:1997nk,Zamolodchikov:1986bd,Fateev:1985mm,EberhardtLectures}. 
In particular, we compare the role of spectral flow in this compact setting as compared to the SL(2,$\R$) case discussed in the main text. 
% In Appendix \ref{sec: appA BTZ} we review some important facts about BTZ black holes and asymptotically AdS$_3$ spacetimes. Appendix \ref{sec: appC Liouville} provides some relevant aspects of Liouville theory and the DOZZ formula \cite{Dorn:1994xn,Zamolodchikov:1995aa}, see also \cite{Teschner:2001rv,Teschner:2003en,Ribault:2014hia,Eberhardt:2023mrq}.
In Appendix \ref{sec: appD SymOrbifolds} we briefly discuss symmetric product orbifold CFTs and the computation of some of their correlators by means of covering map techniques \cite{Lunin:2000yv,Lunin:2001pw,Pakman:2009ab,Pakman:2009zz,Dei:2019iym}.

%%%%%%%%%%%%%%%%%%%%%%%%%%

%%%%%%%%%%%%%%%%%%%%%%%%%%%%%
\newpage
%%%%%%%%%%%%%%%%%%%%%%%%%%%%%%%

%%%%%%%%%%%%%%%%%%%%%%%%%%%%%

 \section{Geometry, spectrum and spacetime symmetries}
\label{sec: bosonic AdS3 spectrum}

Let us start by motivating the study of strings propagating in AdS$_3$. We consider type II superstring theory on 
\begin{equation}\mathrm{Mink}^{1,4}\times S^1_y \times T^4,
\end{equation}
and further include $n_5$ NS5-branes wrapping the $S^1_y$ (often called the $y$-circle) and the torus, together with $n_1$ (fundamental) F1-strings  wrapped on $S^1_y$, smeared along the $T^4$ directions. For this supersymmetric configuration, which preserves 8 supercharges, the (string frame) geometry takes the form  \cite{Dabholkar:1990yf,Callan:1991at,Tseytlin:1996as} 
\begin{equation}
\label{NS5F1 metric}
    ds^2 = f_1^{-1}(r) (-dt^2 + dy^2) + f_5(r)(dr^2 + r^2 d\Omega^2_3) + dz_i dz^i, 
\end{equation}
where $y \sim y + 2\pi R_y$, the $z_i$ with $i=1,\dots,4$ are coordinates on the $T^4$, $r$ is the radial coordinate, and $f_{1,5}(r)$ are the transverse harmonic functions 
\begin{equation}
\label{NS5F1 harmonic functions}
  f_{1,5}(r) = 1+ \frac{r_{1,5}^2}{r^2}\,, \qquad
  r_5^2= \alpha' n_5 \,,
  \quad 
 r_1^2= \frac{g_s^2 \alpha'n_1}{v_4}.  
\end{equation}
Here $g_s$ is the string coupling at infinity, $\alpha'$ is the inverse string tension, and $v_4$ is the volume of the $T^4$ in string units%, $\mathrm{Vol}(T^4)=v_4 \alpha'^2 $
. For this to be a string vacuum we also need to have a non-trivial dilaton $\Phi$ and an NSNS 3-form field-strength $H$, namely 
\begin{equation}
\label{NS5F1 dilaton and Bfield}
    e^{2\Phi} = g_s^2 \frac{f_5}{f_1} \,, \quad 
    H = 2 \left(   n_5 + \frac{g_s^2 n_1}{v_4} \frac{f_5} { f_1}\, *_6\right) \ep_{S^3}, 
\end{equation}
with $\ep_{S^3}$ the volume form on  the transverse $S^3$ and $*_6$ the six-dimensional Hodge dual. Assuming the torus is roughly square, the smearing approximation is valid for $r^2 \gg v_4^{1/2} \alpha'$. 

This background is S-dual to the D1D5 system, considered in \cite{Maldacena:1997re} as one of the original incarnations of the holographic duality, i.e.~the AdS/CFT correspondence relating the near-horizon AdS$_3$ regime to the dual conformal field theory on the branes. In our NS frame, the decoupled regime is obtained by zooming to small $r$ or taking $n_1$ and $n_5$ to be very large, so that one can drop the "1+" factors in the harmonic functions \eqref{NS5F1 harmonic functions}. 
At the practical level, this scaling limit is reached as a two-step procedure. We first take $r,g_s \to 0$ with $r/g_s$ fixed (or equivalently, $n_1 \to \infty$ and $g_s\to 0$ with $g_s^2n_1$ fixed). This intermediate step leads to the so-called five-brane decoupling limit, where the geometry is asymptotically flat but with a dilaton that runs linearly\footnote{In this setup, one can study a non-AdS version of holography \cite{Asrat:2017tzd}. We will come back to this in Sec.~\ref{sec: 3 roads to LST}.} with $\log r$.  The second step corresponds to taking $r, \alpha' \to 0$ with $r/\alpha'$ fixed, which is known as the string decoupling limit. This leads to  
\begin{equation}
\label{AdS3S3T4 metric}
    ds^2 = n_5 \left[
    \frac{r^2}{\alpha'} (-dt^2 + dy^2) + 
    \frac{\alpha'}{r^2}dr^2 + \alpha' d\Omega_3^2  
    \right] + dz_i dz^i \, , 
\end{equation}
where we have further rescaled $r$ with a factor $\sqrt{n_1 n_5 / v_4}$ for convenience, while 
\begin{equation}
\label{D1D5 AdS B-field and dilaton}
    H = 2 n_5  (1 + *_6 ) \ep_{S^3} \, , \quad 
    e^{2\Phi} = \frac{v_4 n_5}{n_1}. 
\end{equation}
The near-horizon geometry is then 
\begin{equation}
\label{AdS3S3T4}
    \mathrm{AdS}_3 \times S^3 \times T^4 , 
\end{equation}
the radii of both AdS$_3$ and the $S^3$ in string units being set by $n_5$. On the other hand, $n_1$ only appears in the constant dilaton, such that the effective six-dimensional string coupling is $g_6^2 = g_s^2/v_4 = n_5/n_1$. A weakly-coupled, low-energy supergravity description is  thus accurate as long as $n_1 \gg n_5 \gg 1$. Finally, as the  $H$-field becomes self-dual, the number of preserved supercharges is enhanced from 8 to 16.  

This instance of the AdS$_3$/CFT$_2$ correspondence is quite special. Indeed, all sources are of the NSNS type, and moreover the geometry \eqref{AdS3S3T4} consists of a product of Lie groups, namely
\begin{equation}
    \mathrm{SL}(2,\R) \times \mathrm{SU}(2) \times \mathrm{U}(1)^4 \,, 
\end{equation}
where, to be precise, we need to work with the universal cover of $\mathrm{SL}(2,\R)$, for which the time direction is non-compact\footnote{We will omit the distinction in the remainder of these notes.}. This means that the we can quantize  the  worldsheet theory \cite{Polchinski:1998rq,Polchinski:1998rr}, which takes the form of a (super) WZW model, and can be treated exactly in $\alpha'$, i.e.~all the way down to  small values of $n_5$. As we will discuss later on, the smallest possible value $n_5=1$ is actually rather exceptional since the AdS$_3$ becomes string-size and the fundamental strings become tensionless \cite{Gaberdiel:2018rqv,Giribet:2018ada}. Of course, we still need $n_1$ to be very large in order for the topological expansion in powers of $g_s\sim e^{\Phi}$ to be under control.   
 
In the remainder of this section and several of the following ones we focus on the AdS$_3$ sector, and consider bosonic  strings for simplicity  \cite{Maldacena:2000hw}. We now describe some of the main aspects of the corresponding SL(2,$\R$) sigma model, starting with the classical particle and string geodesics, then moving to the spectrum of the model and its free field description \cite{Giveon:1998ns}, and finally introducing the partition function. Along the way we highlight the crucial role played by the spectral flow automorphisms of the underlying affine symmetry algebra. We end with a brief discussion of correlation functions in the unflowed sector of the theory.

\subsection{Geometric aspects and semiclassical considerations}

\subsubsection*{Global AdS$_3$. Particle and string geodesics}

 In global coordinates, the AdS$_3$ line  element and $B$-field (with $H=dB$, at least locally) read 
\begin{equation}
\label{AdS3 metric and B field}
    ds^2 = k\left(- \cosh^2 \rho \, dt^2 + d\rho^2 + 
    \sinh^2 \rho \, dy^2 \right) \ , \quad B = k \sinh^2 \rho dt \wedge dy.  
\end{equation}
The geometry has constant negative curvature, and its isometries are generated by the six vector fields 
\begin{subequations}
    \begin{eqnarray}
    \zeta^1_{L/R} &=&
\frac{1}{2} \sin (x^\pm) \der_\rho + \cos(x^\pm) \left[ \tanh^{-1}(2\rho) \der_\pm - \sinh^{-1}(2\rho) \der_{\mp} \right]\, , \\
\zeta^2_{L/R} &=&
\frac{1}{2} \cos (x^\pm) \der_\rho - \sin(x^\pm) \left[ \tanh^{-1}(2\rho) \der_\pm - \sinh^{-1}(2\rho) \der_{\mp} \right]\, , \\ [1ex]
\zeta^3_{L/R} &=&  \der_{\pm}
\, ,
\end{eqnarray}
\end{subequations}
with $x^\pm = t\mp y$. These vector fields satisfy $[\zeta^a_{L},\zeta^b_R] = 0$ for $a,b=1,2,3$. After defining $J^a = i \zeta^a$ and $J^\pm = J^1 \pm i J^2$  we obtain two copies of the SL(2,$\R$) Lie algebra,
\begin{equation}
    [J^3,J^{\pm}]= \pm J^\pm\,, \quad 
    [J^+,J^{-}]= -2 J^3 \,.
\end{equation}
The coordinates $t$ and $y$ paramterize the (topological) cylinder that constitutes the conformal boundary of AdS$_3$. More precisely,  we  identify $J_R^3 + J_L^3$ and $J_R^3 - J_L^3$ as generating translations along the timelike and circular directions at fixed radial distance. 
% correction (2) The corresponding eigenvalues,
The eigenvalues of $J^3_R$ and $J^3_L$, respectively 
denoted as $m$ and $\bar{m}$ in what follows (as opposed to $m_R$ and $m_L$) are then associated to the spacetime energy $E$ and the quantized angular momentum $n_y$, respectively, i.e. 
\begin{equation}
\label{mmb E and L}
    m = \frac{E + n_y}{2} \, \quad \mb = \frac{E - n_y}{2}.
\end{equation} 
In the Lorentzian theory, which is what we are interested in, the spectrum is built out of states with $m,\mb \in \R$. 

Each point on the AdS$_3$ manifold defines an element $g$ in SL(2,$\R$). Parametrising the latter via its isomorphism\footnote{A generic group element $
    \left(\begin{smallmatrix}
        a & b \\
        c & d
    \end{smallmatrix}\right)$ with $
a,b,c,d \in \R$ and $ad-bc=1$ can be parametrized in terms of an SU(1,1) element $   \left(\begin{smallmatrix}
        \alpha & \beta \\
        \bar{\beta} & \bar{\alpha}
    \end{smallmatrix}\right) $, with 
 $\alpha,\beta \in \mathbb{C}$ and $ |\alpha|^2-|\beta|^2=1$. The explicit map is $
2 \alpha = a+d+i(b-c)$, $ 
2\beta = b+c+i(a-d)$.} with SU$(1,1)$, we can use the Pauli matrices $\sigma_a$ to write 
\begin{equation}
    g = e^{i v \sigma_3 } e^{\rho \sigma_1} e^{i u \sigma_3}\,,
\end{equation}
with $u = \frac{1}{2}(t+y)$ and $v = \frac{1}{2}(t-y)$. 
We now use this (elliptic) parametrization to describe the SL(2,$\R$) WZW-model. The key aspects of this type of two-dimensional conformal field theories are reviewed in Appendix \ref{sec: appB WZW models}; see \cite{DiFrancesco:1997nk} for more details. The action takes the form 
\begin{equation}
    S_{\rm WZW} = \frac{k}{8\pi} \int d^2\sigma \Tr \left[ g^{-1} \der_\alpha g g^{-1}\der^{\alpha} g \right] + k \Gamma\,     
\end{equation}
where $\Gamma$ stands for the Wess-Zumino term discussed around Eq.~\eqref{def WZ term}. The parameter $k$, known as the level, sets the scale of the target space geometry in units of $\alpha'$, which has been set to 1. Defining the generators 
\begin{equation}
    t^1 = \frac{1}{2}\sigma_1 \,, \qquad 
    t^2 = \frac{1}{2}\sigma_2 \,, \qquad 
    t^3 = -\frac{i}{2}\sigma_3 \, ,
\end{equation}
we obtain the conserved currents 
\begin{equation}
    J^a_\alpha = k \Tr \left[ t^a \der_\alpha g g^{-1} \right] \, , \qquad 
    \Jb^a_\alpha = k \Tr \left[ t^a g^{-1} \der_\alpha g \right] \, .
\end{equation}
In terms of the right- and left-moving worldsheet coordinates $\tau^\pm = \tau \pm \sigma$, the equations of motion read simply 
\begin{equation}
    \der_- (\der_+ g g^{-1}) = 0 \, .
\end{equation}
This implies that the conserved currents of the model  only depend on one of the light-cone coordinates. More precisely, we have $J^a = J^a (\tau^+)$ and $\Jb^a = \Jb^a (\tau^-)$. Combining $J^{1,2}$ into the ladder operators $J^\pm = J^1 \pm i J^2$ (and similarly for $\Jb^{\pm}$) we find that, in terms of the target space fields,  
\begin{equation}
    J^3 = k \left[ \der_+ u + \cosh (2\rho) \der_+ v \right] \, , \qquad 
    J^\pm = k \, e^{\mp 2 i u} \left[ \der_+ \rho \pm i \sinh (2\rho) \der_+ v\right] \, , 
\end{equation}
while 
\begin{equation}
    \Jb^3 = k \left[ \der_- v + \cosh (2\rho) \der_- u \right] \, , \qquad 
    \Jb^\pm = k \, e^{\mp 2 i v} \left[ \der_- \rho \pm i \sinh (2\rho) \der_- u\right] \, . 
\end{equation}
The general solution to the equation of motion can be written as $g = g_+(\tau^+) g_-(\tau^-)$, and the closed string boundary conditions impose $g_+(\tau^+ + 2\pi) = g_+(\tau^+) M$ and $g_-(\tau^- - 2\pi) = M^{-1} g_+(\tau^-)$ for some constant monodromy matrix  $M \in $ SL(2,$\R$). 

\begin{figure}
    \centering
    \includegraphics[scale=0.7]{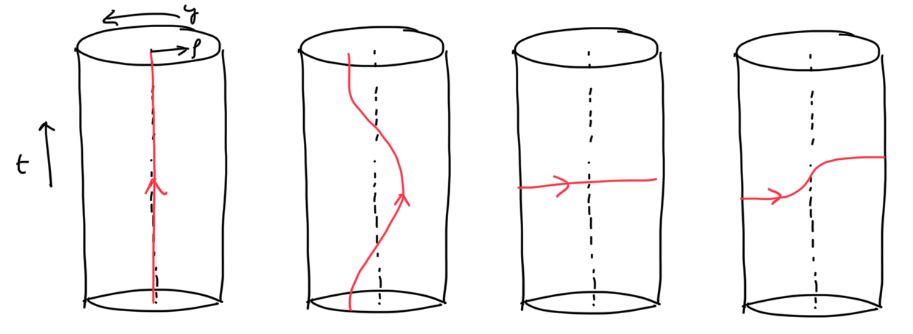}
    \caption{From left to right: Particle geodesics (a) timelike solution with $\rho_0=0$ (b) more general timelike configuration (c) basic spacelike solution (d) more general spacelike configuration. }
    \label{fig: particle geodesics}
\end{figure}

For bosonic strings propagating in AdS$_3\times M_{\rm int}$ there will be an internal CFT, and one should also include the usual $bc$-system of reparametrization ghosts \cite{Polchinski:2012nh,Blumenhagen:2013fgp}. Quantization is then implemented by means of the BRST procedure, such that, focusing on the right-moving sector, physical states must be in the cohomology of the BRST charge 
\begin{equation}
\label{QBRST}
    \Qq = \oint \frac{d\tau^+}{2\pi} \left( c \, T_{\rm tot} + \text{ghosts} \right) \:\! ,
\end{equation}
where normal ordering is implied. Here $T_{\rm tot}(\tau^+) = T(\tau^+)  + T_{\rm int} (\tau^+) $, where we reserve the symbol $T(\tau^+) $ for the contribution coming from the AdS$_3$ sector. In the classical limit the latter takes the form $T = k^{-1} \eta_{ab}J^a J^b$, where the Killing metric has non-zero entries $\eta_{+-} = \eta_{-+} = \frac{1}{2}$ and $\eta_{33} = -1$. Physical solutions must satisfy the Virasoro constraints, that is, the corresponding $T_{\rm tot}$ and $\bar{T}_{\rm tot}$ must vanish.

The simplest solutions are the time-like geodesics with $\rho = 0$ and $t = \alpha \tau$, corresponding to $g = e^{i \alpha \tau \sigma_3}$,  and the space-like  geodesics with $t = y= 0$,  and $\rho = \alpha \tau$, which give $g = e^{ \alpha \tau \sigma_1}$. These geodesics and some other ones discussed below are depicted in Fig.~\ref{fig: particle geodesics}.  The constant parameter $\alpha$ sets either the energy or the radial momentum, respectively. Both solutions depend only on $\tau$, hence they are effectively one-dimensional and correspond to particle geodesics. Assuming that $T_{\rm int} = h_{\rm int}\geq 0$, as expected for compact $M_{\rm int}$, only the time-like solution with  $J^\pm=0$ and $J^3 = k\alpha/2$ and hence $T = - \frac{k}{4}\alpha^2$ can solve the Virasoro constraint for real values of $\alpha$, namely by setting $k \alpha^2 = 4h_{\rm int}$. For the space-like case, $J^3=0$ and $J^+ = J^- = k \alpha/2$, which leads to $T = +\frac{k}{4}\alpha^2$, hence there is no non-trivial solution for $\alpha \in \R$ and $h_{\rm int}\geq 0$. More general particle geodesics can be obtained by conjugation with constant matrices $h,\hb \in $ SL(2,$\R$), i.e.~by taking $g \to h g \bar{h}$. For instance, one can easily obtain a time-like geodesics of the form 
\begin{equation}
\label{timelike geodesic w=0}
    y = 0 \,, \qquad \sinh\rho = \sinh \rho_0 \sin \alpha \tau
    \, , \qquad \tan t = \frac{\tan \alpha \tau}{
    \cosh \rho_0 
    } \, . \qquad (\tau>0)
\end{equation}

\begin{figure}
    \centering
    \includegraphics[scale=0.7]{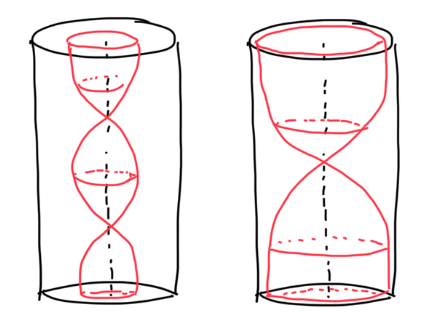}
    \caption{Short string (left) and long string (right) solutions with spectral flow $\w=1$.}
    \label{fig: string geodesics}
\end{figure}

Starting from a given classical solution $g = g_+ g_-$, one can generate new configurations that solve the equations of motion  by implementing $g_+ \to h_+(\tau^+) g_+$ and $g_- \to g_- h_-(\tau^-) $ for any $h_\pm \in $ SL(2,$\R$). The more interesting cases for us are obtained by  using 
\begin{equation}
    h_\pm = e^{\frac{i}{2} \w_\pm \tau^\pm \sigma_3} \, \qquad \w_\pm \in \mathbb{Z} \, . 
\end{equation}
In our parametrization, these shift the time and angular coordinates as 
\begin{equation}
\label{t and y spectral flow}
    t \to t + \w \tau \,, \quad 
    y \to y + \w \sigma\,. 
\end{equation}
Here we have chosen $\w_-=\w_+\equiv \w$ as required by the periodicity of the string worldsheet $\sigma \sim \sigma + 2\pi$ (recall that our time coordinate is non-compact). Configurations with $\w \neq 0$ are called spectrally flowed solutions. The resulting geodesics now depend explicitly on $\sigma$, hence we see them as describing the trajectory of a string. Roughly speaking, they correspond to rotating the original ones around the vertical $\rho=0$ axis, while boosting them at the same time. As a result, they wind around the angular direction of the target space $\w$ times, leading to the configurations depicted in Fig.~\ref{fig: string geodesics}. However, one should keep in mind that the $y$-circle becomes contractible in the interior of AdS$_3$, hence $\w$ is \textit{not} a conserved charge.

At the level of the currents, we get 
\begin{equation}
\label{Lorentzian classical flow currents}
    J^3 = \tilde{J}^3 + \frac{k}{2}\w \,, \qquad J^\pm =  e^{\mp i \w \tau^\pm} \tilde{J}^\pm \, , \qquad T = \tilde{T} - \w \tilde{J}^3 - \frac{k}{4}\w^2\, ,
\end{equation}
where $\tilde{J}^a$ and $\tilde{T}$ are those of the original unflowed geodesic. 
When the latter is a time-like geodesic, the Cartan current zero-mode and physical state conditions read 
\begin{equation}
    J_0^3 = \tilde{J}^3_0 + \frac{k}{2}\w \, , \qquad
    \tilde{T} - \w \tilde{J}^3 - \frac{k}{4}\w^2 + h_{\rm int} 
    %=- \frac{k}{4}\left(\w + \alpha\right)^2 + h_{\rm int}
    = 0
\end{equation}
where $J_0^3 = \int_0^{2\pi} \frac{d\tau^+}{2\pi} J^3$, and similarly for the left-moving sector. Assuming $h_{\rm int} = \hb_{\rm int}$, and using $\tilde{T} = -\frac{k}{4}\alpha^2$, the energy gives
\begin{equation}
    E = J_0^3 + \Jb_0^3 = \frac{k}{2} \w + \frac{2}{\w}\left( h_{\rm int} - \frac{k}{4}\alpha^2\right) \, ,
\end{equation}
It is thus bounded from above by $E_{\rm max} = \frac{k\w}{2} + \frac{2h_{\rm int}}{\w}$, and unbounded from below. An explicit example is given by the solution 
\begin{equation}
    y = \w \sigma \,, \qquad \sinh\rho = \sinh \rho_0 \sin \alpha \tau
    \, , \qquad \tan t = \frac{\tan \alpha \tau+\cosh \rho_0 \tan \w \tau}{
    \cosh \rho_0 - \tan \w \tau \tan \alpha \tau
    } \, ,
\end{equation}
for some constant $\rho_0$, such that $\alpha$ sets the radial momentum here as well. In general, this represents a string bound to the interior of AdS$_3$ which oscillates around the center. These are the short string solutions. Moreover, one can see that for $h_{\rm int} = \frac{k}{4}\w^2$ we need to set $\alpha = 0$, giving solutions with  
\begin{equation}
    y = \w \sigma \, , \qquad 
    \rho = \rho_0 \, , \qquad 
    t = \w \tau \, ,
\end{equation}
such that the string stays at constant radial distance, while the energy comes only from the winding part, $E= E_{\rm max} =  k\w $. 

So far, it appears that we cannot get string geodesics with $E>E_{\rm max}$. Here is where the second type of solutions come into play, i.e.~those obtained by spectrally flowing the original space-like geodesics. When spectral flow is included one can actually solve the Virasoro condition even though $\tilde{T}$ changes sign. The resulting energy reads 
\begin{equation}
    E =  \frac{k}{2} \w + \frac{2}{\w}\left( h_{\rm int} + \frac{k}{4}\alpha^2\right) \, .
\end{equation}
It is now bounded from below by $E_{\rm min} = \frac{k\w}{2} + \frac{2h_{\rm int}}{\w}$, and unbounded from above. The threshold is thus at $E_{\rm min} = \frac{k\w}{2}$, corresponding to $h_{\rm int}= 0$. For instance, we may have 
\begin{equation}
    y = \w \sigma \, , \qquad 
    \rho = \alpha |\tau| \, , \qquad 
    t = \w \tau \, ,
\end{equation}
These are the long strings, also known as scattering states, which come from radial infinity at $t \to - \infty$, then approach the center of AdS$_3$, and finally go back to the near-boundary region of the geometry as $t \to +\infty$.  
A slightly more general example corresponds to the spectrally flowed solution originating from that of Eq.~\eqref{timelike geodesic w=0}, namely 
\begin{equation}
\label{Lorentzian Classical long string}
    y = \w \sigma \,, \qquad \sinh\rho = \cosh \rho_0 \sinh \alpha \tau
    \, , \qquad \tan t = \frac{\tan \w \tau  + \tanh \alpha \tau \sinh \rho_0 }{
    1 - \tan \w \tau \tanh \alpha \tau \sinh \rho_0 
    } \, . 
\end{equation}

We conclude that for fixed $h_{\rm int} \neq \frac{k}{4}\w^2$ the minimal energy state is always in the short string sector. On the other hand, at $h_{\rm int} = \frac{k}{4}\w^2$ the short string solution becomes identical to a long string configuration.  The latter can thus become unbounded and turn into a long string. These solutions are allowed to stay at constant $\rho$ because the gravitational attraction is exactly compensated by the repulsion generated by the $B$-field. In this way, even very long strings located near the asymptotic boundary can remain at finite energy.

\subsubsection*{Euclidean AdS$_3$ and normalizability conditions}

\begin{figure}
    \centering
    \includegraphics[scale=0.8]{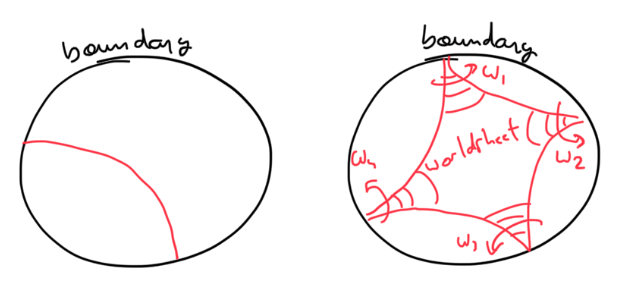}
    \caption{Geodesic for an unflowed two-point function (left) and classical configuration associated with a four-point function involving operators with non-trivial spectral flow charges (right).}
    \label{fig: geodesics H3+}
\end{figure}

For holographic applications it will be useful to consider the Euclidean version of the model, achieved by  implementing the Wick rotation $t \to -i t_E$. The resulting hyperbolic manifold is known as $H_3^+$. It is \textit{not} a group manifold,  but can be understood as the coset space SL(2,$\mathbb{C}$)/SU(2). Changing coordinates to\footnote{Note that this maps the boundary (Euclidean) cylinder to the complex plane parametrized by $\gamma$ and $\gammab$ at $\rho \to \infty$. In the Lorentzian setting, a similar change of variables gives the Poincaré metric on AdS$_3$. This covers only part of the original manifold, that is, the diamond-shaped region known as the Poincaré patch. This is different from the Euclidean case, where $\gamma,\gammab$ and $\phi$ cover the full $H_3^+$.} 
\begin{equation}
    \gamma = e^{t_E -iy} \tanh \rho \, \quad 
    \phi = - t_E + \log (\cosh \rho), 
\end{equation}
the metric and $B$-field (up to an exact 2-form) become
\begin{equation}
\label{H3 metric and B-field}
    ds^2 = k \left(
    d\phi^2 + e^{2\phi}d\gamma d\bar{\gamma} \right)\, , \quad 
    B = -\frac{k}{2} e^{2\phi} d\gamma \wedge d\bar{\gamma} \, .
\end{equation}
Here the Killing vectors give 
\begin{equation}
\label{Ja gammaphi H3}
   k^{-1}  J^+ = -\der_\gamma \, \quad k^{-1} J^3 =  - \gamma \der_\gamma + \frac{1}{2}\der_\phi 
    \,,\quad 
    k^{-1} J^- = - \gamma^2 \der_\gamma + \gamma \der_\phi + e^{-2\phi} \der_{\bar{\gamma}}. 
\end{equation}
Note that the final term in $J^-$ vanishes at large $\phi$. Moreover, $\gamma$ and $\bar{\gamma}$ are now complex coordinates on the  boundary sphere.  Some of the classical solutions discussed above are depicted in the $H_3^+$ setting in Fig.~\ref{fig: geodesics H3+}. 

As in the Lorentzian case, in the $H_3^+$ coset model the dimensionless parameter $k$ setting the size of the geometry in string units corresponds to the level of the affine symmetry algebra. In order to gain some intuition, we start by considering the classical limit $k \gg 1$, where only the zero-mode algebra is relevant. The quadratic Casimir operator $C$, which reads 
\begin{equation}
    2 C = - 2 J^3 J^3 + J^+ J^- + J^- J^+,  
\end{equation}
is then proportional to the corresponding Laplacian. In this limit scalar vertex operators of the model reduce to solutions of the Klein-Gordon equation. This is known in the literature as the mini-superspace limit \cite{Teschner:1997fv}. 
A useful basis of solutions parametrized by a complex variable $x$ is given by the functions 
\begin{equation}
\label{Vjx semiclassical}
    V_j(x) = \frac{2j-1}{\pi} \left(
    e^{\phi} |\gamma - x|^2 + e^{-\phi}
    \right)^{-2j}.  
\end{equation}
The corresponding masses, i.e.~the eigenvalues of $C$, can be read off from the Klein-Gordon equation associated to the metric \eqref{H3 metric and B-field}, 
\begin{equation}
\label{KG equation AdS3}
\left[\nabla_{H_3^+}^2 - \frac{4j(j-1)}{k}\right] V_j(x) = 0. 
\end{equation}
The operators $J^a$ act on $V_j(x)$ as  
\begin{equation}
\label{Ja Diff ops x}
    J^+ \sim  \der_x \, \quad J^3 \sim   x \der_x + j 
    \,,\quad 
    J^- \sim  x^2 \der_x + 2 j x \, . 
\end{equation}
We identify $j$ with the spin of the corresponding representation. The functions $V_j(x)$ can be thought of as the components of a spacetime tensor of weights $(j,j)$.
It is interesting to look at their behaviour at large $\phi$,  
\begin{equation}
\label{Vh large phi}
    V_j(x) \sim e^{2(j-1)\phi} \delta^{(2)}(\gamma - x) + \frac{(2j-1)}{\pi} e^{-2j\phi} |\gamma-x|^{-4j}\,,
\end{equation}
where we have used the following representation of the complex Dirac delta function\footnote{Since we omit most of the anti-holomorphic factors, from now on we will simply write $\delta(x)$.}: %holaa
\begin{equation}
    \delta^{(2)}(x) = \frac{2j-1}{\pi} \lim_{\vep\to 0} \frac{\vep^{2(2j-1)}}{(\vep^2 + |x|^2)^{2j}}  \, , \qquad 
    \re j > \frac{1}{2} \,.
\end{equation} 
In Eq.~\eqref{Vh large phi} we have omitted a series of subleading corrections accompanying each of the two leading terms. We see from the first term that there is a leading divergence centered at $\gamma = x$, while the second term sets the decay rate at large $\phi$ for more general values of $\gamma$. 

We would now like to identify the two contributions described above as the non-normalizable and  normalizable terms that appear in the usual holographic computations. 
For this, we employ an alternative description in terms of energy eigenstates, obtained by applying the Mellin-type transform 
\begin{equation}
\label{Mellin}
    V_{jm\mb} \equiv \int_\mathbb{C} d^2x\, x^{j-m-1} \xb^{j-\mb-1} V_j(x).  
\end{equation}
The integral is to be performed over the full complex plane, although when $m-j \in \mathbb{N}_0$ and $\mb-j \in \mathbb{N}_0$ ($-m-j \in \mathbb{N}_0$ and $-\mb-j \in \mathbb{N}_0$) it coincides with a double contour integral around $x=0$ ($x = \infty$) up to an overall factor. By means of the integral \cite{Maldacena:2001km}
\begin{equation}
\label{integral Maldacena AppC}
\frac{1}{\pi}\int_\mathbb{C} d^2x \, x^{a}\xb^{\bar{a}} (1-x)^{b}(1-\xb)^{\bar{b}} = 
\frac{\Gamma(a+1)
\Gamma(b+1)
\Gamma(-\bar{a}-\bar{b}-1)}{
\Gamma(-\bar{a})
\Gamma(-\bar{b})
\Gamma(a+b+2)
}
    \,,
\end{equation}
we obtain
\begin{equation}
\label{Vjm large phi}
V_{jm\mb} \sim e^{2(j-1)\phi} \gamma^{j-m-1}\bar{\gamma}^{j-\mb-1} + 
    \frac{ \Gamma(1-2j)
\Gamma(j+m)
\Gamma(j-\mb)}{
\Gamma(2j)
\Gamma(1+m-j)
\Gamma(1-\mb-j)
}  
e^{-2j\phi} \gamma^{-j-m}
    \bar{\gamma}^{-j-\mb}
\end{equation}
at large $\phi$. Note that one must have $m-\mb \in \mathbb{Z}$ for the wave-function to be single-valued. Under this condition, the factor in front of the second term in \eqref{Vjm large phi} is actually symmetric under the exchange $m \leftrightarrow \mb$, as follows from the identity $\Gamma(z)\Gamma(1-z) = \pi/ \sin \pi z$. 
The inner product we consider is the one associated to the $L^2$ norm. In terms of the coordinates ($\phi,\gamma,\bar{\gamma}$) it takes the form 
\begin{equation}
    \langle j,j'\rangle \equiv \frac{1}{\pi^3 k^{3/2}} \int d\phi d\gamma d\bar{\gamma} \, e^{2\phi} \overline{V}_j(x) V_{j'}(x) \,. 
\end{equation}
With this norm,  one can ask which values of $j$ correspond to states that should be included in the physical spectrum. The answer to this question is two-fold. 

For generic values of $m$ and $\mb$ and $\re j \neq \frac{1}{2}$, either the first or the second term in  Eq.~\eqref{Vjm large phi} lead to a divergent norm upon integrating over $\phi$. However, when $j = \frac{1}{2} + i s$ with $s \in \R$ the solutions are delta-function normalizable. These states, which share many features with those appearing in Liouville theory \cite{Zamolodchikov:1995aa}, are those originally considered in the Euclidean model \cite{Teschner:1999ug}. In the Lorentzian theory, they constitute the AdS$_3$ analogues of plane waves in flat space. They are associated to particles that come from the boundary, then approach interior of AdS$_3$ where they interact with the NS5 and F1 sources, and finally bounce back to infinity. The parameter $s$ gives the radial momentum. Upon implementing spectral flow, they will correspond to the long string states discussed above. The   relative coefficient appearing in \eqref{Vjm large phi} can be interpreted as the (classical limit of the) reflection coefficient
\begin{equation}
 R_{jm\mb}  = \frac{ \Gamma(1-2j)
\Gamma(j+m)
\Gamma(j-\mb)}{
\Gamma(2j)
\Gamma(1+m-j)
\Gamma(1-\mb-j)
},  
\end{equation}
which amounts to a pure phase for $m,\mb \in \R$ (as long as $m-\mb \in \mathbb{Z}$). 

On the other hand, something special happens for real  $j$: there are certain values of $m$ and $\mb$ for which $R_{jm\mb}$ develops simple poles. These occur at $m = \pm(j + n)$ with $n \in \mathbb{N}_0$, and similarly for $\mb$. If we were interested in the $H_3^+$ model \textit{per se}, it would be natural to consider only complex values for $m$ and $\mb$ \cite{Teschner:1997fv}. However, we are ultimately interested in the analytic continuation to Lorentzian (global) AdS$_3$, hence, as discussed above, it makes more sense to take $m,\mb \in \R$ since these quantum numbers parametrize the energy and angular momentum, see Eq.~\eqref{mmb E and L}. This allows to construct normalizable states by performing the following rescaling: 
\begin{equation}
    V_{jm\mb} \to \tilde{V}_{jm\mb} = \lim_{\mb \to \pm (j+\bar{n})} \lim_{m \to \pm (j+n)} (j+n \mp m)(j+\bar{n} \mp \mb) V_{jm\mb} \, \qquad n,\bar{n} \in \mathbb{N}_0\,.  
\end{equation}
This has the effect of removing the first term in \eqref{Vjm large phi}, while the second one remains finite. Hence, at least at the classical level, for $j > \frac{1}{2}$ such configurations are normalizable, and should therefore correspond to physical states. 

We will see below that in the quantum theory the reflection coefficient appears in the two-point functions. The poles described above then indicate the quantum numbers for which the external operators go on-shell. More precisely, the $m$-basis employed here is the analogue of the usual Fourier basis, and these singularities play the same role as the usual LSZ-type poles in QFT \cite{Peskin:1995ev}. This is consistent with our decision to include the corresponding states in the physical spectrum.
 
\medskip
%%%%%%%%%%%%%%%%

In order to go beyond the classical limit we need to study the coset model based on $H_3^+$. It follows from \eqref{H3 metric and B-field} that the action (including the WZ term) reads
\begin{equation}
\label{Swzw H3}
    S = \frac{k}{\pi} \int d^2z \left(\der \phi \derb \phi + e^{2\phi} \der \bar{\gamma}
    \derb \gamma \right), 
\end{equation}
where the target space coordinates are promoted to fields that depend on the complex worldsheet coordinates $z,\bar{z}$. This can also be derived from Eq.~\eqref{SWZW} by an appropriate  parametrization of the coset elements, namely 
\begin{equation}
  g = \left( \begin{array}{ccc}
       e^{-\phi} + e^{\phi} \gamma \gammab 
       &&  e^{\phi} \gamma \\
        e^{\phi} \gammab && e^{\phi}
    \end{array}
   \right) \, , \qquad 
    g \in H_3^+ \simeq \frac{\mathrm{SL}(2,\mathbb{C})}{\mathrm{SU}(2)} 
    \, .
\end{equation}
Focusing on the holomorphic sector, the  explicit expressions of the conserved currents read
\begin{equation}
\label{Ja classical}
    k^{-1}J^+ =  e^{2\phi} \der\gammab 
    \,, \quad 
    k^{-1}J^3 = e^{2\phi} \gamma \der \gammab - \der \phi
    \,, \quad 
    k^{-1}J^- = e^{2\phi} \gamma^2 \der \gammab - 2 \gamma \der \phi -\der \gamma\,. 
\end{equation}
Roughly speaking, these expressions can be understood  as the one-forms dual to the killing vectors in Eq.\eqref{Ja gammaphi H3}. Similar to the Lorentzian case, classical solutions are given by configurations that factorize as 
$g(z,\zb) = g(z)\bar{g}(\zb)$. They can be parametrized as \cite{deBoer:1998gyt}  
\begin{equation}
\label{Classical sol gamma beta}
    \gamma = a(z) + \frac{\bar{b}(\zb) e^{-2c(z)}}{1+b(z)\bar{b}(\zb)} \,, \quad 
    \gammab = \bar{a}(z) + \frac{b(z) e^{-2\bar{c}(\zb)}}{1+b(z)\bar{b}(\zb)}\,,
\end{equation}
\begin{equation}
\label{Classical sol phi}
    \phi = c(z) + \bar{c} (\zb) + \log [1+b(z)\bar{b}(\zb)] \, , 
\end{equation}
where the holomorphic functions $a(z)$, $b(z)$ and $c(z)$ and their anti-holomorphic counterparts are arbitrary. As will be justified below, the boundary conditions corresponding to the insertion of an operator characterized by a given value of $j$ at a point $x=x_0$ on the asymptotic sphere and at a point $z= z_0$ on the worldsheet are of the form  
\begin{equation}
\label{OPE phi gamma with Vjx}
    \phi(z \sim z_0) \sim - \frac{j}{k} \log |z-z_0|^2
    \, , \quad 
    \gamma(z \sim z_0) \sim x_0 + {\cal{O}}\left(|z-z_0|^{2j/k}\right)\,.
\end{equation}
The simplest classical configuration associated with a two-point function with operators $V_j(x_i,z_i)$ inserted at $z_1=x_1 = 0$ and $z_2 = x_2 \to \infty$ thus reads
\begin{equation}
\label{basic solution j}
    \phi(z,\zb) = - \frac{j}{k} \log |z|^2
    \, , \quad 
    \gamma(z) = 0\,, 
\end{equation}
which corresponds to $a(z) = b(z) = 0$ and $c(z) = - \frac{j}{k} \log z$ in Eqs.~\eqref{Classical sol gamma beta} and \eqref{Classical sol phi}. 
This describes a geodesic joining the two boundary points through the interior of $H_3^+$. In the Lorentzian case, this corresponds to $\rho= 0$ and $t = 2 j \tau/k$, i.e.~the timelike geodesic considered around Eq.~\eqref{timelike geodesic w=0}, with $\alpha = 2j/k$. 

We now consider small perturbations on top of the solution \eqref{basic solution j}, following \cite{Maldacena:2001km}. One can for instance perturb $\gamma$ by turning on either one of the functions $a(z)$ and $\bar{b}(\zb)$ in \eqref{Classical sol gamma beta}. At first order, this gives 
\begin{equation}
\label{2 perturbations gamma}
\delta_{a,n}\gamma = \vep z^n \, \quad  \text{or} \, \quad \delta_{b,n}\gamma = \vep \zb^n |z|^{4j/k} \,, \quad  n \in \mathbb{Z}\,,   
\end{equation}
which correspond to the action of the current modes $J^+_n$ and $J^-_n$, respectively. Indeed, in our parametrization the currents \eqref{Ja classical} are computed from the formula $J^a = k \Tr \left[t^a \der g g^{-1}\right]$ with generators 
\begin{equation}
t^+ = \left(\begin{array}{cc}
    0 & 1 \\
    0 & 0
\end{array}\right) \, \qquad    
t^3 = -\frac{1}{2}\left(\begin{array}{cc}
    1 & 0 \\
    0 & -1
\end{array}\right) \, \qquad    
t^- = \left(\begin{array}{cc}
    0 & 0 \\
    -1 & 0
\end{array}\right) \, .
\end{equation}
We are interested in perturbations giving non-trivial modifications at infinity, corresponding to solutions associated with the boundary insertion of operators of the form $(J^{\pm}_n V_j)$, and should thus only consider perturbations that are non-normalizable. Here the norm is defined using the target space metric, 
% Answer (3)
namely\footnote{The factor $|z|^{-2}$ comes from the conformal transformation relating the worldsheet cylinder and the flat complex coordinates \cite{Maldacena:2001km}. Indeed, we start by considering the (euclidean) string worldsheet as an infinite cylinder --consistent, for instance, with our interpretation of the solution in Eq.~\eqref{basic solution j} -- and then map it to the complex $z$-plane as usual.} 
\begin{equation}
\label{norm ws perturbations}
    \int \frac{d^2z}{|z|^2} \left[ 
    (\delta\phi)^2 + e^{
    2\phi
    } \delta\gamma \delta\gammab
    \right]. 
\end{equation}
For spacetime weights satisfying  $\frac{1}{2} < j < \frac{k}{2}$, one finds that the non-normalizable modes correspond to $n=0,-1,-2,\dots$ for $J^+_n$ and $n=-1,-2,\dots$ for $J^-_n$. Conversely, the normalizable modes can be thought of as annihilating the original operator $V_j$. For $J^\pm_n$ with $n>0$ this is precisely what we expect for an affine primary state, see Appendix \ref{sec: appB WZW models}. Additionally we expect $V_j(x_1=0,z_1=0)$ to be annihilated also by $J^-_0$ since the Mellin transform of Eq.~\eqref{Mellin} tells us that we are dealing with the wave-function of a state which has the lowest possible value of $m$. In other words, it is, the lowest-weight state in the corresponding representation of the zero-mode algebra, and has $m=j$.

However, the story is different for when $j\geq \frac{k}{2}$. Repeating the perturbative analysis for states with $j = \tilde{\jmath} + \frac{k}{2} \w$, where $0 < \tilde{\jmath} < \frac{k}{2}$ and $\w \in \mathbb{Z}$, we now find that the these states are annihilated by the modes  $J^+_n$ with $n= \w + 1, \w + 2, \dots$ and $J^-_n$ for $n= -\w,-\w + 1, -\w + 2, \dots$. This means that we effectively shift the modes of $J^{\pm}$ by $\pm \w$ units. Looking back at Eq.~\eqref{Lorentzian classical flow currents}, which in Euclidean signature reads $J^\pm \to z^{\pm \w} J^\pm$, we see that this is precisely what we expect from a state obtained by including $\w$ units of spectral flow. These are the short string states. More explicitly, we will show below that they corresponds to lowest-weight states of spin $\tilde{\jmath}$ with respect to a different SL(2,$\R$) subalgebra generated by 
\begin{equation}
    \left\{\tilde{J}_0^+,\tilde{J}_0^-,\tilde{J}_0^3\right\} \equiv \left\{J^+_\w, J^-_{-\w},J_0^3 - \frac{k}{2}\w \right\}\,.  
\end{equation}
 
Finally, we need to consider the cases with Re$(j) = \frac{k}{2} \w$ separately. Here the two perturbations in \eqref{2 perturbations gamma} with $n=\pm \w$ become exactly the same. On the other hand, a new solution appears; it takes the form
\begin{equation}
    \phi = - \frac{\w}{2} \log |z|^2 \,, \quad 
    \gamma = \vep z^\w \log |z|^2. 
\end{equation}
This is precisely the (analytic continuation of the Lorentzian) solution for a long string  at small radial momentum, obtained by setting $\alpha \sim \vep$ in Eq.~\eqref{Lorentzian Classical long string}. One finds that these states are annihilated by $J^\pm_{n}$ with $n > \pm \w $. As will be discussed below, these spectrally flowed states are built from \textit{unflowed} states which belong to the continuous representations of SL(2,$\R$), for which $m$ is unbounded.

\subsection{Wakimoto fields and spacetime symmetries}
\label{sec: wakimoto syms}

The current algebra is defined by the OPEs
\begin{equation}
    J^a(z)J^b(w) \sim \frac{\eta^{ab} k/2}{(z-w)^2} + \frac{ f^{ab}_{\phantom{ab}c} J^c(w)}{z-w} \, ,
    \label{OPEjSL2}
\end{equation}
where $k$ is the level of the affine algebra, while
\begin{equation}
-2\eta^{33} = \eta^{+-} = 2\, , \qquad f^{+-}_{\phantom{+-}3}=-2 \,, \qquad 
    f^{3+}_{\phantom{3+}+}=-
    f^{3-}_{\phantom{3-}-}=1
\,. 
\end{equation} 
The energy-momentum tensor and the central charge follow from the Sugawara construction, reviewed in Appendix \ref{sec: appB WZW models}, and are given by
\begin{equation}
    T_{\mathrm{sl}}(z) = \frac{1}{2(k-2)} \left[-2 J^3 (z) J^3(z) + J^+(z) J^-(z) + J^-(z) J^+(z) \right]\,,  
    \label{TSL2}    
\end{equation}
and 
\begin{equation}
    c_{\mathrm{sl}} = \frac{3k}{k-2}\,.
\end{equation}
Note that the "-2" factors in the denominators are due to quantum effects, and were invisible in the classical large $k$ limit, see for instance Eq.~\eqref{KG equation AdS3}.  Identical expressions hold for the anti-holomorphic sector.

At the classical level, we can introduce a pair of Lagrange multipliers $\beta,\betab$ in order to  rewrite the action \eqref{Swzw H3} as 
\begin{equation}
\label{Swzw H3 beta}
    S = \frac{1}{2\pi}\int d^2z \left(\der \phi \derb \phi + \betab \der \bar{\gamma} + \beta
    \derb \gamma  - 
    e^{-\sqrt{\frac{2}{k}}\phi} \beta \betab \right). 
\end{equation}
Here we have rescaled $\phi \to \phi/ \sqrt{2k}$ for normalization purposes. 
At large $\phi$ the interaction term can be neglected, hence the theory becomes that of a free scalar field and a so-called $\beta\gamma$-system. Moreover, in this regime the EOMs ensure that $\beta$, $\gamma$ and $\der \phi$ are holomorphic. 
At the quantum level, there is indeed a free-field description valid in the above regime, that is, near the AdS$_3$ boundary. This is known as the Wakimoto representation \cite{Wakimoto:1986gf}. However, one must take into account the non-trivial transformation of the path integral measure\footnote{See \cite{Hosomichi:2000bm,Ishibashi:2000fn} for an explicit derivation, and also the earlier references \cite{Gawedzki:1991yu,Tseytlin:1993my}.}. This results in a small but crucial modification in the exponent of the interaction term, and further generates a (spacetime) dilaton linear in $\phi$, leading to 
\begin{equation}
\label{Swzw H3 beta quantum}
    S = \frac{1}{2\pi} \int d^2z \left(\frac{1}{2}\der \phi \derb \phi  + \frac{Q_\phi}{4 \sqrt{2}} R^{(2)}\phi + \betab \der \bar{\gamma} + \beta
    \derb \gamma  - 
    e^{-\sqrt{\frac{2}{k-2}}\phi} \beta \betab \right),      
\end{equation}
where $R^{(2)}$ is the worldsheet curvature scalar and $Q_\phi = -\frac{1}{\sqrt{k-2}}$.    
Near the asymptotic boundary, the interaction can be treated perturbatively, and the theory can be studied in terms of the free field OPEs, namely 
\begin{equation}
\label{OPE beta gamma and phi}
    \phi(z,\zb)\phi(0) \sim - \log |z|^2\,, \quad 
    \beta(z) \gamma(0) \sim - \frac{1}{z} \,.
\end{equation}
The holomorphic currents are realized as 
\begin{equation}
\label{Ja Wakimoto}
    J^+ = \beta \,, \quad 
    J^3 = (\beta \gamma) - \sqrt{\frac{k-2}{2}} \der \phi
    \,, \quad 
    J^- = (\beta \gamma^2) - \sqrt{2(k-2)}\gamma \der\phi - k \der \gamma. 
\end{equation}
We see from the OPE \eqref{OPE beta gamma and phi} that $\beta$ acts as (minus) the derivative w.r.t.~$\gamma$, hence for large $k$ these expressions reproduce those given in Eq.~\eqref{Ja gammaphi H3} above, as they should. On the other hand, by solving the EOMs for $\beta,\betab$ we go back to the currents of Eq.~\eqref{Ja classical}.
We also have 
\begin{equation}
    T(z) = -\beta \der \gamma  - \frac{1}{2} \der \phi \der \phi - \frac{Q_\phi}{\sqrt{2}} \der^2 \phi  \, , 
\end{equation}
with central charge 
\begin{equation}
\label{centralcharge Wakimoto}
    c = 2 + 1 + 6Q_\phi^2 = \frac{3k}{k-2}\, ,
\end{equation}
justifying the appearance of the charge $Q_\phi$. 

The free field representation of the SL(2,$\R$) vertex operators reads
\begin{equation}
\label{Vjm Wakimoto}
    V_{jm\mb} = \gamma^{j-m-1}\gammab^{j-\mb-1} e^{(j-1)\sqrt{\frac{2}{k-2}} \phi},  
\end{equation}
which should be compared with the first term in \eqref{Vjm large phi}, and the corresponding worldsheet weight is 
\begin{equation}
\label{def delta w=0}
    \Delta = -\frac{j(j-1)}{k-2},  
\end{equation}
consistent with Eq.~\eqref{KG equation AdS3}.
By using the OPEs in \eqref{OPE beta gamma and phi} we find that, in our conventions\footnote{The reader might prefer to work with a slightly more natural convention in which the coefficients generated by the action of $J^{\pm}_0$ vanish for $m = \pm j$, analogously to what is usually done in the SU(2) case. However, for AdS$_3$ this actually introduces a number of inconvenient factors in different formulas. Our notation follows that of Refs.~\cite{Maldacena:2001km,Dei:2021xgh,Dei:2021yom,Bufalini:2022toj}.}, the currents act as follows:  
\begin{equation}
    J^{\pm}(z) V_{jm}(0) \sim \frac{m \pm (1-j)}{z} V_{j,m\pm 1}(0) \, , \quad 
    J^{3}(z) V_{jm}(0) \sim \frac{m }{z} V_{jm}(0) \, .
\end{equation}
for $m \neq \pm j$, while $J^{\pm}(z) V_{j,\mp j}(0) \sim 0$. Here we have suppressed higher-order terms. We have also omitted the anti-holomorphic indices, and shall continue doing so. 

The operators \eqref{Vjm Wakimoto} form the so-called $m$-basis, where the Cartan current $J^3$ acts diagonally. It will prove extremely useful to construct the dual $x$-basis, involving vertex operators whose wave functions are localized near $\gamma \sim x$ at the asymptotic boundary, see Eq.~\eqref{Vjx semiclassical}. For this we need to invert the transform defined in Eq.~\eqref{Mellin}. Roughly speaking, this can be written as 
\begin{equation}
    V_{j}(x,z) = \sum_{m,\bar{m}}  x^{m-j}
    \bar{x}^{\bar{m}-j} \, V_{jm}(z) \, . 
    \label{ExpSL2}
\end{equation}
To be precise, \eqref{ExpSL2} holds for states in the discrete sector, where $j \in \R$. Hence, poles in the integrand of \eqref{Mellin} coming from the expansion around  $x=0$ ($x=\infty$) are associated to states in the ${\cal D}_j^+$ (${\cal D}_j^-$) representation. 
For states in the continuous sector, \eqref{ExpSL2} is modified to account for the fact that, although $m-\bar{m}$ is an integer number, $m+\bar{m}$ can take arbitrary real values, which must be integrated over, leading to   
\begin{equation}
    V_{j}(x,z) = \frac{i}{(2\pi^2)}\sum_{m-\bar{m}}
    \int_{-\infty}^\infty d(m+\bar{m}) \, x^{m-j}
    \bar{x}^{\bar{m}-j} \, V_{jm}(z)\,.
    \label{ExpSL2Cont}
\end{equation} 

As in the classical limit, the zero-modes of the SL(2,$\R$) currents act as differential operators on such states, namely\footnote{Here our convention differs from that used in \cite{DiFrancesco:1997nk} by a sign. } 
\begin{equation}
\label{OPE Ja Vjx}
    J^a(z) V_{j}(x,0) \sim \frac{1}{z} \, (D_x^a V_j)(x,0) \,,
\end{equation}
with 
\begin{equation}
\label{Da Diff ops x}
    D_x^+ =  \der_x \, \quad D_x^3 =   x \der_x + j 
    \,,\quad 
    D_x^- =  x^2 \der_x + 2 j x.  
\end{equation}
This shows that $J_0^+$ generates translations in $x$-space, hence, at least for discrete states, $V_j(x,z)$ can be understood as a lowest-weight operator translated from the origin, i.e. 
\begin{equation}
    V_j(x,z) = e^{x J_0^+ + \bar{x} \bar{J}_0^+} V_{jj} (z) e^{-x J_0^+ -\bar{x} \bar{J}_0^+} \, ,
\end{equation}
where we have identified $V_j(x=0,z) = V_{jj}(z)$. Conversely, $V_j(x\to \infty,z)$ is directly related to $V_{j,-j}(z)$. In the Lorentzian setting, states in lowest-weight (highest-weight) representations, i.e.~with $m = j + n$ ($m = - j - n$) with $n\in \mathbb{N}_0$ are interpreted as in-going (out-going) particle excitations. 
It is instructive to compute the OPEs 
\begin{equation}
    \phi(z,\zb) V_j(x,0) \sim -(j-1)\sqrt{\frac{2}{k-2}} \log |z|^2 V_j(x,0) \, , \quad 
    \gamma(z) V_j(x,0) \sim x V_j(x,0) \, .  
\end{equation}
The first expression justifies \textit{a posteriori} the boundary conditions employed in Eq.~\eqref{OPE phi gamma with Vjx}, which should be valid in the limit of large $k$ and large $j$ (and was written in terms of the unrescaled $\phi$). 

Having the holographic correspondence in mind, we would like to interpret the (integrated version of the) worldsheet vertex operators $V_j(x,z)$  as local operators of a boundary conformal field theory. Eqs.~\eqref{OPE Ja Vjx} and \eqref{Da Diff ops x} show that the global generators of the corresponding \textit{spacetime} Virasoro algebra (not to be confused with the \textit{worldsheet} Virasoro algebra) can be identified with 
\begin{equation}
    \L_{-1} = \oint dz J^+(z) \,, \quad 
        \L_{0} = \oint dz J^3(z) \,, \quad 
            \L_{1} = \oint dz J^-(z) \,.
\end{equation}
The Wakimoto representation allows for a simple construction of the rest of the spacetime Virasoro modes, namely \cite{Giveon:1998ns} 
\begin{equation}
\label{Ln Wakimoto}
    \L_n = \oint dz \left[(n+1) \gamma^n J^3 - n \gamma^{n+1} J^+\right]\,.
\end{equation}
This expression is valid in the large $\phi$ region where the free field description is accurate. The  modes $\L_n$ are indeed BRST-invariant. They are also unique up to BRST exact states, i.e.~those with integrands proportional to $2 \gamma^n J^3 -\gamma^{n+1}J^+- \gamma^{n-1}J^- $, obtained by acting with the BRST charge $\Qq$ on $\gamma^n$. The current OPEs combined with the fact that 
\begin{equation}
    J^3(z) \gamma^n(0) \sim -\frac{n}{z} \gamma^n(0) \,, \quad 
    J^-(z) \gamma^n(0) \sim -\frac{n}{z} \gamma^{n+1}(0)\,,
\end{equation}
imply that
\begin{equation}
\label{Virasoro alg spacetime}
    [\L_n,\L_m] = (n-m) \L_{n+m} + \Ii \frac{k}{2}  (n^3-n) \delta_{n+m,0} \,, 
\end{equation}
where 
\begin{equation}
\label{Def identity op wakimoto}
    \Ii \equiv \oint \gamma^{-1} \der \gamma \,.
\end{equation}
It can be seen that the operator $\Ii$ commutes with all $\L_n$. In a classical solution, $\Ii$ is an integer that counts how many times the string worldsheet winds around the boundary circle. (In the Lorentzian framework we have $\gamma \sim e^{t- i y}$ at fixed $\rho$, hence $\Ii$ keeps track of how many times we go around the spatial circle upon shifting $\sigma \to \sigma + 2\pi$.) This quantity is interpreted as  counting the number of fundamental strings sources present in our configuration\footnote{We should stress that this is  formally infinite from the perturbative worldsheet point of view when working at first order in the topological expansion.}, namely $n_1$ \cite{Giveon:1998ns}. By replacing $\Ii \to n_1$ in \eqref{Virasoro alg spacetime} we read off the spacetime central charge
\begin{equation}
    c_{\rm st} = 6 k n_1.  
\end{equation}
However, it should be noted that in the quantum theory the role of the operator $\Ii$ is quite subtle; we will come back to it later on when discussing spectral flow.

Suppose that there is an additional set of conserved Kac-Moody currents $K^a(z)$ in our string background, satisfying 
\begin{equation}
    [K^a_n,K^b_m] = i g^{ab}_{\phantom{ab}c} K^c_{n+m} +
    \frac{k'}{2}n \delta^{ab}\delta_{n+m,0}
\end{equation}
for some integer level $k'$ and structure constants $g^{ab}_{\phantom{ab}c}$. This is the case, for instance when the background is AdS$_3\times S^3 \times T^4$ (or K3), where there are currents $K^a(z)$ which generate the $S^3$ isometries. The former are identified with the R-symmetry generators of the holographic SCFT. The Wakimoto fields allow us  to translate this into spacetime modes, 
\begin{equation}
\label{def K(x) modes Wakimoto}
    \Kk^a_n = \oint dz\, K^a(z) \gamma^n(z). 
\end{equation}
This leads to following extension of the spacetime symmetry algebra: 
\begin{eqnarray}
    [\Kk^a_n,\Kk^b_m] &=& i f^{ab}_{\phantom{ab}c} \Kk^c_{n+m} +
    \Ii \, \frac{k'}{2}n \delta^{ab}\delta_{n+m,0} \, , \\
\left[\L_n ,\Kk^a_m\right] &=& -m \Kk^a_{n+m}  \, .
\end{eqnarray}
Since $\Ii$ is the same operator that appeared in the Virasoro case, we can read off $k_{\rm st}' \equiv  n_1 k'$.

We can also compute the action of the modes $\L_n$ on the vertex operators. By using Eq.~\eqref{Vjm Wakimoto}, we get\footnote{The odd minus sign in \eqref{Ln st on Vjm} comes from our conventions for the expansion in terms of the boundary coordinate $x$, see Eq.~\eqref{ExpSL2}.}  
\begin{equation}
\label{Ln st on Vjm}
    [\L_n,V_{j,-m}(z)] = 
    [n(j-1)-m] V_{j,-(m+n)}(z)
    \,. 
\end{equation}
This tells us that, up to the integration over the worldsheet coordinates $z$ and $\zb$, the $m$-basis operators $V_{jm\mb}(z)$ can be interpreted as the (spacetime) Virasoro modes of a local operator $V_j(x,z)$ of holomorphic weight $j$ of the boundary theory. 

Let us discuss in more detail the allowed values for the spin quantum number $j$. Our semiclassical analysis indicated the presence of two types of physical modes:  those with $j \in \frac{1}{2} + i \R$, which were only delta-function normalizable, and those with  $j > \frac{1}{2}$. However, we must be careful since this was done in the $k \to \infty$ limit, hence our conclusions could -- and will -- get modified in the quantum theory, i.e.~at finite $k$. 

The broad picture is consistent with the well-known 
representation theory of the zero-mode SL(2,$\R$) algebra. 
The relevant irreducible representations are either of the discrete
lowest/highest-weight type, or of the continuous type. The principal discrete series of lowest-weight are built from the states $|j,j\rangle$, annihilated by $J_0^-$, by acting with $J_0^+$, thus spanning 
\begin{equation}
    \label{D+rep}
        {\cal{D}}_j^+ =
        \Big\langle \, 
        |j m\rangle \ , \ m=j
        ,j+1,j+2,\cdots
        \Big\rangle,
    \end{equation}
where $J_0^3|j m\rangle = m|j m\rangle$. Here $|j m\rangle$ represents  the state created by inserting $V_{jm} (z)$ at the origin $z=0$. The representations ${\cal{D}}_j^+$ are unitary  for any $j > 0$. This also holds for their conjugate representations, denoted as ${\cal{D}}_j^-$, which are highest-weight representations defined analogously.  
% For consistency one must restrict to 
% \begin{equation}
%         \frac{1}{2} < j < \frac{k-1}{2},
%         \label{Djrange}
% \end{equation}
% as follows from $L^2(AdS_3)$ normalisation conditions, no-ghost theorems and spectral flow considerations to be discussed below.
On the other hand, the principal continuous series are given by 
\begin{equation}
\label{Contrep}
    {\cal{C}}_j^\alpha = 
        \Big\langle \,
        |j  m \rangle \ , \ 0 \leq \alpha < 1 \ , \  m=\alpha
        ,\alpha\pm 1,\alpha \pm 2,\cdots
        \Big\rangle.
    \end{equation}
These are unitary if $j=\frac{1}{2} + i s$ with $s \in \R$. (Strictly speaking, the representation with $s = \alpha - \frac{1}{2}=0$ is reducible.) Note that, as opposed to what happens for discrete states, in the continuous case the allowed values of $m$ are not related to that of $j$.

\begin{figure}
    \centering
    \includegraphics[scale=0.7]{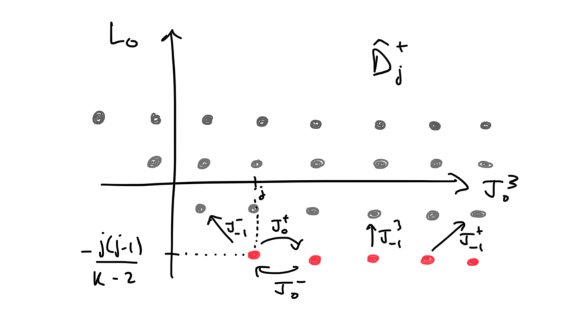}
    \caption{Weight diagram for the lowest-weight representation $\hat{{\cal{D}}}_j^+$}
    \label{fig: weight diagram D+ w=0}
\end{figure}

The above analysis was restricted to the zero-mode algebra. For WZW models the number of symmetry generators is enhanced to the affine extension of the SL(2,$\R$) algebra at level $k$. The commutation relations of the generators $J^a_n$ with $n \in \mathbb{Z}$ are contained in the OPEs \eqref{OPEjSL2}. Each of the above states defines an affine primary state, satisfying 
\begin{equation}
    J_n^a |jm\rangle = 0 \, , \qquad \forall \,\, n > 0, 
\end{equation}
from which the full affine module is generated by acting freely with all modes $J^a_{n}$ with $n<0$. As an example, we show the weight diagram for the $\hat{{\cal{D}}}_j^+$ representation in the plane of eigenvalues of $L_0$ and $J_0^3$ in Fig.~\ref{fig: weight diagram D+ w=0}.

However, as in the SU(2) case, there exist null descendants of the form 
\begin{equation}
    \left(J_{-1}^-\right)^{k-2j+1} |j \, j \rangle = 0\,.  
\end{equation}
For instance, the simplest null state is given by the $J_{-1}^-|\frac{k}{2} \frac{k}{2} \rangle$, since 
\begin{equation}
\label{Null state SL2}
    \left| J_{-1}^-\left|\frac{k}{2} \frac{k}{2} \right\rangle \right|^2 = \left\langle \frac{k}{2} \frac{k}{2} \right| J_{1}^+ J_{-1}^- \left| \frac{k}{2} \frac{k}{2} \right\rangle = 
    \left\langle \frac{k}{2} \frac{k}{2} \right| k - 2 J_{0}^3 \left| \frac{k}{2} \frac{k}{2} \right\rangle = 0 \, .
\end{equation}
As discussed in Appendix \ref{sec: appB WZW models}, this can be used to show that (unflowed) states with $j > \frac{k}{2}$ actually decouple \cite{Zamolodchikov:1986bd}. Moreover, we have argued above that, at least semiclassically, states where $j$ is an integer multiple of $\frac{k}{2}$ are rather special, as this is the value for which the long strings appear. This suggests that we should restrict to discrete representations with $\frac{1}{2} < j < \frac{k}{2}$.  Actually, this intuition turns out to be slightly off: the correct range being given by  
\begin{equation}
\label{Djrange}
    \frac{1}{2} < j < \frac{k-1}{2}\,. 
\end{equation}
It was shown in \cite{Maldacena:2001km} that only such states satisfy the requisite no-ghost theorem in the bosonic string context.
We will shortly provide an argument for the upper bound in \eqref{Djrange} based on spectral flow.

We now come back to the full bosonic string theory on AdS$_3\times M_{\rm int}$. One finds that states belonging to the continuous representations are problematic. Indeed, the mass-shell condition reads
\begin{equation}
    -\frac{j(j-1)}{k-2} + h_{\rm int} - 1 + N = 0\,,
\end{equation}
where $N$ is the level of (worldsheet Virasoro) 
%descendence ANSWER (5)
descendants. 
For $j = \frac{1}{2} + i s$ we get $-j(j-1) = \frac{1}{4} + s^2$, hence this Virasoro constraint cannot be satisfied for $N > 0$ assuming $h_{\rm int} \geq 0$.  Moreover, Eq.~\eqref{KG equation AdS3} shows that any such state with $s\neq 0$ has a spacetime mass which violates the Breitenlohner-Freedman bound for AdS$_3$: 
\begin{equation}
    m^2 = 4 j(j-1) = - 1 - 4 s^2 < m_{BF}^2 = -1\,.
\end{equation}
Here we are working in units of the AdS$_3$ radius  and in the large $k$ regime. We conclude that these modes are tachyonic, and should be excluded. This is actually a relief: we argued above that $j$ should be identified with the holomorphic weight in the boundary theory, and for this the latter to be a unitary CFT we ought to avoid including operators with complex weights. 

However, the above discussion raises two very important problems. First, in our worldsheet description we have not found physical operators corresponding to the short and long string solutions observed at the classical level, only those associated with particle-like solutions. Relatedly,  we have not been able to construct any operator with a spacetime weight larger than $\frac{k}{2}$. We expect a large number of states above this value since the spacetime central charge is $c_{\rm st} = 6 k n_1$, and we are working at large $n_1$. The solution to both issues was found in \cite{Maldacena:2001km}: we are missing the spectrally flowed representations. 

\subsection{Spectrally flowed representations}

Algebraically, spectral flow refers to a family of authomorphisms of the affine SL(2,$\R$) algebra at level $k$, namely the transformation $J^a_n \to \tilde{J}^a_n$ with  
\begin{equation}
\label{J modes spectral flow}
    \tilde{J}^{3}_n  = J^3_n - \frac{k}{2} \w \delta_{n,0} \,, \quad 
    \tilde{J}^\pm_n = J^\pm_{n\pm \w} \, , \qquad 
    \w \in \mathbb{Z} \, .
\end{equation}
The effect on the worldsheet Virasoro modes is given by 
\begin{equation}
\label{Ln spectral flow}
    L_n \to \tilde{L}_n = L_n + \w J^3_n - \frac{k}{4} \w^2 \delta_{n,0} \, . 
\end{equation}
All commutation relations are preserved by these transformations, and of course the affine level $k$ and central charge $c = \frac{3k}{k-2}$ also remain  unchanged. Nevertheless, the effect on the different states is highly non-trivial. More precisely, we now show that a  whole new family of physical string states is generated. These are distinct from the unflowed states that we have considered so far, which is in contrast with what happens for WZW models based on compact groups such as SU(2), where spectral flow merely reshuffles the different primaries and descendants without producing any new states, see Appendix \ref{sec: appB WZW models}. 

Let us see how this works for SL(2,$\R$). Consider a state defined as an affine primary of spin $j$ and projection $m$ with respect to the unflowed currents $\tilde{J}^a_n$,  for a given $\w$. This state is denoted $|jm\w\rangle$, and it  satisfies
\begin{equation}
\label{Jtilde state jmw 1}
    \tilde{J}^3_0 |jm\w\rangle = m |jm\w\rangle \, , \quad 
    \tilde{J}^{\pm}_{\pm 0} |jm\w\rangle = 
    [m \pm (1-j)]
    |j,m\pm 1,\w\rangle \, .
\end{equation}
Restricting to $\w\geq 0$ without loss of generality, the following properties hold as well:
\begin{equation}
\label{Jtilde state jmw 2}
    \tilde{J}^3_n |jm\w\rangle = J^3_n |jm\w\rangle = 0 \,, \quad 
    \tilde{J}^\pm_n |jm\w\rangle = J^\pm_{n\pm \w} |jm\w\rangle = 0 \,, \qquad 
    \forall \, \, n > 0 \, .
\end{equation}
All other modes act non-trivially, except when dealing with a lowest/highest weight state with $m=\pm j$, for which we additionally have
\begin{equation}
   \tilde{J}^\mp_0 |j, \pm j,\w\rangle = J^\mp_{\mp\w} |j, \pm j,\w\rangle = 0\,. 
\end{equation}
This shows that for $\w>0$ the state $|jm\w\rangle$ is \textit{not} an affine primary with respect to the action of the physical currents -- by which we mean $J^a_n$ -- since generically it is not annihilated by the positive modes $J^+_{n}$ with $n= 1, \dots\, \w$. 

However, recall that physical states need not be  affine primaries: in string theory we only require that they are built by combining (worldsheet) Virasoro primaries in a BRST-invariant way. Although in the context of WZW models being an affine primary implies being a Virasoro primary as well, the converse statement is not true \cite{DiFrancesco:1997nk}. 
Indeed, from Eq.~\eqref{Ln spectral flow} we see that $|jm\w\rangle$ \textit{is} a Virasoro primary, i.e.~it is an eigenstate of $L_0$ and further satisfies $L_{n} |jm\w\rangle = 0$ for all $n>0$. Hence, such states (and their descendants) should be included in the spectrum, provided they satisfy the mass-shell condition. In order to identify the relevant quantum numbers we compute the action of the zero modes $J^a_0$ and $L_0$. By using Eq.~\eqref{J modes spectral flow} in combination with Eqs.~\eqref{Jtilde state jmw 1} and \eqref{Jtilde state jmw 2}, we derive 
\begin{equation}
    J^3_0|jm\w\rangle = \left(\tilde{J}^3_0 + \frac{k}{2} \w \right) |jm\w\rangle = \left(m + \frac{k}{2}\w\right) |jm\w\rangle \,, 
\end{equation}
and 
\begin{equation}
    J^-_0|jm\w\rangle = \tilde{J}^-_\w  |jm\w\rangle = 0 \, .
\end{equation}
We thus identify $|jm\w\rangle$ as the lowest-weight state in discrete representation of the physical zero-mode algebra with spin 
\begin{equation}
\label{h spectral flow}
    h = m + \frac{k}{2} \w \, .
\end{equation}
Note that we have landed on a discrete lowest-weight representation \textit{regardless of the unflowed representation we started with}. In other words, we need not impose any condition on whether $j$ is real or not: we always end up with $h \in \R$.   
The highest-weight flowed primary from which the  corresponding conjugate representation is generated is of the form $|j,-m,-\w\rangle$. Finally, from Eq.~\eqref{Ln spectral flow} we get 
\begin{equation}
    L_0 |j,m, \w \rangle = \left(\tilde{L}_0 - \w \tilde{J}^3_0 - \frac{k}{4}\w^2\right) |j, m, \w \rangle
    \, , \quad 
\end{equation}
from where we read off the physical worldsheet weights of such spectrally flowed states,  
\begin{equation}
\label{Delta spectral flow}
    \Delta = - \frac{j (j-1)}{k-2} - m \w - \frac{k}{4} \w^2 = - \frac{j (j-1)}{k-2} - h \w + \frac{k}{4} \w^2 \, .
\end{equation}

\begin{figure}
    \centering
    \includegraphics[scale=0.7]{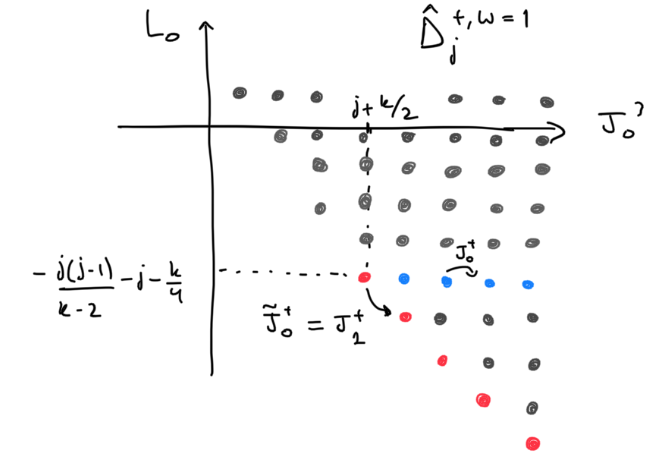}
    \caption{Weight diagram for the lowest-weight representation $\hat{{\cal{D}}}_j^{+,\w}$ with $\w=1$. }
    \label{fig: weight diagram D+ w=1}
\end{figure}

We now argue that states belonging to the spectrally flowed representations, which will be denoted as $\Dd_j^{\pm,\w}$ and $\Cc_{j}^{\alpha,\w}$, are indeed new Virasoro primaries, as neither them nor their descendants can be obtained as a linear combination of states belonging to the unflowed representations. This can be seen for instance by looking at the allowed values for the corresponding worldsheet weights $\Delta$. Recall that in each \textit{unflowed} affine module, defined by the value of the corresponding spin $j$, the primary weight in Eq.~\eqref{def delta w=0} constitutes a lower bound for the eigenvalue of $L_0$. For  (physical) discrete states we thus have an overall lower bound of the form $\Delta > \Delta(j=\frac{k-1}{2})$, while for the continuous representations, which are tachyonic anyway, we get $\Delta > -\frac{1}{4(k-2)}$. In contrast, from Eq.~\eqref{Delta spectral flow} we immediately find that in generic spectrally flowed sectors we can obtain arbitrarily negative  values of $\Delta$ simply by increasing the absolute value of $m$, see Fig.~\ref{fig: weight diagram D+ w=1}. It follows that these must be new physical states.

\begin{figure}
    \centering
    \includegraphics[scale=0.7]{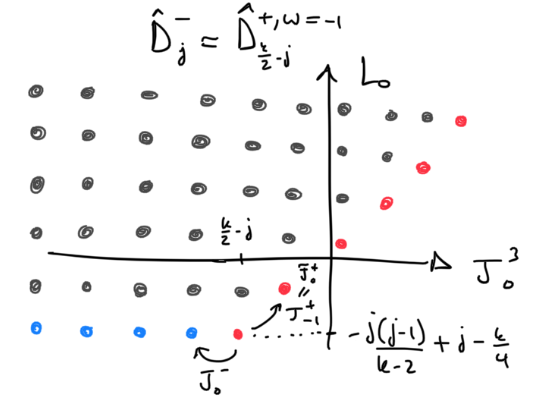}
    \caption{Equivalence between the representations $\hat{{\cal{D}}}_{j}^{-}$ and $\hat{{\cal{D}}}_{\frac{k}{2}-j}^{+,\w}$ with $\w=-1$. }
    \label{fig: weight diagram SeriesId}
\end{figure}

The only exceptions correspond to the affine modules  $\hat{\Dd}_j^{\pm,\w=\mp 1}$. The reason for this can be understood in terms of a more general phenomenon  known as the SL(2,$\R$) series identifications, which we now describe. From Eqs.~\eqref{h spectral flow} and \eqref{Delta spectral flow} we see that 
that the states $|j,-j,\w\rangle$ and $|\frac{k}{2}-j,\frac{k}{2}-j,\w-1\rangle$  have exactly the same quantum numbers. Hence, they should be identified (up to a normalization constant, whose precise form will be obtained in due course). As these states can be used to generate their full affine modules, we get the following set of isomorphims: 
\begin{equation}
\label{series Id SL2 1}
    \hat{\Dd}_{j}^{\pm,\w} \simeq \hat{\Dd}_{\tilde{\jmath}}^{\mp,\w \pm 1} 
    \,, \quad \tilde{\jmath} =\frac{k}{2}-j \, .
\end{equation}
In particular, $\hat{\Dd}_j^{\pm,\w = \mp 1}$ are identified with unflowed representations. This is shown for a particular example in Fig.~\ref{fig: weight diagram SeriesId}. 
We also note that this explains why we had to impose $j < \frac{k-1}{2}$ in Eq.~\eqref{Djrange}, as opposed to the classical upper bound $j<\frac{k}{2}$. Indeed, states in $\hat{\Dd}_j^{+,\w=-1}$ with $j > \frac{k-1}{2}$ would be identified with unflowed states in a highest-weight representation of spin $\tilde{\jmath} < \frac{1}{2}$, which are not normalizable.

Due to the series identifications \eqref{series Id SL2 1}, we should be careful and avoid overcounting when describing the spectrum of the model. This can be achieved for example by only including the discrete representations $\hat{\Dd}_{j}^{+,\w\geq 0}$ and $\hat{\Dd}_{j}^{-,\w \leq 0}$. Equivalently, we could work only with the $\hat{{\cal{D}}}_j^{+,\w}$ representations and include all $\w \in \mathbb{Z}$.  This fits nicely with the fact that, as in the unflowed sector, from the spacetime point of view discrete representations of the highest-weight type are interpreted as out-going modes while their lowest-weight counterparts are the in-going ones. In other words, the operators that create the states $|jm\w\rangle$ and $|j,-m,-\w \rangle$ in the worldsheet theory contribute to the same $x$-basis  vertex operator, denoted as $V_{jh}^w(x,z)$ with $h=m+\frac{k}{2}\w$. This discussion holds for spectrally flowed representations of the zero-mode algebra regardless of whether they were built from unflowed states in the discrete or continuous sectors. 
Hence, we conclude that the physical spectrum of the SL(2,$\R$) WZW model is characterized by a Hilbert space of the form \cite{Maldacena:2000hw,Maldacena:2000kv,Maldacena:2001km} 
\begin{equation}
    \Hh = \Hh_{\w=0}^{\rm disc}\oplus_{\w \in \mathbb{Z}_{\neq 0}} \left[\Hh_\w^{\rm disc} \oplus \Hh_\w^{\rm cont}\right] \,,
\end{equation}
with 
\begin{equation}
\Hh_\w^{\rm disc} = 
    \int_{\frac{1}{2}}^{\frac{k-1}{2}} dj \, \hat{\Dd}_j^{+,\w} \otimes
    \hat{\Dd}_j^{+,\w} \,,
\end{equation}
and 
\begin{equation}
    \Hh_\w^{\rm cont} = \int_{\frac{1}{2}+i \R}
    dj \int_0^1 d\alpha \, 
    \hat{\Cc}_j^{\alpha,\w} \otimes
    \hat{\Cc}_j^{\alpha,\w}. 
\end{equation}  
It was shown in \cite{Maldacena:2000kv} that this leads to a modular invariant torus partition function.

We now discuss in more detail the main properties of the spectrally flowed vertex operators. In the $m$-basis we denote them as $V_{jm}^\w(z)$. For $\w\geq0$, they are formally defined by the following OPEs: 
\begin{subequations}
\label{flowedOPESL2}
\begin{eqnarray}
    J^{+}(z)V_{jm }^{\w}(w) &=& \frac{(m+1-j)V_{j,m+1 }^{\w}(w)}{(z-w)^{\w+1}} + 
    \sum_{n=1}^\w 
    \frac{(J^+_{n-1} V_{jm }^\w)(w)}{(z-w)^n} + \dots \, , \\ [1ex]
    J^{3}(z)V_{jm }^{\w}(w) &=& 
    \frac{\left(m+ \frac{k}{2}\w \right)V_{jm }^{\w}(w)}{(z-w)}
    + \dots \, , \\ [1ex]
    J^{-}(z)V_{jm }^{\w}(w) &=& (z-w)^{\w-1} (m-1+j) V_{j,m-1 }^{\w}(w) + \dots \, , 
\end{eqnarray}
\end{subequations}
where the ellipsis indicate higher order terms. Note that the leading terms in the OPEs of $V_{jm}^\w(z)$ with $J^\pm$ are known since they correspond to the action of the modes $J^\pm_{\pm \w} = \tilde{J}^\pm_0$. In contrast,  there is no simple expression available for the following orders. Similar equations hold for $\w<0$ with the roles of $J^+$ and $J^-$ inverted. Even though we omit the anti-holomorphic dependence, it should be clear that the operators $V_{jm}^\w(z)$ have analogous OPEs with the currents $\Jb^a$. 

We can also provide an $x$-basis construction for vertex operators with a non-trivial spectral flow charge. Indeed, for $\w>0$ all spectrally flowed $m$-basis primaries obtained above are annihilated by $J_0^-$. For each of them, the remaining  states in the corresponding zero-mode representation are obtained by acting freely with $J_0^+$. Spectrally flowed $x$-basis operators are defined in analogy with those of the unflowed sector, see Eq.~\eqref{ExpSL2}. More precisely, we have 
\begin{equation}
\label{def Vjhw x-basis}
    V_{jh\bar{h}}^\w(x,\xb,z,\zb) =  
    e^{x J_0^+ + \bar{x} \bar{J}_0^+} V_{jm \mb}^\w (z) e^{-x J_0^+ -\bar{x} \bar{J}_0^+} \, .
\end{equation}
Note that the holomorphic and anti-holomorphic spacetime weights $h = m + \frac{k}{2}\w$ and $\bar{h} = \mb + \frac{k}{2} \w$ could be different. Nevertheless, in what follows we will mostly focus on spacetime scalars, that is, operators with $h = \bar{h}$. Hence, from now on we shall omit anti-holomorphic quantities and simply use the shorthand $V_{jh}^\w(x,z)$. 

In Eq.~\eqref{def Vjhw x-basis} we have identified $V_{jh}^\w(x=0,z) = V_{jm}^\w(z)$ and $V_{jh}^\w(x\to \infty,z) \sim  V_{j,-m}^{-\w}(z)$, so that when working in the $x$-basis it makes sense to restrict to $\w \geq 0$. The defining OPEs \eqref{flowedOPESL2} imply that for $V_{jh}^\w(x,z)$ we have 
\begin{subequations}
\begin{eqnarray}
   J^+(w)  V^{\w}_{jh}(x,z) & = & 
   \sum_{n=2}^{\w+1} \frac{\left(J_{n-1}^+ 
    V_{jh}^\w\right) (x,z) }{(w-z)^n}
    +\frac{\left[J_{0}^+, 
    V_{jh}^\w (x,z)\right]  
      }{(w-z)}  + \cdots \, , \\ [1ex]
    J^3(w)  V^{\w}_{jh}(x,z) & = & 
    x \sum_{n=2}^{\w+1} \frac{\left(J_{n-1}^+ 
    V_{jh}^\w\right) (x,z) }{(w-z)^n}
    +\frac{\left[J_{0}^3, 
    V_{jh}^\w (x,z)\right]  
      }{(w-z)}  + \cdots
    \, , \\ [1ex]
    J^-(w)  V^{\w}_{jh}(x,z) & = &  
    x^2 \sum_{n=2}^{\w+1} \frac{\left(J_{n-1}^+ 
    V_{jh}^\w\right) (x,z) }{(w-z)^n}
    +\frac{\left[J_{0}^-, 
    V_{jh}^\w (x,z)\right]  
      }{(w-z)}  + \cdots
      \, , 
\end{eqnarray}
\label{JVxOPE}
\end{subequations}
As in the unflowed sector, the zero modes act as differential operators in $x$, 
\begin{subequations}
\begin{eqnarray}
    \left[J_{0}^+, 
    V_{jh}^\w (x,z)\right] &=& \der_x V_{jh}^\w (x,z) \, , 
    \label{J0+dx}\\[1ex]
    \left[J_{0}^3, 
    V_{jh}^\w (x,z)\right] &=& (x\der_x+h) V_{jh}^\w (x,z) \, , \\[1ex]
    \left[J_{0}^-, 
    V_{jh}^\w (x,z)\right] &=& (x^2\der_x + 2hx) V_{jh}^\w (x,z) \, ,
\end{eqnarray}    
\label{diffopsx}
\end{subequations}
\hspace{-0.1cm}while 
\begin{equation}
    \left(J_{\pm \w}^\pm V_{jh}^\w\right) (x,z) = \left[h-\frac{k}{2} \w \pm (1-j)\right] V_{j,h\pm 1}^\w (x,z)  \, . 
    \label{JwVx}
\end{equation}
Importantly, in terms of the currents\footnote{$J^-(x,z)$ is often written simply as $J(x,z)$ in the literature. }
\begin{equation}
    J^3(x,z) = J^3(z) - x J^+(z) \, , \quad J^-(x,z) = J^-(z) - 2 x J^3(z)+x^2 J^+(z)
\label{defJx}
\end{equation} 
we get
\begin{subequations}
\begin{eqnarray}
    J^3(x,w)  V^{\w}_{jh}(x,z) & = & 
    \frac{h 
      }{(w-z)} V_{jh}^\w(x,z) + \cdots
    \, , \label{J3xOPE}\\ [1ex]
    J^-(x,w)  V^{\w}_{jh}(x,z) & = &  
    (w-z)^{\w-1} \left(J_{-w}^-V_{jh}^\w\right) (x,z) + \cdots
    \, . 
    \label{JmxOPE}
\end{eqnarray}
\end{subequations}
This shows that when the currents are inserted at the same point in $x$ as the vertex operators, the corresponding OPEs become analogous to the $m$-basis ones.

Let us now discuss the physical spectrum of the full string worldsheet theory. We restrict to the Virasoro primary sector for simplicity, and consider operators of the form 
\begin{equation}
    \Vv (x,\xb,z,\zb) \equiv V_{jh}^\w(x,\xb,z,\zb) V_{\rm int}(z,\zb). 
\end{equation}
Focusing on the holomorphic sector, we impose the Virasoro constraint 
\begin{equation}
    \label{Virasoro condition with w}
    -\frac{j(j-1)}{k-2}
    - h \w + \frac{k}{4}\w^2+ h_{\rm int} - 1 = 0. 
\end{equation}
For states in flowed lowest-weight representations we have $h = m + \frac{k}{2}\w$ with $m = j + n $ for some non-negative integer $n$. We can thus solve for $j$ in Eq.~\eqref{Virasoro condition with w}, and also compute the spacetime weight, giving 
\begin{equation}
\label{h physical discrete w}
   h =  n+\w + \frac{1}{2} + 
   \sqrt{(k-2) \left( h_0 - n \w - \frac{\w( \w+1)}{2} \right)} \,,
\end{equation}
with 
\begin{equation}
\label{def h0}
   h_0 =  \frac{1}{4(k-2)} + h_{\rm int}-1.
\end{equation}
The allowed range for $j$ given in Eq.~\eqref{Djrange} further  imposes the bounds  
\begin{equation}
\label{bound on h0 w}
    \frac{k}{4}\w^2 + \frac{w}{2} \leq h_{0}\leq  
       \frac{k}{4}(\w+1)^2 - \frac{w+1}{2}.
\end{equation}
For a given $h_0$ in this range, setting $n=0$ in \eqref{h physical discrete w} gives the lowest allowed value of $h$. In particular, when the bound is saturated (for a fixed $\w$) we can write this as \begin{equation}
    h^{\rm min} = \frac{k}{4}\w + \frac{h_0^{\rm min}}{\w} = \frac{1}{2} + \frac{k}{2}\w\, ,
\end{equation}
and 
\begin{equation}
    h^{\rm min}  = \frac{k}{4}(\w+1) + \frac{h_0^{\rm max}}{\w+1} = 
    -\frac{1}{2} + \frac{k}{2}(\w+1). 
\end{equation}
We now switch to states belonging to the continuous sector. Here we interpret \eqref{Virasoro condition with w} as an equation for $h$ instead of $j = \frac{1}{2} + i s$, which gives 
\begin{equation}
    h = \frac{k}{4}\w + \frac{1}{\w}\left[ \frac{s^2}{(k-2)} 
    + h_{0}\right]. 
\end{equation}
The lowest possible value for $h$  clearly corresponds to $s=0$, i.e.~ 
\begin{equation}
    h^{\rm min}=  \frac{k}{4}\w + \frac{h_0}{\w} 
\end{equation}
for any $h_0$.
For a given value of $\w$, we thus find that, in the limit $j\to \frac{1}{2}$ or $j\to \frac{k-1}{2}$, that is, when we saturate the bound \eqref{Djrange} for the discrete states,  the minimal spacetime energy $E = h + \bar{h}$ for the corresponding short strings becomes exactly that of a continuous one with a winding charge given by either $\w$ or $\w+1$. Said differently, whenever they exist, i.e.~for a given choice of $h_{\rm int}$ such that $h_0$ is in the range \eqref{bound on h0 w}, the corresponding state with the lowest possible $h$ always corresponds to a short string. On the other hand, when we instead have  
\begin{equation}
\label{quantum long string range}
    \frac{k}{4}\w^2 - \frac{w}{2} \leq h_{0}\leq  
       \frac{k}{4}\w^2 + \frac{w}{2}\,,
\end{equation}
the lowest energy eigenvalue is realized in the long string sector. 
This is to be compared with our analysis of the classical string geodesics in global AdS$_3$, where we found that long strings can only give the lowest-energy states when we sit precisely at $h_{\rm int} = \frac{k}{4}\w^2$. In the classical regime, i.e.~at large $h$ and $k$, we have $h_0 \approx h_{\rm int}$, see Eq.~\eqref{def h0}. We see from Eq.~\eqref{quantum long string range} that in the quantum theory this is extended to a strip of width $\w$.

\subsubsection{Wakimoto fields and $x$-basis spectrally flowed vertex operators}
\label{sec: Wakimoto Vjhw}

Before finishing this section, we briefly go back to the Wakimoto free field description. Spectral flow corresponds to the transformation\footnote{This gives the correct transformation for $J^3$ thanks to the normal ordering constant $\w$ appearing when flowing the product $(\beta \gamma)$, see Eq.~\eqref{Ja Wakimoto}.  }  
\begin{equation}
\label{spectral flow wakimoto}
    \gamma(z)  \to z^\w\gamma(z) \,, 
    \quad 
    \beta(z)  \to z^{-\w}\beta(z) \,, 
    \quad 
    \phi(z,\zb) \to \phi(z,\zb) - \w \sqrt{\frac{k-2}{2}} \log |z|^2.  
\end{equation}
In the $x$-basis, the zero-mode of 
$J^+ = \beta$ acts as $\der_x$, hence $\gamma$ is translated in $x$-space as $\gamma(x,z) = \gamma (z) - x$, while $\phi(z,\zb)$ is unaffected.
This leads to  OPEs of the flowed vertex operators with the Wakimoto fields of the form 
\begin{equation}
\label{OPE beta Vw}
\beta (z) V_{jh}^\w(x,0) \sim \frac{h-\frac{k}{2}\w+1-j}{z} V_{j,h+1}^\w(0) 
\end{equation}
and 
\begin{equation}
\label{OPE phi Jvx w}
    \phi (z,\zb) V_{jh}^\w(x,0) \sim -\sqrt{\frac{2}{k-2}}\left(j -1 + \frac{k-2}{2}\w\right) \log |z|^2  V_{jh}^\w(x,0), 
\end{equation}
which justifies the boundary conditions used in Eq.~\eqref{OPE phi gamma with Vjx} for spectrally flowed states. On the other hand, the OPE with $\gamma(z)$ is regular, although it must satisfy 
\begin{equation}
\label{OPE gamma Vw}
    \gamma(z) V_{jh}^\w(x,0) = x V_{jh}^\w(x,0) + 
    z^\w V_{j,h-1}^\w(x,0) + \cdots . 
\end{equation}

An explicit expression for the spectrally flowed vertex operators in the free field language was derived quite recently in  \cite{Dei:2023ivl,Knighton:2023mhq}. One has 
\begin{equation}
\label{Vw xbasis Wakimoto}
    V_{jh}^\w(x,z) \equiv e^{(j-1+\frac{k-2}{2}\w)\sqrt{\frac{2}{k-2}} \phi(z)} \left[\frac{\der^\w(\gamma(z)-x)}{\w!}\right]^{\frac{k}{2}\w+j-h-1}
     \delta_\w(\gamma(z)-x) \, ,
\end{equation}
where we have introduced the product of delta-function operators 
\begin{equation}
\label{def deltaw}
     \delta_\w (\gamma(z)-x) \equiv \prod_{n=0}^{\w-1} \delta (\der^n (\gamma(z)-x)) \, . 
\end{equation}
As a sanity check, for $\w=0$ this trivializes and, given that one must also set $h=j$, \eqref{Vw xbasis Wakimoto} reduces to the $x$-basis version of\eqref{Vjm Wakimoto}, namely 
\begin{equation}
\label{V xbasis Wakimoto}
    V_j(x,z) = (\gamma(z)-x)^{-1} e^{(j-1)\sqrt{\frac{2}{k-2}} \phi} \, .
\end{equation}
Note that inside Cauchy integrals $(\gamma(z)-x)^{-1}$ behaves effectively as $\delta(\gamma(z)-x)$. 

Let us now discuss why \eqref{Vw xbasis Wakimoto} is the correct generalization of \eqref{V xbasis Wakimoto} for $\w>0$. The exponential in $\phi$ is dictated by the OPE \eqref{OPE phi Jvx w}, and has conformal weight 
\begin{equation}
    \Delta\left[e^{(j-1+\frac{k-2}{2}\w)\sqrt{\frac{2}{k-2}} \phi}\right] = -\frac{1}{k-2}\left(j+\frac{k-2}{2}\w\right)\left(j-1+\frac{k-2}{2}\w\right) \, .
\end{equation}
Additionally, the operator in \eqref{def deltaw} is designed to enforce that, near $V_{jh}^\w(x,z=0)$, the field $\gamma$ approaches the boundary insertion point $x$, and that all its derivatives up to $\der^{\w-1}\gamma$ vanish, as indicated by the expansion in Eq.~\eqref{OPE gamma Vw}. Such delta-function operators were introduced in \cite{Witten:2012bh}\footnote{The operators $\delta(\gamma(z))$  and $\delta(\beta(z))$ are defined such that, when inserted in a correlation function defined in the path integral formalism, they instruct us to consider configurations where the corresponding field vanishes at the insertion point. In terms of OPEs, this leads to  
\begin{equation}
\label{OPEs delta gamma}
    \gamma(z) \delta (\gamma(w)) \sim (z-w) \der \gamma \delta(\gamma )(w) \, , \qquad
    \beta(z) \delta (\gamma(w)) \sim (z-w)^{-1} \delta'(\gamma (w)) \, ,
\end{equation}
and, analogously, 
\begin{equation}
\label{OPEs delta beta}
    \gamma(z) \delta (\beta(w)) \sim (z-w)^{-1} \delta'(\beta (w)) \, , \qquad
    \beta(z) \delta (\beta(w)) \sim 
     (z-w) \der \gamma \delta(\beta )(w) \, ,
\end{equation}
which can be derived  from the  $\beta(z)\gamma(0)$ OPE combined with the integral representation $\delta(\gamma) = (2\pi)^{-1}\int d\sigma \exp(i \sigma \gamma)$. As will be done in Sec.~\ref{sec: AdS3xS3xT4} for the fermionic sector in the superstring case, spectral flow in the $\beta\gamma$ sector is more easily discussed in a bosonized description, see \cite{Knighton:2023mhq}.
%Furthermore, at leading order, the fusion of delta function operators further gives 
% \begin{equation}
%     \delta (\gamma(z))\delta (\gamma(w)) \sim - (z-w)^{-1} \delta (\gamma(w))\delta (\der \gamma(w)) \, .
% \end{equation}
}.
In particular, since $\gamma(z)$ is dimensionless, the worldsheet conformal dimension of $\delta_\w(\gamma(z)-x)$ comes solely from the derivatives, giving 
\begin{equation}
    \Delta[\delta_\w(\gamma)] = - \frac{\w(\w-1)}{2} \, .
\end{equation}
Finally, the power of $\der^\w(\gamma(z)-x)$ appearing in \eqref{OPE gamma Vw} has weight 
\begin{equation}
    \Delta \left[(\der^\w \gamma)^{\left(\frac{k}{2}-h+j-1\right)}\right] = \w \left(\frac{k}{2}-h+j-1\right) \, ,
\end{equation}
and it guarantees that the first regular term in the OPE of $\gamma(z)-x$ and $V_{jh}^\w(x,z)$ has order $z^\w$ and simply shifts $h \to h-1$, consistent with \eqref{OPE gamma Vw}. It also gives the correct OPE with $\beta$, and completes the appropriate total weight 
\begin{equation}
    \Delta\left[e^{(j-1+\frac{k-2}{2}\w)\sqrt{\frac{2}{k-2}} \phi}\right] + \Delta \left[(\der^\w \gamma)^{\left(\frac{k}{2}-h+j-1\right)}\right] +  \Delta[\delta_\w(\gamma)] 
    %= - \frac{j(j-1)}{k-2}-h\w+\frac{k}{4}\w^2 
    = \Delta \left[V_{jh}^w\right].
\end{equation}
We will make extensive use of the definition \eqref{Vw xbasis Wakimoto} when computing correlation functions in the free field limit in Sec.~\ref{sec: holography}. In particular, we will show that the presence of the delta-function operator \eqref{def deltaw} has crucial consequences from the holographic point of view.

Finally, we also comment relation between the operator $\Ii$ defined in \eqref{Def identity op wakimoto} when discussing the worldsheet realization of the  spacetime Virasoro algebra. Eq.~\eqref{spectral flow wakimoto} gives 
\begin{equation}
\label{Ii shift w Wakimoto}
    \Ii \to \oint z^{-\w}\gamma^{-1} \der \left(z^\w\gamma \right) = \Ii + \w.
\end{equation}
Hence, strictly speaking, $\Ii$ cannot be interpreted as the identity operator of the spacetime theory, as it acts differently in each spectral flow sector. Now, although the spectral 
flow charge behaves effectively as a winding charge when the (long) strings are near the AdS$_3$ boundary, it is not a conserved quantity since the $y$-circle is contractible in the bulk interior. In particular, this leads to the presence of  dynamical processes where a long string is emitted or absorbed. Given that $\Ii$ is supposed to count the number of fundamental string sources in the background, this suggests that $n_1$ should be thought of as a dynamical quantity. We will come back to this later on.

\subsection{Comments on the partition function}

We now consider the partition function. Our treatment will follow the qualitative analysis of \cite{Maldacena:2000hw}, Appendix B. Although the resulting expressions for the characters and partition function are useful form the intuitive point of view, they are somewhat ill-defined due to various divergencies. For a more precise treatment, which is outside of the scope of these notes, we refer the reader to \cite{Maldacena:2000kv,Hanany:2002ev,Israel:2003ry,Eberhardt:2020bgq,Ashok:2020dnc}. 

We study the characters of the discrete and continuous sectors separately. 
Let us first consider the discrete sector. We already know that, once spectral flow is taken into account, it is enough to work only with representations of the lowest-weight type, together with their spectrally flowed images. Naively, the computation should be somewhat similar to that of the SU(2) WZW model (see Appendix \ref{sec: appB WZW models}) except for the range of allowed values for the spin $j$, which is $\frac{1}{2} < j < \frac{k-1}{2}$ instead of $0 \leq j \leq \frac{k}{2}$. However, it turns out that the computation is actually simplified. This is because in the SL(2,$\R$) case the expression analogous to Eq.~\eqref{norm proto null states} reads  
\begin{equation}
\label{norm proto null states SL2}
    || (j_{-1}^-)^{N}|j,j\rangle ||^2 = \prod_{n=1}^N n (k-2j-1+n)\,,  
\end{equation}
implying that a null state similar to those found in the SU(2) context can only be obtained for SL(2,$\R$) by choosing $j = \frac{k-1+N}{2}$ for some positive integer $N$ . As a consequence, no null states appear when imposing the physical restriction in Eq.~\eqref{Djrange}! In the unflowed sector, this leads to simple global characters of the form 
\begin{equation}
    \label{GlobalCharacter-j-SL2}
    \chi_j^+(z) = \sum_{n=0}^\infty y^{j+n} = \frac{
    y^{j}}{
    1-y}, 
\end{equation}
where $z$ is the chemical potential related to the charge under $J_0^3$ and $y = e^{2\pi i z}$. 
Including the affine descendants then gives 
\begin{equation}
\label{WrongCharacter-l-SL2}
    \chi_j^+(\tau,z) = \Tr_j^+\left[
    q^{L_0 - \frac{c}{24}} y^{J_0^3}
    \right]
    = \frac{q^{-\frac{j(j-1)}{k-2}-\frac{k}{8(k-2)}} \chi_j^+(z)}
    {
    \prod_{n=1}^{\infty} (1-q^n) (1-y q^n) (1- y^{-1}q^n)
    } \,, 
\end{equation}
with $c=\frac{3k}{k-2}$, $\tau$ the modular parameter, and  $q=e^{2\pi i \tau}$.

Let us now consider the spectrally flowed short string sector. Here the characters can be written as 
\begin{equation}
    \chi_j^{+,\w}(\tau,z) = \Tr_j^+\left[
    q^{L_0 - \w J_0^3 - \frac{k}{4} \w^2 - \frac{c}{24}} y^{J_0^3+\frac{k}{2}\w}
    \right] = q^{-\frac{k}{4}\w^2} y^{\frac{k}{2}\w} \chi_j^+(\tau,z - \w \tau), 
\end{equation}
which gives 
\begin{eqnarray}
   \chi_j^{+,\w}(\tau,z) &=&  \frac{q^{-\frac{j(j-1)}{k-2}-\w j- \frac{k}{4}\w^2-\frac{k}{8(k-2)}} y^{j + \frac{k}{2}\w}}
    {
    (1-y q^{-\w})\prod_{n=1}^{\infty} (1-q^n) (1-y q^{n-\w}) (1- y^{-1}q^{n+\w})
    } \nn \\
    &=& \frac{q^{-\frac{j(j-1)}{k-2}-\w j- \frac{k}{4}\w^2-\frac{k}{8(k-2)}} y^{j + \frac{k}{2}\w}}
    {
    (1-y q^{-\w})\prod_{n=1}^{\infty} (1-q^n) (1-y q^{n}) (1- y^{-1}q^{n})
    } \prod_{n=1}^\w \frac{1-y^{-1}q^n}{1-y q^{n-\w}}  \nn \\
    &=& \frac{q^{-\frac{j(j-1)}{k-2}-\w j- \frac{k}{4}\w^2-\frac{k}{8(k-2)}} y^{\frac{k}{2}\w} \chi_j^+(z) }
    {\prod_{n=1}^{\infty} (1-q^n) (1-y q^{n}) (1- y^{-1}q^{n})
    } \prod_{n=1}^\w \frac{1-y^{-1}q^n}{1-y q^{-n}}.  
\end{eqnarray}
Here we recognize the overall shifts in the worldsheet weight and the $J_0^3$ charge produced by spectral flow. From the final expression we also find that the last factor removes the modes associated to the action of  $J^-_{-1},\dots,J^-_{-\w}$ and adds those obtained by acting with $J^+_1,\dots,J^+_{\w}$, as expected. We can simplify this further using 
\begin{equation}
    \prod_{n=1}^\w \frac{1-y^{-1}q^n}{1-y q^{-n}} = 
    \prod_{n=1}^\w \left(-y^{-1}q^n\right) = (-1)^\w y^{-\w} q^{\frac{1}{2}\w (\w+1)}, 
\end{equation}
so that 
\begin{equation}
\label{chi_j+w}
    \chi_j^{+,\w}(\tau,z) = \frac{q^{-\frac{1}{8}
    %-\frac{1}{k-2}\left(j-\frac{1}{2}+\frac{k-2}{2}\w\right)
    }\left(y q^{-\frac{1}{k-2}}\right)^{j-\frac{1}{2}+\frac{k-2}{2}\w}}
    {(y^{-\frac{1}{2}}-y^{\frac{1}{2}})\prod_{n=1}^{\infty} (1-q^n) (1-y q^{n}) (1- y^{-1}q^{n})
    }.
\end{equation}

In order to obtain the full contribution from discrete states to the torus partition function\footnote{To be precise, here we also perform an analytic continuation both in the worldsheet and in spacetime, which allows us to take $\tau$ and $z$ to be complex, and $\bar{\tau}$ and $\zb$ as their complex conjugates. } we consider the diagonal combination of holomorphic and anti-holomorphic characters, integrate over the unflowed spins $j$ and sum over the spectral flow charges $\w \in \mathbb{Z}$. Remarkably, Eq.~\eqref{chi_j+w} shows that the characters depend only on the combination $t = j - \frac{1}{2} + \frac{k-2}{2}\w$, hence the integral in $j$ over $\frac{1}{2}<j<\frac{k-1}{2}$ and the sum over $\w \in \mathbb{Z}$ combine to give an integral in $t$ over the full real axis. This gives 
\begin{equation}
    Z^{\rm disc}(\tau,z) = e^{\pi k \frac{(\im z)^2}{\im \tau}} |\Theta_1(\tau,z)|^{-2}\int_{-\infty}^{+\infty} dt \, e^{\frac{4 \pi \im \tau}{k-2}  t^2- 4\pi   \im z \, t} \sim \frac{e^{2\pi \frac{(\im z)^2}{\im \tau}}}{\sqrt{\im \tau} \,   |\Theta_1(\tau,z)|^{2}}, 
\end{equation}
where $\Theta_1$ is the elliptic theta function. The final  expression is  manifestly modular invariant, modulo a small number of subtleties discussed in  \cite{Maldacena:2000hw}.

We will be less precise regarding the continuous representations ${\cal{C}}_j^\alpha$, although we include a few interesting comments for completeness. The (unflowed) global character reads 
\begin{equation}
    \chi_j^\alpha(z) = y^{\alpha} \sum_{n\in \mathbb{Z}} e^{2\pi i z n} = 2\pi y^\alpha \sum_{m \in \mathbb{Z}} \delta(z+m).   
\end{equation}
Since this is non-zero only when $z$ takes an integer value, we can set $y \to 1$ in various factors of the full affine character, leading to 
\begin{eqnarray}
    \chi_j^{\alpha}(\tau,z) &=&  \frac{q^{\frac{1+4s^2}{4(k-2)}-\frac{k}{8(k-2)}} \chi_j^\alpha(z)}
    {
    \prod_{n=1}^{\infty} (1-q^n) (1-y q^{n}) (1- y^{-1}q^{n})
    } \nn \\
    &=& \frac{2\pi}{\eta^{3}(q)} q^{\frac{s^2}{k-2}} \sum_{m \in \mathbb{Z}} e^{2 \pi i m \alpha} \delta(z+m),   
\end{eqnarray}
with $\eta(q)$ the Dedekind eta function. 
Upon implementing spectral flow, we obtain the physical characters 
\begin{equation}
    \chi_j^{\alpha,\w}(\tau,z) 
    = 
    \frac{2\pi}{\eta^{3}(q)} q^{\frac{s^2}{k-2}+\frac{k}{4}\w^2} \sum_{m \in \mathbb{Z}} e^{2 \pi i m \left(\alpha-\frac{k}{2}\w\right)} \delta(z-\tau \w +m).
\end{equation}
To obtain the full contribution from continuous states we combine the holomorphic and anti-holomorphic characters, and integrate over $s \in \R$ and over $\alpha \in [0,1)$. The latter integral forces $m$ to be the same on both sides, while the former is a gaussian integral, assuming a uniform density of states. When the dust settles, one finds 
\begin{equation}
\label{Zcont Sl2}
    Z^{\rm cont}(\tau,z) = \frac{1}{\sqrt{\im \tau} |\eta(\tau)|^6} \sum_{\w,m = - \infty}^{+\infty}
    \delta (z-\w \tau + m)\,, 
\end{equation}
which is formally modular invariant. 

We finish by discussing the interpretation of the factor $\delta (z-\w \tau + m)$. If we were to perform an orbifold identification and consider strings propagating in a locally (Euclidean) AdS$_3$ space with a periodic euclidean time coordinate such that the boundary became a torus, the relation between the Cartan generator $J_0^3$ on the worldsheet and the spacetime Virasoro mode $\Ll_0$ would suggest that $z$ should be interpreted as the spacetime modular parameter, i.e.~that of the \textit{boundary} torus. The worldsheet spectrum would also include twisted sectors, labelled by two integers $c$ and $d$. In this context, the long-string partition function would end up being proportional to \cite{Maldacena:2000kv,Eberhardt:2020bgq}
\begin{equation}
    \sum_{a,b,c,d \in \mathbb{Z}}\delta(z (c \tau + d)-a \tau - d) \,, \qquad ad-bc=1 \, . 
\end{equation}
This localizes on configurations where there is a holomorphic map between the modular parameters of the worldsheet and boundary tori, i.e. 
\begin{equation}
    z = \frac{a \tau + d}{c \tau + d}. 
\end{equation}

The continuous sector is analyzed much more precisely in \cite{Maldacena:2000kv} by studying strings on thermal AdS$_3$. The main issue is that, as it turns out, the assumption of a uniform density of states $\rho(s)$ is too strong. However, the rough discussion we have included here is consistent with our classical intuition: long strings can reach the near-boundary region, but we know from \eqref{Swzw H3} that the corresponding contribution to the path integral are highly suppressed unless $\gamma$ is  holomorphic. 
Given that $\gamma$ can be thought of as parametrizing the AdS$_3$ boundary, the leading contributions would then come from configurations for which there exists a holomorphic covering map from the worldsheet to the boundary (or several of them). Roughly speaking, in Eq.~\eqref{Zcont Sl2} we have picked up the contributions of all such maps that are in the (worldsheet) untwisted sector $(c,d)=(0,1)$ and wind exactly $\w$ times around the boundary spatial circle. These types of maps will play a prominent role in Sec.~\ref{sec: bosonic correlators} below when computing correlators involving spectrally flowed insertions.   

\newpage

\section{Exact description of the unflowed sector}
\label{sec: Exact theory, unflowed sector}

So far we have discussed the spectrum of the theory, using both the classical limit of the model and the Wakimoto free-field approach for an approximate account of its quantized version. We now present the exact description of the unflowed sector of the SL(2,$\R$) WZW model, including correlation functions \cite{Kutasov:1999xu,Maldacena:2001km}. In this sector both the vertex operators and their correlators can be understood by analytic continuation from the Euclidean model on $H_3^+$ \cite{Teschner:1997fv,Teschner:1997ft,Teschner:1999ug,Ribault:2005wp,Ishibashi:2000fn,Hosomichi:2000bm}.  We leave the computation of spectrally flowed correlators for Section \ref{sec: bosonic correlators}.  

\subsection{Reflection symmetry. Two- and three-point functions}

Until now we have neglected the interaction term in \eqref{Swzw H3 beta quantum}, which is, strictly speaking, only valid at the  AdS$_3$ boundary. In the near-boundary region we can of course still use perturbation theory \cite{Knighton:2023mhq,Knighton:2024qxd}. However, the fact that we are dealing with a WZW model allows for a much greater computational power. One can use both conformal and SL(2,$\R$) Ward Identities to derive various differential equations in $z$ and $x$ that must be satisfied by the correlators. Moreover,  the Sugawara construction also implies the well-known Knizhnik-Zamolodchikov (KZ) equation. Even though they only appear in representations that are not part of the physical spectrum, further constraints can be derived from the existence of null vectors in SL(2,$\R$) affine modules\footnote{Several details related the derivation of these differential equations and their consequences are reviewed in Appendix \ref{sec: appB WZW models} for the SU(2) case.}. One can also obtain additional  constraints by imposing crossing symmetry. In the SU(2), $H_3^+$ and SL(2,$\R$) models it turns out that this is enough to derive two- and three-point functions (on the sphere)  \textit{exactly}! The same goes for several specific higher-point functions.

In the exact theory, vertex operators $V_j(x)$ are defined by their OPEs with the currents, see Eq.~\eqref{OPE Ja Vjx}. One starts by  constructing the  \textit{macroscopic} operators of \cite{Teschner:1999ug}, i.e.~those belonging to the continuous representations. The \textit{microscopic} operators associated with the discrete representations are  then defined by analytic  continuation to real values of $j$. Since the conformal weight $(k-2)\Delta  = -j(j-1)$ is symmetric under the exchange $j \to 1-j$, there is direct relation between operators with spins $j$ and $1-j$. This is known as the reflection symmetry, and takes the following form: \begin{equation}
\label{ReflectionUnflowed}
    V_{j}(x,z) = B(j) \int d^{2}x' \, |x-x'|^{-4j}V_{1-j}(x',z). 
\end{equation}
The physical meaning of the reflection coefficient 
\begin{equation}
    B(j) = \frac{(k-2) \nu^{1-2j}}{\pi \gamma(\frac{2j-1}{k-2})} \, , \qquad 
    \gamma(x) \equiv \frac{\Gamma(x)}{\Gamma(1-\xb)}\,,
\end{equation}
will become clear shortly.  Here we have included   a constant $\nu$, which must be independent of $j$.

The global Ward identities imply that  correlation functions 
\begin{equation}
    \langle V_{j_1}(x_1,z_1) \dots 
    V_{j_n}(x_n,z_n)
    \rangle
\end{equation}
must be annihilated by the differential operators 
\begin{equation}
    \sum_{i=1}^n \der_{x_i} \,, \quad 
    \sum_{i=1}^n x_i \der_{x_i} + j_i \,, \quad 
    \sum_{i=1}^n x^2_i\der_{x_i} + 2j_i x_i \, ,
\end{equation}
and 
\begin{equation}
    \sum_{i=1}^n \der_{z_i} \,, \quad 
    \sum_{i=1}^n z_i \der_{z_i} + \Delta_i \,, \quad 
    \sum_{i=1}^n z^2_i\der_{z_i} + 2\Delta_i z_i \,. 
\end{equation}
This fixes the dependence of two- and three-point functions on the insertion points $z_i$ and $x_i$. For the two-point function, an important subtlety is that, as a consequence of the reflection symmetry \eqref{ReflectionUnflowed}, one must consider not only the usual solution $\langle V_{j_1} V_{j_2}\rangle \sim |x_{12}|^{-4j_1}$ for $j_1=j_2$ (and with $x_{12} = x_1-x_2$),  but also a second one, given by the distributional solution of the form $ \langle V_{j_1} V_{j_2}\rangle \sim \delta(x_{12})$, valid when $j_2 = 1-j_1$. This comes from the distributional identity $x\delta'(x) = -\delta(x)$. Consequently,  the two-point function reads 
\beq
\langle V_{j_1}(x_1,z_1)V_{j_2}(x_2,z_2)\rangle
=\frac{1}{|z_{12}|^{4\Delta_{1}}}\off{\delta(x_1-x_2)\delta(j_1+j_2-1)+ B(j_1)\frac{\delta(j_1-j_2)}{|x_{12}|^{4j_1}}}\label{SL2 bosonic 2point} \, .  
\eeq
These two terms precisely match what we would expect from the product of two asymptotic expressions of the form given in Eq.~\eqref{Vh large phi}. Indeed, for large $k$ we have 
\begin{equation}
\label{def B(j)}
    B(j) = \nu^{1-2j} \frac{2j-1 }{\pi} \frac{\Gamma(1- \frac{2j-1}{k-2})}{
    \Gamma(1+ \frac{2j-1}{k-2}) 
    } 
    \approx \frac{2j-1}{\pi}
\end{equation}
as long as $\nu(k\to \infty) \to 1$, which reproduces the relative coefficient in \eqref{Vh large phi}. The exact value of $\nu$ is not fixed by internal consistency of the WZW model and can be chosen conveniently. In \cite{Teschner:1999ug} this was taken to be 
\begin{equation}
\label{nu in Teschner}
    \nu = (k-2)\gamma\left(\frac{k-1}{k-2}\right) \,, 
\end{equation}
which effectively trivializes $B(j=1)$ for all $k$. 

On the other hand, three-point functions take the form 
\begin{equation}
\braket{V_{j_1}(x_1,z_1)V_{j_2}(x_2,z_2)V_{j_3}(x_3,z_3)} = C(j_1,j_2,j_3) \Bigg|\frac{
    x_{12}^{j_3-j_1 -j_2 }
    x_{23}^{j_1-j_2 -j_3 }
    x_{13}^{j_2-j_1 -j_3 }
    }{
    z_{12}^{\Delta_1+\Delta_2-\Delta_3}
    z_{23}^{\Delta_2+\Delta_3-\Delta_1}
    z_{13}^{\Delta_1+\Delta_3-\Delta_2}}\Bigg|^2\,. 
\end{equation}
The structure constants $C(j_1,j_2,j_3)$ were computed exactly in \cite{Teschner:1997ft}. This was done by combining the constraints coming from null vectors belonging to representations with \cite{Dei:2021yom}
\begin{equation}
\label{degenerate reps SL2}
    j = \frac{1}{2} \left[r + s(k-2)\right] \, \quad \, \text{where} \, \quad 
    \left\{ 
    \begin{array}{c}
        r \leq 0 \, , \, s \leq 0 \, ,\\
        r >2  \, , \, s > 0 \, ,
    \end{array}\right.
\end{equation}
with consistency conditions associated with crossing symmetry. The method closely follows the analogous computation performed in the context of Liouville theory in \cite{Dorn:1994xn,Zamolodchikov:1995aa}. Both CFTs share a number of important features: Liouville theory is the simplest non-compact model with  Virasoro symmetry, while the $H_3^+$ model is the simplest one with a non-abelian Kac-Moody current algebra. As shown in \cite{Ribault:2005wp}, $H_3^+$ correlators can be derived from Liouville correlators. At the path integral level, this can be understood from the fact that,  working with the Wakimoto fields \eqref{Swzw H3 beta quantum}, one can integrate out the $\beta \gamma$-system explicitly, which leads to an effective action Liouville-type action for the non-compact scalar field $\phi$, albeit with a modified background charge \cite{Hikida:2007tq}.  

The computation of $C(j_1,j_2,j_3)$ is beyond the scope of these notes; see \cite{Ribault:2014hia,Eberhardt:2023mrq} for a pedagogical derivation. The final result (known as the DOZZ formula for the Liouville case) takes the form \cite{Maldacena:2001km}
\begin{equation}
\label{C123}
    C(j_1,j_2,j_3) = - \frac{G(1-J) }{
    2 \pi^2  \nu^{J-1} \gamma\left(\frac{k-1}{k-2}\right) G(-1)
    } \prod_{i=1}^3 \frac{G(2j_i-J)}{G(1-2j_i)}, 
\end{equation}
with $J = j_1 + j_2 + j_3$, and where $G(j)$ is defined in terms of the Barnes double Gamma function, closely related to the function $\Upsilon(j)$ originally used in \cite{Zamolodchikov:1995aa}. The following properties of the function $G(j)$ will be especially relevant for us:
\begin{itemize}
    \item For generic values of $k$, $G(j)$ has simple poles at 
    \begin{equation}
    \label{C123 Gj poles}
        j = n + m(k-2) 
        \, \quad \, \text{where} \, \quad 
    \left\{ 
    \begin{array}{c}
        n \geq 0 \, , \, m \geq 0 \, ,\\
        n \leq -1  \, , \, m \leq -1 \, .
    \end{array}\right.
    \end{equation}
    When $k$ is integer some of these become double poles. 

    \item Moreover,  $G(j)$ satisfies the shift identities 
    \begin{equation}
        G(j+1) = \gamma \left(-\frac{j+1}{k-2}\right) G(j) \, , \qquad 
        G(j - (k-2)) = \frac{\gamma(j+1)}{(k-2)^{2j+1}} G(j) \, .
    \end{equation}
\end{itemize}

As a check, we can recover the two-point function \eqref{SL2 bosonic 2point} by taking one of the insertions to be the identity operator. This is done by setting $j_3 = \vep$ and carefully taking the limit $\vep \to 0$. 
The $x$-dependence of the second term is easily obtained. In general, the structure constant vanishes for $j_3 \sim \vep \to 0$ due the divergence  of the factor $G(1-2j_3)$ in the denominator, although for $j_1=j_2$ two terms in the numerator diverge as well. One obtains  
\begin{eqnarray}
    &&\frac{G(2j_1-J)G(2j_2-J)}{G(-1)G(1-2j_3)} = 
    \frac{G(j_{12}-\vep)G(-j_{12}-\vep)}{G(-1)G(1-2\vep)} \nn \\
    &=& 
    \frac{G(-1+j_{12}-\vep)G(-1-j_{12}-\vep)}{G(-1)G(-1-2\vep)} 
    \frac{
    \gamma\left(\frac{\vep + j_{12}}{k-2}\right)
    \gamma\left(\frac{\vep - j_{12}}{k-2}\right)
    }{
    \gamma\left(\frac{2\vep -1}{k-2}\right)
    \gamma\left(\frac{2\vep}{k-2}\right)
    } \nn \\
    &\approx &
    (k-2) 
    \gamma\left(\frac{k -1}{k-2}\right) 
    \frac{2\vep}{\vep^2 -j_{12}^2} \to 
    2\pi (k-2) 
    \gamma\left(\frac{k -1}{k-2}\right) \delta(j_{12}) \,,   
\end{eqnarray}
with $j_{12} = j_1-j_2$. The contact term is a bit more subtle. It can be picked up by noting that $x_{12}^{j_3-j_1-j_2}$ is meromorphic in $j_3$ \textit{except} near $j_3 = j_1 + j_2 -1$ when $x_{1}=x_2$. This translates into the distributional identity  \cite{Teschner:1999ug,Gelfand1964} 
\begin{equation}
\label{pole x12 contact term}
 |x_{12}|^{2(\vep-j_1-j_2)} \sim \frac{\pi \delta (x_{12}) }{1-j_1-j_2+\vep} \,, 
\end{equation}
which, in combination with the structure constant (which is regular), leads to the first term in \eqref{SL2 bosonic 2point}. 

\subsection{OPE structure and four-point functions}

From the three-point function one can derive the following (primary) OPE between vertex operators in continuous representations: 
\begin{eqnarray}
\label{OPE Teschner cont}
    && V_{j_1}(x_1,z_1)V_{j_2}(x_2,z_2)   \\
    && = \int_{\frac{1}{2}+i\R_+}  \frac{dj_3}{|z_{12}|^{2(\Delta_1+\Delta_2-\Delta_3)}} \int_{\mathbb{C}} d^2x_3  \frac{ C(j_1,j_2,j_3)}{
    |x_{12}^{j_1+j_2-j_3} x_{23}^{j_2+j_3-j_1} x_{31}^{j_3+j_1-j_2}|^2} V_{1-j_3}(x_3,z_3) \,. \nn
\end{eqnarray}
The integrand is actually symmetric under $j_3 \to 1-j_3$, as follows from Eqs.~\eqref{ReflectionUnflowed} and \eqref{integral Maldacena AppC}, which allows one to extend the integration region to the full line $j_3 = \frac{1}{2}+i \R$. Strictly speaking, Eq.~\eqref{OPE Teschner cont} is valid only for spins satisfying \cite{Teschner:1999ug}  
\begin{equation}
\label{Teschner OPE range}
|\re \left(j_1-j_2 \right)| < \frac{1}{2} \,  \quad \text{and } \quad     
|\re \left(j_1+j_2-1\right)| < \frac{1}{2} \, .
\end{equation}
Nevertheless, one can use it compute the OPE for real values of $j_1$ and $j_2$ by analytic continuation. One must take into account, however, that the integrand has many poles in $j_3$, and, as we move $j_1$ and $j_2$ in the complex plane, some of these poles might cross the integration contour. When this happens, the OPE picks up a discrete set of additional contributions from the corresponding residues. Let us stress that these poles come not only from the structure constants but also from the powers of $x_{ij}$, in a similar fashion as what we saw above. 

In special cases, more precisely when $j_1$ and/or $j_2$ belong to the set of degenerate representations in Eq.~\eqref{degenerate reps SL2}, these manipulations can lead to situations where the contribution from the integral vanishes exactly. In such situations the OPE becomes a sum over a finite set of contributions coming from the relevant residues. As discussed in \cite{Ribault:2014hia}, this allows one to derive the structure constants and operator product expansions for related models with finite, discrete spectra, such as minimal models and the SU(2) WZW model, see Eq.~\eqref{3pt-SU2-zudep}.  

The full set of contributions to the $V_{j_1}V_{j_2}$ OPE is described in detail in \cite{Teschner:1999ug}. For reasons that will become clear shortly, a particularly important case to consider is that with $j_1=1$ and $j_2=j$ in the physical range for the discrete series in AdS$_3$, see Eq.~\eqref{Djrange}. In the limit $z_{12} \to 0$, the leading term is regular and comes from the pole at $j_3 = j_1+j_2-1$, i.e.~exactly that of Eq.~\eqref{pole x12 contact term}. This pole sits at $\re j_3 = 0$ when $\re j_1 = \re j_2 = \frac{1}{2}$, but for $j_1 = 1$ and $j_2 = j > \frac{1}{2}$ it moves to $\re j_3 > \frac{1}{2}$, thus crossing the integration line in the process. This leads to 
\begin{equation}
\label{OPE V1 Vj}
     V_{1}(x_1,z_1)V_{j}(x_2,z_2) \sim \int_{\mathbb{C}} d^2x_3 \, \delta(x_{12}) |x_{23}|^{2(2j+1)}
     |x_{31}|^{-2} V_{1-j}(x_3,z_2) = 
     \delta(x_{12}) V_{j}(x_2,z_2)\, ,
\end{equation}
where we have used  \eqref{ReflectionUnflowed}. (The fact that we have a unit coefficient on the RHS follows from the choice in Eq.~\eqref{nu in Teschner}.)  This OPE can be checked by considering the semi-classical expressions \eqref{Vh large phi} near the boundary : 
\begin{equation}
    V_{1}(x_1) V_j(x_2) \sim \delta(\gamma - x_1) 
    e^{2(j-1)\phi} \delta(\gamma - x_2) \sim  \delta(x_{12}) V_j(x_2). 
\end{equation}
This will be important in our discussion of the spacetime symmetries. 

Finally, we consider the four-point functions of the $H_3^+$ WZW model. As usual, global conformal invariance implies
\begin{equation}
    \langle 
    V_{j_1}(x_1,z_1)
    V_{j_2}(x_2,z_2)
    V_{j_3}(x_3,z_3)
    V_{j_4}(x_4,z_4)
    \rangle = |Z[z_i,\Delta_i]
    X[x_i,j_i]\Ff(x,z)|^2\, ,
\end{equation}
where 
\begin{eqnarray}
    Z[z_i,\Delta_i]
    &=& 
    z_{21}^{-\Delta_1-\Delta_2
    +\Delta_3-\Delta_4}
    z_{31}^{-\Delta_1+\Delta_2- \Delta_3+\Delta_4}
    z_{32}^{\Delta_1-\Delta_2-\Delta_3+\Delta_4}
    z_{34}^{-2\Delta_4} \,, \\
    X[x_i,j_i]
    &=& 
    x_{21}^{-j_1-j_2+j_3-j_4}
    x_{31}^{-j_1+j_2-j_3+j_4}
    x_{32}^{j_1-j_2+j_3-j_4}
    x_{34}^{-2j_4} \,, 
    \end{eqnarray}
while the worldsheet and spacetime cross-ratios are defined as 
\begin{equation}
    z = \frac{z_{32} z_{14}}{z_{12}z_{34}} \, , \qquad
    x = \frac{x_{32} x_{14}}{x_{12}x_{34}} \, .
\end{equation}
The function $\Ff(x,z)$ is the holomorphic conformal block. It is constrained by the KZ equation \eqref{KZequation-WZW}, which in the $x$-basis mixes differential operators in the variables $z$ and $x$. More explictely, this reads \cite{Teschner:1999ug}
\begin{equation}
\label{KZ equation SL2}
    \der_z \Ff = \frac{1}{k-2} \left(\frac{P_x}{z} + \frac{Q_x}{z-1}\right) \,,
\end{equation}
with 
\begin{equation}
P_x = x^2(x-1)\der_x^2 -  \left[(\kappa-1)x + 2j_1 - 2j_4 (x-1)\right]x\der_x - 2 \kappa j_4 x - 2 j_1 j_4
\end{equation}
and 
\begin{equation}
Q_x = -x(x-1)^2\der_x^2 + \left[(\kappa-1)(x-1) - 2j_2  - 2j_4x \right] (x-1)\der_x 
+ 2 \kappa j_4(x- 1) - 2 j_2 j_4, 
\end{equation}
where $\kappa = j_3 - j_1 -j_2 -j_4$. The solution is not known in closed form for generic SL(2,$\R$) spins. Nevertheless, by inserting the OPE \eqref{OPE Teschner cont} one obtains a series expansion in powers of $z$ related to the factorization of the four-point function in the chanel $14 \to 23$. More precisely, one has 
\begin{equation}
    |\Ff(x,z)| = \int_{\frac{1}{2}+i \R} dj \frac{C(j_1,j_4,j)C(j,j_2,j_3)}{B(j)} |\Ff_j(x,z)|^2
    \label{factorization unflowed}
\end{equation}
where 
\begin{equation}
    \Ff_j(x,z) = z^{\Delta_j - \Delta_1 - \Delta_4} 
    x^{j-j_1-j_4} \sum_{n=0}^\infty f_{j,n}(x) z^n\, . 
\end{equation}
The KZ equation then implies that the zero-th order solution corresponds to the hypergeometric function 
\begin{equation}
f_{j,0}(x) = 
{}_2F_1 (j-j_1+j_2,
j+j_3-j_4,2j,x) \,,  
\end{equation}
suitably combined with a second solution corresponding to the exchange $j \to 1-j$ in order to give a monodromy-invariant result \cite{Maldacena:2001km}. The rest of the functions $f_{j,n}(x)$ can in principle be determined iteratively from the KZ equation as well. 

As it comes from the OPE \eqref{OPE Teschner cont}, this expansion is strictly speaking only valid for external spins in the range \eqref{Teschner OPE range} (replacing $j_2 \to j_4$), while a similar restriction holds for $j_{2}$ and $j_3$. 
For other values of $j_i$ one must proceed by analytic continuation. This is necessary in order to compute, for instance, the four-point  function for  unflowed states in the discrete representations of AdS$_3$, namely the supergravity modes. As discussed for the OPE expansion, the procedure is much more interesting than it sounds. It was famously shown in \cite{Maldacena:2001km} by studying the factorization limit of such four-point functions that the $s$-channel includes contributions associated with the propagation of long string states with $\w=1$. The precise expressions are consistent with the three-point functions which violate spectral flow conservation, to be discussed in Sec.~\ref{sec: bosonic correlators} below. This shows, once again, that the AdS$_3$ model would be inconsistent without the inclusion of spectrally flowed states.    

\subsection{Spacetime symmetry generators}

Let us now go back to string theory on AdS$_3\times M_{\rm int}$. A generic physical vertex operator takes the form 
\begin{equation}
\Phi = P(J,\dots) 
\bar{P}(\bar{J},\dots) V_j(x,\xb,z,\zb) V_{\rm int} (z,\zb)\,,
\end{equation}
where $P$ and $\bar{P}$ are polynomials with overall scaling degree $N$ and $\bar{N}$, such that the physical state conditions  read 
\begin{equation}
\label{VirasoroWS}
    -\frac{j(j-1)}{k-2} + h_{\rm int} -1 + N = 
    -\frac{j(j-1)}{k-2} + \bar{h}_{\rm int} -1 + \bar{N} = 0.  
\end{equation}
At the very least, we should be able to build the spacetime  Virasoro and Kac-Moody generators without relying on the free-field approximation. Moreover, the construction should be independent of the details of the internal geometry \cite{Kutasov:1999xu}. 

As it turns out, the key ingredient is the SL(2,$\R$) primary vertex operator with $j=1$. This is an $(h,\hb) = (1,1)$ tensor in spacetime, which has  vanishing worldsheet weight. Although $V_1$ is not the worldsheet identity, i.e.~the operator with $j=0$, it is related to it through the analytic continuation of the $H_3^+$ reflection formula \eqref{ReflectionUnflowed} away from $j \in \frac{1}{2}+i \R$, namely  $
    V_{0}(x,z) \sim  \int d^{2}x' \, V_{1}(x',z).$ 
Using the operator $V_1$ allows us to solve the Virasoro conditions \eqref{VirasoroWS} simply by setting $h_{\rm int} = \bar{h}_{\rm int} = 0$ and $N = \bar{N} = 1$. The proposal for the worldsheet avatars of the local operators generating the (holomorphic) spacetime symmetry algebra are then as follows. For  the operator $\Ii$ and the Kac-Moody currents ${\cal{K}}^a$ we have 
\begin{equation}
\label{def I(x)}
    \Ii = \frac{1}{(k-2)^2}\int d^2z \, \left(J\Jb V_1 \right) (x,\xb,z,\zb)  \, , 
\end{equation}
where $\left(J\Jb V_1 \right)$ stands for the normal-ordered product $J^-(x,z) \Jb^-(\xb,\zb) V_1(x,\xb,z,\zb)$,  and 
\begin{equation}
\label{def K(x)}
    \Kk^a(x) = - \frac{1}{k-2}\int d^2z \, K^a(z) \left(\Jb V_1 \right)(x,\xb,z,\zb) \,,
\end{equation}
respectively, while the spacetime energy-momentum tensor is written as 
\begin{equation}
\label{def T(x)}
    \Tt(x) = \frac{1}{2(k-2)}\int d^2z \, \left(\left[
    (\der_x J)\der_x + 2 (\der_x^2 J)
    \right] \Jb V_1 \right) (x,\xb,z,\zb) 
    \,. 
\end{equation}
The SL(2,$\R$) currents are inserted so that $(J\Jb V_1)$ and $(\Jb V_1)$ transform as operators of spin (0,0) and (1,0) under the action of the left-handed and right-handed zero-mode algebras.  We now show that these operators satisfy the expected OPEs in spacetime, namely
\begin{subequations}
\label{KT st OPE1}
    \begin{eqnarray}
     \Kk^a(x) \Kk^b(y) &\sim& \frac{\delta^{ab} k' \Ii}{2 (x-y)^2} + \frac{i f^{ab}_{\phantom{ab}c} K^c(y)}{x-y} \,, \\
     \Tt(x) \Tt(y) &\sim &
     \frac{c_{\rm st}^{(1)} \, \Ii}{2(x-y)^4} + \frac{2 T(y)}{(x-y)^2} + 
     \frac{\der_y T(y)}{x-y}, 
    \end{eqnarray}
\end{subequations}
with $c_{\rm st}^{(1)} = 6k$ and $\Ii$ acting as the spacetime identity up to a constant, at least when working in the unflowed sector of the theory, and 
\begin{subequations}
\label{KT st OPE2}
    \begin{eqnarray}
     \Kk^a(x) \Vv_h(y,\yb) &\sim& \frac{ t^a \Vv_h(y,\yb)}{(x-y)}\,, 
     \label{KT st OPE2 a}\\
     \Tt(x) \Vv_h(y,\yb) &\sim&  \frac{ h\Vv_h(y,\yb)}{(x-y)^2}
     + \frac{ \der_y\Vv_h(y,\yb)}{(x-y)} \, . 
    \end{eqnarray}
\end{subequations}
Here we have introduced the vertex operators of the boundary theory, obtained by integrating the worldsheet ones, i.e.
\begin{equation}
    \Vv_j(x,\xb) \equiv \int d^2z \, V_j (x,\xb,z,\zb) V_{\rm int} (z,\zb)\,,
\end{equation} 
and set $h=\hb=j$.

The main technical obstacle is that the definitions in Eqs.~\eqref{def K(x)} and \eqref{def T(x)} are given in terms of integrals in $z$ over the full complex plane, as opposed to contour integrals. In order to bypass this issue, we use the following trick. Since we are only interested in the singular terms, the relevant information is still available upon applying $\der_{\xb}$ on both sides of all OPEs in Eqs.~\eqref{KT st OPE1} and \eqref{KT st OPE2}, and using the distributional identity 
\begin{equation}
    \der_{\xb} \left[(x-y)^{-1}\right] = \pi \delta (x-y) \,.  
\end{equation}
Indeed, since $\Kk^a(x)$ and $\Tt(x)$ must be holomorphic only the contact terms should survive.  
On the worldsheet side, we use the 
the following identity for $V_1$:  
\begin{equation}
    \der_{\xb} \left(\bar{J}^-V_1\right) 
    (x,\xb,z,\zb)= (k-2) \, \der_{\zb} V_1(x,\xb,z,\zb)\,, 
\end{equation}
which is a direct consequence of the Sugawara construction, see Appendix \ref{sec: appB WZW models}. This leads to 
\begin{equation}
\label{def K(x) contour}
    \der_{\xb}\Kk^a(x) = \oint_{\Cc} dz \, K^a(z)  V_1(x,\xb,z,\zb) \,, 
\end{equation}
and 
\begin{equation}
   \der_{\xb} \Tt(x) = \frac{1}{2} \oint_{\Cc} dz \, \left[
    (\der_x J^-)\der_x + 2 (\der_x^2 J^-)
    \right] V_1(x,\xb,z,\zb) \,, 
\end{equation}
which hold inside generic correlation functions. More precisely, upon applying $\der_{\xb}$ on the original definitions the integrands become total derivatives \textit{up to} contact terms from the (derivatives of the) OPEs with other vertex operators a given correlator, hence the integration contour $\Cc$ should be understood as a series of small circles surrounding those additional insertions. In other words, we get for example 
\begin{equation}
\label{WS ope Kax V}
    \langle \der_{\xb}\Kk^a(x) \Vv_j(y,\yb) \dots\rangle \sim \int d^2w \oint_w dz \langle  V_1(x,\xb,z,\zb) V_{j}(y,\yb,w,\wb) 
     K^a(z) V_{\rm int}(w,\wb)  \dots \rangle + \cdots \,,
\end{equation}
where we have focused on the contribution from a small circle around one of the additional insertions. A key property of the operator $V_1$ is the OPE obtained in Eq.~\eqref{OPE V1 Vj}. 
It implies that \eqref{WS ope Kax V} becomes  
\begin{equation}
\label{WS ope Kax V 2}
    \langle \der_{\xb}\Kk^a(x) \Vv(y,\yb) \dots\rangle \sim \pi \delta(x-y) t^a \langle \Vv(y,\yb) \dots  \rangle + \cdots \,,
\end{equation} 
where we have used that $K^a(z)V_{\rm int}(w,\bar{w}) \sim (z-w)^{-1}t^aV_{\rm int}(w,\bar{w})$. This confirms \eqref{KT st OPE2 a}. Similarly, when there are additional current insertions we get contributions of the form  
\begin{equation}
\label{WS ope KKax 1}
    \langle \der_{\xb}\Kk^a(x) \Kk^b(y) \dots\rangle \sim \int d^2w \oint_w dz \langle  K^a(w) V_1(x,\xb,z,\zb) 
    K^b(z) \left(\bar{J}V_1\right)(y,\yb,w,\wb)  \dots \rangle + \cdots \,. 
\end{equation}
This contains single- and double-pole contributions from the OPE of the worldsheet currents. Proceeding as before and using 
\begin{eqnarray}
   (k-2) \lim_{z\to w} \der_z V_1(x,\xb,z,\zb) 
 \left(\bar{J}V_1\right)(y,\yb,w,\wb) &=& \lim_{z\to w} \der_x \left(JV_1\right)(x,\xb,z,\zb) 
 \left(\bar{J}V_1\right)(y,\yb,w,\wb) \nn \\
 &=& \der_x \delta(x-y) (J\bar{J} V_1)(y,\yb,w,\wb) 
\end{eqnarray}
for the double pole residue, we get
\begin{equation}
    \langle \der_{\xb}\Kk^a(x) \Kk^b(y) \dots\rangle \sim \pi  \langle \left[\delta(x-y) i f^{ab}_{\phantom{ab}c} \Kk^c(y) + \der_x \delta(x-y) (k'/2) \delta^{ab} \Ii \right]\dots  \rangle \,,
\end{equation}
as expected from Eq.~\eqref{KT st OPE1}. A similar set of computations involving the proposed definition for $\Tt(x)$ confirms the rest of the spacetime symmetry algebra. For a more detailed study of the properties of $\Ii$, see \cite{Kutasov:1999xu,Porrati:2015eha}.

As a sanity check, we now make contact with the free field description. At large $\phi$, we combine the currents of Eq.~\eqref{Ja classical} into 
\begin{equation}
\Jb (\xb,\zb) \approx k \left[
(\gammab-\xb)^2 e^{2\phi} \derb \gamma 
-2 (\gammab-\xb) \derb \phi - \derb \gammab
\right]\,. 
\end{equation}
Starting from \eqref{def K(x)} and using the leading semi-classical expression for $V_1$ near the boundary, $V_1(x)\approx \delta(x-\gamma)$, combined with the fact that $\gamma$ must be holomorphic in this regime, we obtain 
\begin{equation}
    \Kk^a(x) = \int d^2z \, K^a(z) \derb \gammab \, \delta(x-\gamma) = \int d^2z \, K^a(z) \derb \left(\frac{1}{x-\gamma}\right) =  \oint dz \, 
    \frac{K^a(z)}{x-\gamma}\,,
\end{equation}
so that, upon formally expanding in powers of $x$ we find 
\begin{equation}
    \Kk^a(x) = \sum_n x^{-n-1}\oint dz 
    \, K^a(z) \gamma^n \, , 
\end{equation}
thus reproducing \eqref{def K(x) modes Wakimoto}. Similarly, for the operator $\Ii$ we have  
\begin{equation}
\label{I def leading}
    \Ii = \int d^2z \, \der \gamma \derb \gammab \, \delta(x-\gamma) = 
    \oint dz \, 
    \frac{\der \gamma}{x-\gamma} = 
    \sum_n x^{-n} 
    \oint dz \, 
    \gamma^{n-1} \der \gamma = \oint 
    dz \, \gamma^{-1}\der \gamma \, .
\end{equation}
Finally, for the case of $\Tt(x)$ we have 
\begin{equation}
    \Tt(x) = \frac{1}{2}\int d^2 z 
    \, 
    \left[(\der_x J^-)\der_x  + 2 (\der_x^2 J^-)\right]
    \derb \left(\frac{1}{x-\gamma}\right)
    = 
    \oint dz \, 
\left[\frac{J^3(x,z)}{(x-\gamma)^2} 
    + \frac{2 J^+(z)}{x-\gamma}\right] \, , 
\end{equation}
where in the last step we have used that $x \delta'(x) = -\delta(x)$. This leads to 
\begin{equation}
    \Tt(x) = \sum_n x^{-n-2} \oint dz\,  \left[(n+1) \gamma^n J^3(z) - n \gamma^{n+1} J^+(z)\right] \, ,
\end{equation}
in accordance with Eq.~\eqref{Ln Wakimoto}. 

Although there are many other important aspects one could include, we choose to end our discussion of the unflowed sector of the theory at this point. We now move to the main subject of these notes: correlation functions involving spectrally flowed operators.  

%%%%%%%%%%%%%%%%%%%%%%%%%%%%%
%%%%%%%%%%%%%%%%%%%%%%%%%%%%%
\newpage
%%%%%%%%%%%%%%%%%%%%%%%%%%%%%%%

\section{Correlation functions and spectral flow}
\label{sec: bosonic correlators}

In this section we study correlators with spectrally flowed insertions. We start by considering a small number of  particular cases, which can be computed by using $m$-basis methods, and discuss the corresponding the spectral flow violation rules \cite{Fateev,Maldacena:2001km,Cagnacci:2013ufa,McElgin:2015eho}. We then describe the constraints imposed by the so-called local Ward identities, which ultimately imply a set of recursion relations between primary correlators. Upon introducing the $y$-transform, the analogue of the $x$-basis for spectrally flowed operators \cite{Dei:2021xgh,Dei:2021yom}, we show how one can recast these recursion relations in terms of differential equations. The latter can be solved in full generality for three-point functions \cite{Iguri:2022eat,Bufalini:2022toj}. Finally, we briefly discuss the application of similar techniques to four-point functions \cite{Dei:2021yom,Dei:2022pkr,Iguri:2024yhb,Barone:2025vww}.

\subsection{$m$-basis flowed correlators and parafermions}

The SL(2,$\R$) WZW model is believed to be exactly solvable. By this, we mean that it should be possible to compute all the relevant CFT data, namely the spectrum, the OPE coefficients (or equivalently the three-point functions), and the dynamics as captured by the four-point functions, which ought to be  crossing symmetric. One of the main interests of these notes is to describe the state-of-the-art in what has, for a long time, delayed the possibility of saying it  \textit{is} actually solved: the computation of correlation functions involving spectrally flowed insertions. 

For pedagogical reasons, we choose to follow the historical timeline, and first discuss the original computations of \cite{Fateev,Maldacena:2001km}, see also \cite{Giribet:2001ft,Hofman:2004ny,Iguri:2007af,Giribet:2011xf,Cagnacci:2013ufa}. We thus focus on the first non-trivial cases for two- and three-point functions.
The main tool comes from what is known as the parafermionic decomposition, where one factorizes the SL(2,$\R$) fields into an operator belonging to the SL(2,$\R$)/U(1) coset model, and an exponential operator associated to the remaining U(1) factor. The former coset model is usually called the (semi-infinite) \textit{cigar} theory due to shape of the resulting geometry.
It actually corresponds to the Euclidean version of the two-dimensional black-hole, see \cite{Witten:1991yr}.
At the level of unflowed vertex operators, this decomposition reads 
\begin{equation}
V_{jm\mb} = \Psi_{jm} e^{m \sqrt{\frac{2}{k}} \varphi} \,,  
\end{equation}
where, as usual, we have omitted the anti-holomorphic components and labels. Here $\varphi$ is a canonically normalized free (holomorphic) scalar, which bosonizes the Cartan current, i.e.
\begin{equation}
    J^3(z) = - \sqrt{\frac{k}{2}} \der \varphi(z), \qquad \varphi(z)\varphi(w) \sim - \log(z-w),  
\end{equation}
while $\Psi_{jm\mb}$ is a coset operator satisfying
\begin{equation}
\label{parafermion weight}
    \Delta[\Psi_{jm\mb}] = -\frac{j(j-1)}{k-2} + \frac{m^2}{k} \, . 
\end{equation}
This is the parafermion operator, whose OPEs with the currents contain fractional powers of $z$ \cite{Fateev:1985mm}. 
In geometric terms, one can interpret the quantum numbers of $\Psi_{jm\mb}$ as follows. The momentum along the asymptotic circle (which has radius $\sqrt{k}$) is given by $m-\mb \in \mathbb{Z}$, while the  winding corresponds to $m+\mb \in k \mathbb{Z}$. The important point is that, while the former is conserved, there is no reason to expect that winding is conserved as well since the circle is contractible at the tip of the cigar \cite{Fateev}.  
Similarly, the currents factorize as 
\begin{equation}
J^\pm= \sqrt{k} \,  \psi^\pm e^{\pm \sqrt{\frac{k}{2}}\varphi}\,,  
\end{equation}
since they have charge $\pm 1$, and where $\Delta[\psi^\pm] = 1+\frac{1}{k}$.  

This coset decomposition is quite useful for describing spectrally flowed ($m$-basis) primary states. Indeed, one can easily check that the expression 
\begin{equation}
    V_{jm}^\w(z) = \Psi_{jm}(z) e^{\left(m + \frac{k}{2}\w \right) \sqrt{\frac{2}{k}} \, \varphi(z)} \, 
    \label{VwMbasisSL2}
\end{equation}
satisfies all the defining properties of the flowed operator, namely its worldsheet weight and OPEs with the SL(2,$\R$) currents. Importantly, the parafermionic sector remains untouched. 
Let us first use this to compute the $x$-basis two-point functions with spectral flow. As usual, the global Ward identities imply that\footnote{In principle, the spacetime Ward identities allow for a contact term with $h_2 = 1-h_2$, but this clashes with the worldsheet Ward identities \cite{Dei:2021xgh}.} 
\begin{equation}
    \langle V_{j_1h_1}^{\w_1} (x_1,z_1) V_{j_2h_2}^{\w_2} (x_2,z_2) \rangle = 
    \frac{\delta(h_1-h_2) D(h_i,j_i,\w_i) }{|x_{12}|^{4h_1}|z_{12}|^{4\Delta_1}} \,, 
\end{equation}
and can only be non-vanishing for either $j_2 = j_1$ or $j_2 = 1-j_1$. Our goal is to compute the structure constant $D(h_i,j_i,\w_i)$. For this, we can take $x_1 = z_1 = 0$ and $x_2=z_2 = \infty$, in which case the $x$-basis flowed operators become $m$-basis flowed primaries. More explicitly, we have 
\begin{eqnarray}
    && \delta(h_1-h_2) D(h_1,j_i,\w_i) = \langle V_{j_1h_1}^{\w_1} (0,0) V_{j_2h_2}^{\w_2} (\infty,\infty) \rangle = \langle V_{j_1m_1}^{\w_1}(0) V_{j_2,-m_2}^{-\w_2}(\infty) \rangle \nn \\
    && \qquad = \langle \Psi_{j_1 m_1}(0) \Psi_{j_2,-m_2}(\infty)\rangle \langle e^{\left(m_1 + \frac{k}{2}\w_1 \right) \sqrt{\frac{2}{k}} \, \varphi(0)} e^{-\left(m_2  + \frac{k}{2}\w_2 \right) \sqrt{\frac{2}{k}} \, \varphi(\infty)} \rangle \,,
\end{eqnarray}
with $m_i = h_i - \frac{k}{2}\w_i$, and where we have made use of \eqref{VwMbasisSL2}. The point is that the parafermionic correlator is \textit{the same} as the one appearing in the unflowed two-point function, which in particular imposes $m_1 = m_2$. Since the $z$-dependence has been stripped off, the only role of the free-field correlator is to impose the additional charge conservation condition $\w_1 = \w_2$, consistent with the requirement $h_1 = h_2$. Had we not set the $z$-dependence beforehand, the free-field factors would have provided the additional terms in the weights as compared to the unflowed operators, namely  $\Delta_i - \Delta_i(\w_i=0) =  - m_i \w_i - \frac{k}{4} \w_i^2$. Consequently, using the Mellin transform \eqref{Mellin}, we can express the unflowed two-point function \eqref{SL2 bosonic 2point} in the $m$-basis, and read off the desired result  
\cite{Maldacena:2001km}
\begin{eqnarray}
    && \langle V_{j_1 h_1}^{\w_1} (x_1,z_1) V_{j_2 h_2}^{\w_2} (x_2,z_2) \rangle = \nn \\
    &&\qquad \frac{\delta_{\w_1,\w_2}\delta(h_1-h_2)}{|z_{12}|^{4\Delta_1}|x_{12}|^{4h_1}}\left[\delta (j_1+j_2-1) +  \frac{\pi \delta(j_1-j_2) B(j_1) \gamma\left(j_1+h_1-\frac{k}{2}\w_1\right)}{\gamma(2j_1) \gamma\left(1-j_1+h_1-\frac{k}{2}\w_1\right)}\right].
    \label{SL2 2point bosonic with flow}
\end{eqnarray}
The computation of three-point functions which conserve spectral flow, i.e.~those for which $\w_1 + \w_2 = \w_3$, follows analogously, although when all three $\w_i$ are non-zero charge the method imposes an additional constraint on the $j_i$. The explicit  result, which can be constructed from the unflowed $m$-basis three-point functions obtained in \cite{Satoh:2001bi} in terms of hypergeometric functions ${}_3F_2$, is not particularly illuminating.  

Are we done? If the spectral flow charge was a conserved quantity, the answer would be yes. However, this need not be the case. Consider the simplest possible example, namely the three-point function with $\boldsymbol{\w} = (\w_1,\w_2,\w_3)= (1,0,0)$:  
\begin{equation}
\label{3pt 100 a}
\langle V_{j_1 h_1}^1 (0,0) V_{j_2} (1,1) V_{j_3} (\infty,\infty) \rangle = \langle V_{j_1m_1}^{1}(0) V_{j_2}(1,1) V_{j_3,-j_3}(\infty) \rangle\,. 
\end{equation}
Here the vertex operator in the middle is still in the $x$-basis. Indeed, when inserted at a generic point, say $x_2=1$, this is a linear combination of an infinite number of $m$-basis operators, see Eq.~\eqref{ExpSL2Cont}. Charge conservation implies that only a mode with $m_2 = m_3 - m_1 - \frac{k}{2}$ can contribute, giving  
\begin{equation}   
\eqref{3pt 100 a}=\langle \Psi_{j_1 m_1}(0) \Psi_{j_2,m_2}(1) \Psi_{j_3,-m_3}(\infty)\rangle \langle e^{\left(m_1 + \frac{k}{2}\right) \sqrt{\frac{2}{k}} \, \varphi(0)} e^{m_2 \sqrt{\frac{2}{k}} \, \varphi(1)} 
    e^{-m_3 \sqrt{\frac{2}{k}} \, \varphi(\infty)}\rangle \,.
    \nn
\end{equation}
An analogous charge conservation condition holds in the anti-holomorphic sector, which we have omitted. Hence, the parafermion correlator is unlike any appearing in the unflowed sector: momentum is conserved, as it should, but the total winding is not (by one unit). As argued above, we do not expect this to vanish.

In order to compute \eqref{3pt 100 a}, we take advantage of the fact that, as can be infered from the parafermionic decomposition, the state with $\w=1$ can be constructed explicitly from the fusion of unflowed operators \cite{Maldacena:2001km}. More explicitly, it follows from Eq.~\eqref{VwMbasisSL2} that
\begin{equation}
   V_{jm}^{1}(z) = \lim_{\vep\to 0}\vep^{m}V_{jm}(z+\vep)V_{\frac{k}{2}  \frac{k}{2}}(z)\,. \label{OPEw1}
\end{equation}
For this reason,  
\begin{equation}
    V_{\frac{k}{2}\frac{k}{2}}(z) =  e^{\sqrt{\frac{k}{2}} \varphi} 
\end{equation}
is known as the \textit{spectral flow operator}. Even though it is outside of the physical range \eqref{Djrange}, this is a very interesting operator for the following  reasons: 
\begin{itemize}
\item its parafermionic component $\Psi_{\frac{k}{2}\frac{k}{2}}$ is simply the identity, as can be seen from Eq.~\eqref{parafermion weight}, 

\item it has a null descendent at level 1, namely 
\begin{equation}
\label{nullstate 0 SL2}
    J_{-1}^- \Big|\frac{k}{2},\frac{k}{2}\Big\rangle =0
    \,,
\end{equation}
as discussed around Eq.~\eqref{Null state SL2}, and 

\item the worldsheet Virasoro mode $L_{-1}$ acts very simply on it,  
\begin{equation}
\label{KZ for null state SL2}
   L_{-1} \Big|\frac{k}{2},\frac{k}{2}\Big\rangle = -J_{-1}^3 \Big|\frac{k}{2},\frac{k}{2}\Big\rangle.
\end{equation}
\end{itemize}

Let us stress that this approach can  be extended to larger values of  $\w$ \cite{Iguri:2022eat}. In the $x$-basis, the generalization of Eq.~\eqref{OPEw1} to arbitrary $\w>0$ reads
\begin{equation}
    V_{jh}^{\w}(x,z) = \lim_{\vep,\bar{\vep}\to 0}  \vep^{m \w}
    \bar{\vep}^{\bar{m} \w} \int d^2y \,  y^{j-m-1} \bar{y}^{j-\bar{m}-1} V_{j}(x+y,z+\vep) V_{\frac{k}{2}\, \frac{k}{2}\w}^{\w-1} (x,z),  
    \label{proposalWxbasis}
\end{equation}
where $V_{\frac{k}{2}\, \frac{k}{2}\w}^{\w-1} (x,z) =  e^{\w\sqrt{\frac{k}{2}} \varphi} $
To understand how this works, we can for instance evaluate it at $x=0$, giving 
\begin{eqnarray}
    V_{jh}^{\w}(0,z) &=& \lim_{\vep,\bar{\vep}\to 0}  \vep^{m \w}
    \bar{\vep}^{\bar{m} \w} \int d^2y \,  y^{j-m-1} \bar{y}^{j-\bar{m}-1} V_{j}(y,z+\vep) V_{\frac{k}{2},\frac{k}{2}\w}^{\w-1} (0,z) \nn \\
    &=& \lim_{\vep,\bar{\vep}\to 0}  \vep^{m \w}
    \bar{\vep}^{\bar{m} \w}  V_{jm}(z+\vep) V_{\frac{k}{2}\frac{k}{2}}^{\w-1} (z) = V_{jm}^\w (z) \, ,
\end{eqnarray}
as expected. Another simple check comes from studying the action of the zero-mode currents on \eqref{proposalWxbasis}. For $J_0^+$, this gives  $\der_y$ when acting on the spectrally unflowed vertex $V_{j}(x+y,z+\vep)$ in the integrand, while on the spectral flow operator it acts as  $\der_x-\der_y$, such that the total action is characterized by $\der_x$. The action of $J_0^3(x) \equiv J_0^3- x J_0^+$, is slightly more interesting to derive. Using $J_0^3(x) = J_0^3(x+y) + y J_0^+$, we get 
\begin{eqnarray}
   \left( J_0^3 V_{jh}^{\w}\right)(x,z) &= & \lim_{\vep,\bar{\vep}\to 0}  \vep^{m \w} \bar{\vep}^{\bar{m} \w}
      \int d^2y \,  y^{j-m-1} \bar{y}^{j-\bar{m}-1} \times \nn \\
      && \qquad 
      \left(y \der_y + j +  \frac{k}{2}\w\right)V_{j}(x+y,z+\vep) V_{\frac{k}{2},\frac{k}{2}\w}^{\w-1} (x,z) \nn \\
      &=& \left(m + \frac{k}{2}\w \right)V_{jh}^{\w}(x,z) = hV_{jh}^{\w}(x,z),
\end{eqnarray}
consistent with \eqref{flowedOPESL2}. A similar computation can be performed for  $J_0^-(x)$. 

Coming back to the particular case of $\w=1$, we can now relate the spectral flow violating three-point function \eqref{3pt 100 a} to an unflowed four-point function. Moreover, this is a special four-point function, which can be derived exactly due to the fact that one of the insertions, namely the spectral flow operator $V_{\frac{k}{2}}(x)$, has a null descendant. This provides a differential constraint which, when combined with the usual KZ equation, completely fixes the dependence on the spacetime and worldsheet cross-ratios. More explicitly, we consider 
\begin{align}
    \begin{aligned}
    &\langle V_{j_1}(x_1,z_1)  
    V_{j_2}(x_2,z_2)
    V_{j_3}(x_3,z_3) 
    V_{\frac{k}{2}}(x_4,z_4)
    \rangle =  \\ 
    &\tilde{C}(j_1,j_2,j_3) |{\cal{F}}(x,z)|^2 \left|\frac{
    x_{32}^{(j_1+j_4-j_2-j_3)}
    x_{34}^{-2j_4}
    x_{31}^{(j_4+j_2-j_1-j_3)}
    x_{21}^{(j_3-j_1-j_4-j_2)}}{
    z_{32}^{(\Delta_2+\Delta_3-\Delta_1-\Delta_4)}
    z_{34}^{2\Delta_4}
    z_{31}^{(\Delta_1+\Delta_3-\Delta_4-\Delta_2)}
    z_{21}^{(\Delta_1+\Delta_4+\Delta_2-\Delta_3)}}\right|^2\,,
    \label{4ptconserving}
    \end{aligned}
\end{align}
where 
\begin{equation}
    z = \frac{z_{32}z_{14}}{z_{12}z_{34}} \, , \quad 
    x = \frac{x_{32}x_{14}}{x_{12}x_{34}} \, , \quad 
    \Delta_4 = -\frac{k}{4} \, , 
    \quad 
    j_4 = \frac{k}{2}.
\end{equation}
Eqs.~\eqref{nullstate 0 SL2} and \eqref{KZ for null state SL2} imply that an extra insertion of the operators 
\begin{equation}
   O^{\mathrm{NS}} \equiv  \oint_{z_4} dz' \frac{J^-(x_4,z')}{(z'-z_4)}\, , 
      \label{ONSOKZ1}
      \end{equation}
   \begin{equation}
   O^{\mathrm{KZ}} \equiv  \der_{z_4} + \oint_{z_4} dz' \frac{J^3(x_4,z')}{(z'-z_4)}, 
   \label{ONSOKZ2}
\end{equation}
annihilates the correlator. By inverting the contours one derives the following two conditions for the conformal block: 
\begin{equation}
    \left[\frac{x}{z} - \frac{x-1}{(z-1)}\right] x (x-1) \der_x {\cal{F}}(x,z) = 
     \left[ \kappa \left(\frac{x^2}{z}-\frac{(x-1)^2}{(z-1)}
     \right) + \frac{2j_1 x}{z} + \frac{2j_2 (x-1)}{(z-1)}\right] {\cal{F}}(x,z) \, ,
     \label{NSww}
\end{equation}
and  
\begin{equation}
    -\der_z {\cal{F}}(x,z) = \frac{x (x-1)}{z(z-1)} \der_x {\cal{F}}(x,z)
    + \left[\frac{j_1}{z} + \frac{j_2}{z-1}
    +\kappa \left(\frac{x}{z}-\frac{x-1}{z-1}\right)\right] {\cal{F}}(x,z)\, ,
    \label{KZww}
\end{equation}
where $\kappa=j_3-j_1-j_2-j_4$. These are solved exactly by 
\begin{equation}
    \label{Pw111}
    {\cal{F}}(x,z) = z^{j_1}(z-1)^{j_2}x^{2j_2+\kappa}
    (x-1)^{2j_1+\kappa} (z-x)^{j_4-j_1-j_2-j_3}.  
\end{equation}
We can also fix the constant $\tilde{C}$ in \eqref{4ptconserving} by considering the factorization limit, see the discussion around Eq.~\eqref{factorization unflowed}. Indeed, by considering the null-state equation, this time for a three-point function of the form  $\langle V_{j_1} V_{j_2}V_{\frac{k}{2}}\rangle$, one finds that it vanishes unless $j_2 = \frac{k}{2}-j_1$. The precise statement can be derived directly from the structure constants \eqref{C123}, giving 
  \begin{equation}  
    \label{3ptwithk2}
    C\left(j_1,\frac{k}{2},j\right) \sim  \, \delta \left(j_1+j-\frac{k}{2}\right)\, ,
\end{equation}
where we have omitted a $j$-independent albeit $k$-dependent factor of the form $\sqrt{B(j_1) B(\tilde{\jmath}_1)}$ with $\tilde{\jmath}_1 = \frac{k}{2}-j_1$, see \eqref{def B(j)}. Hence, there is a single intermediate channel, leading to 
\begin{equation}
    \tilde{C}
    \left(j_1,j_2,j_3\right) \sim 
    B\left(\frac{k}{2}-j_1\right)^{-1}     
    C\left(\frac{k}{2}-j_1,j_2,j_3\right) \sim B(j_1)     
    C\left(\frac{k}{2}-j_1,j_2,j_3\right).  
\end{equation}
We have thus computed \eqref{4ptconserving} exactly. 

We can now use this to compute the three-point function \eqref{3pt 100 a} by means of Eq.~\eqref{proposalWxbasis}. Employing purely holomorphic notation (and ignoring  a couple of overall signs) we have 
\begin{eqnarray}
&& \langle V_{j_1 h_1}^1 (0,0) V_{j_2 h_2} (1,1) V_{j_2 h_2} (\infty,\infty) \rangle   \\
    && \quad = \lim_{\vep\to 0}  \vep^{m_1}
    \int dy \,  y^{j_1-m_1-1} \langle V_{j_1}(y,\vep)  
    V_{j_2}(1,1)
    V_{j_3}(\infty,\infty) 
    V_{\frac{k}{2}}(0,0)
    \rangle  \,.
    \nn \\
    && \quad \approx \tilde{C}(j_1,j_2,j_3)\lim_{\vep\to 0}  \vep^{j_1 + m_1}
    \int dy \,  y^{j_2+j_3-m_1-\frac{k}{2}-1} (y-\vep)^{\frac{k}{2}-j_1-j_2-j_3}
    \label{y integral 100}
\end{eqnarray}
where we have used that for these insertion points $x_{12} = y-1$, $z_{21}= \vep-1$, $x = \frac{y}{y-1}$ and $z = \frac{\vep}{\vep-1}$, and that we are interested in the $\vep \to 0$ limit. The latter trivializes once we change variables $y \to y \vep$, so the correlator does not vanish in general. By using \eqref{integral Maldacena AppC} we finally obtain 
\begin{equation}
    \langle V_{j_1 h_1}^1 (0,0) V_{j_2} (1,1) V_{j_2} (\infty,\infty) \rangle =      \frac{
C\left(\frac{k}{2}-j_1,j_2,j_3\right)B(j_1) \gamma \left(j_1 + h_1 - \frac{k}{2}\right) \gamma(j_2 + j_3 - h_1)}{
    \gamma \left(j_1 + j_2 + j_3 - \frac{k}{2}\right)
    }. 
    \label{W100 final}
\end{equation}

This is an important result, and we pause here in order to emphasize its physical  interpretation, which had a strong impact on the study of strings in AdS$_3$. We started by considering a curved background sourced by $n_5$ NS5-branes (with $n_5 \sim k$ in the supergravity regime) and $n_1$ fundamental strings. In the near-horizon region the geometry becomes AdS$_3\times M_{\rm int}$. Naively, one would expect the charges $n_1$ and $n_5$ to be fixed quantities. Nonetheless, upon taking $\frac{1}{2}<j_2,j_3 < \frac{k-1}{2}$ and $j_1 \in \frac{1}{2}+ i \R$, we find that the correlator in Eq.~\eqref{W100 final} describes a non-zero probability for a process where two short strings join to form a long one. As discussed in earlier sections, this long string can then travel all the way to the boundary in finite proper time, and escape from the AdS$_3$ region. In other words, the system can dynamically emit an F1-string, hence $n_1$ can actually change! In this case it changes by one unit because the long string in question is singly wound. 

This interpretation raises an interesting question regarding the holographic CFT. Our discussion of the spacetime Virasoro algebra led us to conclude that the corresponding central charge was given by $c_{\rm st} = 6 k n_1$. What do we mean then, when we say that $n_1$ should be considered dynamical? 

In passing, we note that there have been some clues along the way pointing to the fact that taking this as a fixed central charge for the spacetime theory might be a bit too quick. This is related to the appearance of the operator $\Ii$ (see Eqs.~\eqref{Def identity op wakimoto} and \eqref{def I(x)}) in the spacetime OPEs involving $\Tt(x)$ and $\Kk^a(x)$. As we briefly discussed early on, $\Ii$ is not exactly proportional to the identity operator of the boundary theory: it only  behaves as such when working in a fixed spectral flow sector. Roughly speaking, its expectation value counts the number of strings in the background. 

The resolution to this puzzle was given in \cite{Kim:2015gak} where the authors argued that we are actually working in the grand canonical ensemble. By this we mean that on the boundary we are combining an infinite number of individual holographic CFTs with many different central charges, and fixing an associated chemical potential. In order to fix $n_1$ instead, one should perform a Legendre transformation by adding a source for $\Ii$. The computational consequences of this realization at the level of correlation functions were then recently studied in \cite{Eberhardt:2021vsx}, while an explicit, very precise check was obtained in \cite{Eberhardt:2020bgq} by considering the partition function in the tensionless limit. 

We will have more to say about this later on. For now, we go back to our study of spectrally flowed three-point functions. Note that, from Eq.~\eqref{W100 final}, we can deduce the parafermion correlator with one unit of winding violation. We can then use it to compute all SL(2,$\R$) correlation functions which, so to speak,  violate total spectral flow conservation by one unit, namely those with $\w_1 + \w_2 - \w_3 = \pm 1$. Thus, we ask again (at least for three-point functions): are we done?

\subsection{Spectral flow selection rules}

This section is not called \textit{Conclusions and outlook}, so the reader might suspect the  answer is no. We are actually quite far from being done. In order to see this, we now study the spectral flow selection rules more systematically \cite{Maldacena:2001km,McElgin:2015eho}. 
We focus on correlators of Virasoro primary operators since the descendant case can be treated similarly \cite{McElgin:2015eho}. Moreover, we start from the knowledge that, for two point functions, spectral flow \textit{is} conserved\footnote{Two independent  derivations can be found in Refs.~\cite{McElgin:2015eho} and \cite{Dei:2021xgh}.}. This extends to the full affine modules, such that, up to the series identifications discussed around Eq.~\eqref{series Id SL2 1}, the different spectral flow sectors are orthogonal to each other. 

The distinction between correlators of $m$-basis vertex operators and those involving the $x$-basis ones will be important in what follows.   Let us first work in the $m$-basis, where $\w$ can be any integer number. Consider the state 
\begin{equation}
\label{generic state}
    \prod_{i=1}^{n-1}V_{j_i m_i}^{\w_i}(z_i)|0\rangle \, .
\end{equation}
The defining OPEs \eqref{flowedOPESL2} imply that it is annihilated by 
\begin{equation}
    {\cal J}^+=
    \oint dz \, J^+(z) \prod_{i=1}^{n-1} (z-z_i)^{\w_i+1},   
\end{equation}
where the contour encircles all of the $z_i$.
Since the different spectral flow sectors are orthogonal to each other, this means that, upon using the $VV$ OPEs to  write the generic state \eqref{generic state} as a sum of individual operators with definite spectral flow charges, denoted as $\w_n$, each of them inserted at the origin, they must all be annihilated separately. Now, in terms of modes at the origin we have ${\cal J}^+ = J^+_p + b_+ J^+_{p-1} + \dots$, where the highest mode number is $p = \sum_{i=1}^{n-1} (\w_i + 1)$, and $b_+$ is some constant. Since there are no null states in these representations, this implies an upper bound on $\w_n$, namely $\w_n \leq p- 1$. One can play a similar game with the operator 
\begin{equation}
    {\cal J}^-=
    \oint dz \, J^-(z) \prod_{i=1}^{n-1} (z-z_i)^{-\w_i+1} \, ,  
\end{equation}
such that ${\cal J}^- = J^-_q + b_- J^-_{q-1} + \dots$,  with $q = \sum_{i=1}^{n-1}(1- \w_i)$. This leads to the lower bound $\w_n \geq 1-q$, hence we have
\begin{equation}
    2-n \leq \w_n- \sum_{i=1}^{n-1}\w_i \leq   n-2. 
\end{equation}
Consequently, if we consider the inner product with a conjugated operator, which amounts to flipping $\w_n \to -\w_n$, we obtain the $m$-basis selection rules for flowed correlators\footnote{Slightly more stringent selection rules can be obtained by specifying how many of the involved operators are in the continuous and discrete representations, respectively, or by working with highest/lowest weight states, which can be always achieved by acting with the currents \cite{Maldacena:2001km,McElgin:2015eho}.  
}:
\begin{equation}
    \Big|\sum_{i=1}^n \w_i\Big| \leq n-2. 
\end{equation}
Hence, for flowed three-point functions in the $m$-basis we were actually already done, as spectral flow can only be violated by one unit. 

However, somewhat surprisingly at first, the story is different for the correlation functions in the $x$-basis. Consider the state 
\begin{equation}
\label{generic state x}
    \prod_{i=1}^{n-1}V_{j_i h_i}^{\w_i}(x_i,z_i)|0\rangle \, ,
\end{equation}
where we now take $\w_i \geq 0$ $\forall \, i$.
The current $J^+(z) = J^+(x,z)$ is the same at any point in $x$-space, hence the first part of our previous derivation involving ${\cal J}^+$ goes through. However, this is not so for $J^-$ since $J^-(x_i,z) \neq J^-(x_j,z)$ when $x_i \neq x_j$, see Eq.~\eqref{defJx}. As a consequence, the operator ${\cal J}^-$ does \textit{not} annihilate the state \eqref{generic state x}. This means that  we end up with a single bound instead of two, and consequently obtain a much weaker restriction\footnote{Although $x$-basis correlators can be expanded in terms of $m$-basis ones (with descendant insertions), there is in principle no contradiction since this sum contains an infinite number of terms. Nevertheless, it would be useful to derive this explicitly.}: non-vanishing $x$-basis spectrally flowed $n$-point functions must satisfy 
\begin{equation}
\label{x basis selection rules w}
    \w_i - \sum_{j \neq i} \w_j \leq n-2 \,. 
\end{equation}

One way to understand this is as follows. The translation from $x=0$ to an arbitrary point in $x$-space can be seen as an automorphism of the current algebra. In other words, the modes $J^a_n(x)$ -- all of them evaluated at the \textit{same} value of $x$ -- satisfy exactly the same algebra as the original modes $J^a_n = J^a_n(0)$ for any $x$. Hence, one could think about performing the spectral flow operation along any of the resulting Cartan generators. Now for an $m$-basis correlator we always sit at the origin and consider spectral flow along a fixed direction in \textit{isospin} space, namely the one defined by the Cartan current $J^3$ at $x = 0$. (Equivalently, for  $x \to \infty$ we simply flip the sign of $\w$). This is different in the $x$-basis, as for each individual operator $V_{j_i h_i}^{\w_i}(x_i,z_i)$ one considers spectral flow along the direction of $J^3(x_i)$, defined by the corresponding boundary insertion point. Here lies the fundamental difference between both types of correlation functions \cite{Eberhardt:2018ouy}.

\subsection{Spectrally flowed correlators and covering maps}

The analysis of the previous section shows that there are many more non-trivial correlators that we need to compute. The problem is that the tools we have employed so far are not really useful in this context. On the one hand, $m$-basis techniques cannot be used as we do not expect generic correlators to have a well-defined limit where we send all $x_i$ to $0$ or $\infty$. On the other hand, while the formula \eqref{proposalWxbasis} works for arbitrary spectral flows, its application quickly becomes cumbersome. For instance, for a three-point function with all $\w_i >0$ we would need to compute a complicated six-point function with three unflowed insertions and three \textit{generalized} spectral flow operators. 

\subsubsection{The recursion relations}

These issues where first addressed in \cite{Eberhardt:2019ywk,Dei:2021xgh}. The authors noticed that, similar to the usual global Ward identities, consistency of the correlators with the OPEs between vertex operators and currents implies a series of constraints, which are in principle powerful enough to compute all spectrally flowed correlation functions in terms of the unflowed ones. These were dubbed \textit{local} Ward identities, and as we now describe, they take the form of recursion relations in the $h_i$ quantum numbers. 

Let us define the shorthands
\begin{equation}\label{defF}
    F =\left\langle \prod_{j=1}^n V_{j_j h_j}^{\w_j}(x_j,z_j)\right\rangle\,, 
\end{equation}
and 
\begin{equation}
    F_m^i =\left\langle 
    \left(J_m^+V_{j_i h_i}^{\w_i}\right)(x_i,z_i)
    \prod_{j\neq i} V_{j_j h_j}^{\w_j}(x_j,z_j)\right\rangle.  
    \label{defFin}
\end{equation}
Although we know a couple of these quantities a bit more explicitly, namely $F_0^i = \der_{x_i} F$ and 
\begin{equation}
    F_{\w_i}^i = \left(h_i-\frac{k}{2} \w_i + 1-j_i\right) \left\langle V_{j_i, h_i+1}^{\w_i}(x_i,z_i) \prod_{j\neq i} V_{j_j h_j}^{\w_j}(x_j,z_j)\right\rangle\,,
    \label{Fwi}
\end{equation}
which follows from \eqref{JwVx}, the rest of the $F^i_m$ with $m=1,\dots, \w_i-1$ are yet to be determined. Counting only the $F_m^i$ with $m>0$, for a given $n$-point function we thus have a total of $\sum_{i=1}^n (\w_i-1)$ unknowns. For later convenience, we also define  
\begin{eqnarray}
    F_{-\w_i}^{i,-} &=& \left\langle 
    \left(J_{-\w_i}^-V_{j_i h_i}^{\w_i}\right)(x_i,z_i)
    \prod_{j\neq i} V_{j_j h_j}^{\w_j}(x_j,z_j)\right\rangle \nn \\
    &=& \left(h_i-\frac{k}{2} \w_i - 1+j_i\right) \left\langle V_{j_i, h_i-1}^{\w_i}(x_i,z_i) \prod_{j\neq i} V_{j_j h_j}^{\w_j}(x_j,z_j)\right\rangle\,.
\end{eqnarray}
The OPEs in Eq.~\eqref{JVxOPE} allow us to expand  correlators involving a current insertion  as
\begin{subequations}
\begin{eqnarray}
    \left\langle J^+(z) \prod_{j=1}^n V_{j_j h_j}^{\w_j}(x_j,z_j)\right\rangle &=& \sum_{i=1}^n \left[
    \frac{\der_{x_i} F}{z-z_i} + \sum_{m=1}^{\w_i} 
    \frac{F_{m}^i}{(z-z_i)^{m+1}}
    \right] + \cdots \, , \\[1ex]
    \left\langle J^3(z) \prod_{j=1}^n V_{j_j h_j}^{\w_j}(x_j,z_j)\right\rangle &=& \sum_{i=1}^n \left[
    \frac{(h_i+x_i\der_{x_i}) F}{z-z_i} + \sum_{m=1}^{\w_i} 
    \frac{x_i F_{m}^i}{(z-z_i)^{m+1}}
    \right] + \cdots\, , \\[1ex]
    \left\langle J^-(z) \prod_{j=1}^n V_{j_j h_j}^{\w_j}(x_j,z_j)\right\rangle &=& \sum_{i=1}^n \left[
    \frac{(2h_i x_i + x_i^2\der_{x_i} )F}{z-z_i} + \sum_{m=1}^{\w_i} 
    \frac{x_i^2 F_{m}^i}{(z-z_i)^{m+1}}
    \right] + \cdots
\end{eqnarray}
\end{subequations}
The above expressions can be combined in order to compute 
\begin{equation}
\label{Gjdef}
    G_j(z) \equiv \left\langle J^-(x_j,z) \prod_{l=1}^n V_{j_l h_l}^{\w_l}(x_l,z_l)\right\rangle = 
    \sum_{i\neq j} \left[
    \frac{(2h_i x_{ij} + x_{ij}^2\der_{x_i} )F}{z-z_i} + \sum_{n=1}^{\w_i} 
    \frac{x_{ij}^2 F_{n}^i}{(z-z_i)^{n+1}}
    \right] + \cdots
\end{equation}
for each $j=1,\dots,n$, where $x_{ij} = x_i - x_j$ and $J^-(x_j,z) = J^-(z)-2x_j J^3(z)+2x_j^2 J^+(z)$. The relevance of these quantities comes from the fact that  Eq.~\eqref{JmxOPE} imposes stringent restrictions on the behavior of $G_j(z)$ when we take $z \to z_j$. Indeed, they must satisfy 
\begin{equation}
    \lim_{z\to z_j} (z-z_j)^{1-\w_j}G_j(z) =  F_{-\w_j}^{j,-} \,. 
    \label{Gjconditions}
\end{equation}
Even though $G_j(z)$ as given in \eqref{Gjdef} is clearly regular in the limit $z\to z_j$, this is much more restrictive, as we have found that the first $\w_j-1$ regular terms in the Taylor expansion  around $z_j$  must vanish, while the term of order $(z-z_j)^{\w_j-1}$ must have coefficient $F_{-\w_i}^{j,-}$. After using this for every $j=1,\dots,n$, we end up with a total of $\sum_{i=1}^n \w_i$ \textit{linear} constraints, which involve the primary correlators as well as all of the $F^i_m$ with $m\geq 0$. 

This provides, first, a complicated linear system which allows one to solve for the unknown $F^i_m$ with $m=1,\dots, \w_i-1$. We are then left with exactly $n$ recursion relations, involving the original correlator,  its $x_i$ derivatives, and also other primary $n$-point functions with the same spectral flow charges but where one of the $h_i$ is shifted upwards or downwards by one unit. The latter are precisely the correlators we have denoted as $F^i_{\w_i}$ and $F^{i,-}_{-\w_i}$. Hence, these relations are recursive. They fully  characterize the $h_i$-dependence of the flowed correlators. 

However, in practice it is quite difficult to solve this system in full generality. As we now describe, this has recently been achieved for the case of three-point functions \cite{Dei:2021xgh,Iguri:2022eat,Bufalini:2022toj}. For higher point functions it is not known how to derive the general expressions of the recursion relations, although important progress was achieved in \cite{Dei:2021yom} for $n=4$.  

\subsubsection{The $y$-basis: from recursion relations to differential equations}
\label{sec: conjecture}

As will be discussed in Sec.~\ref{sec:tensionless} below, these recursion relations admit a very simple solution in the tensionless string limit, which corresponds to $k=3$ in the bosonic set-up \cite{Eberhardt:2019ywk,Eberhardt:2025sbi}. For more general values of $k$, however, the derivation and structure of the recursion relations derived from the local Ward identities is much more involved. 

Nevertheless, we note that there is an important similarity with the case of (spacetime) global Ward identities. Indeed, the latter imply somewhat analogous recursions between correlators with shifted eigenvalues $m_i \to m_i \pm 1$. 
The solutions are the complicated expressions obtained in \cite{Satoh:2001bi} in terms of hypergeometric functions of the ${}_3F_2$ type, namely the Clebsch-Gordan coefficients for SL(2,$\R$). However, we know that these $m$-basis recursions are transformed into differential equations when working with $x$-basis operators, on which the zero modes $J^a_0$ act as the differential operators \eqref{Da Diff ops x}. This drastically simplifies the analysis, and allows one to quickly deduce the relevant $x$-dependence. The idea put forward in \cite{Dei:2021xgh,Dei:2021yom} was to proceed analogously in the flowed case, that is, to define yet another complex variable $y$ conjugate to the quantum number $h$, in terms of which the modes $\tilde{J}^3_0 = J^3_0 - \frac{k}{2}\w$ and $\tilde{J}^\pm_{0} = J^\pm_{\pm \w}$ act as differential operators. 

How should we define the relevant (linear combinations of) spectrally flowed vertex operators? The answer is already hidden in Eq.~\eqref{proposalWxbasis}. Indeed, by rescaling $y \to y \vep^\w$, we obtain 
\begin{equation}
\label{proposalWxybasis}
    V_{jh}^{\w}(x,z) = 
    \int d^2y \,  y^{\frac{k}{2}\w + j-h-1}  \bar{y}^{\frac{k}{2}\w + j -\hb-1} V_j^\w (x,y,z)\, ,
\end{equation}
where we have defined\footnote{For $\w=0$ we simply have $y=x$.}  
\begin{equation}
    V_j^\w (x,y,z) \equiv \lim_{\vep,\bar{\vep}\to 0}  |\vep|^{2j \w} V_{j}(x+y\vep^\w,z+\vep) V_{\frac{k}{2} \frac{k}{2}\w}^{\w-1} (x,z).
    \label{defVWxyz}
\end{equation}
Eq.~\eqref{proposalWxybasis} is the exact analogue of the Melling-type transform \eqref{Mellin} for spectrally flowed operators. As in the unflowed case, for vertex operators built upon the lowest-weight representations we can invert this and write 
\begin{equation}
    V_j(x,y,z) = \sum_{h,\bar{h}} y^{h-\frac{k}{2}\w-j}\bar{y}^{\bar{h}-\frac{k}{2}\w-j} V_{jh{\bar{h}}}^\w(x,z) \, ,
\end{equation}
where, as usual, we have suppressed some of the anti-holomorphic variables. This should be compared with Eq.~\eqref{ExpSL2}. Similar expressions hold for operators coming from the highest-weight states and the continuous sector.  We should stress that these linear combinations are quite non-standard from the CFT point of view, as the different contributions have distinct spacetime and worldsheet weights!

Let us state two important properties.  From our definition of $y$-basis operators one can also deduce the reflection property for long strings, namely 
\begin{equation}
\label{reflection y basis}
    V_{1-j}^\w (x,y,z) = B(1-j) \int d^{2}y' \, |y-y'|^{4j-4}V_{j}^\w(x,y',z) \,.
\end{equation}
Moreover, while the zero modes still act as in Eq.~\eqref{diffopsx}, we now have 
\begin{subequations}
\label{JwVy}
\begin{eqnarray}
    \left(J_{\w}^+V_{j}^\w\right) (x,y,z) &=& \der_y V_{j}^\w (x,y,z) \, , 
    \\[1ex]
    \left(J_{0}^3V_{j}^\w\right) (x,y,z) &=& \left(y\der_y + j + \frac{k}{2}\w \right) V_{j}^\w (x,y,z) \, , 
    \\[1ex]
    \left(J_{-\w}^- V_{j}^\w\right) (x,y,z) &=& \left(y^2\der_y + 2jy \right) V_{j}^\w (x,y,z) \, .
    \label{diffopsyMinus}
\end{eqnarray}    
\label{diffopsy}
\end{subequations}
We now showcase how to use the $y$-basis language in practice.

As a warm-up, let us see how this leads to an alternative method for computing the flowed two-points $\langle V_{j_1h_1}^\w(x_1,z_1)V_{j_2h_2}^\w(x_2,z_2)\rangle$ without going through the $m$-basis and the parafermionic decomposition. We start by considering the corresponding $y$-basis correlator $\langle V_{j_1}^\w(x_1,y_1,z_1)V_{j_2h_2}^\w(x_2,y_2,z_2)\rangle$, and inserting an operator involving the currents of the model: 
\begin{equation}
    \oint_{z_1} dz \frac{(z-z_2)^\w}{(z-z_1)^\w}\langle J^-(x_1,z)V_{j_1}^\w(x_1,y_1,z_1)V_{j_2h_2}^\w(x_2,y_2,z_2) \, .
\end{equation}
The overall powers ensure that we pick up the residue at $z_1$ where the current acts on $V_{j_1}^{\w_1} (x_1,y_1,z_1)$ as $\left(J_{-\w_1}^- V_{j_1}^{\w_1}\right) (x_1,y_1,z_1)$. On the other hand, we can also turn the contour around and act on $V_{j_2}^{\w_2} (x_2,y_2,z_2)$. Using that $J^-(x_1,z) = J^-(x_2,z) - 2 x_{12} J^3(x_2,z) + x_{12}^2 J^+(z)$, we see that there we pick up the action of the mode $J^+_{\w_2}$. More explicitly, this gives the differential equation 
\begin{equation}
 \left(
 z_{12}^\w (y_1^2 \der_{y_1} + 2 j_1 y_1) + \frac{x_{12}^2}{z_{21}^\w} \der_{y_2}
 \right)  \langle V_{j_1}^\w(x_1,y_1,z_1)V_{j_2h_2}^\w(x_2,y_2,z_2) \rangle  = 0 \, .
\end{equation}
An analogous equation with $1 \leftrightarrow 2$ can be obtained by starting with $J^-(x_2,z)$. We now combine this with the constraints coming from the global Ward identities, which imply that 
\begin{equation}
    \langle V_{j_1}^\w(x_1,y_1,z_1)V_{j_2h_2}^\w(x_2,y_2,z_2) \rangle = x_{12}^{-h_1^0 - h_2^0} z_{12}^{-\Delta_1^0 - \Delta_2^0} 
    F \left( y_1 \frac{z_{12}^\w}{x_{12}},y_2 \frac{z_{12}^\w}{x_{12}}\right) \, , 
\end{equation}
where 
\begin{equation}
    h^0_i = j_i + \frac{k}{2}\w_i
    \, , \qquad \Delta^0_i 
    = - \frac{j_i(j_i-1)}{k-2} - j_i \w_i - \frac{k}{4}\w_i^2\, , 
\end{equation}
and  
\begin{equation}
    F \left( y_1 ,y_2 \right) \equiv \langle V_{j_1}^\w(0,y_1,0)V_{j_2h_2}^\w(\infty,y_2,\infty) \rangle \, .
\end{equation}
This can be understood by noting that $V_{j_i}(x_i,y_i=0,z_i) = V_{j_ih_i^0}^{\w_i}(x_i,z_i)$ and thinking in terms of a power series expansion around $y_i=0$, such that, due to  Eqs.~\eqref{proposalWxybasis} and \eqref{Delta spectral flow}, each additional power of $y_i$ shifts $h_i \to h_i + 1$ and $\Delta_i \to \Delta_i - \w_i$. 
As a result, we reinterpret the above constraints as two differential equations for $F(y_1,y_2)$, namely  
\begin{equation}
    \left[
y_1^2 \der_{y_1} + 2 j_1 y_1 +  (-1)^\w \der_{y_2}
 \right] F(y_1,y_2) = 0 = 
 \left[
y_2^2 \der_{y_2} + 2 j_2 y_2 + (-1)^\w \der_{y_1}
 \right] F(y_1,y_2) \,.
\end{equation}
Up to constant rescalings (and omitting the anti-holomorphic sector), this has two independent solutions, one of them distributional. They take the form
\begin{equation}
    F(y_1,y_2) = \delta(j_1-j_2) \left[(-1)^w + y_1 y_2\right]^{-j_1-j_2} \,
,
\end{equation}
and 
\begin{equation} 
    F(y_1,y_2) = \delta(j_1+j_2-1) y_1^{-j_1}y_2^{-j_2} \delta \left((-1)^w + y_1 y_2 \right) \, . 
\end{equation}
As in the unflowed case, the overall normalization for the solution is conventional, but the relative coefficient between the two terms is physical. The latter can be fixed by means of the reflection identity \eqref{reflection y basis}.
Upon including the anti-holomorphic contributions and integrating over $y_1$ and $y_2$ as in  \eqref{proposalWxybasis}, this leads precisely to the two terms in Eq.~\eqref{SL2 2point bosonic with flow}.   

Let us now move to three-point functions. Working with $y$-basis operators will allow us to rewrite the recursion relations derived formally above as differential equations for correlators of the form  
\begin{equation}
    F_y 
    \equiv \langle V_{j_1 }^{\w_1}(x_1,y_1,z_1)V_{j_2 }^{\w_2}(x_2,y_2,z_2)V_{j_3 }^{\w_3}(x_3,y_3,z_3)\rangle\, .
    \label{V1V2V3xy}
\end{equation}
The original $x$-basis correlators in Eq.~\eqref{defF} can then be obtained from these by means of \eqref{proposalWxybasis}, although the integrals are somewhat non-trivial and have not been worked out in full generality, see \cite{Dei:2021xgh}. It is useful to fix $x_1 = z_1=0$, $x_2 = z_2=1$ and $x_3 = z_3=\infty$, and consider 
\begin{equation}
    \hat{F}_y = 
    \langle V_{j_1 }^{\w_1}(y_1)V_{j_2 }^{\w_2}(y_2)V_{j_3 }^{\w_3}(y_3)\rangle\equiv \langle V_{j_1 }^{\w_1}(0,y_1,0)V_{j_2 }^{\w_2}(1,y_2,1)V_{j_3 }^{\w_3}(\infty,y_3,\infty)\rangle\,
    .\label{Vw1w2w301infty}
\end{equation}  
In this case the global Ward identities imply 
\begin{align}
\begin{aligned}
   &\langle V_{j_1 }^{\w_1}(x_1,y_1,z_1)V_{j_2 }^{\w_2}(x_2,y_2,z_2)V_{j_3 }^{\w_3}(x_3,y_3,z_3)\rangle =  \frac{x_{21}^{h^0_3-h^0_1-h^0_2}
    x_{31}^{h^0_2-h^0_1-h^0_3}
    x_{32}^{h^0_1-h^0_2-h^0_3}}{
    z_{21}^{\Delta^0_1+\Delta^0_2-\Delta^0_3}
    z_{31}^{\Delta^0_1+\Delta^0_3-\Delta^0_2}
    z_{32}^{\Delta^0_2+\Delta^0_3-\Delta^0_1}}
    \times \,  \\[1ex]
    &      \left\< V_{j_1}^{\w_1} \left(0,y_1 \frac{x_{32} z_{21}^{\w_1}z_{31}^{\w_1}}{x_{21}x_{31}z_{32}^{\w_1}},0\right)V_{j_2}^{\w_2} \left(1,
    y_2 \frac{x_{31} z_{21}^{\w_2}z_{32}^{\w_2}}{x_{21}x_{32}z_{31}^{\w_2}}
    ,1\right)V_{j_3}^{\w_3}\left(\infty,
    y_3 \frac{x_{21} z_{31}^{\w_3}z_{32}^{\w_3}}{x_{31}x_{32}z_{21}^{\w_3}},\infty\right)\right\> \, .
\end{aligned}
\label{ybasisx1x2x3fixing}
\end{align}
The procedure outlined above implies that, whenever the system is compatible and the $F_n^i$ can be solved for, the $y$-basis correlator $\hat{F}_y$ will satisfy recursion relations of the following schematic form \cite{Bufalini:2022toj}: 
\begin{equation}
    \left[y_i(y_i\der_{y_i} + 2j_i) + \sum_{j=1}^3(A_{ij} y_j -B_{ij})\der_{y_j} 
    + C_i \right]\langle V_{j_1}^{\w_1}(y_1)
    V_{j_2}^{\w_2}(y_2)
    V_{j_3}^{\w_3}(y_3) \rangle = 0 \, , 
    \label{GenEqsEven}
\end{equation}
where $A_{ij}$, $B_{ij}$ and $C_i$ are constants, but depend on the spins $j_i$ and the charges $\w_i$. 
Moreover, generically the way in which these equations are derived only depends on the charges $\w_i$ involved in the correlator, hence the recursion relations do \textit{not} dependent on whether the corresponding vertex operators belong to spectrally flowed discrete or continuous representations. Interestingly, using the definition given in Eq.~\eqref{defVWxyz}, these differential equations can be shown to be equivalent to the null-state conditions associated to the generalized spectral flow operators $V_{\frac{k}{2} \frac{k}{2}\w}^{\w-1} (x,z)$ \cite{Iguri:2022eat}.

Before discussing the general case we provide three simple examples for clarity: the three-point functions with $\boldsymbol{\w}=(1,0,0)$, $\boldsymbol{\w}=(1,1,0)$ and $\boldsymbol{\w}=(2,1,1)$. In the first case there is only one constraint, as only one of the operators has a non-trivial spectral flow charge, and its OPEs with the currents contain no unknown contributions. The latter only appear for correlators involving insertions with $\w_i\geq 2$ for some $i$. Hence, in the $x$-basis the constraint reads 
\begin{equation}
   \sum_{i\neq 1} 
    \frac{(2h_i x_{i1} + x_{i1}^2\der_{x_i} )}{z_{1i}}F  = F^{1,-}_{-1} \, ,
\end{equation}
where $h_{2}=j_2$ and $h_3=j_3$.
Moving to the $y$-basis and making use of the global Ward identities and of \eqref{ybasisx1x2x3fixing}, we find that $\hat{F}_y = 
\langle V_{j_1 }^{1}(0,y_1,0)V_{j_2}(1,1)V_{j_3}(\infty,\infty)\rangle$ must satisfy 
\begin{equation}
    \left[y_1(y_1-1) \der_{y_1} + (2y_1-1)j_1 + j_2 + j_3 - \frac{k}{2}\right]\hat{F}_y^{\boldsymbol{\w}=(1,0,0)} = 0, 
\end{equation}
hence it must be of the form 
\begin{equation}
\label{Fy result 100}
    \hat{F}_y^{\boldsymbol{\w}=(1,0,0)} \sim y_1^{j_2+j_3-j_1-\frac{k}{2}} (y_1-1)^{\frac{k}{2}-j_1-j_2-j_3} \, ,
\end{equation}
up to a factor that is independent of $y_1$ (and will be fixed below in the general case). Finally, \eqref{proposalWxybasis} tells us that in order to go back to the $x$-basis and derive the dependence in $h_1$ we must insert an additional factor of $y_1^{j_1+\frac{k}{2}-h_1}$ and integrate over $y_1$. This gives exactly the same integral as that obtained around Eq.~\eqref{y integral 100} from $m$-basis methods, showing that both approaches indeed give the same result.  

For the correlator with $\boldsymbol{\w}=(1,1,0)$ we still have no unknowns to deal with, but there are now two differential equations, needed for deriving the dependence in  $y_1$ and $y_2$. These conditions come from inserting the current operators $J^-(x_1,z)$ and $J^-(x_2,z)$ and constraining the behavior of the resulting expressions as $z \to z_1$ and $z \to z_2$, respectively. They take the form 
\begin{subequations}
\begin{eqnarray}
     \sum_{i\neq 1} 
    \frac{(2h_i x_{i1} + x_{i1}^2\der_{x_i} )}{z_{1i}}F + \frac{x_{21}^2}{z_{12}^2} F_1^2   &=& F^{1,-}_{-1} \, , \\
    \sum_{i\neq 2} 
    \frac{(2h_i x_{i2} + x_{i2}^2\der_{x_i} )}{z_{2i}}F + \frac{x_{12}^2}{z_{21}^2} F_1^1  &=& F^{2,-}_{-1} \, . 
\end{eqnarray}
\end{subequations}
Proceeding as before, these lead to 
\begin{subequations}
\begin{eqnarray}
\left[ y_1(y_1-1)\der_{y_1} + (y_2-1)\der_{y_2}
+ (2y_1-1)j_1+j_2+j_3
\right]\hat{F}_y^{\boldsymbol{\w}=(1,1,0)} &=& 0 \, , \\
\left[ y_2(y_2-1)\der_{y_2} + (y_1-1)\der_{y_1}
+ (2y_2-1)j_2+j_2+j_3
\right]\hat{F}_y^{\boldsymbol{\w}=(1,1,0)} &=& 0 \, , 
\end{eqnarray}
\end{subequations}
hence the solution is 
\begin{equation}
    \hat{F}_y^{\boldsymbol{\w}=(1,1,0)} \sim (y_1-1)^{j_2-j_1-j_3}(y_2-1)^{j_1-j_2-j_3}
    (y_1y_2-1)^{j_3-j_1-j_2} \, .
\end{equation}
From the different factors and their powers we see that the structure of the result resembles that of the $x$-dependence in an unflowed three-point function. In fact, in order to derive the $h$-dependence of the correlator we need to compute the integral 
\begin{equation}
\int_{\mathbb{C}} dy_1 \, dy_2 \, y_1^{j_1+\frac{k}{2}-h_1-1} y_2^{j_2+\frac{k}{2}-h_2-1}    |y_1-1|^{2(j_2-j_1-j_3)}|y_2-1|^{2(j_1-j_2-j_3)}
    |y_1y_2-1|^{2(j_3-j_1-j_2)} \, ,
\end{equation}
which is exactly the same integral needed for computing an $m$-basis three-point function in the unflowed sector (after integrating out $x_3$, and with $y_1$ and $y_2^{-1}$ playing the role of $x_1$ and $x_2$). This agrees with our previous computations: for correlators where the total spectral flow is conserved -- and here it is, in the sense that if we take $x_2 \to \infty$ the second insertion becomes and $m$-basis flowed primary with spectral flow charge $-1$ -- the flowed $x$-basis structure constants are nothing but the unflowed ones multiplied by the corresponding Clebsch-Gordan coefficients, with $m_i = h_i-\frac{k}{2}\w_i$. As will become clear shortly, the structural similarities between $y$-basis flowed correlators and the $x$-basis unflowed ones are no coincidence.

Finally, we look at the correlator with $\boldsymbol{\w}=(2,1,1)$. This is the first time we have to deal with one of the unknowns, namely $F_1^1$, and with all three variables $y_i$. Since $\w_1=2$, we get two conditions from the insertion of $J^-(x_1,z)$, plus one more from each of the other current insertions at $x_2$ and $x_3$: 
\begin{subequations}
\begin{eqnarray}
     \sum_{i\neq 1 } 
    \frac{(2h_i x_{i1} + x_{i1}^2\der_{x_i} )}{z_{1i}}F + \frac{x_{21}^2}{z_{12}^2} F_1^2 + \frac{x_{31}^2}{z_{13}^2} F_1^3   &=& 0 \, , \\
    -\sum_{i\neq 1} 
    \frac{(2h_i x_{i1} + x_{i1}^2\der_{x_i} )}{z_{1i}^2}F - \frac{2x_{21}^2}{z_{12}^3} F_1^2 - \frac{2x_{31}^2}{z_{13}^3} F_1^3   &=& F^{1,-}_{-1} \, , \\
    \sum_{i\neq 2} 
    \frac{(2h_i x_{i2} + x_{i2}^2\der_{x_i} )}{z_{2i}}F + \frac{x_{12}^2}{z_{21}^2} F_1^1 + \frac{x_{12}^2}{z_{21}^3} F_2^1 + \frac{x_{32}^2}{z_{23}^2} F_1^3  &=& F^{2,-}_{-1} \, , \\
    \sum_{i\neq 3} 
    \frac{(2h_i x_{i3} + x_{i3}^2\der_{x_i} )}{z_{3i}}F + \frac{x_{13}^2}{z_{31}^2} F_1^1 + \frac{x_{13}^2}{z_{31}^3} F_2^1 + \frac{x_{23}^2}{z_{32}^2} F_1^2  &=& F^{3,-}_{-1} \, . 
\end{eqnarray}
\end{subequations}
As described above, we simply use either one of the latter two equations to get an expression for $F_1^1$, and insert it into the other. After using the global Ward identities once again, we obtain 
\begin{align}
\begin{aligned}
\hspace{-0.5cm} \left[ (2 y_2-y_2^2-1)\der_{y_2} + (y_3^2-2y_3+1)\der_{y_3} +  \der_{y_1}
 -2(y_2-1)j_2 + 2(y_3-1)j_3
\right]\hat{F}_y &= 0 \, , \\[1ex]
\left[ y_1(y_1-1)\der_{y_1} + (y_2-2)\der_{y_2}+  y_3\der_{y_3}
+ (2y_1-1)j_1+j_2+j_3
\right]\hat{F}_y &= 0 \, , \\[1ex]
 \left[ y_1\der_{y_1}- (y_2-1)\der_{y_2} - (y_3-1)\der_{y_3}
+j_1-j_2-j_3
\right]\hat{F}_y &= 0 \, ,
\end{aligned}
\end{align}
where $\hat{F}_y\equiv \hat{F}_y^{\boldsymbol{\w}=(2,1,1)}$. Here the solution takes the form  
\begin{equation}
    \hat{F}_y^{\boldsymbol{\w}=(2,1,1)} \sim (y_1 y_3 -y_1+1)^{j_2-j_1-j_3}(2-y_2-y_3)^{j_1-j_2-j_3}
    (y_1y_2-y_1-1)^{j_3-j_1-j_2} \, , 
\end{equation}
such that the powers are the same as in the previous case, but the individual factors start becoming more complicated. Note, however, that upon rescaling $y_1 \to \vep y_1$, $y_2 \to \vep^{-1}y_2$ and $y_3 \to \vep^{-1}y_3$, and further taking the limit $\vep \to 0$, we recover the same structure as in the previous example,  up to some unimportant signs.  
 The motivation for considering this limit is that, as can be seen from Eq.~\eqref{Vw1w2w301infty}, it actually corresponds to unfixing $x_2$ and taking it to infinity. More precisely, we take $(x_1,x_2,x_3)\to (0,\infty,\infty)$, hence all three insertions become $m$-basis flowed primaries. Since $\w_1-\w_2-\w_3=0$ the total spectral flow is again conserved, so the result necessarily had to be the same.

Clearly the above method quickly becomes cumbersome as one increases the spectral flow charges, and we need a more systematic way to deal with these systems of linear constraints. We now state the general solution, first conjectured in \cite{Dei:2021xgh}, and then proved in \cite{Iguri:2022eat,Bufalini:2022toj}.
For the so-called \textit{odd parity} correlators, namely when $\w_1+\w_2+\w_3 \in 2 \mathbb{Z}+1$, one has 
\begin{align}
\left\langle V^{\w_1}_{j_1}(0, y_1, 0) \,  V^{\w_2}_{j_2}(1, y_2, 1) \, V^{\w_3}_{j_3}(\infty, y_3, \infty) \right\rangle = 
C_{\boldsymbol{\w}} (j_i)
X_{123}^{\frac{k}{2}-j_1-j_2-j_3} \prod_{i=1}^3 X_i^{-\frac{k}{2}+j_1+j_2+j_3-2j_i} \ ,
\label{3pt-odd-parity}
\end{align}
while for the \textit{even parity} case, i.e.~when $\w_1+\w_2+\w_3 \in 2 \mathbb{Z}$, 
\begin{align}
\left\langle V^{\w_1}_{j_1}(0, y_1, 0) \,  V^{\w_2}_{j_2}(1, y_2, 1) \, V^{\w_3}_{j_3}(\infty, y_3, \infty) \right\rangle = 
C_{\boldsymbol{\w}} (j_i) X_\emptyset^{j_1+j_2+j_3-k}\prod_{i<\ell} X_{i \ell}^{j_1+j_2+j_3-2j_i-2j_\ell} \  .
\label{3pt-even-parity}
\end{align}
Here $\boldsymbol{\w}=(\w_1, \w_2, \w_3)$, and for any subset $I \subset \{ 1,2,3 \}$, 
\be 
X_I(y_1,y_2,y_3)\equiv \sum_{i \in I:\ \varepsilon_i=\pm 1} P_{\boldsymbol{\w}+\sum_{i \in I} \varepsilon_i e_{i}} \prod_{i\in I} y_i^{\frac{1-\varepsilon_i}{2}} \ . 
\label{X_I-3pt}
\ee
with $e_1 = (1,0,0)$, $e_2 = (0,1,0)$ and $e_3 = (0,0,1)$.
The numbers $P_{\boldsymbol{\w}}$ are defined as  
\be 
P_{\boldsymbol{\w}} = 0 \qquad \text{for} \qquad \sum_j \w_j < 2 {\rm Max}(\w_i) \quad \text{or}\quad \sum_i \w_i \in 2\mathds{Z}+1
\ee
and 
\be
P_{\boldsymbol{\w}} =S_{\boldsymbol{\w}} \frac{G\left(\frac{-\w_1+\w_2+\w_3}{2} +1\right) G\left(\frac{\w_1-\w_2+\w_3}{2} +1\right) G\left(\frac{\w_1+\w_2-\w_3}{2} +1\right) G\left(\frac{\w_1+\w_2+\w_3}{2}+1\right)}{G(\w_1+1) G(\w_2+1) G(\w_3+1)}  \ , 
\label{Pw-definition}
\ee
otherwise, where $G(n)$ is the Barnes $G$ function, $G(n)=\prod_{i=1}^{n-1} \Gamma(i)$, while $S_{\boldsymbol{\w}}$ is a phase which will not be too important for our purposes. Finally, the $y_i$-independent normalizations are defined in terms of the unflowed ones,  and read
\begin{equation}
\label{consdei}
    C_{\boldsymbol{\w}}(j_1,j_2,j_3) = \left\{\begin{array}{cc}
    C(j_1,j_2,j_3),     & \text{if} \quad  \w_1 + \w_2 + \w_3 \in 2\mathbb{Z}, \\
    {\cal{N}}(j_1) C\left(\frac{k}{2}-j_1,j_2,j_3 \right),     & \qquad  \text{if} \quad  \w_1 + \w_2 + \w_3 \in 2\mathbb{Z}+1, 
    \end{array} \right.
\end{equation}  
with 
\begin{equation}
    {\cal{N}}(j) = \sqrt{\frac{B(j)}{B\left(
    \frac{k}{2}-j\right)}} \, .     \label{Ndef}
\end{equation}
It can be seen from the two-point function that $\N(j)$ is precisely the relative coefficient discussed above Eq.~\eqref{series Id SL2 1} in relation to  the series identifications. 

\subsubsection{Holomorphic covering maps}
\label{sec: covering maps}

We would like to turn the constraints \eqref{Gjconditions} into differential equations for the correlators $\hat{F}_y$ for arbitrary values of the $\w_i$. The usual method for this type of computation involves writing the current insertions in terms of contour integrals, and then inverting the contour to make the currents act on the rest of the vertex operators. As for the selection rules, the problem is that $J^-$ acts differently near each of the insertion points $x_i$.

In order to deal with this, it would be extremely useful to have a (holomorphic) function $\Gamma(z)$ capable of making the operator 
\begin{equation}
    J^-(\Gamma(z),z) = J^-(z)-2\Gamma(z) J^3(z)
    +\Gamma^2(z) J^+(z) \, 
    \label{JGamma}
\end{equation}
act as the mode $J^-_{-\w_i}(x_i)$ on each of the spectrally flowed insertions. More explicitly, we need $\Gamma(z)$ to satisfy
\begin{equation}
\Gamma(z) \sim x_i + a_i (z-z_i)^{\w_i} + \cdots\qquad \text{near} \quad 
z = z_i\,, \qquad i=1,2,3 \,,
\label{coveringmapexp}
\end{equation}
where the ellipsis indicates higher order terms in $(z-z_i)$. We also need \eqref{JGamma} to act trivially at infinity (if there is no operator inserted there), which imposes $\Gamma(z\to \infty) = \Gamma_\infty$ for some constant $\Gamma_\infty$. As it turns out, in many cases -- not all -- such functions do exist. They are known as holomorphic covering map, and are the subject of Hurwitz theory, see for instance  \cite{cavalieri_miles_2016}. 

For a given three-point function, the covering map exists whenever
\begin{equation}
    \sum_{i=1}^3 \w_i \in 2\mathbb{Z}+1 
    \qqquad 
    \sum_{i=1}^3 \w_i > 2 {\rm Max}(\w_i) - 1 
    \qqquad \w_i> 0\,, \,\, \forall \, i,  \label{conditionsmapwi}
\end{equation}
and, in the genus zero case, it is unique.
The second of these conditions fits almost perfectly with the selection rule in Eq.~\eqref{x basis selection rules w}, i.e.~it only excludes the cases where the latter inequality is saturated. These \textit{edge} cases will be treated separately. However, the first condition in \eqref{conditionsmapwi} tells us that, without including any further ingredients, the derivation we discuss below works only for the odd parity cases. 

The explicit construction of the covering map is carried out for instance in \cite{Lunin:2000yv}, and we will review it below. 
Before that, let us highlight the main properties of these maps that will be relevant for the computation of genus zero worldsheet correlators: \begin{itemize}
    \item $\Gamma(z)$ is a rational function. 
    
    \item The coefficients $a_i$ appearing in Eq.~\eqref{coveringmapexp} take the form  
\begin{equation}
    a_i =  
    \left(
    \begin{array}{c}
        \frac{\w_i+\w_{i+1}+\w_{i+2}-1}{2}  \\
        \frac{-\w_{i}+\w_{i+1}+\w_{i+2}-1}{2}
    \end{array}\right)
   \left(
    \begin{array}{c}
        \frac{-\w_{i}+\w_{i+1}-\w_{i+2}-1}{2} \\
        \frac{\w_{i}+\w_{i+1}-\w_{i+2}-1}{2}
    \end{array}\right)^{-1}
    \frac{x_{i,i+1}x_{i+2,i}}{
    x_{i+1,i+2}}
    \left(\frac{z_{i+1,i+2}}{
    z_{i,i+1}z_{i+2,i}}\right)^{\w_i}\,, 
    \label{coveringmapcoeffs}
\end{equation}
where the subscripts are understood mod 3. Note that the last two factors trivialize upon setting $x_1=z_1= 0$, $x_2=z_2= 1$ and $x_3=z_3= \infty$. In a slight abuse of notation, later on we will use the same notation "$a_i$" for the resulting purely numerical coefficients.  

\item At generic points $\Gamma(z)$ is an $N$-to-1 map, where $N$ is computed from the Riemann-Hurwitz formula, namely  
\begin{equation}
    N = 1+\frac{1}{2} \sum_{i=1}^n\left(\w_i-1\right) \, . 
\end{equation}  

\item $\Gamma(z)$ has $N$ simple poles, whose locations will be denoted $\lambda_a$, with $a=1,\dots,N$. 

\item The derivative of the map $\der \Gamma(z)$ takes the simple form 
\begin{equation}
\label{der Gamma}
    \der \Gamma(z) \sim \frac{\prod_{i=1}^3 (z-z_i)^{\w_i-1}}{
    \prod_{a=1}^N (z-\lambda_a)^{2}
    } \,, 
\end{equation}
up to an overall factor independent of $z$, which will, for now,  be irrelevant for us. 
\end{itemize}

Let us now discuss how these covering maps are obtained \cite{Lunin:2000yv}. For simplicity, we  fix the insertions points at $(z_1,z_2,z_3) = (x_1,x_2,x_3) = (0,1,\infty)$. This singles out the vertex operator at infinity, where we must  require that 
\begin{equation}
    \Gamma(z\to \infty) = (-1)^{\w_3+1} a_3^{-1} z^{\w_3} + \cdots\,, 
\end{equation}
as follows from the transformations $z\to -1/z$ and $x\to -1/x$. Moreover, we focus on genus zero covering maps since we work with the worldsheet theory on the sphere, 
hence the number of preimages at a generic point is given by $N$, defined in Eq.~\eqref{Ndef}. The covering map is then a ratio of polynomials with known degrees, i.e. 
\begin{equation}
   \Gamma(z)=\frac{p(z)}{q(z)} \, , \qquad {\rm deg}(p) = N \, , \qquad {\rm deg}(q) = N-\w_3 \, . 
\end{equation}
One can construct a second order differential equation whose independent solutions, which we denote as $f(z)$, are linear combinations of $p(z)$ and $q(z)$. This reads 
\begin{equation}
\label{diff eq Gamma 1}
  0 = - {\rm det} \left( \begin{array}{ccc}
        f & f' & f'' \\
        p & p' & p'' \\
        q & q' & q'' \\
    \end{array}\right) = (p' q - p q') f'' - (p'' q -p q'') f' + (p'' q' -p' q'') f \, .
\end{equation}
Now, consider the function $q^2 \der \Gamma = p'q - p q'$. This is a polynomial of degree ${\rm deg}(p) + {\rm deg}(q)-1 = \w_1 + \w_2 -2$, which we require to have zeros of order $\w_1-1$ and $\w_2-1$ at $z=0$ and $z=1$, respectively, see Eq.~\eqref{der Gamma}. Hence, it must take the form 
\begin{equation}
    p' q - p q' =  A z^{\w_1-1}(z-1)^{\w_2-1} \,, 
\end{equation}
for some constant $A$. This also tells us that 
\begin{equation}
p'' q -p q'' = (p' q - p q' )' = A z^{\w_1-2}(z-1)^{\w_2-2} [ (\w_1+\w_2-2)z-\w_1 +1]
    \,.
\end{equation}
It then follows from the fact that $p$ and $q$ are solutions of \eqref{diff eq Gamma 1} -- which we can assume to have no common roots -- that 
\begin{equation}
 p'' q' -p' q'' = B z^{\w_1-2} (z-1)^{\w_2-2} \,, 
\end{equation}
where   $B$ is some other constant. Inserting these expressions into \eqref{diff eq Gamma 1} and dividing by $A z^{\w_1-2} (z-1)^{\w_2-2}$ thus gives 
\begin{equation}
   z (z-1)f'' + [(\w_1-1)- (\w_1 + \w_2 - 2)z]f' + C f = 0 \, ,
\end{equation}
with $C = - B/A$. This is a differential equation of the hypergeometric type. At large $z$, we can write $f \sim z^n$ for some $n>0$. We then get the cuadratic condition 
\begin{equation}
n(n-1)-n(\w_1+\w_2-2)+ C =0 \, .    
\end{equation}
Imposing that the two solutions $n_\pm$ satisfy $n_+ - n_- = \w_3$ then fixes $C = N (N-\w_3)$. Consequently, the general solution reads 
\begin{equation}
    f = \alpha \, {}_2F_1 (-N,\w_3-N,1-\w_1;z) + \beta z^{\w_1} {}_2F_1 (\w_1-N,\w_1+\w_3-N,1+\w_1;z) \, , 
\end{equation}
where $\alpha,\beta$ are arbitrary constants. the second term must be proportional to the polynomial $p$ since it vanishes at $z=0$, while the first one has the correct degree to be identified with $q$. Fixing the normalization by imposing $\Gamma(1)=1$ finally gives\footnote{The symbol $\Gamma$ appearing on the LHS of \eqref{Map solution} denotes the covering map, while on the RHS it refers to the usual Gamma-function. The distinction should be clear from the context.} 
\begin{equation}
\label{Map solution}
    \Gamma(z) = \frac{N!(N-\w_3)!\Gamma(1-\w_1)}{w_1!f(N-\w_1)!\Gamma(N+1-\w_1 - \w_3)} \frac{z^{\w_1} {}_2F_1 (\w_1-N,\w_1+\w_3-N,1+\w_1;z)}{{}_2F_1 (-N,\w_3-N,1-\w_1;z)} \, .
\end{equation}

Before coming back to the correlators, let us look at a few examples. When all $\w_i$ are equal to $1$ we simply have the identity map $\Gamma(z)=z$. The case with $\w_1 = \w_3 = \w$ and $\w_2=1$ should give the map relevant for computing the spectrally flowed  two-point function. Indeed, at the level of the covering map, having $\w_2=1$ implies that nothing special happens at $z=z_2=1$, as should be the case if we insert the worldsheet identity operator, which can be identified $V_{00}(z)\sim V_{0}(x,z) \sim V_{\frac{k}{2} \frac{k}{2}}^{-1}(x,z)$ by means of the series identifications, see Eq.~\eqref{series Id SL2 1}. In this case the map takes the form $\Gamma(z) = z^{\w}$. As for a more non-trivial example, when $\w_1=\w_2=\w_3=3$ we get 
\begin{equation}
    \Gamma(z) = \frac{z^3 (2-z)}{2z-1}\,, 
\end{equation}
which, as expected, satisfies 
\begin{equation}
    \Gamma(z\sim 0) = -2 z^3 + \cdots \, , \quad 
    \Gamma(z\sim 1) = 1-2 (z-1)^3 + \cdots \, , \quad 
    \Gamma(z\to \infty) \sim - \frac{1}{2} z^3  \, .
\end{equation}

As pointed out above, the existence of a covering map with the above properties is related with the spectral flow selection rules of the worldsheet correlators\footnote{In general, the non-existence of a map does not imply that the flowed correlators vanish. As will be discussed in section \ref{sec:tensionless} below, the relation becomes more precise in the tensionless string limit. }. These covering maps are precisely those used in the computation of correlators in symmetric orbifold theories \cite{Lunin:2000yv}, where the charges $\w_i$ should be identified with the \textit{twists} of the corresponding operators, see Appendix \ref{sec: appD SymOrbifolds}. In such theories, three-point functions can only be non-zero if an appropriate covering map exists, i.e.~when the conditions given in Eq.~\eqref{conditionsmapwi} are satisfied. If the holographically dual theory to our worldsheet model was a symmetric orbifold, the same would be true for the string correlators in AdS$_3$. However, we find that the conditions \eqref{conditionsmapwi} are similar, but not identical to the selection rules obtained in Eq.~\eqref{x basis selection rules w}, with $n=3$. More precisely, the latter allow for correlators with $\w_i \geq 0$ as long as $w_i+w_j \geq \w_k-1$ (for all $i,j,k$), while covering maps exist only when the strict versions of both inequalities are satisfied  and the parity constraint $\sum_{i=1}^3 \w_i \in 2\mathbb{Z}+1$ holds. A similar discussion holds for higher-point functions, although in those cases the conditions on the spectral flow charges are only necessary (but not sufficient) for the existence of the corresponding covering maps.

Actually, we have already computed  the simplest examples of non-zero correlators in situations where the covering map does not exist, i.e.~correlators with unflowed insertions such as those with $\boldsymbol{\w}=(0,0,0)$ and $\boldsymbol{\w}=(1,0,0)$, and also even parity flowed correlators such as those with $\boldsymbol{\w}=(2,1,1)$.  One of the reasons for the existence of these additional correlators comes from the fact that, as we know, in the SL(2,$\R$) WZW model the spectral flow charges are not uniquely defined, at least for short strings.  We conclude that the holographic CFT is \textit{not} a symmetric orbifold, although in some sense it appears to be quite close. We will come back to this discussion several times in the remainder of these notes.

\subsubsection{Derivation for odd parity correlators}
\label{sec:oddcases}

We now show how the existence of the covering maps $\Gamma(z)$ allows us to derive the $y$-basis differential equations satisfied by the corresponding odd parity correlators $F_y$ without having to worry about all the unknown $F_m^i$  \cite{Eberhardt:2019ywk,Dei:2021xgh,Bufalini:2022toj}. Consider the following correlator: 
\begin{equation}
    G(z) = \left\langle 
    J^-(\Gamma(z),z)\,
    V_{j_1}^{\w_1}(x_1,y_1,z_1)
    V_{j_2}^{\w_2}(x_2,y_2,z_2)
    V_{j_3}^{\w_3}(x_3,y_3,z_3) \right\rangle \, .
\end{equation}
We will compute the contour integral 
\begin{equation}
    \oint_{z_i} \frac{dz \, G(z)}{(z-z_i)^{\w_i}}
        \label{ointJgammazi}
\end{equation}
in two different ways. On the one hand, by means of \eqref{coveringmapexp}, near $z=z_i$ we can write 
\begin{equation}
    J^-(\Gamma(z),z) \sim J^-(x_i,z) 
    - 2 a_i (z-z_i)^{\w_i} J^3(x_i,z)
    +a_i^2 (z-z_i)^{2\w_i} J^+(z) + \cdots.
\end{equation}
Eqs.~\eqref{JmxOPE} and \eqref{diffopsy} then give  
\begin{equation}
       \oint_{z_i} \frac{dz \, G(z)}{(z-z_i)^{\w_i}} =  \left[ 
    \left(2j_i y_i + y_i^2 \der_{y_i}\right) 
    -2 a_i  \left( j_i+\frac{k}{2}\w_i + y_i \der_{y_i}\right)
    +a_i^2 \der_{y_i}
    \right] 
    F_y\, ,
\label{ointJgammazi1}
\end{equation}
where $F_y$ was defined in Eq.~\eqref{V1V2V3xy}. On the other hand, by combining the definition of $J^-(\Gamma(z),z)$ and the OPEs in Eq.~\eqref{JVxOPE} we get  
\begin{equation}
    G(z) = \sum_{j=1}^{3}     \left\{
    -\frac{2 [\Gamma(z)-x_j](y_j \der_{y_j} + j_j + \frac{k}{2}\w_j)}{z-z_j} F_y + \sum_{m=0}^{\w_j}
    \frac{[\Gamma(z)-x_j]^2}{(z-z_j)^{m+1}} F_{y,m}^j
    \right\} \,,
    \label{ointJgammazi2}
\end{equation}    
where $F_{y,m}^i$ stands for the $y$-basis version of the $F_{m}^i$  in Eq.~\eqref{defFin}. Let us look carefully at the RHS of \eqref{ointJgammazi2}. This is a rational function of $z$ which, as implied by \eqref{coveringmapexp} in combination with the constraint equations -- the fact that the RHS of Eqs.~\eqref{Gjdef} must satisfy \eqref{Gjconditions} -- has zeros of order $\w_i-1$ at all $z_i$. It also has double poles at the $N$ simple poles of $\Gamma(z)$ due to the $\Gamma(z)^2$ factors. Moreover,  it further goes to zero as $z^{-2}$ for $z\to \infty$ since the coefficient of the putative ${\cal O}\left(z^{-1}\right)$ contribution is given by 
\begin{eqnarray}
    &&\sum_{j=1}^{3}  \left\{
    - \left[2 [\Gamma_\infty-x_j]\left(y_j \der_{y_j} + j_j + \frac{k}{2}\w_j\right)\right] F_y+ 
    [\Gamma_\infty-x_j]^2 F_{y,0}^j
    \right\} \\
    && \hspace{-0.3cm} =  \sum_{j=1}^{3}  \left\{
    - \left[2 [\Gamma_\infty-x_j]\left(y_j \der_{y_j} + j_j + \frac{k}{2}\w_j\right)\right]  + 
    [\Gamma_\infty-x_j]^2 \der_{x_j}
    \right\} F_y \nn \\
    && \hspace{-0.3cm} =  \sum_{j=1}^{3}  \left\{ \Gamma_{\infty}^2 \der_{x_j} -2
    \Gamma_{\infty} \left( y_j \der_{y_j} + j_j + \frac{k}{2}\w_j + x_j\der_{x_j} \right) +
    x_j^2 \der_{x_j} + 2 \left( y_j \der_{y_j} + j_j + \frac{k}{2}\w_j  \right) x_j 
    \right\} F_y \nn 
\end{eqnarray}
which vanishes due to the global Ward identities (recall that $\left( y_j \der_{y_j} + j_j + \frac{k}{2}\w_j \right)$ is the $y$-basis analogue of $h_j$). These are exactly the properties that, up to a multiplicative factor,  uniquely define the derivative of the covering map in Eq.~\eqref{der Gamma}. Hence, we find 
\begin{equation}
    G(z) = \alpha \, \der \Gamma(z) \, , 
\end{equation}
where $\alpha$ must be independent of $z$. 
Since $\der \Gamma(z \sim z_i) = a_i \w_i (z-z_i)^{w_i-1}$, it follows that  
\begin{equation}
\oint_{z_i} \frac{dz \, G(z)}{(z-z_i)^{\w_i}} = \alpha \, a_i \w_i  \,.   \label{RHSodd}
\end{equation}
The coefficient $\alpha$ can be computed explicitly as follows \cite{Eberhardt:2019ywk}. The function $-\Gamma(z)^{-1}\der \Gamma(z)$ clearly has  unit residues at all $\lambda_a$. This implies that
\begin{equation}
    \alpha N = -\sum_{a=1}^N \oint_{\lambda_a} dz \frac{G(z)}{\Gamma(z)} = -\sum_{a=1}^N \sum_{j=1}^3 \sum_{m=0}^{\w_j}\oint_{\lambda_a} dz \frac{\Gamma(z)}{(z-z_i)^{m+1}} F_{y,m}^j\,, 
\end{equation}
where we have used that only the terms proportional to $\Gamma(z)^2$ in \eqref{ointJgammazi2} contribute. By turning the contour around and using the defining properties of the covering map (there is no residue at infinity since $\Gamma^{-1} G (z\to \infty) \sim z^{-2}$), we get
\begin{equation}
    \alpha N = \sum_{j=1}^3 \sum_{m=0}^{\w_j} 
    \left[ \oint_{z_j} dz \frac{x_j + a_j (z-z_j)^{\w_j}+\cdots}{(z-z_j)^{m+1}}F_{y,m}^j\right] 
 = \sum_{j=1}^3     \left[
    x_j  F_{y,0}^j + a_j F_{y,\w_j}^j
    \right] \, .
\end{equation}
Consequently, using the global Ward identities once more we find  
\begin{equation}
    \alpha = \frac{1}{N} \sum_{j=1}^3 \left(x_j \der_{x_j} + a_j\der_{y_j}\right) F_y = \frac{1}{N}\sum_{j=1}^3\left[
    (a_j - y_j)\der_{y_j} - j_j - \frac{k}{2}\w_j \right] F_y
    \, . 
\end{equation}
By combining Eqs.~\eqref{ointJgammazi1} and \eqref{RHSodd} and fixing the insertion points as in \eqref{Vw1w2w301infty} we finally conclude that 
\begin{equation}
 \left\{    (y_i-a_i)^2 \der_{y_i} 
    +2 j_i (y_i-a_i) +
    \frac{a_i \w_i}{N}\left[ 
    \sum_{j=1}^3\left(
    (y_j-a_j)\der_{y_j} + j_j \right) + \frac{k}{2} \right] \right\} \hat{F}_y = 0,
\label{recursionywithmap}
\end{equation}
for $i=1,2,3$, and where now the $a_i$ are purely numerical. 

This system of first order equations, which fits the structure given in Eq.\eqref{GenEqsEven}, is easily solved and fixes the dependence of $\hat{F}_y$ on $y_1$, $y_2$ and $y_3$ completely. Ignoring the overall $y_i$-independent normalization for now, we obtain 
\begin{align}
\langle V_{j_1}^{\w_1}(y_1)
    V_{j_2}^{\w_2}(y_2)
    V_{j_3}^{\w_3}(y_3) \rangle_{\rm odd} & \sim (y_1-a_1)^{-2j_1}
    (y_2-a_2)^{-2j_2}
    (y_3-a_3)^{-2j_3} \label{oddfinal2} \\
    &\quad \times \left(
    \w_1\frac{y_1+a_1}{y_1-a_1}
    +\w_2\frac{y_2+a_2}{y_2-a_2}
    +\w_3\frac{y_3+a_3}{y_3-a_3} -1
    \right)^{\frac{k}{2}-j_1-j_2-j_3}. \nn 
\end{align}  
To make the connection with Eq.~\eqref{3pt-odd-parity} we note that, while the $P_{\boldsymbol{\w}}$ defined in \eqref{Pw-definition} are somewhat complicated, their ratios are actually very simple. For instance, consider the $X_1$ term appearing in \eqref{3pt-odd-parity}. From Eqs.~\eqref{X_I-3pt} and \eqref{Pw-definition}, and up to an overall sign, we have 
\begin{equation}
    X_1 = P_{\boldsymbol{\w}-e_1} y_1 + P_{\boldsymbol{\w}+e_1} = P_{\boldsymbol{\w}-e_1} \left(y_1+\frac{P_{\boldsymbol{\w}+e_1}}{P_{\boldsymbol{\w}-e_1}}\right) = 
    P_{\boldsymbol{\w}-e_1} \left(y_1-a_1\right) \, .
\end{equation}
A similar result holds for $X_2$ and $X_3$, while $X_{123}$ can also be simplified in this manner. The $y$-dependence of the expression in \eqref{oddfinal2} is thus precisely that of Eq.~\eqref{3pt-odd-parity}. 

We note that, as is evident from Eq.~\eqref{oddfinal2}, $y$-basis three-point functions diverge whenever a variable $y_i$ approaches the corresponding $a_i$. In this sense, the coefficients of the covering map define very special points in the $y$-hyperplane. We will come back to this in Sections \ref{sec:tensionless} and \ref{sec: holography} when discussing the comparison with the correlators of the putative holographically dual theory.

%%%%%%%%%%%%%%%%%%%5
\subsubsection{Series identifications and even parity correlators}
\label{sec:evenparitycases}

Let us now consider even parity correlators, satisfying 
\begin{equation}
    \sum_{i=1}^3 \w_i \in 2\mathbb{Z} 
    \qqquad 
    \sum_{i=1}^3 \w_i > 2 {\rm Max}(\w_i) - 1 
    \qqquad \w_i> 0\,, \,\, \forall \, i \, ,  \label{wiEvenCases}
\end{equation}
for which it is not possible to construct a holomorphic covering map satisfying all the relevant properties. For concreteness, in the following we also take $\w_3 \geq \w_{1,2}$. 

As discussed below Eq.\eqref{GenEqsEven}, for deriving the $y$-basis differential equations it is enough to focus on the discrete sector of the theory. As it turns out, in this sector even and odd parity cases are very closely related. This is due to the series identifications among flowed representations with charges shifted upwards or downwards by one unit. Indeed, from Eq.~\eqref{series Id SL2 1} one can derive the following identities for $y$-basis operators  \cite{Iguri:2022eat}
\begin{equation}
 V_j^\w(x,y=0,z) =  {\cal{N}}(j)
 \lim_{y\to \infty}  y^{k-2j} V_{\frac{k}{2}-j}^{\w+1}(x,y,z),
 \label{seriesidybasis1}
\end{equation}
and 
\begin{equation}
 \lim_{y\to \infty}  y^{2j} V_j^\w(x,y,z) = {\cal{N}}(j)
 V_{\frac{k}{2}-j}^{\w-1}(x,y=0,z), 
 \label{seriesidybasis2}
\end{equation} 
with ${\cal{N}}(j)$ defined in Eq.~\eqref{Ndef}.  
This means that all even parity correlators can be linked to (at least) three \textit{different} situations where a covering map \textit{does} exists. More explicitly, given a set of charges  ($\w_1,\w_2,\w_3$) satisfying \eqref{wiEvenCases}, it follows that the adjacent assignments $(\w_1+1,\w_2,\w_3)$, $(\w_1,\w_2+1,\w_3)$ and $(\w_1,\w_2,\w_3-1)$ satisfy all conditions in \eqref{conditionsmapwi}. We denote the corresponding covering maps as follows:
\begin{equation}
    \Gamma_1^+ \equiv \Gamma[\w_1+1,\w_2,\w_3] \qqquad 
    \Gamma_2^+ \equiv \Gamma[\w_1,\w_2+1,\w_3] \qqquad 
    \Gamma_3^- \equiv \Gamma[\w_1,\w_2,\w_3-1]\,.
    \label{3maps1}
\end{equation}
Thus, we have  
\begin{subequations}
\begin{align}
&\langle V_{j_1}^{\w_1}(0)
    V_{j_2}^{\w_2}(y_2)
    V_{j_3}^{\w_3}(y_3) \rangle = \lim_{y_1\to \infty} y_1^{k-2j_1} {\cal{N}}(j_1) 
    \langle V_{\frac{k}{2}-j_1}^{\w_1+1}(y_1)
    V_{j_2}^{\w_2}(y_2)
    V_{j_3}^{\w_3}(y_3) \rangle, \\[1ex]
&\langle V_{j_1}^{\w_1}(y_1)
    V_{j_2}^{\w_2}(0)
    V_{j_3}^{\w_3}(y_3) \rangle = \lim_{y_2\to \infty} y_2^{k-2j_2} {\cal{N}}(j_2) 
    \langle V_{j_1}^{\w_1}(y_1)
    V_{\frac{k}{2}-j_2}^{\w_2+1}(y_2)
    V_{j_3}^{\w_3}(y_3) \rangle, \\[1ex]
&\lim_{y_3\to \infty} y_3^{2j_3}\langle        V_{j_1}^{\w_1}(y_1)
    V_{j_2}^{\w_2}(y_2)
    V_{j_3}^{\w_3}(y_3) \rangle = {\cal{N}}(j_3) 
    \langle V_{j_1}^{\w_1}(y_1)
    V_{j_2}^{\w_2}(y_2)
    V_{\frac{k}{2}-j_3}^{\w_3-1}(0) \rangle.
\end{align}
\label{3mapsCorrs1}
\end{subequations}
In many cases we can also shift the spectral flow charges in the opposite direction, leading to the maps  
\begin{equation}
    \Gamma_1^- \equiv \Gamma[\w_1-1,\w_2,\w_3] \qqquad 
    \Gamma_2^- \equiv \Gamma[\w_1,\w_2-1,\w_3] \qqquad 
    \Gamma_3^+ \equiv \Gamma[\w_1,\w_2,\w_3+1]\,, 
       \label{3maps2}
\end{equation}
so that 
\begin{subequations}
\begin{align}
&\lim_{y_1\to \infty} y_1^{2j_1}\langle V_{j_1}^{\w_1}(y_1)
    V_{j_2}^{\w_2}(y_2)
    V_{j_3}^{\w_3}(y_3) \rangle = {\cal{N}}(j_1) 
    \langle V_{\frac{k}{2}-j_1}^{\w_1-1}(0)
    V_{j_2}^{\w_2}(y_2)
    V_{j_3}^{\w_3}(y_3) \rangle \\[1ex]
&\lim_{y_2\to \infty} y_2^{2j_2}\langle V_{j_1}^{\w_1}(y_1)
    V_{j_2}^{\w_2}(y_2)
    V_{j_3}^{\w_3}(y_3) \rangle = {\cal{N}}(j_2) 
    \langle V_{j_1}^{\w_1}(y_1)
    V_{\frac{k}{2}-j_2}^{\w_2-1}(0)
    V_{j_3}^{\w_3}(y_3) \rangle,  \\[1ex]
&\langle V_{j_1}^{\w_1}(y_1)
    V_{j_2}^{\w_2}(y_2)
    V_{j_3}^{\w_3}(0) \rangle = \lim_{y_3\to \infty} y_3^{k-2j_3} {\cal{N}}(j_3) 
    \langle V_{j_1}^{\w_1}(y_1)
    V_{j_2}^{\w_2}(y_2)
    V_{\frac{k}{2}-j_3}^{\w_3+1}(y_3) \rangle. 
\end{align}
\label{3mapsCorrs2}
\end{subequations}
All expressions on the RHS of Eqs.~\eqref{3mapsCorrs1} and \eqref{3mapsCorrs2} are limits of correlators we have already computed (up to an overall constant) in the previous section. In particular, we know that they satisfy the appropriate limits of the differential equations given in \eqref{recursionywithmap}. For instance, $\langle V_{\frac{k}{2}-j_1}^{\w_1-1}(0)
V_{j_2}^{\w_2}(y_2) V_{j_3}^{\w_3}(y_3) \rangle$ is annihilated by the differential operators
\begin{align}
    \begin{aligned}
    & y_2 (y_2 \der_{y_2}+2j_2) + (\w_1 - \w_2 - \w_3)^{-1} \big\{ (\w_1 + \w_2 - \w_3)a_2[\Gamma_1^-]^2 \der_{y_2} \\ 
    &  - 2 a_2[\Gamma_1^-] \left[ (\w_1-\w_3)(j_2+y_2 \der_{y_2}) + \w_2(j_3+y_3 \der_{y_3} - a_3[\Gamma_1^-] \der_{y_3} + j_1+j_3) \right]\big\} 
    \end{aligned}
    \label{ex1a}
\end{align}
and
\begin{align}
    \begin{aligned}
    &y_3 (y_3 \der_{y_3}+2j_3) + (\w_1 - \w_2 - \w_3)^{-1} \big\{ (\w_1 - \w_2 + \w_3)a_3[\Gamma_1^-]^2 \der_{y_3} \\ 
    &- 2 a_3[\Gamma_1^-] \left[ (\w_1-\w_2)(j_3+y_3 \der_{y_3}) + \w_3(j_2+y_2 \der_{y_2} - a_2[\Gamma_1^-] \der_{y_2} + j_1+j_2) \right] \big\},  
    \end{aligned}
    \label{ex1b}
\end{align}
where $a_i[\Gamma_j^\pm]$ denotes the coefficient $a_i$ of the map $\Gamma_j^\pm$. The main point is that these operators must coincide with the appropriate limits of those provided in Eqs.~\eqref{GenEqsEven} for the original even parity correlator. In our example, these read 
\begin{align}
    \begin{aligned}
    y_2 (y_2 \der_{y_2} + 2j_2) + A_{22} y_2 \der_{y_2}
    + A_{23} y_3 \der_{y_3} - B_{22} \der_{y_2} - B_{23} \der_{y_3} - 2 A_{21} j_1 + C_2, \\
    y_3 (y_3 \der_{y_3} + 2j_3) + A_{33} y_3 \der_{y_3}
    + A_{32} y_2 \der_{y_2} - B_{33} \der_{y_3} - B_{32} \der_{y_2} - 2 A_{31} j_1 + C_3.
    \end{aligned}
\end{align}
Proceeding similarly with the rest of the 12 relations, one finds a total of 60 conditions.  21 of these 60 conditions can be used to solve explicitly for all the coefficients $A_{ij}$, $B_{ij}$ and $C_i$ in \eqref{GenEqsEven}. The fact that the remaining 39 identities are then  satisfied provides a highly non-trivial check of procedure, and follows from the identities relating the $a_i$ coefficients of the different covering maps involved.

There are many equivalent ways to write the resulting coefficients. The simplest one leads to the following differential equations for even parity correlators:  
\begin{align}
    \begin{aligned}
   &\Big\{ \left(y_1-a_1[\Gamma_3^-]\right)^2 \der_{y_1} + 
    2 j_1 \left(y_1-a_1[\Gamma_3^-]\right) 
    + \frac{2 a_1[\Gamma_3^-] \w_1}{\w_1 + \w_2 - \w_3} \Big[
    (y_1 - a_1[\Gamma_3^-]) \der_{y_1} + j_1  \\
    &\, + \,  (y_2 - a_2[\Gamma_3^-]) \der_{y_2} + j_2
    -(y_3 - a_3[\Gamma_2^-]) \der_{y_3} - j_3
    \Big] \Big\} \, \langle V_{j_1}^{\w_1}(y_1)
    V_{j_2}^{\w_2}(y_2)
    V_{j_3}^{\w_3}(y_3) \rangle_{\rm even} = 0\, ,      
    \end{aligned}
    \label{Eq1even}
\end{align}
\begin{align}
    \begin{aligned}
   &\Big\{ \left(y_2-a_2[\Gamma_3^-]\right)^2 \der_{y_2} + 
    2 j_2 \left(y_2-a_2[\Gamma_3^-]\right) 
    + \frac{2 a_2[\Gamma_3^-] \w_2}{\w_1 + \w_2 - \w_3} \Big[
    (y_1 - a_1[\Gamma_3^-]) \der_{y_1} + j_1  \\
    &\, + \,  (y_2 - a_2[\Gamma_3^-]) \der_{y_2} + j_2
    -(y_3 - a_3[\Gamma_1^-]) \der_{y_3} - j_3
    \Big] \Big\} \, \langle V_{j_1}^{\w_1}(y_1)
    V_{j_2}^{\w_2}(y_2)
    V_{j_3}^{\w_3}(y_3) \rangle_{\rm even} = 0\,,     
    \end{aligned}
    \label{Eq2even}
\end{align}
\begin{align}
    \begin{aligned}
   &\Big\{ \left(y_3-a_3[\Gamma_1^-]\right)^2 \der_{y_3} + 
    2 j_3 \left(y_3-a_3[\Gamma_1^-]\right) 
    + \frac{2 a_3[\Gamma_1^-] \w_3}{\w_1 - \w_2 - \w_3} \Big[
    (y_1 - a_1[\Gamma_2^-]) \der_{y_1} + j_1  \\
    &\, - \,  (y_2 - a_2[\Gamma_1^-]) \der_{y_2} - j_2
    -(y_3 - a_3[\Gamma_1^-]) \der_{y_3} - j_3
    \Big] \Big\} \, \langle V_{j_1}^{\w_1}(y_1)
    V_{j_2}^{\w_2}(y_2)
    V_{j_3}^{\w_3}(y_3) \rangle_{\rm even} = 0 \,.     
    \end{aligned}
    \label{Eq3even}
\end{align}
Up to an overall constant, the general solution then takes the form 
\begin{eqnarray}
\label{eq: general even solution}
    \langle V_{j_1}^{\w_1}(y_1)
    V_{j_2}^{\w_2}(y_2)
    V_{j_3}^{\w_3}(y_3) \rangle_{\rm even} & \sim &
    \left(
    1-\frac{y_2}{ a_2[\Gamma_3^+]} -\frac{y_3}{a_3[\Gamma_2^+]}+
    \frac{y_2 y_3}{a_2[\Gamma_3^-] a_3[\Gamma_2^+] }\right)^{j_1-j_2-j_3} \nn \\
    & \times &   \left(
    1-\frac{y_1}{a_1[\Gamma_3^+]}-\frac{y_3}{a_3[\Gamma_1^+]} + \frac{y_1 y_3}{ a_1[\Gamma_3^-]  a_3[\Gamma_1^+]}
    \right)^{j_2-j_3-j_1} \\
    &\times &  \left(
    1-\frac{y_1}{a_1[\Gamma_2^+]} -\frac{y_2}{a_2[\Gamma_1^+]} +\frac{y_1 y_2}{a_1[\Gamma_2^+]a_2[\Gamma_1^-]} \right)^{j_3-j_1-j_2}. \nn
\end{eqnarray}
As in the odd case, one can check that this exactly reproduces the $y$-dependence of \eqref{3pt-even-parity}.

%%%%%%%%%%%%%%%%%%%%%%%%%%%%%%%%%%%%%%%%%%%%%%%
\subsubsection{Edge cases, unflowed insertions, and normalization}
\label{sec: edge cases}

Given that all non-zero correlation functions must satisfy the selection rules in Eq.~\eqref{x basis selection rules w}, it remains to consider the edge cases, namely the three-point functions with \begin{eqnarray}
\label{wiedgecases}
    \w_3 = \w_1+\w_2 \quad \text{or} \quad  \w_3 = \w_1+\w_2 +1\, ,   \qquad \w_i\geq 1 \, , \, \forall \, i\, , 
\end{eqnarray}
where we have assumed $\w_3 \geq \w_{1,2}$ for simplicity. We refer to these as the even and odd edge cases, respectively. Below we will also discuss correlators with unflowed insertions. 

Edge cases must be treated carefully as several of the $a_i$ coefficients either vanish or diverge in this limit. Fortunately, alternative techniques involving current insertions are available in these situations \cite{Bufalini:2022toj}. 
In order to see how this works, we start by unfixing the middle insertion. By means of the global Ward identities \eqref{ybasisx1x2x3fixing} we have 
\begin{align}
\label{eq: Mobiusunfix}
    & \braket{
    V^{\w_1}_{j_1}(0,y_1,0)
    V^{\w_2}_{j_2}(x,y_2,1)
    V^{\w_3}_{j_3}(\infty,y_3,\infty)  } \nn\\
     & = x^{-j_1 - j_2 + j_3 + \frac{k}{2}(-\w_1-\w_2 + \w_3)}
     \braket{
     V^{\w_1}_{j_1}\left(0,\frac{y_1}{x},0\right)
     V^{\w_2}_{j_2}\left(1,\frac{y_2}{x},1\right)
     V^{\w_3}_{j_3}\left(\infty,y_3 x,\infty \right)  } \, . 
\end{align}
In the limit $x\to 0$, all vertex operators become $m$-basis flowed primaries. Given that we have $    |\w_1 + \w_2 - \w_3| \le 1$, the edge cases correspond precisely to the configurations for which we expect to have a finite, non-zero $m$-basis correlator! 
This corresponds to the three-point functions studied in \cite{Maldacena:2001km, Cagnacci:2013ufa}, where they were denoted as spectral flow \textit{conserving} and spectral flow \textit{violating} three-point functions, depending on the overall parity of the total spectral flow charge. 
We will derive some of the relevant differential equations satisfied by the edge correlators in the above collision limit, and only recover the full correlators at the end. 

The connection can be made more explicit by deriving a constraint that will be satisfied in both edge cases. We set $x_1 = x_2 = x$ and consider the integral 
\begin{equation}
\oint_{{\cal{C}}} \braket{
J^3(x,z)
V^{\w_1}_{j_1}(x,y_1,z_1)
V^{\w_2}_{j_2}(x,y_2,z_2)
V^{\w_3}_{j_3}(x_3,y_3,z_3)  }dz \, ,  
\end{equation}
where ${\cal{C}}$ denotes a contour encircling all three insertion points. This vanishes since there is no residue at infinity. Turning the countour around, using the OPEs \eqref{JwVy}, fixing the worldsheet insertions to $(0,1,\infty)$, and further consider the limit $(x,x_3)\to (0,\infty)$, we find 
\begin{align}
\begin{aligned}
\label{eq: charge conservation general}
    &\Big[
    y_1 \del_{y_1}   + y_2 \del_{y_2}  - y_3 \del_{y_3} + j_1 + j_2 - j_3  \\ 
    & \qquad + \frac{k}{2} (\w_1 + \w_2 - \w_3)
    \Big] \braket{
V^{\w_1}_{j_1}(0,y_1,0)
V^{\w_2}_{j_2}(0,y_2,1)
V^{\w_3}_{j_3}(\infty,y_3,\infty)  } =0 \, . 
\end{aligned}
\end{align}
This is nothing but the usual $m$-basis charge conservation condition written $y$-basis language. 

For the even edge cases, we can proceed similarly with the integrals 
\begin{equation}
\label{eq: Jminusinsertedeveneasy}
\oint_{{\cal{C}}} \braket{J^-(x_3,z)  
V^{\w_1}_{j_1}(x_1,y_1,z_1)
V^{\w_2}_{j_2}(x_2,y_2,z_2)
V^{\w_3}_{j_3}(x_3,y_3,z_3)  } \frac{(z-z_1)^{\w_1} (z-z_2)^{\w_2}}{(z-z_3)^{\w_3}} dz 
\end{equation}
and
\begin{equation}
    \label{eq: Jminusinsertedevencollision}
    \oint_{{\cal{C}}}  \braket{J^-(x,z) 
    V^{\w_1}_{j_1}(x,y_1,z_1)
    V^{\w_2}_{j_2}(x,y_2,z_2)
    V^{\w_3}_{j_3}(x_3,y_3,z_3)  } \frac{(z-z_3)^{\w_3}}{(z-z_1)^{\w_1}(z-z_2)^{\w_2} }  dz \, . 
\end{equation}
Setting $(z_1,z_2,z_3)=(0,1,\infty)$ while sending $ x \to 0$ and $x_3 \to \infty$ as before, we obtain 
\begin{align}
    \begin{aligned}
    \label{eq: three eq even}
      0 &=  \Big[  (-1)^{\w_1}\del_{y_1}  + (-1)^{\w_3}\del_{y_2} + (y_3^2 \del_{y_3} + 2j_3 y_3)    \Big] \braket{\dots}  \, , \\
      0 &=  \Big[(-1)^{\w_1} (y_1^2 \del_{y_1} + 2j_1 y_1) + (-1)^{\w_3}(y_2^2 \del_{y_2} + 2j_2 y_2) +  \del_{y_3}   \Big] \braket{\dots} \, .  
    \end{aligned}
\end{align}
where $\braket{\dots}$ stands for $\braket{
V^{\w_1}_{j_1}(0,y_1,0)
V^{\w_2}_{j_2}(0,y_2,1)
V^{\w_3}_{j_3}(\infty,y_3,\infty)  }$ with $\w_3 = \w_1+\w_2$.
Up to an overall constant, the solution reads
\cite{Dei:2021xgh}
\begin{align}
    &\braket{
    V^{\w_1}_{j_1}(0,y_1,0)
    V^{\w_2}_{j_2}(0,y_2,1)
    V^{\w_3= \w_1+\w_2}_{j_3}(\infty,y_3,\infty)  }  \\
    & \sim  ((-1)^{\w_1} y_1 -  (-1)^{\w_3}y_2)^{j_3 -j_1 - j_2 }
    (1 + (-1)^{\w_3}y_2 y_3 )^{j_1 - j_2 - j_3} 
    (1 + (-1)^{\w_1}y_1 y_3)^{ j_2  - j_1  - j_3} \, . \nn 
\end{align}
For the odd edge cases, we make use of the integrals  \begin{equation}
\label{eq: Jminusinsertedodd1}
\oint_{{\cal{C}}} \braket{J^-(x_3,z)  V^{\w_1}_{j_1}(x_1,y_1,z_1)
V^{\w_2}_{j_2}(x_2,y_2,z_2)
V^{\w_3}_{j_3}(x_3,y_3,z_3)  } \frac{(z-z_1)^{\w_1+1} (z-z_2)^{\w_2}}{(z-z_3)^{\w_3}} dz \, , 
\end{equation}
and
\begin{equation}
\label{eq: Jminusinsertedodd2}\oint_{{\cal{C}}} \braket{J^-(x_3,z)  V^{\w_1}_{j_1}(x_1,y_1,z_1)
V^{\w_2}_{j_2}(x_2,y_2,z_2)
V^{\w_3}_{j_3}(x_3,y_3,z_3)  } \frac{(z-z_1)^{\w_1} (z-z_2)^{\w_2+1}}{(z-z_3)^{\w_3}} dz \, , 
\end{equation}
leading to 
\begin{align}
    \begin{aligned}
    \label{eq: three eq odd}
    0 & = \Big[ (-1)^{\w_3} \del_{y_2} +  (y_3^2 \del_{y_3} + 2 j_3 y_3  )   \Big] \braket{\dots}  \, , \\
    0 & = \Big[ (-1)^{\w_1} \del_{y_1} + (y_3^2 \del_{y_3} + 2 j_3 y_3  )   \Big] \braket{\dots}\, .  
    \end{aligned}
\end{align}
where $\braket{\dots}= \braket{
V^{\w_1}_{j_1}(0,y_1,0)
V^{\w_2}_{j_2}(0,y_2,1)
V^{\w_3}_{j_3}(\infty,y_3,\infty)  }$ with $\w_3 = \w_1+\w_2+1$. In this case we find 
\begin{align}
    &\braket{
    V^{\w_1}_{j_1}(0,y_1,0)
    V^{\w_2}_{j_2}(0,y_2,1)
    V^{\w_3= \w_1+\w_2+1}_{j_3}(\infty,y_3,\infty)  } \\
    & \sim
    y_3^{j_1 + j_2 - j_3 - \frac{k}{2}} 
    (1 + (-1)^{\w_1}y_1 y_3 + (-1)^{\w_3}y_2 y_3 )^{\frac{k}{2} - j_1 - j_2 - j_3} \, , \nn 
\end{align}
The precise form of the edge correlators away from the collision limit can be obtained by using the global Ward identities. In the odd edge cases both equations in Eq.~\eqref{eq: three eq odd} hold away for arbitrary $x_2$, but  \eqref{eq: charge conservation general} receives corrections. On the other hand, for the even edge cases only the first constraint in \eqref{eq: three eq even} remains valid. One obtains
\begin{eqnarray}
\label{eq: even edge solution}
& \langle 
    V_{j_1}^{\w_1}(y_1)
    V_{j_2}^{\w_2}(y_2)
    V_{j_3}^{\w_3}(y_3) \rangle_{\rm even}  & \sim   \left(
    1 + (-1)^{\w_1}\frac{(\w_1 + \w_2-1)!}{(\w_1-1)!\w_2!}y_3
    +
    (-1)^{\w_3}y_2 y_3 \right)^{j_1-j_2-j_3} 
    \nn \\
    &\times & \hspace{-2cm} 
     \left(
    1 + (-1)^{\w_1+1} \frac{(\w_1 + \w_2-1)!}{\w_1!(\w_2-1)!}
    y_3 + (-1)^{\w_1}y_1 y_3
    \right)^{j_2-j_3-j_1} \\
    &\times &  \hspace{-2cm}
   \left(
    1 + (-1)^{\w_1+1} \frac{\w_1 !\w_2!}{(\w_1+\w_2)!}((-1)^{\w_1} y_1 - (-1)^{\w_3} y_2) \right)^{j_3-j_1-j_2} \, , \nn 
\end{eqnarray}
and 
\begin{align}
\label{eq: general odd edge}
    &\braket{
    V^{\w_1}_{j_1}(y_1)
    V^{\w_2}_{j_2}(y_2)
    V^{\w_3}_{j_3}(y_3)  }_{\rm odd} \sim y_3^{j_1 + j_2 - j_3 - \frac{k}{2}} \\
    &\qquad \times 
    \left(1 + (-1)^{\w_1+1} \frac{(\w_1 + \w_2)!}{\w_1! \w_2!}  y_3 + (-1)^{\w_1}y_1 y_3 + (-1)^{\w_3}y_2 y_3 \right)^{\frac{k}{2} - j_1 - j_2 - j_3} \, . \nn 
\end{align}
One can check that these expressions match the $y$-dependence given Eqs.~\eqref{3pt-even-parity} and \eqref{3pt-odd-parity}. 

\medskip

We have thus derived the $y$-dependence of all correlators where all three insertions are spectrally flowed. By using the series identifications once more, we can also derive three-point functions with unflowed insertions. For instance, for $\w_1=0$ we have 
\begin{equation} 
\langle V_{j_1}
    V_{j_2}^{\w}(y_2)
    V_{j_3}^{\w + 1}(y_3) \rangle =  
    {\cal{N}}(j_1)
     \lim_{y_1\to \infty}
     y_1^{k-2j_1}
     \langle 
     V_{\frac{k}{2}-j_1}^1(y_1)
    V_{j_2}^{\w}(y_2)
    V_{j_3}^{\w + 1}(y_3) \rangle \, ,
    \label{seriesid0ww}
\end{equation}
and 
\begin{equation} 
\langle V_{j_1}
    V_{j_2}^{\w}(y_2)
    V_{j_3}^{\w}(y_3) \rangle =  
    {\cal{N}}(j_1)
     \lim_{y_1\to \infty}
     y_1^{k-2j_1}
     \langle 
     V_{\frac{k}{2}-j_1}^1(y_1)
    V_{j_2}^{\w}(y_2)
    V_{j_3}^{\w}(y_3) \rangle
    \, ,   
    \label{seriesid0ww+1}
\end{equation}
where we have abbreviated $V_{j_1}(0,0)\equiv V_{j_1}$. Hence, up to the overall constant we find 
\begin{equation}
\label{eq: oneunflowed1}
    \langle V_{j_1}
    V_{j_2}^{\w}(y_2)
    V_{j_3}^{\w+1}(y_3) \rangle \sim  y_3^{j_1 + j_2 - j_3 - \frac{k}{2}} \left( 1 - y_3 - (-1)^\w y_2 y_3 \right)^{\frac{k}{2} - j_1 - j_2 - j_3} \, . 
\end{equation}
and 
\begin{equation}
    \langle V_{j_1}
    V_{j_2}^{\w}(y_2)
    V_{j_3}^{\w }(y_3) \rangle \sim 
    \left( y_2 + (-1)^\w  \right)^{j_3 - j_1 - j_2}
    \left( y_3 - 1 \right)^{j_2 - j_1 - j_3}
    \left(  (-1)^{\w} +  y_2 y_3 \right)^{j_1 - j_2 - j_3} \, . 
\end{equation}
Finally, for correlators with two unflowed insertions we simply recover  Eq.~\eqref{Fy result 100}.
% \begin{align}
%     \langle V_{j_1}
%     V_{j_2}
%     V_{j_3}^{1}(y_3) \rangle & =  
%     {\cal{N}}(j_2)
%      \lim_{y_2\to \infty}
%      y_2^{k-2j_2}
%      \langle 
%      V_{j_1}
%     V_{\frac{k}{2}-j_2}^{1}(y_2)
%     V_{j_3}^{1}(y_3) \rangle \nn \\
% & \sim y_3^{j_1 + j_2 - j_3 - \frac{k}{2}} (y_3 - 1)^{\frac{k}{2}-j_1 - j_2 - j_3}
% \, , 
% \end{align}
Upon integrating over $y_i$, one finds that all of these results reproduce the original computations of \cite{Maldacena:2001km,Cagnacci:2013ufa}. Hence, we have established the $y$-dependence for all non-trivial spectrally flowed three-point functions.  

\medskip

We end this section by  fixing  the overall $y$-independent normalizations \cite{Iguri:2022eat}, which confirms the validity of the formula conjectured in \cite{Dei:2021xgh}. Once again, the argument relies on the SL(2,$\R$) series identifications. Indeed, the relations given in Eqs.~\eqref{3mapsCorrs1} and \eqref{3mapsCorrs2} must hold including the overall factors. This allows us to  determine the structure constants recursively, starting from the unflowed three-point functions of \cite{Teschner:1999ug,Maldacena:2001km}. For instance, we first focus on the identity  
    \begin{equation}
     \lim_{y_3\rightarrow \infty} y_3^{2j_3} \left\langle V^{\w_1}_{j_1}(y_1) \,  V^{\w_2}_{j_2}(y_2) \, V^{\w_3}_{j_3}(y_3) \right\rangle = 
    {\cal{N}}(j_3)\left\langle V^{\w_1}_{j_1}(y_1) \,  V^{\w_2}_{j_2}(y_2) \, V^{\w_3-1}_{\frac{k}{2}-j_3}(0) \right\rangle\, .
    \label{identifC}
\end{equation}
Since we are interested in the $y$-independent factors we are free to set $y_1=y_2=0$. According to Eqs.\eqref{3pt-odd-parity} and \eqref{3pt-even-parity}, the product of $X_I$ factors on the left- an right-hand sides of \eqref{identifC} then give either 
\begin{equation}
    P_{\boldsymbol{\w}}^{j_1+j_2+j_3-k} \, P_{\boldsymbol{\w}+e_1+e_2}^{j_3-j_1-j_2}
    \,
    P_{\boldsymbol{\w}+e_2-e_3}^{j_1-j_2-j_3}
    \,
    P_{\boldsymbol{\w}+e_1-e_3}^{j_2-j_3-j_1},
\end{equation}
or 
\begin{equation}
    P_{\boldsymbol{\w}+e_1+e_2-e_3}^{\frac{k}{2}-j_1-j_2-j_3} \,     
    P_{\boldsymbol{\w}+e_1}^{-j_1+j_2+j_3-\frac{k}{2}}
    \, 
    P_{\boldsymbol{\w}+e_2}^{j_1-j_2+j_3-\frac{k}{2}}
    \,
    P_{\boldsymbol{\w}-e_3}^{j_1+j_2-j_3-\frac{k}{2}}\, ,
\end{equation}
depending on the overall parity. In both cases, this implies 
\begin{equation}
    C_{\boldsymbol{\w}}(j_1,j_2,j_3) =
    {\cal{N}}(j_3) C_{\boldsymbol{\w}-e_3}\left(j_1,j_2,\frac{k}{2}-j_3\right). 
\end{equation}
Analogous constraints can be derived by shifting $\w_1$ and $\w_2$ instead. These are consistent with the identities 
\begin{equation}
    {\cal{N}}(j_1) C\left(\frac{k}{2}-j_1,j_2,j_3 \right) = 
    {\cal{N}}(j_2) C\left(j_1,\frac{k}{2}-j_2,j_3 \right) = 
    {\cal{N}}(j_3) C\left(j_1,j_2,\frac{k}{2}-j_3 \right) \, ,
    \label{B123}
\end{equation}
which follow from Eqs.~\eqref{C123} and \eqref{def B(j)}. Since ${\cal{N}}(j) {\cal{N}}(\frac{k}{2}-j) = 1$, we conclude that the normalization is $C_{\boldsymbol{\w}}(j_1,j_2,j_3) = C(j_1,j_2,j_3)$ for all even parity three-point functions, and $C_{\boldsymbol{\w}}(j_1,j_2,j_3) = {\cal{N}}(j_1) C(\frac{k}{2}-j_1,j_2,j_3)$ for the odd parity ones. This argument is valid for discrete representations, although we expect it to hold
also for the continuous series by analytic continuation in $j$ \cite{Maldacena:2001km,Zamolodchikov:1995aa}.

\medskip 

In the previous pages we have finally succeeded in deriving all non-zero three-point functions with arbitrary spectral flow charges. For the third time, we ask: are we done now? Well... almost! The more precise statement is that we have derived all $y$-basis three-point functions. In order to obtain the full $h$-dependence of the $x$-basis three-point functions one actually needs to carry out the integration over the  variables $y_i$, the relevant $h_i$ appearing in the exponents of the prefactors, see Eq.~\eqref{proposalWxybasis}. This computation is beyond the scope of these notes. It was carried out in detail in \cite{Dei:2021xgh} for the case of three long-string states. The result can be written in terms of the so-called Lauricella hypergeometric functions of type A \cite{Lauricella}, and then analytically continued in order to describe correlators with short string insertions. 

An important aspect of the integration over the $y$-variables is that it can lead to new poles in the space of unflowed spins $j_i$, which were not present in the unflowed structure constants \eqref{C123}. The simplest ones to spot arise in odd spectral flow three-point functions, and come from the integration near the locus $y_i=a_i$, see Eq.~\eqref{3pt-odd-parity}. As will be discussed later on, the residues of these correlators at such poles are captured by the free field description \cite{Eberhardt:2021vsx,Knighton:2023mhq,Knighton:2024qxd}. They are also particularly important in the tensionless limit of the theory \cite{Eberhardt:2019ywk,Eberhardt:2025sbi}.

\subsection{The conjecture for four-point functions}
\label{sec: conjecture 4pt W}

This analysis was extended to four-point functions in \cite{Dei:2021yom}. However, an important complication comes from the fact that the corresponding conditions on the spectral flow charges $\w_i$ are necessary for the existence of the appropriate covering maps, but, as opposed to the three-point case, they are not sufficient. Indeed, the values of the $\w_i$ fix the orders of the different ramification points, but the putative map $\Gamma(z)$ is then fixed by the conditions \eqref{coveringmapexp} near $z_i$ for $i=1,2,3$. The resulting function will only be a covering map behaving as \eqref{coveringmapcoeffs} for all four insertions if the additional condition $\Gamma(z_4) = x_4$ is satisfied.  

The authors of \cite{Dei:2021yom} put forward the following conjectured solution: 
\begin{subequations}
\begin{align}
&\hspace{-30pt}\left\langle V_{j_1,h_1, \bar h_1}^{w_1}(0;0) V_{j_2,h_2, \bar h_2}^{w_2}(1;1) V_{j_3,h_3, \bar h_3}^{w_3}(\infty;\infty) V_{j_4,h_4, \bar h_4}^{w_4}(x;z) \right \rangle \nonumber\\
& \hspace{-13pt} = \int \prod_{i=1}^4 \mathrm{d}^2 y_i \ y_i^{\frac{kw_i}{2}+j_i-h_i-1}\bar{y_i}^{\frac{kw_i}{2}+j_i-\bar h_i -1} |X_\emptyset|^{2(j_1+j_2+j_3+j_4-k)} \nonumber  \\
& \quad \quad\times |X_{12}|^{2(-j_1-j_2+j_3-j_4)} \, |X_{13}|^{2(-j_1+j_2-j_3+j_4)} |X_{23}|^{2(j_1-j_2-j_3+j_4)} |X_{34}|^{-4j_4} \nonumber \\
& \quad \quad\times\left\langle V_{j_1}^{0}(0;0) V_{j_2}^{0}(1;1) V_{j_3}^{0}(\infty;\infty) V_{j_4}^{0}\left(\frac{X_{23}X_{14}}{X_{12}X_{34}};z\right) \right \rangle 
\label{4ptf-even}
\end{align}
for $\sum_i w_i \in 2 \mathds{Z}$ and
\begin{align}
&\hspace{-21pt} \left\langle V_{j_1,h_1, \bar h_1}^{w_1}(0;0) V_{j_2,h_2, \bar h_2}^{w_2}(1;1) V_{j_3,h_3, \bar h_3}^{w_3}(\infty;\infty) V_{j_4,h_4, \bar h_4}^{w_4}(x;z) \right \rangle \nonumber\\
&\hspace{-5pt}=\mathcal{N}(j_3) \int \prod_{i=1}^4 \mathrm{d}^2 y_i \ y_i^{\frac{kw_i}{2}+j_i-h_i-1}\bar{y_i}^{\frac{kw_i}{2}+j_i-\bar h_i-1} |X_{123}|^{2(\frac{k}{2}-j_1-j_2-j_3-j_4)} \nonumber \\
& \quad \quad \quad\times |X_{1}|^{2(-j_1+j_2+j_3+j_4-\frac{k}{2})} \, |X_{2}|^{2(j_1-j_2+j_3+j_4-\frac{k}{2})} |X_3|^{2(j_1+j_2-j_3+j_4-\frac{k}{2})} |X_{4}|^{-4 j_4} \nonumber \\
&\quad \quad \quad\times\left\langle V_{j_1}^{0}(0;0) V_{j_2}^{0}(1;1) V_{\frac{k}{2}-j_3}^{0}(\infty;\infty) V_{j_4}^{0}\left(\frac{X_{2}X_{134}}{X_{123}X_{4}};z\right) \right \rangle\, 
\label{4ptf-odd}
\end{align}
\label{4ptf}%
\end{subequations}
for $\sum_i w_i \in 2 \mathds{Z}+1$. Here the factors $X_I$ are defined similarly to the three-point case, see Eq.~\eqref{X_I-3pt}. However, now the numbers $P_{\boldsymbol{\w}}$ are promoted to polynomials $P_{\boldsymbol{\w}}(x,z)$ in the worldsheet and spacetime cross-ratios. The precise definitions can be found in \cite{Dei:2021yom}. The values of $x$ and $z$ where one of these polynomials vanish coincide precisely with the situations where a covering map associated to the original correlator or to one of the adjacent ones (where one or more of the $\w_i$ are shifted by one unit) actually exists. 

We see that, as it was the case for three-point functions, the $X_I$ behave in many ways as generalized differences (although $X_i - X_j \neq X_{ij}$!). For instance, in the even case the generalized cross ratios satisfy 
\begin{equation}
    X \equiv \frac{X_{23}X_{14}}{X_{12}X_{34}} = 1-\frac{X_{24}X_{13}}{X_{12}X_{34}}. 
\end{equation}
Actually, to be precise the recursion relations only fix the overall dependence in the $X_I$ up to an undetermined function $F\left(X,z\right)$. This can then be combined with the KZ equation for the flowed correlator in order to show that the former undetermined function must, as a function of $X$ and $z$ actually satisfy the same differential equations as the \textit{unflowed} conformal block, i.e.~$F\left(X,z\right) \sim {\cal{F}}(X,z)$ with $x$ replaced by $X$, leading to the above conjecture.     
This is a direct generalization of what happens for three-point functions, where we have seen that the unflowed structure constants $C(j_1,j_2,j_3)$ appear in the normalization. 

At the moment of writing these notes, the conjecture put forward in \cite{Dei:2021yom} remains unproven, although a number of important aspects were considered in \cite{Dei:2022pkr} from the holographic perspective. It would also be extremely interesting to show that Eqs.~\eqref{4ptf-even} and \eqref{4ptf-odd} factorize properly, and that it  satisfies crossing symmetry, which was only established for $k=3$ in \cite{Eberhardt:2025sbi}. Roughly speaking, this corresponds to the tensionless limit, which we study in some detail in the following section. For generic values of $k$, a first step in this direction was taken in \cite{Iguri:2024yhb}, where a number of non-trivial spectrally flowed factorization channels  where derived from the above formulas, and shown to be consistent with the corresponding structure constants. This was done by studying the small $z$ limit. Moreover, the small $x$ limit, that is, the regime associated with the OPE structure of the spacetime theory, was further considered in the supersymmetric case and for a specific class of correlators in \cite{Barone:2025vww}. Given that, despite all of these important results, the derivation of the spacetime OPE from the worldsheet perspective remains to somewhat obscure\footnote{This issue was recently considered from the free-field perspective in \cite{Sriprachyakul:2025ubx}.}, it would be extremely interesting to generalize this result to all four-point functions.

%%%%%%%%%%%%%%%%%%%%%%%%%%%%%
\newpage
%%%%%%%%%%%%%%%%%%%%%%%%%%%%%%%

\section{The tensionless string limit}
\label{sec:tensionless}

In the previous chapters we have studied the propagation of strings in AdS$_3$ without specifying the value of the parameter $k$, which sets the value of the affine level.  In general, $k$ is not restricted to be integer since the target space of our WZW model is non-compact.  

In geometric terms, $k$ is identified with the AdS$_3$ squared radius in string units. Although we have drawn important lessons from the semiclassical limit of the model, where $k$ is taken to be very large, working in this regime is by no means necessary. The WZW description is \textit{exact} in $k$, at least as long as $k>2$ such that the SL(2,$\R$) central charge  $c = \frac{3k}{k-2}$ remains positive. In the supersymmetric setting with target space AdS$_3\times S^3 \times T^4$ (or K3) the worldsheet anomaly cancellation condition sets $k = n_5+2$ for the bosonic AdS$_3$ sector, where $n_5$ the number of NS5-brane sources, which is of course quantized, hence one must restrict to\footnote{Models with $2<k<3$ are discussed in \cite{Balthazar:2021xeh}.} $k\geq 3$, $k \in \mathbb{Z}$. We now discuss what happens in the limit where $k$ takes the smallest possible value in this context, $k=3$, which, in this context, corresponds to the backgrounds source by $n_1$ fundamental strings and a single NS5-brane. This defines the tensionless limit of the model. We will see that the worldsheet simplifies drastically, which has important consequences from the holographic perspective.

\subsection{Higher spin massless spectrum} 

We start by looking more closely at the massless sector of the theory. Although so far we have concentrated on operators that are spacetime scalars, here we briefly consider fields with arbitrary (spacetime) spin. The mass-squared formulas for symmetric traceless tensors of rank $s$ and $p$-form fields in AdS$_{d+1}$ read  
\begin{eqnarray}
    m^2_{{\rm AdS}_{d+1}} (s) &=& (\Delta_{\rm st} - s +2-d) (\Delta_{\rm st} + s -2) \,, \\
    m^2_{{\rm AdS}_{d+1}} (p) &=& (\Delta_{\rm st} +p-d) (\Delta_{\rm st} - p) \,,
\end{eqnarray}
respectively, in units of the AdS radius. For AdS$_3$, we identify $\Delta_{\rm st} = h + \hb$ and $s$ or $p$ with $|l| = |h - \hb|$, giving 
\begin{equation}
    m^2_{{\rm AdS}_{3}} = (h+\hb - |h-\hb| )(h+\hb + |h-\hb| -2) \,,  
\end{equation}
in both cases. For primary states with $h = \hb = j$, this reproduces the masses of the unflowed scalars studied in Sec.~\ref{sec: bosonic AdS3 spectrum}, namely $m^2_{{\rm AdS}_{3}} = 4 j (j-1)$.  
From the above discussion, we conclude that massless states appear when (1) either $\hb \leq h = 1$ or $h \leq \hb = 1$, or (2) either $h=0$ or $\hb=0$. The latter option is the relevant one if one is looking for massless higher-spin states. As per the AdS/CFT dictionary, they correspond to spin $l$ conserved currents of the boundary theory. Let us stress that, although so far we have  worked mostly with primary operators, more generally the spacetime spin $h$ should also take into account the action of the descendant modes, characterized by two integers $N$ and $\bar{N}$.

Which massless modes can we actually construct from the worldsheet WZW model? Recall that the mass-shell conditions read 
\begin{equation}
    - \frac{j(j-1)}{k-2} - \w h + \frac{k}{4}\w^2 + h_{\rm int} + N - 1 =  - \frac{j(j-1)}{k-2} - \w \hb + \frac{k}{4}\w^2 + \hb_{\rm int} + \Nb - 1 = 0 \, .
\end{equation}
Subtracting both constraints we get the level-matching condition 
\begin{equation}
    \w (h-\hb) = N - \Nb + h_{\rm int} - \hb_{\rm int}\,. 
\end{equation}
The universal solutions describe the supergravity modes, i.e.~the graviton, dilaton and $B$-field fluctuations. These include the operator $\Ii$ \eqref{def I(x)} and the spacetime energy-momentum tensor defined in Eq.~\eqref{def T(x)}, together with its anti-holomorphic counterpart. They are solutions with $j=1$, $\w=0$, $h_{\rm int} = \hb_{\rm int} = 0$ and $N = \Nb = 1$. 
We now prove that, in the bosonic model and for generic values of $k$, these are the only solutions. 

From now on we set $h_{\rm int} = \hb_{\rm int} = 0$ for simplicity, although the arguments below are easily extended to the general case. In the unflowed sector we only need to consider discrete states in lowest-weight representations. In order to get, say, $h=0$, we need to solve 
\begin{equation}
\label{massless unflowed}
    - j(j-1) + (k-2)(N-1) = 0 \, .
\end{equation}
Now, at level $N$, the eigenvalue of $J_0^3$ is restricted to the range $m \geq j-N$. This is because $m \geq j$ for the affine primary, which can be lowered in at most $N$ units by acting with $(J_{-1}^-)^N$. The lowest energy, which gives the minimal spacetime spin, thus corresponds to $h=m=0$, i.e.~$j = N$, for which \eqref{massless unflowed} becomes $- j(j-1) + (k-2)(j-1) = 0$. The only solutions with $j \neq 1$ correspond to $j= k-2$, which is outside of the physical range given in Eq.~\eqref{Djrange}. 

Moving to short strings with $\w>0$, we now ask for solutions with, say, $\hb=0$, for which 
\begin{equation}
    - \frac{j(j-1)}{k-2} = 1-\Nb - \frac{k}{4}\w^2 \, .
\end{equation}
Imposing $j \leq \frac{k}{2}$ gives $- \frac{j(j-1)}{k-2} \geq - \frac{k}{4}$, which translates into 
\begin{equation}
    1-\Nb - \frac{k}{4}(\w^2-1) \geq 0 \, . 
\end{equation}
This only allows for $\w=1$ and $\Nb = 0,1$. As before, the lowest energy state has $\mb \geq j-\Nb \geq \frac{1}{2}-\Nb \geq -\frac{1}{2}$, which is incompatible with $\hb = \mb + \frac{k}{2}= 0$  for $k >2$. 

Finally, we also consider long string states. Upon setting $\hb=0$, level-matching implies $N = \w h + \Nb$, hence the holomorphic mass-shell condition  becomes 
\begin{equation}
    \frac{1}{4}+s^2 = (k-2)\left(1-\Nb - \frac{k}{4}\w^2\right) \, .
\end{equation}
For $k>2$ and $s\in \R$, there is no solution \textit{unless} $s=0$, $\Nb =0$, $\w=1$ and, crucially, $k=3$.

We conclude that for $k>2$, $k\neq 3$ the only massless modes are the usual supergravity states. On the other hand, at $k=3$ we actually  have an infinite tower of higher-spin massless states with $l = h = N$, which are singly-wound long strings\footnote{Further references regarding higher spin states in this context include \cite{Gaberdiel:2014cha,Gaberdiel:2015mra,Gaberdiel:2015wpo,Gaberdiel:2015uca,Baggio:2015jxa,Ferreira:2017pgt}. } . The interpretation is that, roughly speaking, this constitutes the AdS$_3$ version of the $\alpha'\to \infty$ limit for strings in flat space, where excited string states become massless. As a consequence, the AdS$_3 \times S^3 \times T^4$ model with $n_5=1$ is understood as describing the \textit{tensionless} string limit\footnote{
Note that, for $k=3$ the physical range for the discrete sector becomes 
\begin{equation}
\frac{1}{2} < j < 1,     
\end{equation}
so that the $j=1$ supergravity modes naively drop out. Nevertheless, by means of the SL(2,$\R$) series identifications \eqref{series Id SL2 1} we can reinterpret them as states with $\w=1$ and unflowed spin $\tilde{\jmath} = \frac{3}{2}-1 = \frac{1}{2}$, which merge with the massless long strings described above.}.

\subsection{Alternative interpretation of the long string spectrum}
\label{sec: sym orb spectrum 1}

Although we work mostly with the bosonic SL(2,$\R$) WZW model, in this section we are ultimately interested in type II superstrings propagating in AdS$_3\times S^3 \times T^4$ with a single NS5-brane source.  Since the level of the bosonic SL(2,$\R$) subsector of the corresponding is given by $k=n_5+2 = 3$ \cite{DiVecchia:1984nyg,Kac:1985wdn,Kiritsis:1986wx}, when $n_5=1$ we are in the tensionless limit, for which a tower of higher-spin fields becomes massless.  Although the supersymmetric worldsheet theory for larger values of $n_5$ is studied in detail in Sec.~\ref{sec: AdS3xS3xT4} below, for the case of $n_5=1$ we refer the interested reader to \cite{Eberhardt:2018ouy,Eberhardt:2019qcl,Eberhardt:2020bgq,Dei:2020zui,Knighton:2020kuh,Dei:2023ivl}.

As it turns out, an even more drastic simplification occurs in the supersymmetric setting a the tensionless point. This was discussed in \cite{Eberhardt:2019qcl}, where the authors showed that 
\begin{itemize}
    \item All short-string states drop out of the spectrum.

    \item Out of all long-string states, only those with $j=\frac{1}{2}$ survive.   
\end{itemize}
These statements are not that easy to derive in our language since the RNS formalism breaks down at $n_5=1$. This is because the bosonic SU(2) sector has level $k' = n_5-2$, which becomes negative at $n_5=1$. Due to this technical complication one needs to work with the so-called hybrid formalism \cite{Berkovits:1999im}, which provides an alternative free field realization \cite{Gaberdiel:2021njm,Dei:2020zui,Naderi:2022bus}. Here we simply quote the result that will be most relevant for us,  the above restrictions on the spectrum, which follow from the representation theory of the relevant supergroup, namely PSU(1,1|2) at level 1, for which the maximal bosonic subgroup is SL(2,$\R$)$_1\times$SU(2)$_1$ \cite{Eberhardt:2018ouy}. Nevertheless, we provide the following heuristic motivation: for $n_5=1$ the corresponding AdS space is string-size, and as a consequence there is simply no space for the discrete bound states, while long strings with non-zero radial momentum have nowhere to go!   

We now describe how these additional restrictions affect the physical spectrum in holographic terms. As we have discussed at length, for bosonic strings in AdS$_3 \times M_{\rm int}$ the worldsheet model is built by combining the SL(2,$\R$) WZW model at level $k$ and an internal CFT with central charge 
\begin{equation}
    c_{\rm int} = 26-\frac{3k}{k-2}\,,
\end{equation} 
together with the usual $bc$-system of ghosts. For long string (Virasoro) primaries, the physical state condition is solved by 
\begin{equation}
\label{h for long strings}
    h =\frac{1}{\w} \left[-\frac{j(j-1)}{k-2} + 
    \frac{k}{4}\w^2 + h_{\rm int} - 1 
    \right] \, .
\end{equation}
By defining 
\begin{equation}
\label{Q alpha Eberhardt}
    Q = \frac{k-3}{\sqrt{k-2}} \, , \qquad 
    %\beta = \frac{\frac{k}{2}-1-j}{\sqrt{k-2}}
    \alpha = \frac{j+\frac{k}{2}-2}{\sqrt{k-2}}
    \, , 
\end{equation}
we can rewrite this as 
\begin{equation}
\label{h for bosonic sym orbifold HCFT}
    h = \frac{k (\w^2-1)}{4\w} + \frac{h_{\rm int}}{\w} + \frac{\alpha (Q-\alpha)}{\w} \, .
\end{equation}
Comparing with Eq.~\eqref{SymOrb spectrum generic}, we find that \eqref{h for bosonic sym orbifold HCFT} reproduces exactly the spectrum of a symmetric orbifold theory
\begin{equation}
\label{bosonic sym orbifold HCFT}
    {\rm Sym}^{N}\left(\R_{Q} \times M_{\rm int}\right) \, ,
\end{equation}
where $\R_{Q}$ denotes a Liouville-type factor, i.e.~a linear dilaton theory with background charge $Q$, such that 
\begin{equation}
\label{cseed}
    c_{\rm seed} = 1 + 6 Q^2 + c_{\rm int} = 1+ \frac{6(k-3)^2}{k-2}+ 26 - \frac{3k}{k-2} = 6k \, .
\end{equation}
Some useful details of symmetric orbifold models are reviewed in appendix \ref{sec: appD SymOrbifolds}. As discussed there, for an operator living in a sector of twist $\w$, the first factor on the RHS of Eq.~\eqref{h for bosonic sym orbifold HCFT} gives the weight of the corresponding twist operator $\sigma_\w$, while the second and third factors give that  of the relevant operator of the seed theory as viewed form the associated covering space. This leads us to identify the worldsheet spectral flow sector with the holographic twist sector. It is then tempting to identify $N$, the number of copies, with $n_1$, the number of fundamental strings in the background. Eq.~\eqref{bosonic sym orbifold HCFT}, constitutes our first concrete proposal for the holographic CFT, although we should of course check that the dynamics are indeed the same.

Even though the long string spectrum reproduces that of a symmetric orbifold model, we cannot conclude that this \textit{is} precisely the holographic CFT at the same point in moduli space where the worldsheet theory is defined, at least not for generic values of $k$. This cannot be the case, since the structure of the holographic correlators as computed from the worldsheet is not that of a symmetric orbifold. For instance, in Sec.~\ref{sec: bosonic correlators} we have obtained a full family of correlators which are non-zero even when the corresponding covering maps do not exist, namely the even parity correlators. It was proposed recently in \cite{Balthazar:2021xeh,Eberhardt:2021vsx} that the correct CFT can be defined, at least perturbatively, by starting from the above symmetric orbifold and including a particular marginal deformation by a twist-two operator, which moreover is trivial in the internal sector but contains a Liouville-type potential for the non-compact scalar. We will come back to this in Sec.~\ref{sec: holography} below, when discussing non-protected observables and their holographic interpretation away from the tensionless point. 

For now, we stick to the $k=3$ case, and note that at this point the background charge $Q$ vanishes, and we get  $\alpha = \frac{1}{2}-j$. Hence, upon restricting to long strings with $j= \frac{1}{2}$ we trivialize the sector of the non-compact scalar, and are only left with  the states of a holographic CFT of the form
\begin{equation}
    {\rm Sym}^{N}\left(M_{\rm int}\right). 
\end{equation}
In particular, at $k=3$, integrated operators with  $j = \frac{1}{2}$ and $h_{\rm int}=0$, represent the worldsheet avatars of  the spacetime twist operators, which create the ground states of the corresponding twisted sectors. In the following sections we describe how the corresponding worldsheet correlators simplify drastically in this limit.

Even though will continue using the bosonic language, this analysis is particularly relevant for the supersymmetric setting of AdS$_3 \times S^3 \times T^4$ at $n_5=1$, where it was shown in \cite{Eberhardt:2018ouy,Eberhardt:2019ywk,Dei:2019osr,Eberhardt:2020bgq,Eberhardt:2020akk,Dei:2020zui,Bertle:2020sgd,Dei:2023ivl} that the full HCFT collapses to the following symmetric orbifold: 
\begin{equation}
    {\rm Sym}^{N}\left(T^4\right) \qquad {\rm with} \qquad \N = (4,4) 
    \quad {\rm supersymmetry} \, .
\end{equation}
Here the $S^3$ provides the geometrical realization of the R-symmetry group \cite{Schwimmer:1986mf}, as usual in holography.  
More generally, one could include long string states with $j=\frac{1}{2}+is$. As stated above, $s$ can be interpreted as the radial momentum. In this case one should of course include the non-compact scalar in the seed theory of the holographic CFT.

\subsection{Localization of worldsheet correlators}

Let us now discuss what happens with string correlators in the tensionless limit of the theory\footnote{See the recent publications \cite{Dei:2023ivl,Knighton:2023mhq} for a related discussion using the path integral formalism involving the Wakimoto free fields.}. As argued above, we are ultimately  interested in computing $n$-point functions of spectrally flowed operators with $j_i=\frac{1}{2}$ for $i=1,\dots,n$. For any value of $n$, this is a particular case of a the more general condition given by 
\begin{equation}
\label{j cond loc}
    \sum_{i=1}^n j_i = \frac{k}{2}(n-2)-(n-3)\,. 
\end{equation}
As shown in \cite{Eberhardt:2019ywk}, when this condition is fulfilled the  constraint equations and recursion relations among spectrally flowed correlators derived from the local Ward identities in Sec.~\ref{sec: bosonic correlators} admit a generic solution of the following form: 
\begin{equation}
\label{Loc solution}
    \langle \prod_{i=1}^n 
    V_{j_i h_i}^{\w_i}(x_i,z_i)\rangle = \sum_{\Gamma} \prod_{i=1}^n a_i^{-h_i}
    \bar{a}_i^{-\hb_i} \prod_{j=4}^n \delta\left(x_j-\Gamma(z_j)\right) W_{\Gamma}(z_4,\dots,z_n) \,, 
\end{equation}
where we have fixed $(z_1,z_2,z_3) = (x_1,x_2,x_3) = (0,1,\infty)$ for simplicity (otherwise the coefficients $a_i$, $\bar{a}_i$ depend explicitly on these insertion points). 

Before showing that Eq.~\eqref{Loc solution} solves the recursion relations, let us briefly unravel this formula. For a given set of spectral flow charges $\w_i$, worldsheet insertion points $z_i$, and boundary insertion points $x_i$, one must ask if an appropriate  holomorphic covering map $\Gamma(z)$ exists. When the $\w_i$ fail to satisfy either of the conditions 
\begin{equation}
\label{wi map conditions n-point} 
\sum_{i=1}^n w_i \geq 2 \,{\rm Max}(\w_i) + n - 2 \, , 
    \qquad 
    \sum_{i=1}^n (w_i-1) \in 2\mathbb{Z} \,, 
\end{equation}
the answer is negative. In such situations the solution \eqref{Loc solution} simply vanishes. On the other hand, if the conditions in Eq.~\eqref{wi map conditions n-point} hold, the situation is similar to that of the generic four-point function considered above: one can always find a holomorphic map with appropriate branchings, namely 
\begin{equation}
    \Gamma(z)-\Gamma(z_i) = a_i (z-z_i)^{\w_i} + \cdots \, ,
    \qquad \forall \quad i = 1,\dots,n \,.
\end{equation}
However, further enforcing $\Gamma(z_i) = x_i$ for $i=1,2,3$ is as far as we can go \cite{Pakman:2009zz}. Indeed, this fixes the function $\Gamma(z)$, which will only constitute a covering map needed to define a non-trivial contribution to the correlator if the remaining pairs $(x_i,z_i)$ are such  that 
\begin{equation}
    \Gamma(z_i) = x_i \, ,
    \qquad \forall \quad i = 4,\dots,n \,.
\end{equation}
In other words, for the $x$-basis worldsheet correlators localize on the locus defined by the existence of such covering maps. Although for the three-point case the map is unique (at genus zero), for $n\geq 3$ one could have discrete set of possible maps, hence the sum over $\Gamma$ in \eqref{Loc solution}. 

When at least one covering map exists, the correlator is non-vanishing, and the dependence on $h_i$, $\bar{h}_i$ and $x_{i\geq 4}$ is captured quite simply by the overall prefactor involving the corresponding coefficients $a_i$. Let us stress that the function $W_\Gamma(z_4,\dots,z_n)$ must be independent of the spacetime weights $h_i$ and of all of the $x_i$, but is in principle otherwise arbitrary. Nevertheless, one should keep in mind that there are additional constraints coming from KZ-type equations and other consistency conditions such as crossing symmetry. An explicit expression for $W_{\Gamma}$ was derived recently in \cite{Eberhardt:2025sbi}. It is valid for $n$-point functions $j_i=\frac{1}{2}+i s_i$ for which the momentum is conserved, in the sense that $\sum_{i=1}^n s_i = 0$, consistent with \eqref{j cond loc}. The authors showed that the resulting $k=3$ worldsheet CFT is consistent, and that, upon integrating over the worldsheet moduli, it reproduces the symmetric orbifold correlators exactly.  

Let us discuss how Eq.~\eqref{Loc solution} comes about. We start with the simpler case of three-point functions, for which there are no delta functions, although we still require \eqref{wi map conditions n-point}. We already know from the analysis of Sec.~\ref{sec: bosonic correlators} that using the covering maps in combination with the $JV^w$ OPEs leads to recursion relations of the form 
\begin{equation}
    \sum_{i=1}^3 \left(a_i 
    F^i_+ - h_i F\right) = \frac{N}{a_j \w_j} \left(F^j_- -2 a_j h_j + a_j^2 F^j_+\right) \, \qquad  j = 1,2,3\,,
\end{equation}
with  
$F = \langle \prod_{i=1}^3 
    V_{j_i h_i}^{\w_i}(x_i,z_i)\rangle$ and  $F^j_\pm = \langle \left( J^\pm_{\pm\w_j}V_{j_j h_j}^{\w_j}\right)(x_j,z_j)
    \prod_{i\neq j}
    V_{j_i h_i}^{\w_i}(x_i,z_i)\rangle$, while $2N = 2 + \sum_{i=1}^3 (\w_i-1)$, see Eq.~\eqref{recursionywithmap}. 
It not difficult to see that, when $\sum_{i=1}^3j_i = \frac{k}{2}$, this is satisfied by Eq.~\eqref{Loc solution} since the $F^j_\pm$ are proportional to the analogous correlator with $h_j$ replaced by $h_j \pm 1$, which simply adds or removes a power of $a_j$. 

The slightly more non-trivial statement is that this structure actually generalizes to higher-point functions. Indeed, since we are sitting at the locus where a covering map is available, the recursion relations can be derived directly by a procedure almost analogous to what we did for parity-odd three-point functions in Sec.~\ref{sec: bosonic correlators}. Nevertheless, one must be careful when replacing $\Gamma(z_i)=x_i$ in several places since the correlators with current insertions might involve derivatives of $F$ with respect to $x_i$. When this derivative hits the factor  $\delta\left(x_j-\Gamma(z_j)\right)$, one must use the distributional identity $x \delta'(x) = -\delta(x)$, see \cite{Eberhardt:2019ywk} for details. The only subtlety comes towards the end of the computation of the coefficient $\alpha$ appearing in the relation $G(z) =  \langle J^-\left(\Gamma(z)\right)\prod_{i=1}^n V_{j_i h_i}^{\w_i}(x_i,z_i)\rangle = \alpha \der \Gamma(z)$. We get 
\begin{eqnarray}
    \alpha N &=& \sum_{i=1}^n \left[a_i 
    F^i_+ + \Gamma (z_i) \der_{x_i} F\right] \nn \\ 
    &=&  
    \sum_{i=1}^n \left[a_i 
    F^i_+ + x_i \der_{x_i} F\right] + (n-3) F =
    \sum_{i=1}^n \left[a_i 
    F^i_+ - h_i F\right] + (n-3)F \, ,
\end{eqnarray}
where in the middle step we have used the Ansatz \eqref{Loc solution}, and, in the last one, the global Ward identities. The extra term proportional to $(n-3)$, which was of course absent for the case of three-point functions, leads to recursion relations of the form 
\begin{equation}
    \sum_{i=1}^n \left(a_i 
    F^i_+ - h_i F\right) + (n-3)F = \frac{N}{a_j \w_j} \left(F^j_- -2 a_j h_j + a_j^2 F^j_+\right) \, \qquad  j = 1,2,3\,,
\end{equation}
which are indeed solved by \eqref{Loc solution}. 

Finally, in order to compute the holographic CFT correlator from the string theory perspective, we are instructed to integrate the insertion points $z_{i}$ with $i \geq 4$ over the full complex plane (at genus zero). In general this is a very complicated procedure, see for instance \cite{Maldacena:2001km,Iguri:2024yhb}. Nevertheless, the integration can easily be carried out for the solution in Eq.~\eqref{Loc solution} thanks to the presence of the delta functions $\delta\left(x_i - \Gamma(z_i)\right)$! We simply get a sum over the preimages $\Gamma^{-1}(x_i)$. This is exactly the behavior we expect for the correlation functions of a symmetric orbifold theory on the boundary, see \cite{Lunin:2000yv,Dei:2019iym}. 

\subsection{Localization in the $y$-basis and holographic matching}

In this section we provide the $y$-basis picture of the discussion related to the localization of worldsheet correlators. We also present some details about the precise matching with the symmetric orbifold results for a specific set of three- and four-point functions.

Let us start with three-point functions. For a given odd parity correlator at $k=3$ and $j_i=\frac{1}{2}$ for $i=1,2,3$, 
the $y$-basis differential equations become 
\begin{equation}
\label{3pt y-eq tensionless}
    \left\{(y_i-a_i)^2 \der_{y_i} +  (y_i-a_i) + \frac{a_i \w_i}{N} \left[
    3+\sum_{j=1}^3 (y_j-a_j) \der_{y_i} 
    \right] \right\} F_y = 0 \, .
\end{equation}
In the original solution \eqref{3pt-odd-parity}, the power of $X_{123}$ vanishes, hence the putative expression for the $y$-dependence is of the form 
\begin{align}
\begin{aligned}
\label{original sol 3pt}
    \left\langle V^{\w_1}_{\frac{1}{2}}(0, y_1, 0) \,  V^{\w_2}_{\frac{1}{2}}(1, y_2, 1) \, V^{\w_3}_{\frac{1}{2}}(\infty, y_3, \infty) \right\rangle &\sim  
|X_1 X_2 X_3|^{-2} \\
&\sim |
    \left(y_1-a_1\right)\left(y_2-a_2\right)\left(y_3-a_3\right) |^{-2} \, ,
\end{aligned}
\end{align}
where we have ignored the overall numerical coefficients and structure constants. However, in this special case there exist an additional distributional solution to \eqref{3pt y-eq tensionless}, given by the replacements 
\begin{equation}
    |y_i-a_i|^{-2} \to \delta(y_i-a_i) \,, \qquad i=1,2,3\, .
\end{equation}
It turns out that this is the correct solution for the bosonic SL(2,$\R$) subsector of the tensionless string worldsheet theory \cite{Dei:2022pkr}. More generally, when the condition \eqref{j cond loc} holds, the differential equations decouple, giving simply 
\begin{equation}
    \left(\der_{y_i} + \frac{2j_i}{y_i-a_i}\right) F_y = 0 \,, \qquad i = 1,2,3.
\end{equation}
which be solved by a product of factors of the form $|y_i-a_i|^{-2j_i}$, or, when Re$(2j_i)=1$ for all $i=1,2,3$, by a product of Dirac delta functions
\footnote{
Interestingly, the two solutions are related by the reflection symmetry \eqref{reflection y basis} since 
\begin{equation}
    \int d^2y_i' |y_i-y_i'|^{4j_i-4}|1-a_i^{-1}y_i'|^{-4j_i} = |a_i|^{4j_i} \delta(y_i-a_i) \, .
\end{equation}} localized at $y_i=a_i$.

The existence of the additional distributional solution  suggests that, there are in fact two consistent CFTs with a $k=3$ SL(2,$\R$) Kac-Moody symmetry algebra: one is simply the limit of the more general WZW model at level $k$, where the local ward identities are solved by correlators with factors $(y_i-a_i)^{-2j_i}$, and a second one, where the spectrum only contains long strings, and the correlators localize. In latter case, the exact $x$-basis correlators are nothing but residues of those of the former one, computed at $y_i=a_i$. This possibility was explored recently in \cite{Eberhardt:2025sbi}, and the self-consistency of the alternative $k=3$ CFT was demonstrated. In particular, an explicit expression for the flowed conformal blocks was obtained.  

Let us be a bit more explicit, setting $j_i=\frac{1}{2}$ for simplicity. For three-point functions, we have found that the $y$-variables must be evaluated precisely at the values of the covering map coefficients. While generically the SL(2,$\R$)$_k$ correlators have divergences coming from the integrals over the $y$-variables associated to contributions near the locus where some of the generalized differences $X_I$ vanish, for instance near $y_i = a_i$, the situation is different for the distributional solution we have just found. Here the $y$-integrals leading to the $x$-basis correlators can easily be carried out explicitly, and lead to a finite answer, with all factors of $y_i$ replaced by $a_i$. More explicitly, these integrals take the form 
\begin{equation}
    \int d^2y_i \, y_i^{\frac{k}{2}\w_i + j_i - h_i - 1}\bar{y}_i^{\frac{k}{2}\w_i + j_i - \bar{h}_i - 1} \delta (y_i-a_i) = 
    a_i^{-h_i -\hb_i + 3\w_i-1} \,. 
\end{equation}
This reproduces the $h_i$-dependence anticipated in Eq.~\eqref{Loc solution}. We can also fix the dependence on the spectral flow charges $\w_i$ and the covering map data completely. The normalization $\N \left(\frac{1}{2}\right)C\left(1,\frac{1}{2},\frac{1}{2}\right)$ trivializes, and we are left with  
\begin{equation}
\langle 
    V_{h_1}^{\w_1}(0,0)V_{h_2}^{\w_2}(1,1)V_{h_3}^{\w_3}(\infty,\infty)\rangle_{k=3}
    \sim  \Big|\prod_{i=1}^3 a_i^{-h_i + \frac{3}{2}\w_i-1}  P_{\boldsymbol{\w}-e_i}\Big|^2 \,,
\end{equation}
where we have omitted the $j_i$ since they are all set to $\frac{1}{2}$. As it turns out, by using the definitions of the numbers $P_{\boldsymbol{\w}}$ as well as various identities related to the covering maps, and after including appropriate normalization factors, one can rewrite this as
\begin{equation}
\langle 
    V_{h_1}^{\w_1}(0,0)V_{h_2}^{\w_2}(1,1)V_{h_3}^{\w_3}(\infty,\infty)\rangle_{k=3}
    =
\Big|\Pi^{-\frac{3}{2}} \prod_{i=1}^3 a_i^{-h_i + \frac{3}{4}(\w_i-1)} \w_i^{-\frac{3}{4}(\w_i+1)+1}\Big|^2  \,,
\end{equation}
where $\Pi = \prod_{a=1}^N r_a$ stands for the product of all the residues of the covering map $\Gamma(z)$, namely $r_a = \lim_{z\to \lambda_a} (z-\lambda_a) \Gamma(z)$, which was computed in \cite{Lunin:2000yv}. For states with $j=\frac{1}{2}+i s$, one simply gets an additional momentum conservation delta function $\delta( s_1+s_2+s_3)$ \cite{Eberhardt:2025sbi}.

This result precisely reproduces the symmetric orbifold three-point functions of twist operators   \cite{Lunin:2000yv,Eberhardt:2021vsx} when $c_{\rm seed} = 6k = 18$, whose computation is briefly reviewed in 
Appendix \ref{sec: appD SymOrbifolds}. 
Such correlators are computed by using covering-space methods, where one goes to a covering of the physical space -- here the boundary of AdS$_3$ -- where the boundary conditions enforced by the twist operator insertions trivialize. Our discussion of the worldsheet computation shows that, in the tensionless limit, the string worldsheet \textit{is} precisely this  covering space. As anticipated above, the worldsheet spectral flow charges must be identified with the twists of the corresponding boundary insertions. 

We now move to the more interesting case of four-point functions. For $j_i=\frac{1}{2}$  and $k=3$ the conjecture of \cite{Dei:2021yom} for the $y$-basis correlator reads 
\begin{eqnarray}
&& 
\langle 
    V_{h_1}^{\w_1}(0,y_1,0)V_{h_2}^{\w_2}(1,y_2,1)V_{h_3}^{\w_3}(\infty,y_3,\infty)V_{h_4}^{\w_4}(x,y_4,z)\rangle_{k=3}  \nn \\[1ex]
&& \qquad \qquad = 
|X_{\emptyset}
X_{12} X_{34}|^{-2} 
\langle 
V_{\frac{1}{2}}(0,0)
V_{\frac{1}{2}}(1,1)
V_{\frac{1}{2}}(\infty,\infty)
V_{\frac{1}{2}}(X,z)\rangle_{k=3} \,,
\label{4pt tensionless}
\end{eqnarray}
where the $X_I$ are the generalized differences introduced in Sec.~\ref{sec: conjecture 4pt W}, while $X$ is the generalized cross-ratio. In particular, the factor $X_\emptyset$ is proportional to the polynomial $P_{\boldsymbol{\w}} (x,z)$. Now, the necessary condition for the existence of the relevant covering map is $P_{\boldsymbol{\w}} (x,z) = 0$ since $P_{\boldsymbol{\w}} (x,z)$ vanishes whenever $\Gamma(z)=x$. Here this polynomial appears with power $-1$, signaling that  there is once again an alternative solution where we replace 
\begin{equation}
    |X_\emptyset|^{-2} \to \delta(X_\emptyset) \propto  
    \delta\left(x-\Gamma(z)\right) \,.
\end{equation}
Moreover, at the locus $P_{\boldsymbol{\w}} (x,z)=0$ the rest of Eq.~\eqref{4pt tensionless} simplifies considerably since \cite{Dei:2022pkr} 
\begin{equation}
    X_{12} \sim  (1-a_1^{-1}y_1)(1-a_2^{-1}y_2) \, , \quad \
    X_{34} \sim (1-a_3^{-1}y_3)(1-a_4^{-1}y_4) \, , 
    \quad X \to  z \, ,
\end{equation}
where we have ignored some overall constants. This is consistent with the fact that, when the covering maps exists,  the $y$-basis differencial equations decouple, and are solved by either $(y_i-a_i)^{-2j_i}$ or $\delta(y_i-a_i)$ for four-point functions as well. Here we could have derived this from first principles -- that is, without starting from the general form conjectured in \cite{Dei:2021yom} -- since we work on the locus where a covering map exists. 

The  behavior of the unflowed correlator in the limit $X \to z$ can be derived directly from the KZ equation \eqref{KZ equation SL2} by inserting ${\cal F} = |X-z|^{2c}$, leading to two possibilities: either  $c=0$ or $c= k-\sum_{i=1}^4 j_i = 1$. We keep only the former solution, otherwise the divergence of the prefactor would cancel. Hence, the dependence on the spacetime cross-ratio $x$ is reduced simply to that of the overall Dirac delta function. 
One can also derive the powers of $z$ and $z-1$ in this limit analogously. In fact, for this particular choice of spins the unflowed correlator can even be computed exactly from Liouville theory \cite{Dei:2022pkr} or using free field techniques. We refer the reader interested in more details to the original publications \cite{Dei:2021yom,Dei:2022pkr,Eberhardt:2025sbi}.
When the dust settles, one finds that the contribution from a given covering map $\Gamma$ to the $x$-basis four-point function takes the form  
\begin{equation}
\langle 
    V_{h_1}^{\w_1}(0,0)V_{h_2}^{\w_2}(1,1)V_{h_3}^{\w_3}(\infty,\infty)V_{h_4}^{\w_4}(x,z)\rangle = \delta \left(z-z_\Gamma\right) 
\Big|\Pi^{-\frac{3}{2}} \prod_{i=1}^4 \w_i^{-\frac{3}{4}(\w_i+1)}a_i^{-h_i + \frac{3(\w_i-1)}{4}} \Big|^2 \,,   
\end{equation}
where $\Gamma(z_\Gamma)=x$, and at $k=3$. 

This method can be extended to string theory higher point functions AdS$_3\times M_{\rm int}$ involving flowed states in the continuous representations of SL(2,$\R$) at $k=3$, namely with  $j_i= \frac{1}{2}+i s_i$, further combined with internal excitations. After carrying out the integration over worldsheet moduli by using the factors $\delta(x_i-\Gamma(z_i))$  which ensure the existence of the covering maps, one gets 
\begin{align}
\begin{aligned}
    &\int \text{d}^2 z_4 \cdots \text{d}^2 z_n \ \left\langle \prod_{i=1}^n V_{s_i h_i}^{\w_i}(x_i,z_i)V_{{\rm int},i}(z_i) \right\rangle  = \\
    & \qquad \delta\Big(\sum\nolimits_i s_i \Big)\sum_{\Gamma}   \prod_{i<j}\Big|\left(z_{ij}^\Gamma \right)^{2s_is_j} \Pi^{-\frac{3}{2}} 
 \prod_{i=1}^n  \w_i^{-\frac{3}{4}(\w_i+1)} a_i^{\frac{3(\w_i-1)}{4}-h_i}\Big|^2 \left\langle \prod_{i=1}^n V_{{\rm int},i}(z_i^\Gamma) \right\rangle\ . \label{eq:correlator matching}
\end{aligned}
\end{align}
Here we have inserted the appropriate normalization factors. This  was originally derived from the free-field perspective in \cite{Dei:2023ivl,Knighton:2023mhq}, and also recently from an exact analysis in \cite{Eberhardt:2025sbi}. It shows that, using this alternative 2d CFT for the AdS$_3$ sector of the worldsheet model at $k=3$ gives bosonic string model which is tachyon free, and holographically dual to 
\begin{equation}
\label{bosonic sym orbifold HCFT Q=0}
    {\rm Sym}^{n_1}\left(\R \times M_{\rm int}\right) \, ,
\end{equation}
where $\R$ represents a free non-compact scalar, and $M_{\rm int}$ a compact CFT with central charge $c_{\rm int}=17$.

\subsection{Back to  (stringy) geometry}
\label{sec: stringy geometry}

First of all, what geometry? We are working in the tensionless limit, where the AdS$_3$ target space is string-size. In principle, there is no such thing as an interpretation in terms of particles moving on the base manifold subject to weak gravitational interactions.  Nevertheless, we can follow \cite{Eberhardt:2019ywk} to try and use the Wakimoto fields to provide a geometrical intuition for the correlation functions we have computed in the previous sections. 
We expect the Wakimoto free-field description to be accurate, at least near the AdS$_3$ boundary. 

Consider the field $\gamma(z)$. As discussed around Eq.~\eqref{OPE gamma Vw}, near a spectrally flowed vertex operator we have 
\begin{equation}
\label{OPE gamma Vwxz}
    \gamma(z) V_{h_i}^{\w_i}(x_i,z_i) \sim  x_i V_{h_i}^{\w_i}(x_i,z_i) + (z-z_i)^{\w_i} V_{h_i-1}^{\w_i}(x_i,z_i) \,. 
\end{equation}
In terms of correlators, we know that, for the solution \eqref{Loc solution}, shifting $h_i \to h_i - 1$ amounts to multiplying the original $n$-point function by $a_i$, implying that 
\begin{equation}
\label{OPE gamma Vi corr}
    \langle  \gamma(z \sim z_i) \prod_{j=1}^n V_{h_j}^{\w_j}(x_j,z_j) 
    \rangle \sim  \left(x_i + a_i (z-z_i)^{\w_i} + \cdots \right)
    \langle  \prod_{j=1}^n V_{h_j}^{\w_j}(x_j,z_j) 
    \rangle \, ,
\end{equation}
for all $i=1,\dots,n$. Given that $\gamma(z)$ is holomorphic, has weight zero and remains finite in the $z \to \infty$ limit, it is natural to propose that 
\begin{equation}
\label{gamma is Gamma}
     \langle  \gamma(z) \prod_{j=1}^n V_{h_j}^{\w_j}(x_j,z_j) 
    \rangle = 
    \Gamma(z)
     \langle   \prod_{j=1}^n V_{h_j}^{\w_j}(x_j,z_j) 
    \rangle \, .
\end{equation}
At first sight, this makes perfect sense since the semiclassical solutions are located at $\gamma = x$ on the boundary! 
This also justifies, at least heuristically,  choosing the distributional solutions in the previous section. Indeed, the LHS of \eqref{gamma is Gamma} is always well defined, and is a holomorphic function with all the right properties to be a covering map, hence the primary correlator should vanish unless the map $\Gamma (z)$ actually exists. 

Moreover, the identification of the Wakimoto field $\gamma(z)$ with the covering map $\Gamma(z)$ can be justified by considering a generic primary  correlator and repeating the exercise carried out in Sec.~\ref{sec: covering maps} when deriving the recursion relations, but now  making use of the current 
\begin{equation}
    J^-(\gamma,z) = J^-(z) - 2 \gamma J^3(z) + \gamma^2 J^+(z) \, ,
\end{equation}
as opposed to using directly $J^-(\Gamma,z)$. The OPE \eqref{OPE gamma Vwxz} then implies that the correlator must be proportional to the derivative of an appropriate covering map, if it exists, and vanish otherwise. Of course, this argument relies on the Wakimoto field $\gamma$ being holomorphic, which only holds near the boundary.

We can proceed similarly with the non-compact field $\phi$, for which the derivative satisfies 
\begin{equation}
    \der \phi (z) V_{h_i}^{\w_i}(x_i,z_i) \sim - \sqrt{2} \left(\frac{\w_i-1}{2(z-z_i)} \right) V_{h_i}^{\w_i}(x_i,z_i) \, ,
\end{equation}
where we have used Eq.~\eqref{OPE phi Jvx w} with $j_i=\frac{1}{2}$ and $k=3$. Given that $\der \Gamma (z \sim z_i) \sim \w_i a_i  (z-z_i)^{\w_i-1}$, we have 
\begin{equation}
\frac{\der^2\Gamma}{\der \Gamma} = \sum_{i=1}^n \frac{\w_i-1}{z-z_i} - \sum_{a=1}^N \frac{2}{z-\lambda_a} \, ,    
\end{equation}
hence the candidate relation with the covering map reads 
\begin{equation}
\label{phi as log dGamma}
     \langle  \der\phi(z)\prod_{j=1}^n V_{h_j}^{\w_j}(x_j,z_j) 
    \rangle = 
\sqrt{2} \left[-\frac{\der^2\Gamma(z)}{2\der\Gamma(z)} \right]
     \langle   \prod_{j=1}^n V_{h_j}^{\w_j}(x_j,z_j) 
    \rangle \, . 
\end{equation}

In other words, this suggests that inside tensionless string correlators the following identities hold: 
\begin{equation}
\gamma (z) = \Gamma(z) \ , \qquad 
\phi(z,\zb) \sim \phi_0 -  \sqrt{2} \log |\der \Gamma| \, ,
\end{equation}
Here $\phi_0$ is a constant, that we can formally take to be very large in order to be close to the boundary.
This precisely reproduces what we expect from the classical solutions. For instance, we can look at a two-point functions with insertions of spectral flow charge $\w_1 = \w_2 = \w$ at $z_1=x_1=0$ and $z_2 = x_2 \to \infty$, for which the classical long string configuration reads 
\begin{equation}
    \gamma(z) = z^\w \, , \qquad \phi(z) = \phi_0 - \sqrt{2}(\w-1) \log |z| \,, 
\end{equation}
consistent with the corresponding covering map being $\Gamma(z) = z^w$.

However, a word of caution is in order. There is an important subtlety related to the identifications provided in Eqs.~\eqref{gamma is Gamma} and \eqref{phi as log dGamma}. Both the covering map $\Gamma(z)$ and the function $(\der \Gamma)^{-1} \der^2 \Gamma(z)$ have additional simple poles at $z=\lambda_a$, i.e.~away form the original insertion points. More precisely, there are $N = 1+ \frac{1}{2}\sum_{i=1}^n (\w_i-1)$ such poles. How can they be there on the LHS of \eqref{gamma is Gamma} and \eqref{phi as log dGamma} if there are no vertex operators inserted at these worldsheet points on the corresponding RHS? The answer to this puzzle is related to the need to include screening operators when working in the free-field formalism \cite{Dei:2023ivl}. We will come back to this in Sec.~\ref{sec: free fields k>3} below.

\subsection{Comments on tensionless superstrings on AdS$_3\times S^3 \times T^4$}

Before moving on, let us make a few comments on the supersymmetric case. 
If we insist on working with the RNS formalism at $n_5=1$, the model is of the form 
\begin{equation}
    {\rm SL}(2,\R)_3 \times {\rm SU}(2)_{-1}\times {\rm U}(1)^4 \quad 
    + \quad \text{10 free fermions} \, .
\end{equation}
The problem is that the bosonic SU(2) has negative level and central charge, such that quantization becomes problematic. Nevertheless, one can  use the available free-field realization in terms of symplectic bosons which effectively behave as fermionic ghosts \cite{Gaberdiel:2018rqv,Eberhardt:2025sbi}. These then cancel four fermionic degrees of freedom, generating a consistent theory albeit with only four physical  bosons and four physical  fermions. In other words, the only remaining fields are those of the (supersymmetric) $T^4$ model, the internal CFT. Combining this with the discussion of the previous sections,  one concludes that in the tensionless limit the two-dimensional holographic CFT  dual to tensionless strings in AdS$_3\times S^3 \times T^4$ \textit{is} the symmetric orbifold Sym$^{n_1}\left(T^4\right)$ itself.    

Alternatively, one can work directly in the hybrid formalism, i.e.~with the super WZW model based on PSU(1,1|2)$_1$, as was done in originally in \cite{Eberhardt:2018ouy}. This leads to a beautiful story, which unfortunately lies beyond the scope of these notes. We simply mention that, in this context, the shortening conditions on the spectrum and the localization of worldsheet correlators were proven \cite{Eberhardt:2018ouy,Dei:2020zui,McStay:2023thk}. Moreover, the discussion was extended to higher genus, and the matching of the full string partition function was demonstrated explicitly \cite{Eberhardt:2020bgq,Knighton:2022ipy,Eberhardt:2021jvj,Knighton:2020kuh,Gaberdiel:2020ycd,Gaberdiel:2022oeu,Dei:2023ivl}. The connection with twistor geometry was also emphasized recently in \cite{Bhat:2021dez,McStay:2023thk,McStay:2024dtk}.

Finally, we stress that the non-conservation of spectral flow shows that $n_1$, the number of fundamental strings in the background, is strictly speaking not  constant. This is related to the operator $\Ii$ in \eqref{def I(x)} acting differently depending on the spectral flow sector. It was argued in \cite{Porrati:2015eha} that one should think of the worldsheet theory as describing the system in the grand canonical ensemble. Indeed, the string partition function was actually matched precisely with that of the grand canonical ensemble of symmetric orbifold models over $n_1$, see \cite{Eberhardt:2020bgq} for the details of the computation. 

%%%%%%%%%%%%%%%%%%%%%%%%%%%%%
\newpage
%%%%%%%%%%%%%%%%%%%%%%%%%%%%%%%

%%%%%%%%%%%%%%%%%%%%%%%%%%%%%

\section{The exact chiral ring for strings  in  AdS$_3\times S^3\times T^4$} 
\label{sec: AdS3xS3xT4}

We continue with our holographic applications of the spectrally flowed worldsheet correlators derived in the Sec.~\ref{sec: bosonic correlators}. As a first step towards comparing with the holographic CFT beyond the tensionless limit, i.e.~for $k\geq 3$, we describe the propagation of superstrings in AdS$_3 \times S^3 \times T^4$ and focus on the protected BPS sector. We present the worldsheet operators corresponding to chiral primaries of the spacetime theory, and derive their three-point functions. These are constant along the moduli space thanks to the non-renormalization theorem of \cite{deBoer:2008ss,Baggio:2012rr}, and we show that they reproduce exactly the chiral ring structure constants as computed in the D1D5 CFT at the symmetric orbifold point  \cite{Lunin:2001pw}. Our presentation follows \cite{Dabholkar:2007ey,Giribet:2007wp,Iguri:2022pbp,Iguri:2023khc}.

\subsection{Supersymmetric worldsheet model in the RNS formalism}

String propagation in AdS$_3\times S^3\times T^4$ with NSNS fluxes is characterized by the  supersymmetric WZW model based on  SL(2,$\R$)$\times$SU(2)$\times$U(1)$^4$. As briefly mentioned above, a field redefinition allows one to rewrite any $\N=1$ super WZW model at level $k$ in terms of the bosonic WZW model with a shifted level $k-\mathfrak{c}$, where $\mathfrak{c}$ is the dual Coxeter number, and a system of free fermions \cite{DiVecchia:1984nyg,Kac:1985wdn,Kiritsis:1986wx}. 
For the case at hand, the relevant  OPEs read 
\begin{alignat}{4}
&J^a(z)\psi^b(w)&&\sim \frac{i\epsilon^{ab}{}_{c}\psi^c(w)}{z-w},\quad &&K^a(z)\chi^b(w)&&\sim \frac{i\epsilon^{ab}{}_{c}\chi^c(w)}{z-w},\\
&\psi^a(z)\psi^b(w)&&\sim \frac{\frac{n_5}{2}\eta^{ab}}{z-w},\quad &&\chi^a(z)\chi^b(w)&&\sim \frac{\frac{n_5}{2}\delta^{ab}}{z-w},
\end{alignat}
where $K^a$ are the currents of the SU(2) model studied in Sec.~\ref{sec: SU(2) WZW model}, while $\psi^a$ and $\chi^a$ are the  SL(2,$\R$) and SU(2) fermions, respectively.  In both cases, the supersymmetric level is set by $n_5$, the number of NS5-brane sources, hence the AdS$_3$ and $S^3$ factors have identical radii. The supersymmetric currents split as 
\beq
J^a = j^a + \hat{\jmath}^a\qqquad
K^a = k^a + \hat{k}^a,
\eeq
where $j^a$ and $k^a$ generate bosonic  SL(2,$\R$)$_{k}$ and SU(2)$_{k'}$ current algebras with $k = n_5 + 2$ and $k' = n_5-2$,  and
\beq 
\hat{\jmath}^a = -\frac{i}{n_5}\epsilon^{a}{}_{bc}\psi^b \psi^c\qqquad 
\hat{k}^a = -\frac{i}{n_5}\epsilon^{a}{}_{bc}\chi^b \chi^c \,, \label{Fermionic Currents}
\eeq
generate fermionic SL(2,$\R$)$_{-2}$ and SU(2)$_{2}$ algebras. In other words, at the algebraic level the decomposition reads 
\begin{equation}
    \N = 1 \quad \mathfrak{sl}(2,\R)_{n_5}\oplus \mathfrak{su}(2)_{n_5}  \, \simeq \, \mathfrak{sl}(2,\R)_{n_5+2} \oplus  \mathfrak{su}(2)_{n_5-2} \, \oplus \, 6 \, \text{free fermions} \, .
\end{equation}
We also have the free bosons and fermions associated with the torus directions, denoted  $Y^i$ and $\lambda^i$, with $i=6,\dots,9$.
%
%%%%%%%%%%%%%%%%%%%
%
It will be convenient to bosonize the fermions by introducing canonically normalized scalars $H_I$,  $I=1,\dots 5$, such that\footnote{More precisely, one must refine this and keep track of the cocycle factors by replacing $H_I$ with $\hat{H}_I = H_I + \pi \sum_{J<I} N_J$, where $N_J \equiv \oint i\del H_J $,   
so that 
$       e^{i a \hat{H}_I} e^{i b \hat{H}_J} = e^{i b \hat{H}_J} e^{i a \hat{H}_I} \, e^{i \pi a b}$ if $I > J$ \cite{Dabholkar:2007ey}. } 
\begin{subequations}
\begin{gather}
	 \psisl^\pm = \sqrt{n_5} \, e^{\pm iH_1} 
	 \, , \quad 
	\psisu^\pm = \sqrt{n_5} \, e^{\pm i H_2} 
	 \, , \quad
	 \lambda^{6} \pm i \lambda^7 = e^{\pm i H_4} 
	 \, , \quad 
	 \lambda^{8} \pm i \lambda^9 = e^{\pm i H_5} \, ,  \label{psiH1xiH2} \\
	 \qquad \psisl^0 = \frac{\sqrt{n_5}}{2} \, \left(e^{i H_3} - e^{-iH_3} \right)  \qqquad \psisu^0 =  \frac{\sqrt{n_5}}{2}  \, \left(e^{i H_3} + e^{-iH_3} \right) \, .
\end{gather}
\end{subequations}  
with $ H_I^\dagger = H_I$ for $I \ne 3 $ and $ H_3^\dagger = - H_3 $. 
% Conversely, 
% \begin{subequations}
%   \begin{gather}
%         i\del \hat{H}_1 = \frac{1}{n_5} \, \psisl^+\psisl^- \qqquad   i \del \hat{H}_2 =  \frac{1}{n_5} \, \psisu^+\psisu^-  \qqquad
% 	 i \del \hat{H}_3 = \frac{2}{n_5} \, \psisl^0\psisu^0  \, ,  
%     \\  i\del \hat{H}_4 = i\lambda^6 \lambda^7 \qqquad  i\del \hat{H}_5 = i\lambda^8 \lambda^9 \, ,
% 	\end{gather}
% \end{subequations}

%%%%%%%%%%%%%%%%%%%%%%

The stress tensor $T$ and supercurrent $G$ describing the matter sector of the worldsheet CFT are 
\begin{eqnarray}
    T &=& \frac{1}{n_5} \left(j^a j_a - \psisl^a \der \psisl_a + 
    k^a k_a - \psisu^a \der \psisu_a 
    \right) + \frac{1}{2}
    \left(\der Y^i \der Y_i - \lambda^i \der \lambda_i\right),
    \label{TAdS3S3T4def}
    \\
    G &=& \frac{2}{n_5} \left(
    \psisl^a j_a + \frac{2i}{n_5}\psisl^0 \psisl^1 \psisl^2 + 
    \psisu^a k_a - \frac{2i}{n_5}\psisu^0 \psisu^1 \psisu^2
    \right) + i \:\lambda^i \der Y_i \, .
    \label{GAdS3S3T4def}
\end{eqnarray}
hence the matter central charge gives 
\begin{equation}
    c = \frac{3(n_5+2)}{n_5} + \frac{3}{2} + \frac{3(n_5-2)}{n_5} + \frac{3}{2} + 4 + 2 = 15 \, .
\end{equation}
This is cancelled by the contributions from the standard $bc$ and $\beta \gamma$ ghost systems (the latter should not be confused with the Wakimoto fields). The BRST charge  reads 
\begin{equation}
\label{eq:BRSToperator}
    {\cal{Q}} = \oint dz \left[ c \left(T + T_{\beta\gamma}\right) - \gamma \, G + c(\der c) b - \frac{1}{4} b \gamma^2\right] \, .
\end{equation}
Finally, we take 
\begin{equation}
 \beta = e^{-\varphi} \del \xi \, , \qquad  \gamma =  e^{\varphi} \eta   \,, 
\end{equation}
where $\varphi$ is also canonically normalized and has a background charge $Q_\varphi = -2$, while $\xi(z)\eta(w) \sim (z-w)^{-1}$.
The spacetime supercharges can be constructed as \cite{Giveon:1998ns}
\begin{equation}
    Q_\vep = \oint dz \, e^{-\varphi/2} S_\vep \qqquad   S_\vep = \exp \left(\frac{i}{2} 
    \sum_{I=1}^{5}\vep_I H_I\right),
    \qquad \vep_I=\pm 1\,. \label{supercharges}
\end{equation}
BRST-invariance and mutual locality constraints impose $\vep_1 \vep_2 \vep_3 = \vep_4 \vep_5 = 1$, thus giving precisely the spacetime $\Nn=(4,4)$ global supercurrents. Moreover, the zero-modes of the worldsheet SU(2) currents are identified with the spacetime R-symmetry generators. 

\subsection{Spacetime chiral primaries}

Let us focus on the short string sector of the spectrum.  
We first consider the NSNS states. The (holomorphic part of the) vertex operators holographically dual to the low-lying chiral primaries of the holographic CFT is given by \cite{Gaberdiel:2007vu,Dabholkar:2007ey}
\begin{subequations}
\label{defVW0}
\bea
&\cV_{j}(x,u,z) &= e^{-\varphi(z)} \psi(x,z) V_{j}(x,z)W_{j-1}(u,z),\\
&\cW_{j}(x,u,z) &= e^{-\varphi(z)}V_{j}(x,z) \chi(u,z) W_{j-1}(u,z).
\eea 
\end{subequations}
in the canonical $(-1)$ ghost picture. Here $x$ is the holographic coordinate, $u$ is the SU$(2)$ isospin variable defined analogously, $z$ is the worldsheet coordinate, and $W_l(u,z)$ is an SU(2) primary field of spin $l$.  Given the weights $\Delta_j = -j(j-1)/n_5$ and $\Delta'_l = l (l+1)/n_5$, the Virasoro condition $\Delta_j + \Delta_l = 0$ is solved by setting $l = j-1$. As the supersymmetric zero-modes $J_0^a$ are identified the spacetime Virasoro modes at level $0,\pm 1$, the corresponding weight $H$ is identified with the spin of the corresponding SL$(2,\RR)_{n_5}$. Note that it might be  shifted with respect to the bosonic one due to the fermions. The same holds for the supersymmetric SU(2)$_{n_5}$ spin $L$. We have $H =j = l+1 = L$ for $\cW_{j}$, and $H = j-1 = l = L$ for $\cV_{j}$, showing that we are indeed dealing with states belonging to the chiral multiplets of the HCFT. 

On the other hand, states  associated to spacetime chiral primaries with $H>n_5/2$ belong to the spectrally flowed sectors of the worldsheet theory \cite{Giribet:2007wp}. 
%%%%%%%%%%%%%%%%%%
Bosonic flowed primaries where   studied at length in the previous sections. A similar story holds for the SL(2,$\R$) fermions, for which, in the $m$-basis, we can write 
\beq
\frac{\psi^\w (z)}{\sqrt{n_5}}\equiv \frac{\psi^{-,\omega}(z)}{\sqrt{n_5}} = e^{-i(1+ \omega)H_1(z) }\,,\quad \frac{\psi^{+,\omega}(z)}{\sqrt{n_5}} =e^{i(1- \omega)H_1(z) }\,,\quad\psi^{0,\omega}(z) = \psi^0(z) e^{-i\omega H_1(z)}.
\label{flowedfermionsSL2}
\eeq
Their $x$-basis cousins are defined as usual, i.e.~$\psi^\w(x,z) = e^{x \hat{\jmath}_0^+}  
    \psi^{-,\w}(z) e^{-x \hat{\jmath}_0^+}$, and we also have $\hat{\jmath}^\w(x,z) = e^{x \hat{\jmath}_0^+}  
    \hat{\jmath}^{-,\w}(z) e^{-x \hat{\jmath}_0^+}$.

Now, although including spectral flow in the SU(2) sector is, strictly speaking, not necessary, for short string states it is actually quite useful. Hence, we also introduce the  operators $W_{l,q}^\w (u,z)$, $\chi^\w (u,z)$, $\hat{k}^\w (u,z)$, which are analogous to their SL(2,$\R$) counterparts. 
The short-string flowed supersymmetric vertex operators then read 
\begin{subequations}
\label{defVWw}
\bea
&&\cV^{\omega}_{j}(x,u,z) = \frac{1}{\sqrt{n_5}}e^{-\varphi(z)} \psi^{\omega}(x,z) V^{\omega}_{j}(x,z) 
\chi^{\omega-1}(u,z) W^{\omega}_{j-1}(u,z) ,\\
&&\cW^{\omega}_{j}(x,u,z) = \frac{1}{\sqrt{n_5}}e^{-\varphi(z)} \psi^{\omega-1}(x,z) V^{\omega}_{j}(x,z) 
\chi^{\omega}(u,z) W^{\omega}_{j-1}(u,z) .\label{Flowed Fields}
\eea 
\end{subequations}
Indeed, flowing the SU(2) sector allows us to cancel the extra contributions to the worldsheet weights generated by the SL(2,$\R$) spectral flow, so that flowed states with  $l=j-1$ also satisfy the mass-shell condition. Here we have also introduced the shorthands
\begin{equation}
V^{\omega}_{j}(x,z) \equiv V^{\omega}_{j,h=j_\w}(x,z) \,, \quad 
W^{\omega}_{l}(u,z) \equiv W^{\omega}_{l,q=l_\w}(u,z)
\label{defVjw(x)}
\end{equation}
with $j_\w = j+k \w/2$ and $l_\w = l+k' \w/2$. The resulting spacetime weights are of the form  
\begin{equation}
    H \left[\cV^{\omega}_{j}\right] =  j-1+n_5 \w/2 \qqquad H \left[\cW^{\omega}_{j}\right] = 
    j+n_5 \w/2 \,.
    \label{hdefVWw}
\end{equation}

We now move to the Ramond-Ramond (RR) sector. Here vertex operators are constructed from the  spin fields defined in Eq.~\eqref{supercharges}. 
% as 
% \begin{equation}
%     S_{\vep_1,\vep_2,\vep_3,\vep_4,\vep_5} = e^{i\frac{\vep_1}{2}\hat{H}_1+i\frac{\vep_2}{2}\hat{H}_2+i\frac{\vep_3}{2}\hat{H}_3+i\frac{\vep_4}{2}\hat{H}_4+i\frac{\vep_5}{2}\hat{H}_5},
% \end{equation}
%Where the $H_i$ bosonic fields are the bosonized fermions plus the corresponding cocycles factors. Note that the field $\hat{H}_3$ involves naturally both the SL$(2,\RR)$ and SU$(2)$ sectors. The GSO projection is satisfied if
% \begin{equation}
%     \prod_i^{5} \vep_i = 1
% \end{equation}
We denote the AdS$_3\times S^3$ chirality as $\vep = \vep_1\vep_2\vep_3$, such that the  
%We can fix $\vep_3 = \vep \vep_1\vep_2$. 
GSO projection imposes $\vep_4\vep_5=\vep$. The relevant unflowed primaries involve SL(2,$\R$)$_{-2}\times$SU(2)$_2$ fields  
% \begin{equation}
%     S_{\vep_1,\vep_2,\vep_4,\vep} = e^{i\frac{\vep_1}{2}\hat{H}_1+i\frac{\vep_2}{2}\hat{H}_2+i\frac{\vep\vep_1\vep_2}{2}\hat{H}_3+i\frac{\vep_4}{2}(\hat{H}_4+\vep\hat{H}_5)}\equiv s_{\vep_1,\vep_2,\vep} e^{i\frac{\vep_4}{2}(\hat{H}_4+\vep\hat{H}_5)},
% \end{equation}
% The fields $s_{\vep_1,\vep_2,\vep}$ provides two
 of spins $(-1/2,1/2)$
 %product representation of the SL$(2,\RR)_{\hat{k}}$ and SU$(2)_{\hat{k}'}$ algebras at levels $\hat{k}=-2$ and $\hat{k}'=2$. The $\hat{\jmath}^0_0$ and $\hat{k}^0_0$ eigenvalues of each field $\hat{m}$ and $\hat{n}$ are given by 
% \begin{equation}
%     \hat{m} = \frac{\vep_1}{2}\qqquad \hat{n}= \frac{\vep_2}{2}.
% \end{equation}
% Then we can construct and $x$- and $u$-basis fields by shift the lowest-weight field as
, namely
\begin{equation}
    s_{\vep}(x,u,z) = e^{u\hat{k}^0_0 } e^{x\hat{\jmath}^0_0 } e^{\frac{i}{2} \left( -H_1(z)-H_2(z)+\vep H_3(z) \right) } e^{-x\hat{\jmath}^0_0 } e^{-u\hat{k}^0_0 } \qqquad \vep = \pm 1 .
    \label{Rxbasis}
\end{equation} 
% \begin{subequations}
% \bea
%     &&s_{+}(x,v) = s_{--+} - x s_{+-+} + iu s_{-++} - ixu s_{+++}\\
%     &&s_{-}(x,v) = s_{--} + x s_{+--} + iu s_{-+-} + ixus_{++-}
% \eea
% \end{subequations}
BRST-invariant unflowed states have total spins $(H,L) = (j-\frac{1}{2},l+\frac{1}{2})$, and at ghost picture ($-\frac{1}{2}$) they take the form \cite{Kutasov:1998zh,Dabholkar:2007ey} 
\begin{equation}
    \Yy^{\ep}_{j}(x,u,z) = e^{-\frac{1}{2}\varphi(z)} s_{-}(x,u,z)V_j(x,z)W_{j-1}(u,z)e^{i\frac{\ep}{2}\of{H_4(z)-H_5(z)}}, \label{R vertex}
\end{equation}
where we have renamed $\vep_4 \to \ep$. 
% The $\Yy^{\ep}_{j}$ belong are in a spin  representation of the full SL$(2,\RR)$ and SU$(2)$ algebras generated by the full currents $J^a_0$ and $K^a_0$ respectively. 
The corresponding flowed operators read  
% from the $m$- and $n$- basis expression of \eqref{R vertex} given by
% \begin{align}
%     \Yy^\vep_{jmn} =& e^{-\frac{\varphi}{2}}\Bigg\{\off{\of{s_{--}V_{j,m+\frac{1}{2}}+s_{+--}V_{j,m-\frac12}}W_{l,n+\frac12}}+\\
%     &+\off{i\of{s_{-+-}V_{j,m+\frac{1}{2}}+s_{++-}V_{j,m-\frac12}}W_{l,n-\frac12}}\Bigg\}e^{i\frac{\vep}{2}\of{\hat{H}_4-\hat{H}_5}} \nn
% \end{align}
% Applying a spectral flow automorphism we get an operator constructed by the flowed spin fields
% \begin{equation}
%     s^{\w}_{\vep_1,\vep_2,-} = e^{i\of{\frac{\vep_1}{2}-\w}H_1 +i\of{\frac{\vep_2}{2}-\w}H_2 -\frac{i\vep_1\vep_2}{2}H_3 }
% \end{equation}
% And the full vertex operator becomes
% \begin{align}
%     \Yy^{\ep,\w}_{jmn} =& e^{-\frac{\varphi}{2}}\Bigg\{\off{\of{s^\w_{--}V^\w_{j,m+\frac{1}{2}}+s^\w_{+--}V^\w_{j,m-\frac12}}W^\w_{l,n+\frac12}}+\\
%     &+\off{i\of{s^\w_{-+-}V^\w_{j,m+\frac{1}{2}}+s^\w_{++-}V^\w_{j,m-\frac12}}W^\w_{l,n-\frac12}}\Bigg\}e^{i\frac{\vep}{2}\of{\hat{H}_4-\hat{H}_5}} \nn
% \end{align}
% Similar to what happend with the NS flowed vertex operators, the Virasoro condition imposes $l = j-1$ and $m = -n = j-\frac{1}{2}$. Then, the flowed vertex simplify to 
% \begin{equation}
%      \Yy^{\ep,\w}_{j} = e^{-\frac{\varphi}{2}}s^\w_{--}V^\w_{j,j}W^\w_{j-1,-j+1}e^{i\frac{\vep}{2}\of{\hat{H}_4-\hat{H}_5}}
% \end{equation}
% And the $x$-basis version as
\begin{equation}
    \Yy^{\ep,\w}_{j}(x,u,z) = e^{-\frac{1}{2}\varphi(z)}s^\w_{-}(x,u,z)V^\w_{j}(x,z)W^\w_{j-1}(u,z)e^{i\frac{\ep}{2}\of{H_4(z)-H_5(z)}} \label{R flowed vertex},
\end{equation}
where $s^\w_{-}(x,u)$ is built upon the flowed primary $
     s^{\w}_{--} = e^{-i\of{\frac{1}{2}+\w}H_1(z) -i\of{\frac{1}{2}+\w}H_2(z) -\frac{i}{2}H_3(z) } $, i.e.~the extremal state in a spin ($-\frac{1}{2}-\w,\frac{1}{2}+\w$) representation of the fermionic zero-mode algebra. Consequently, the spacetime weights are 
\begin{equation}
    H \left[\Yy^{\ep,\w}_{j}\right] =  j-1/2+n_5 \w/2 .
\end{equation}

Let us now describe what each of these operators corresponds to in terms of the D1D5 CFT at the symmetric orbifold point, namely Sym$^{N}\left(T^4\right)$ with $N=n_1 n_5$, following \cite{Argurio:2000tb,Dabholkar:2007ey,Giribet:2007wp}. Recall that this is not the dual CFT away from $n_5=1$. Nevertheless, it contains all the relevant information when discussing observables protected by supersymmetry, such as the chiral primary spectrum, the  corresponding three-point functions, and also the extremal four-point functions \cite{Pakman:2009ab}. 

At large $N$, we should identify single-string states in the bulk with single-cycle fields of the dual CFT. As discussed in Appendix \ref{sec: appD SymOrbifolds}, each twist sector contains four chiral primary multiplets  \cite{Lunin:2001pw}. We will denote them as $O_n^-(x,u)$, $O_n^\ep(x,u)$ and $O_n^+(x,u)$, respectively. Their holomorphic weights are given by  
\begin{equation}
    H \left[O_n^{-} \right] = \frac{n-1}{2}  \,,\quad
    H \left[O_n^{\ep} \right] = \frac{n}{2} \,,\quad
    H \left[O_n^{+} \right] = \frac{n+1}{2} \,;\quad
 n=1,2,\dots.    \label{D1D5CFTweights3}
\end{equation}
These local operators of the HCFT living on the AdS$_3$ boundary correspond to $x$-basis operators of the worldsheet theory integrated over the worldsheet coordinates $z$ and $\zb$. 
Focusing on chiral primary operators that are spacetime scalars, the worldsheet operators constructed above are promoted to NSNS and RR vertex operators by including the  anti-holomorphic polarizations, giving, for instance, 
\bea
    &&\mathbb{V}^{\omega}_{j}(x,\bar{x},u,\bar{u},z,\bar{z}) \equiv \\ 
    &&\frac{1}{n_5}e^{-\varphi(z)-\bar{\varphi}(\bar{z})} \psi^{\omega}(x,z)\bar{\psi}^{\omega}(\bar{x},\bar{z}) V^{\omega}_{j}(x,\bar{x},z,\bar{z}) 
\chi^{\omega-1}(u,z) \bar{\chi}^{\omega-1}(\bar{u},\bar{z}) W^{\omega}_{j-1}(u,\bar{u},z,\bar{z}).\nn 
\eea
Here we have momentarily reinstated the dependence of the bosonic primaries on the anti-holomorphic variables for clarity.  $\mathbb{W}_j^\w$ and $\mathbb{Y}_j^{\ep,\w}$ are defined analogously. 
Hence, up to the normalization (to be discussed later on), the holographic dictionary reads
\begin{gather}
\begin{aligned}
    O_n^- (x,\bar{x},u,\bar{u}) & \leftrightarrow  \mathbb{V}^{\omega}_{j}(x,\bar{x},u,\bar{u},z,\bar{z}) , \\[1ex]
    O_n^\ep (x,\bar{x},u,\bar{u}) & \leftrightarrow \mathbb{Y}^{\ep,\omega}_{j}(x,\bar{x},u,\bar{u},z,\bar{z}) , \\[1ex]
    O_n^+ (x,\bar{x},u,\bar{u}) & \leftrightarrow  \mathbb{W}^{\omega}_{j}(x,\bar{x},u,\bar{u},z,\bar{z}) ,  
\end{aligned}
\end{gather}
together with the identification 
\begin{equation}
 n \,=\, 2 j - 1 + n_5 \, \w \,. \label{njw}
\end{equation}

From the worldsheet point of view, the allowed ranges are  
\begin{equation}
    j = 1,\frac{3}{2},\dots,
\frac{n_5}{2}
,\qquad
\w = 0,1,\dots \, , 
\label{jwrangecc}
\end{equation}
which shows that the worldsheet theory accounts for all chiral primaries of the holographic CFT, except for those belonging to twisted sectors for which $n$ is a (non-zero) multiple of $n_5$ \cite{Dabholkar:2007ey,Giribet:2007wp}. 
These would sit exactly at the lower boundary of the allowed range for $j$. However, at this point, the spectrum degenerates due to the presence of the zero-momentum states belonging to the continuous representations~\cite{Teschner:1997ft,Giveon:2001up}. 
This indicates that the NS5-F1 model sits at a singular point in moduli space, see  \cite{Seiberg:1999xz}. Upon (perturbatively) including RR fluxes, the long-string sector is lifted, and this issue is resolved \cite{Eberhardt:2018vho}.

Finally, we also need to discuss picture changing. Due to the presence of the ghost background charge, for three-point functions it is necessary to derive the ghost picture $(0)$ version of the NSNS operators constructed above. These are defined as 
\begin{equation}
    \Oo^{(0)}(z) = \lim_{w\rightarrow z}\off{e^{\varphi(w)} G(w)} \Oo^{(-1)}(z)\,.
\end{equation}
We obtain 
\begin{equation}
 \Vv^{\w,(0)}_j(x,u,z) = \Aa^{\w,1}_j(x,u,z) + (-1)^\w \Aa^{\w,2}_j(x,u,z) \,,
    \label{Vpicture0}
\end{equation}
with 
\begin{eqnarray}
   \Aa^{\w,1}_j(x,u,z) &=& \off{j^{-}_{-1-\w}(x,z)- H\hat{\jmath}^{-}_{-1-\w}(x,z)} \hat{\psi}^\w(x,z) V^\w_j(x,z) \hat{\chi}^{\w}(u,z) W^\w_{j-1}(u,z) \quad , \\[1ex]
   \Aa^{\w,2}_j(x,u,z) &=&-\frac{1}{n_5}\off{k^{+}_{\w}(u,z)-H\hat{k}^+_w(u,z)}\psi^{\w}(x,z)V^{\w}_j(x,z)\chi^{\w}(u,z)W^\w_{j-1}(u,z) ,\quad
\end{eqnarray}
and
\begin{equation}
    \Ww^{\w,(0)}_j(x,u,z) = \Bb^{\w,1}_j(x,u,z) + (-1)^\w \Bb^{\w,2}_j(x,u,z) \,,
        \label{Wpicture0}
\end{equation}
with 
\begin{eqnarray}
    \Bb^{\w,1}_j(x,u,z) &=& \off{k^{-}_{-1-\w}(u,z)-H\hat{k}^{-}_{-1-\w}(u,z)} \hat{\psi}^\w(x,z) V^\w_{j}(x,z)\hat{\chi}^{\w} (u,z) W^\w_{j-1}(u,z) , \quad \\
   \Bb^{\w,2}_j(x,u,z) &=& \frac{1}{n_5} \off{j^{+}_{\w}(x,z) + H\hat{\jmath}^{+}_\w(x,z)} \psi^{\w}(x,z)V^\w_j(x,z)\chi^\w(u,z)W^\w_{j-1}(u,z)  . \quad
\end{eqnarray}
In the RR sector, we will also need the field \eqref{R flowed vertex} in its $(-\frac{3}{2})$ ghost picture version for computing two-point functions. This takes the form 
\beq
\Yy^{\ep,\w,(-\frac{3}{2})}_j(x,u,z) =-\frac{\sqrt{n_5}
}{2j-1+n_5\w} e^{-\frac{3}{2}\varphi} s^{\w}_{+}(x,u,z) V^{\w}_j(x,z)W^{\w}_{j-1}(u,z) e^{i\frac{\ep}{2}\of{H_4-H_5}}
\,, \label{R flowed vertex -3/2}
\eeq
where $s^\w_{+}(x,u)$ is the $x$- and $u$-basis version of $
s^\w_{--+} = e^{-i\of{\frac{1}{2}+\w}H_1-i\of{\frac{1}{2}+\w}H_2+\frac{i}{2}H_3}$.
%%%%%%%%%%%%%%%%%%%%%%%%%%%%%%%%%
\subsection{Flowed primary correlators in the SU(2) and fermionic sectors}
\label{sec:primarycorrs}
%%%%%%%%%%%%%%%%%%%%%%%%%%%%%%%

In the SU(2) sector, spectrally flowed correlators are merely complicated linear combinations of primary and descendant unflowed correlators. Nevertheless,  we can use techniques analogous to those of Sec.~\ref{sec: bosonic correlators} to derive them directly. For this, we introduce the $v$-basis operators 
\begin{equation}
    \T{W}_l(u,v,z) = \sum_{n,\bar{n} = -l}^l v^{l+n}\bar{v}^{l+\bar{n}} W_{l,n+\frac{k'}{2}\w,\bar{n}+\frac{k'}{2}\w}(u,z)\,, 
\end{equation}
in analogy with the SL(2,$\R$) $y$-basis construction. 
By using the same covering maps, it is easy to see that three-point functions of flowed SU(2) primaries must  satisfy the same recursion relations as those of the SL(2,$\R$) model, albeit with the replacements 
\begin{equation}
\label{SL2 to SU2 replacements}
    k \to -k' \, , \qquad j_i \to -l_i\, . 
\end{equation}
It follows that, after fixing the overall dependence on $z_i$ and $u_i$, one gets  
\begin{align}
\langle \T{W}_{l_1}^{\w_1}(v_1)
    \T{W}_{l_2}^{\w_2}(v_2)
    \T{W}_{l_3}^{\w_3}(v_3) \rangle & = N_{\rm odd}'\, (v_1-a_1)^{2l_1}
    (v_2-a_2)^{2l_2}
    (v_3-a_3)^{2l_3} \label{oddfinal2su2} \\
    &\quad \times \left(
    \w_1\frac{v_1+a_1}{v_1-a_1}
    +\w_2\frac{v_2+a_2}{v_2-a_2}
    +\w_3\frac{v_3+a_3}{v_3-a_3} -1
    \right)^{-\frac{k'}{2}+l_1+l_2+l_3}, \nn 
\end{align}  
for odd parity correlators, and
\begin{eqnarray}
\label{eq: general even solution su2}
    \langle \T{W}_{l_1}^{\w_1}(v_1)
    \T{W}_{l_2}^{\w_2}(v_2)
    \T{W}_{l_3}^{\w_3}(v_3) \rangle &= & N_{\rm even}'
    \left(
    1-\frac{v_2}{ a_2[\Gamma_3^+]} -\frac{v_3}{a_3[\Gamma_2^+]}+
    \frac{v_2 v_3}{a_2[\Gamma_3^-] a_3[\Gamma_2^+] }\right)^{-l_1+l_2+l_3} \nn \\
    & \times &   \left(
    1-\frac{v_1}{a_1[\Gamma_3^+]}-\frac{v_3}{a_3[\Gamma_1^+]} + \frac{v_1 v_3}{ a_1[\Gamma_3^-]  a_3[\Gamma_1^+]}
    \right)^{-l_2+l_3+l_1} \\
    &\times &  \left(
    1-\frac{v_1}{a_1[\Gamma_2^+]} -\frac{v_2}{a_2[\Gamma_1^+]} +\frac{v_1 v_2}{a_1[\Gamma_2^+]a_2[\Gamma_1^-]} \right)^{-l_3+l_1+l_2}\,, \nn
\end{eqnarray}
for even parity correlators. The normalizations are given by 
\begin{equation}
    N_{\rm even}'(l_i,\w_i) = C'\left(l_1,l_2,l_3\right) \tilde{N}_{\rm even}' \,, \quad 
    N_{\rm odd}'(l_i,\w_i) = C'\left(k'/2-l_1,l_2,l_3\right) \tilde{N}_{\rm odd}'\,,
\end{equation}
where $C'(l_1,l_2,l_3)$ is the unflowed SU(2) three-point functions computed in \cite{Zamolodchikov:1986bd}, while 
\begin{equation}
    \tilde{N}_{\rm even}'(l_i,\w_i) = P_{(\w_1,\w_2,\w_3)}^{-l_1-l_2-l_3+k'}P_{(\w_1+1,\w_2+1,\w_3)}^{-l_3+l_2+l_1} P_{(\w_1,\w_2+1,\w_3+1)}^{-l_1+l_2+l_3}P_{(\w_1+1,\w_2,\w_3+1)}^{-l_2+l_3+l_1}\,, 
\end{equation}
and
\begin{equation}
    \tilde{N}_{\rm odd}'(l_i,\w_i) =   \of{\frac{P_{(\w_1-1,\w_2-1,\w_3-1)}}{\w_1+\w_2+\w_3-1}}^{-\frac{k'}{2}+l_1+l_2+l_3} 
    P_{(\w_1-1,\w_2,\w_3)}^{-l_3-l_2+l_1+\frac{k'}{2}} P_{(\w_1,\w_2-1,\w_3)}^{-l_3+l_2-l_1+\frac{k'}{2}}
    P_{(\w_1,\w_2,\w_3-1)}^{l_3-l_2-l_1+\frac{k'}{2}} \, .
\end{equation} 

We will also need correlators involving flowed fermions and spin fields. They can be deduced by noting that $\psi^{\w}(x,z)$ ($\chi^\w(u,z)$) is the flowed version of a spin $\hat{\jmath}=-1$ ($\hat{l}=1$) unflowed fermion, and hence belongs to a spin $\hat{\jmath}_\w = -1-\w$ ($\hat{l}_\w = 1+\w$)  representation of the zero-mode algebra of SL$(2,\RR)_{-2}$ (respectively SU$(2)_{2}$). Similarly, the fermionic field $\hat{\psi}^\w(x,z)$ ($\hat{\chi}^\w(u,z)$) can be seen as the flowed version of the fermionic identity, with flowed spin $\hat{\jmath}_\w = -\w$ ($\hat{l}_\w = \w$). Finally, the spin fields $s_\pm^\w(x,u,z)$ belong to  SL(2,$\R$)$_{-2}\times$SU(2)$_2$ representations with flowed spins $(\hat{\jmath}_\w,\hat{l}_\w)=(-\frac{1}{2}-\w,\frac{1}{2}+\w)$, obtained by implementing a spectral flow transformation on a spin $(\hat{\jmath},\hat{l})=(-\frac{1}{2},\frac{1}{2})$ state. Consequently, all relevant spectrally flowed fermionic correlators are obtained from the bosonic formulas given above by inserting the corresponding spins and levels. 

%%%%%%%%%%%%%%%%%%%%%%%%%%%%%%%
%%%%%%%%%%%%%%%%%%%%%%%%%%%%%%%

%%%

\subsection{Spectrally flowed correlators involving current insertions}

Let us consider NS-NS-NS short-string three-point functions, and focus on those involving only states polarized along the AdS$_3$ directions for concreteness. These correlators are computed by inserting two vertex operators with ghost picture (-1) and one with ghost picture (0). We see from  Eq.~\eqref{Vpicture0} that it is not enough to know the SL(2,$\R$) primary correlators, as one must also compute descendant correlators of the form 
\beq
\braket{V^{\w_1}_{j_1}(x_1,z_1)V^{\w_2}_{j_2}(x_2,z_2)\of{j^{\w_3}V^{\w_3}_{j_3}}(x_3,z_3)}.
\label{currentcorrelators}
\eeq
Here we continue using the shorthand $V^{\w}_{j}(x,z) \equiv V^{\w}_{j,h}(x,z)$ for $h = j + \frac{k}{2}\w$, which belongs to the $\mathcal{D}_j^{+,\w}$ representation, while the insertion involving the current is
\beq
\of{j^{\w}V^{\w}_{j}}(x,z) \equiv \oint_z \frac{dw}{(w-z)^{\w+1}} e^{x j^+_0}j^-(w) V^{\w}_{j,j}(z)e^{-x j^+_0}.
\eeq
with $V^{\w}_{j,j}(z)$ a flowed primary $m$-basis operator, derived from an unflowed lowest-weight state, as opposed to an $x$-basis operator. Due to the usual complications coming from spectral flow, correlators of the form \eqref{currentcorrelators} were originally identified as one as of the main obstacles for the worldsheet  computation of structure constants of the holographic chiral ring by the authors of \cite{Giribet:2007wp,Iguri:2022pbp}. 

The relevant bosonic three-point functions thus involve the action of the mode  $j^-_{-1-\w}(x)$ on a vertex operator $V_{j}^{\w}(x,z)$, which is a negative mode of the current $j^-(x,z)$ in the corresponding spectrally flowed frame. The problem is that, since these are not affine primary states, one cannot use the usual contour integration techniques to compute this, since one would need to start dealing with the unknown correlators $F_{n}^i$ discussed in Sec.~\ref{sec: bosonic correlators}, see Eq.~\eqref{defFin}. This was the approach considered recently in \cite{Eberhardt:2025sbi}. Nevertheless, this can be bypassed by noting that the series identification \eqref{series Id SL2 1} imply
% \beq
% V^{\w}_{j,j} = \N(j) V^{\w+1}_{\frac{k}{2}-j,-(\frac{k}{2}-j)}.
% \eeq
% Therefore
\beq
\of{j^{\w}V^{\w}_{j}}(x,z) = \N(j) \oint_z \frac{dw }{(w-z)^{\w+1}}e^{x j^+_0} j^-(w) V^{\w+1}_{\frac{k}{2}-j,-(\frac{k}{2}-j)}(z)
e^{-x j^+_0}, 
\eeq
where $V^{\w+1}_{\frac{k}{2}-j,-(\frac{k}{2}-j)}(z)$ is the highest-weight $m$-basis operator of the $\mathcal{D}_{\frac{k}{2}-j}^{-,\w}$ representation. The current mode $j^-_{-1-\w}$ can then be seen as a zero-mode in the spectrally flowed frame associated with an operator of charge $\w+1$, leading to
\beq
\label{eq44}
\of{j^{\w}V^{\w}_{j}}(x,z) = -\N(j) e^{x j^+_0}V^{\w+1}_{\tilde{\jmath},-\tilde{\jmath}-1}(z)
e^{-x j^+_0}=  -\N(j) V^{\w+1}_{\tilde{\jmath},\tilde{h}-1}(x,z),
\eeq
with
\begin{equation}
\tilde{\jmath} = k/2 - j \,,\quad  \tilde{h}= -\tilde{\jmath} + k(\w+1)/2 \,.   
\end{equation}
In Eq.~\eqref{eq44}, $V^{\w+1}_{\tilde{\jmath},-\tilde{\jmath}-1}(z)$ is an $m$-basis operator with unflowed spin projection $m=-\tilde{\jmath}-1$ belonging to the $\mathcal{D}_{\tilde{\jmath}}^{-,\w+1}$, while $V^{\w+1}_{\tilde{\jmath},\tilde{h}-1}(x,z)$ is its $x$-basis counterpart.
We conclude that we can express the relevant correlation functions with current insertions  in terms of flowed primary correlators in the following way: 
\begin{align}
\langle V^{\w_1}_{j_1}(x_1,z_1)V^{\w_2}_{j_2}(x_2,z_2)&\of{j^{\w_3}V^{\w_3}_{j_3}}(x_3,z_3)\rangle=\nn\\ &-\N(j_3) 
\braket{V^{\w_1}_{j_1}(x_1,z_1)V^{\w_2}_{j_2}(x_2,z_2)V^{\w_3+1}_{\tilde{\jmath}_3,\tilde{h}_3-1}(x_3,z_3)}\, .\label{Current Correlator} 
\end{align}
In $y$-basis language, the correlator appearing on the RHS of  \eqref{Current Correlator} reads
\beq
\braket{V^{\w_1}_{j_1}V^{\w_2}_{j_2}V^{\w_3+1}_{\tilde{\jmath}_3,\tilde{h}_3-1}} \hspace{-0.05cm}  = \hspace{-0.2cm} \lim_{y_3\rightarrow \infty} y_3^{k-2j_3}\braket{\T{V}^{\w_1}_{j_1}(y_1=0)\T{V}^{\w_2}_{j_2}(y_2=0)D^{-}_{y_3,\tilde{\jmath}_3}\T{V}^{\w_3+1}_{\tilde{\jmath}_3}(y_3)}, 
\eeq
where we have fixed  the insertions at $(z_1,z_2,z_3)=(x_1,x_2,x_3)=(0,1,\infty)$, and used that for half-BPS states we are dealing with lowest-weight vertex operators. 
By means Eqs.~\eqref{oddfinal2} and \eqref{eq: general even solution} we then obtain
\begin{equation}
\braket{V^{\w_1}_{j_1}V^{\w_2}_{j_2}V^{\w_3+1}_{\tilde{\jmath}_3,\tilde{h}_3-1}}  
    = \alpha_{\Bw} \lim_{y_3\rightarrow \infty} y_3^{k-2j_3}\braket{\T{V}^{\w_1}_{j_1}(y_1=0)\T{V}^{\w_2}_{j_2}(y_2=0)\T{V}^{\w_3+1}_{\tilde{\jmath}_3}(y_3)},
\end{equation}
where the coefficient is given by
\begin{align}
    \alpha_{\Bw} \equiv \begin{dcases}
        & \frac{2 a_3[\Gamma_{13}^{++}]\off{(\w_1-\w_2)(j_1-j_2)+(\w_3+1)(\frac{k}{2}-j_3)}}{\w_1+\w_3-\w_2+1} \,\, \text{if} \,\, 
        \sum_{i=1}^3 \w_i \in 2\mathbb{Z}+1,\\
        & \frac{2a_3[\Gamma_3^+]\off{(1+\w_1+\w_2)j_3-(1+\w_3)(j_1+j_2)-\frac{k}{2}(\w_1+\w_2-\w_3)}}{\w_3-\w_2-\w_1} \,\, \text{if} \,\,
        \sum_{i=1}^3 \w_i \in 2\mathbb{Z}.
    \end{dcases}
\end{align}
Here $a_3[\Gamma^{++}_{13}]$ denotes the coefficient $a_3$ of the covering map $\Gamma[\w_1+1,\w_2,\w_3+1](z)$. Finally,  by using the series identification once more we conclude that 
\beq
\braket{V^{\w_1}_{j_1}V^{\w_2}_{j_2}\of{j^{\w_3}V^{\w_3}_{j_3}}} = \alpha_{\Bw} 
\braket{V^{\w_1}_{j_1}V^{\w_2}_{j_2}V^{\w_3}_{j_3}}.\label{Current Correlator final}
\eeq
where the insertions are Moebius-fixed. 
Recall that this result holds only for the discrete lowest-weight states. Analogous formulas hold for the SL(2,$\R$) fermionic sector, and also for the SU(2) bosons and fermions.

%%%%%%%%%%%
\subsection{Supersymmetric three-point functions}

We now have all the necessary  ingredients for computing the NS-NS-NS short-string correlators with arbitrary spectral flow charges. This includes not only the extremal correlators (in the holographic CFT language, i.e.~those with $H_1 + H_2 = H_3$), some of which were briefly discussed in \cite{Giribet:2007wp}, but also the non-extremal ones. 

We continue with correlators involving three vertex operators polarized in the AdS$_3$ directions, namely $
\langle \Vv^{\w_1}_{j_1} \Vv^{\w_2}_{j_2} \Vv^{\w_3,(0)}_{j_3} \rangle$, and take the picture (0) operator to be the one with the largest spectral flow charge, say  $\w_3$, while also setting $(x_1,x_2,x_3) = (u_1,u_2,u_3) = (z_1,z_2,z_3)  = (0,1,\infty)$. All $y$-basis (and $v$-basis) three-point functions involved in the supersymmetric computation\footnote{The edge cases must be treated separately. As shown in \cite{Iguri:2023khc}, those correlators vanish.} turn out to be regular in the limits  $y_i\to 0$ (and $v_i \to 0$). We can thus select the relevant residues simply by setting $y_i = v_i = 0$. 
The vertex operators involved in $\langle \Vv^{\w_1}_{j_1} \Vv^{\w_2}_{j_2} \Vv^{\w_3,(0)}_{j_3} \rangle$ were given in Eqs.~\eqref{defVWw} and \eqref{Vpicture0}. For even parity correlators only the first term in \eqref{Vpicture0} contributes, while only the second one is relevant in the odd parity cases. In the former case, we have 
\beq
\langle \Vv^{\w_1}_{j_1} \Vv^{\w_2}_{j_2} \Vv^{\w_3,(0)}_{j_3} \rangle = \langle \Vv^{\w_1}_{j_1} \Vv^{\w_2}_{j_2} \Aa^{\w_3,1}_{j_3} \rangle \qqquad 
\sum_i \w_i \in 2\mathbb{Z}. 
\eeq
This factorizes into ghost, bosonic and fermionic correlators, the latter including a current insertion, namely  
\begin{equation}
    \langle \psi^{\w_1}\psi^{\w_2}\of{\hat{\jmath}^{\w_3}\hat{\psi}^{\w_3}}\rangle = - \frac{2a_3[\Gamma^+_3](2+\w_1+\w_2+\w_3)}{\w_3-\w_2-\w_1}\langle\psi^{\w_1}\psi^{\w_2}\hat{\psi}^{\w_3}\rangle, 
\end{equation}
where we have used \eqref{Current Correlator final} with $k \to \hat{k}= -2$, $j_{1,2} \to \hat{\jmath}_{1,2} = - 1$ and $j_3 \to \hat{\jmath}_3=0$. Hence,  
\begin{align}
    &\langle \Vv^{\w_1}_{j_1} \Vv^{\w_2}_{j_2} \Vv^{\w_3,(0)}_{j_3} \rangle = (h_1+h_2+h_3-2)
    \label{VVV0primary}\\
    &\quad \times\frac{2(1+\w_3)a_3[\Gamma_3^+]}{\w_1+\w_2-\w_3}\braket{V^{\w_1}_{j_1}V^{\w_2}_{j_2}V^{\w_3}_{j_3}}\langle\psi^{\w_1}\psi^{\w_2}\hat{\psi}^{\w_3}\rangle\braket{W^{\w_1}_{j_1-1}W^{\w_2}_{j_2-1}W^{\w_3}_{j_3-1}}\braket{\hat{\chi}^{\w_1}\hat{\chi}^{\w_2}\hat{\chi}^{\w_3}},\nn
\end{align}
where $h_i = j_i + n_5\w_i/2 = H_i+1$. The different spectrally flowed primary three-point can be evaluated explicitly. Although the individual factors look somewhat complicated, the final result is remarkably simple:
%that
% \begin{equation}
%     \frac{G[1+n]}{G[n]} = (n-1)!,\quad\quad  \frac{G[2+n]G[n]}{G[n+1]^2} = n
% \end{equation}
% is easy to check that
% \begin{equation}
%     \frac{2(1+\w_3)a_3[\Gamma_3^+]}{\w_1+\w_2-\w_3}\braket{V^{\w_1}_{j_1}V^{\w_2}_{j_2}V^{\w_3}_{j_3}}\langle\psi^{\w_1}\psi^{\w_2}\hat{\psi}^{\w_3}\rangle\braket{W^{\w_1}_{j_1-1}W^{\w_2}_{j_2-1}W^{\w_3}_{j_3-1}} \braket{\hat{\chi}^{\w_1}\hat{\chi}^{\w_2}\hat{\chi}^{\w_3}}= n_5 \Cc(j_i),
% \end{equation}
\begin{align}
    \langle \Vv^{\w_1}_{j_1} \Vv^{\w_2}_{j_2} \Vv^{\w_3,(0)}_{j_3} \rangle = n_5 (h_1+h_2+h_3-2) \Cc(j_i).
    \label{VVV0final}
\end{align}
Here $\Cc(j_i)$ is the product of the SL$(2,\RR)_k$ and SU$(2)_{k'}$ \textit{unflowed} three-point functions. 
It was shown in \cite{Dabholkar:2007ey} that for short strings the relation between the SL(2,$\R$) and SU(2) spins stemming from the Virasoro condition, namely $l_i = j_i-1$, leads to important cancellations for the latter, giving  
\beq
   \Cc(j_i) \equiv  C(j_i)C'(j_i-1) =
%    \label{CCprime1}
 \sqrt{\frac{b^2\gamma(-b^2)}{4\pi \nu }}\prod_{i=1}^3 \sqrt{B(j_i)} \equiv Q\prod_{i=1}^3 \sqrt{B(j_i)}, 
\label{defCproduct}
\eeq
where $b^2 = n_5^{-1}$. 

The computations above show that such cancellations extend non-trivially to the spectrally flowed sectors of the theory. As a result of the structural similarities between spectrally flowed correlators for  SL(2,$\R$) and SU(2), all combinatorial factors coming from the different contributions in the second line of \eqref{VVV0primary} exactly cancel. The only dependence of the final expression \eqref{VVV0final} on the spectral flow charges $\w_i$ is contained in the overall prefactor. 

As it turns out,  Eq.~\eqref{VVV0final} actually holds  for odd parity correlators as well. We have 
\beq
\langle \Vv^{\w_1}_{j_1} \Vv^{\w_2}_{j_2} \Vv^{\w_3,(0)}_{j_3} \rangle = \langle \Vv^{\w_1}_{j_1} \Vv^{\w_2}_{j_2} \Aa^{\w_3,2}_{j_3} \rangle, 
\eeq
and using that 
\begin{align}
    \langle W^{\w_1}_{l_1}W^{\w_2}_{l_2}\of{k^{+}_{\w_3}W^{\w_3}_{l_3}}\rangle = & \, \langle \T{W}^{\w_1}_{l_1}(v_1=0)\T{W}^{\w_2}_{l_2}(v_2=0)\p_{v_3}\T{W}^{\w_3}_{l_3}(v_3)\Large{|}_{v_3=0}\rangle \\
    =& -\frac{2\off{\w_3(l_1+l_2-\frac{k'}{2})-(1+\w_1+\w_2)l_3}}{(1+\w_1+\w_2+\w_3)a_3}\braket{W^{\w_1}_{l_1}W^{\w_2}_{l_2}W^{\w_3}_{l_3}},\nn
\end{align}
where $a_3$ is the coefficient from the map $\Gamma[\w_1,\w_2,\w_3]$, and
\begin{equation}
    \langle \hat{\chi}^{\w_1}\hat{\chi}^{\w_2}\of{\hat{k}^+_{\w_3}\chi^{\w_3}}\rangle=\frac{2}{a_3}\braket{\hat{\chi}^{\w_1}\hat{\chi}^{\w_2}\chi^{\w_3}},
\end{equation}
we find 
\begin{align}
     \langle \Vv^{\w_1}_{j_1} \Vv^{\w_2}_{j_2} \mathcal{A}^{\w_3,2}_{j_3}\rangle =& -\of{h_1+h_2+h_3-2}\\
     &\times 2\frac{\braket{V^{\w_1}_{j_1}V^{\w_2}_{j_2}V^{\w_3}_{j_3}}\braket{W^{\w_1}_{j_1-1}W^{\w_2}_{j_2-1}W^{\w_3}_{j_3-1}}\braket{\psi^{\w_1}\psi^{\w_2}\psi^{\w_3}}\braket{\hat{\chi}^{\w_1}\hat{\chi}^{\w_2}\chi^{\w_3}}}{(1+\w_1+\w_2+\w_3)a_3}\nn.
\end{align}
One readly obtains \eqref{VVV0final} after 
inserting the explicit expressions for each factor. 
A similar computation can be carried out for NS-NS-NS correlators involving one, two or three states polarized in the  SU(2) directions. The final results for these cases will be given in the following section.

Finally, we consider the R-R-NS three-point functions. The correlators
\begin{equation}
    \braket{\Yy^{\ep_1,\w_1}_{j_1} \Yy^{\ep_2,\w_2}_{j_2} \Vv^{\w_3}_{j_3}} \, \qquad  \text{and}  \qquad \braket{\Yy^{\ep_1,\w_1}_{j_1}\Yy^{\ep_2,\w_2}_{j_2}\Ww^{\w_3}_{j_3}}
\end{equation}
are technically simpler since no picture changing is necessary. 
The only new pieces of information that we need are the fermionic correlators involving spectrally flowed spin fields, namely $
    \braket{s_{-}^{\w_1} s_{-}^{\w_2} \psi^{\w_3}\hat{\chi}^{\w_3}}\braket{e^{\frac{i\ep_1}{2}\of{H_4-H_5}}e^{\frac{i\ep_2}{2}\of{H_4-H_5}}}$. 
The different sectors factorize up to an overall phase coming from the cocycle factors, which can be ignored since it will cancel out upon including the contributions from the anti-holomorphic sector.  
The torus correlators involving  $H_4$ and $H_5$ impose $\ep_1 = -\ep_2$. On the other hand, the SL$(2,\RR)$ and SU$(2)$ contributions give a product of a flowed three-point function with SL$(2,\RR)_{-2}$ spins $(\hat{\jmath}_1,\hat{\jmath}_2,\hat{\jmath}_3) = (-\frac{1}{2},-\frac{1}{2},-1)$ and SU$(2)_2$ spins $(\hat{l}_1,\hat{l}_2,\hat{l}_3)=(\frac{1}{2},\frac{1}{2},0)$. 
As for the NS-NS-NS cases, we find that all combinatorial factors related to spectral flow cancel out, giving
\begin{equation}
\braket{\Yy^{\ep_1,\w_1}_{j_1}\Yy^{\ep_2,\w_2}_{j_2}\Vv^{\w_3}_{j_3}} = 
\braket{\Yy^{\ep_1,\w_1}_{j_1}\Yy^{\ep_2,\w_2}_{j_2}\Ww^{\w_3}_{j_3}}  = \sqrt{n_5}\, 
\Cc(j_i) \xi^{\ep_1,\ep_2}\,, \qquad 
\xi =  \left(\begin{smallmatrix}
        0 & 1 \\
        1 & 0
    \end{smallmatrix}\right) \,.
\end{equation}

%%%%%%%%%%%%%%%
\subsection{Normalization and holographic matching}
\label{sec:normcorrelators}
%%%%%%%%%%%%%%%

As argued in \cite{Maldacena:2001km,Giribet:2007wp} and proved in \cite{Iguri:2022pbp}, in order to obtain an exact holographic matching, the worldsheet vertex operators must be properly normalized. 
Local operators of the CFT living on the AdS$_3$ boundary are given by vertex operators such as $\Vv(x,z)$ integrated over their worldsheet insertion point. Importantly, the worldsheet two-point functions contain a divergent factor $\delta(h_1 - h_2)$. This divergence is cancelled by the integration over $z_1$ and $z_2$, or alternatively by fixing the insertion points at $0$ and $\infty$ and dividing by the remaining conformal volume \cite{Maldacena:2001km,Giribet:2007wp}.  However, this cancellation is not exact, and introduces an additional finite but non-trivial multiplicative factor depending on $h$ and $\w$.  

This constant factor can be obtained by using the spacetime Ward identities associated with the R-symmetry currents $\mathcal{K}^a(x)$, see Sec.~\ref{sec: Exact theory, unflowed sector} above. In the supersymmetric version of the model we have 
\beq
\mathcal{K}^a(x) = - \int d^2z \, \frac{1}{n_5 c_\nu} K^a \bar{j}(\bar{x})V_1(x,\bar{x},z,\zb),\quad c_\nu = \frac{\pi\gamma(1-b^2)}{\nu b^2}. 
\eeq
Hence a generic vertex operator of the form $V^\omega_{h} (x) V_{\mathrm{int}}$, where $V_{\mathrm{int}}$ stands for the internal, fermionic and ghost contributions, must satisfy 
\beq
\braket{\mathcal{K}^{3}(x_1)V^\omega_{j}(x_2)V_{\mathrm{int},2}V^\omega_{j}(x_3)V_{\mathrm{int},3}}= \off{\frac{q_2}{x_{12}}+\frac{q_3}{x_{13}}}\braket{V^\omega_{j}(x_2)V_{\mathrm{int},2}V^\omega_{j}(x_3)V_{\mathrm{int},3}},\label{62}
\eeq
where $q_2$ and $q_3 = -q_2$ denote the corresponding R-charges. We can evaluate both sides of Eq.~\eqref{62} independently. Using the three-point functions derived above, the LHS becomes 
\beq
\frac{-q_2}{n_5c_\nu}\frac{\xb_{12}\xb_{13}}{\xb_{23}}\left |\frac{z_{23}}{z_{12}z_{13}}\right |^2\off{2\omega+1-2h}\braket{V_1(x_1)V^\omega_{j}(x_2)V^\omega_{j}(x_3)}\braket{V_{\mathrm{int},2}V_{\mathrm{int},3}},
\eeq
Moreover, \eqref{defCproduct} gives 
\beq
\braket{V_1(x_1)V^\omega_{j}(x_2)V^\omega_{j}(x_3)} = \mathcal{C}(1,j,j) = 2Q^2B(j). 
\eeq
The comparison with the RHS then shows that the string two-point function differs from the spacetime one by a factor 
\beq
\off{2h-1-2\omega}n_5c_\nu^{-1}2Q^2B(j).
\eeq
We thus derive the precise identification 
\begin{equation}
    \mathbb{O}^{\omega}_{j}(x,\bar{x},u,\bar{u},z,\bar{z}) = \frac{\Oo^{\omega}_{j}(x,\bar{x},u,\bar{u},z,\bar{z})}{\sqrt{2c_\nu^{-1}n_5Q^2\of{2h-1}B(j)v_4}},
    \label{ONormNS}
\end{equation}
where $h = j+n_5\w/2$,  $v_4$ is the $T^4$ volume in string units, and $\mathbb{O}$ stands for either $\mathbb{V}$ or $\mathbb{W}$. A similar computation shows that, due to the extra factor appearing in Eq.~\eqref{R flowed vertex -3/2}, for vertex operators in the RR sector of the worldsheet theory we have
\begin{equation}
    \mathbb{Y}^{\ep,\bar{\ep},\w}_j(x,\bar{x},u,\bar{u},z,\bar{z}) = \sqrt{\frac{2h-1}{2c_\nu^{-1}n_5^2Q^2B(j)v_4}} \Yy^{\ep,\bar{\ep},\w}_j(x,\bar{x},u,\bar{u},z,\bar{z}).
    \label{ONormR}
\end{equation}
In these expressions we have reinserted the anti-holomorphic dependence for clarity.  

String three-point functions among spacetime chiral primaries are thus obtained directly from the results obtained in the previous sections by including the explicit normalization factors. One must also include a factor of the string coupling $g_s = \sqrt{\frac{n_5 v_4}{n_1}}$ -- obtained from Eq.~\eqref{D1D5 AdS B-field and dilaton} combined with S-duality -- together with an additional factor of $v_4$.  Consequently, we find that the full set of normalized (spacetime) three-point functions under consideration takes the following form: 
\begin{subequations}
\label{Final3ptspacetimeNS}
\bea
&\braket{\mathbb{V}^{\w_1}_{j_1}\mathbb{V}^{\w_2}_{j_3}\mathbb{V}^{\w_3,(0)}_{j_3}}&= \frac{1}{\sqrt{N}}\off{\frac{\of{h_1 + h_2 + h_3-2}^4}{(2h_1-1)(2h_2-1)(2h_3-1)}}^{1/2},\\
&\braket{\mathbb{W}^{\w_1}_{j_1}\mathbb{V}^{\w_2}_{j_2}\mathbb{V}^{\w_3,(0)}_{j_3}}&=  \frac{1}{\sqrt{N}}\off{\frac{\of{1+ h_1 - h_2 - h_3}^4}{(2h_1-1)(2h_2-1)(2h_3-1)}}^{1/2},\\
&\braket{\mathbb{W}^{\w_1}_{j_1}\mathbb{W}^{\w_2}_{j_2}\mathbb{V}^{\w_3,(0)}_{j_3}}&=  \frac{1}{\sqrt{N}}\off{\frac{\of{h_1+h_2-h_3}^4}{(2h_1-1)(2h_2-1)(2h_3-1)}}^{1/2},\\
&\braket{\mathbb{W}^{\w_1}_{j_1}\mathbb{W}^{\w_2}_{j_2}\mathbb{W}^{\w_3,(0)}_{j_3}}&=  \frac{1}{\sqrt{N}}\off{\frac{\of{h_1+h_2+h_3-1}^4}{(2h_1-1)(2h_2-1)(2h_3-1)}}^{1/2},
\eea
\end{subequations}
and
\begin{subequations}
\label{Final3ptspacetimeR}
\bea
&\braket{\mathbb{Y}^{\ep_1,\bar{\ep}_1,\w_1}_{j_1}\mathbb{Y}^{\ep_2,\bar{\ep}_2,\w_2}_{j_2}\mathbb{V}^{\w_3}_{j_3}}&=  \frac{1}{\sqrt{N}}\off{\frac{\of{2h_1-1}\of{2h_2-1}}{(2h_3-1)}}^{1/2}\delta^{\ep_1,\ep_2}\delta^{\bar{\ep}_1,\bar{\ep}_2},\\
&\braket{\mathbb{Y}^{\ep_1,\bar{\ep}_1,\w_1}_{j_1}\mathbb{Y}^{\ep_2,\bar{\ep}_2,\w_2}_{j_2}\mathbb{W}^{\w_3}_{j_3}}&=  \frac{1}{\sqrt{N}}\off{\frac{\of{2h_1-1}\of{2h_2-1}}{(2h_3-1)}}^{1/2}\xi^{\ep_1,\ep_2}\xi^{\bar{\ep}_1,\bar{\ep}_2},
\eea
\end{subequations}
where $h_i = j_i+n_5 \w_i/2$. Here $
   \delta =  \left(\begin{smallmatrix}
        1 & 0 \\
        0 & 1
    \end{smallmatrix}\right)$ and $
   \xi =  \left(\begin{smallmatrix}
        0 & 1 \\
        1 & 0
    \end{smallmatrix}\right)$. The overall scaling with $N=n_1 n_5$ is obtained from \cite{Dabholkar:2007ey}  
\beq
\frac{1}{\sqrt{N}}= \frac{g_s}{n_5 \sqrt{v_4}} \sqrt{\frac{2\pi^5}{\nu b^4\gamma(1+b^2)}}.
\eeq
Note that the six-dimensional coupling $g_6 = g_s/\sqrt{v_4}$ is independent of the $T^4$ volume.
Finally, the selection rules on the SU(2) spins $l_i$ and spectral flow charges $\w_i$ can be summarized as follows:
\beq
l_i \leq l_k + l_j  \quad \text{and} \quad \w_i \leq \w_k +\w_j 
\quad \forall \quad i,j,k=1,2,3.
\label{finalfusionrules}
\eeq

The fusion rules \eqref{finalfusionrules} and the final expressions for the structure constants of the AdS$_3/$CFT$_2$ chiral ring presented in Eqs.~\eqref{Final3ptspacetimeNS} and \eqref{Final3ptspacetimeR} precisely reproduce the holographic CFT computations at the symmetric orbifold point \cite{Jevicki:1998bm,Lunin:2000yv,Lunin:2001pw,Dabholkar:2007ey,Giribet:2007wp}, see Appendix \ref{sec: appD SymOrbifolds}. 

\subsection{Higher-point functions}

The method reviewed in this section was recently extended to extremal four-point functions in \cite{Barone:2025vww}. This was based on the conjectured formula for flowed bosonic four-point functions of \cite{Dei:2021yom}. The analysis of the flowed conformal blocks builds on \cite{Dei:2022pkr,Iguri:2024yhb}, combined with going to the limit of small spacetime cross-ratio and integrating over the worldsheet one by adapting the analysis of \cite{Maldacena:2001km}. In the end, all results are consistent with the expected structure of the spacetime OPE. Finally, the free-field approach for higher-point functions was also  considered in \cite{Sriprachyakul:2024gyl,Yu:2024kxr,Yu:2025qnw,Knighton:2026wva}.  

%%%%%%%%%%%%%%%%%%%%%%%%%%%%%%%5

%%%%%%%%%%%%%%%%%%%%%%%%%%%%%
\newpage
%%%%%%%%%%%%%%%%%%%%%%%%%%%%%%%

%%%%%%%%%%%%%%%%%%%%%
\section{On the holographic theory for other values of $k$}
\label{sec: holography}

So far we have derived a precise holographic matching for (1) generic correlators at the tensionless point, namely $k=3$, and (2) correlators which are protected by supersymmetry for $k>3$.
In both cases, the dual derivation was performed at the symmetric orbifold point in moduli space. In this section, we discuss what is known about the holographic theory away from the tensionless point, and go beyond protected quantities. Our discussion draws heavily from the proposals of \cite{Balthazar:2021xeh,Eberhardt:2021vsx}, and also from the recent checks performed in \cite{Knighton:2023mhq,Knighton:2024qxd}. 

\subsection{An alternative description for long strings}

We have seen in previous sections that long stings can reach the asymptotic boundary of spacetime while remaining at finite energy. In fact, in the tensionless limit, long strings are, so to speak, forced to lie almost on top of the AdS$_3$ boundary. As a result, the string  worldsheet itself becomes the covering space employed in the computation of a given correlation function of the boundary theory. As argued in section \ref{sec:tensionless},  in this limit the dynamics of worldsheet excitations accurately describe the physics of the holographic CFT, which takes the form of a symmetric orbifold Sym$^{n_1} (\R_Q \times M_{\rm int})$ in the bosonic setting, where $n_1$ is the number of fundamental string sources. It thus makes sense to wonder what long strings can tell us about the nature of the dual theory\footnote{After all, in most examples of AdS/CFT we identify the dual CFT with the low energy limit of the theory on the D-brane sources.} at $k \neq 3$. 

In order to explore this, we briefly shift gears and consider the theory \textit{on} the long strings themselves \cite{Seiberg:1999xz} (see also \cite{Knighton:2024pqh}). More precisely, we study excitations on top of the classical long string solutions, employing light-cone quantization for simplicity. The simplest solution of this type corresponds to a single long string with unit winding sitting at a fixed radial distance; the latter is taken to be very large so that we remain close to the boundary, and are able to make use of the Wakimoto free-field description. This classical configuration takes the form 
\begin{equation}
    \gamma(z) = z \qquad \phi(z) = \phi_0\gg 1 \, . 
\end{equation}
Focusing on the holomorphic sector, we use the gauge freedom to fix $\gamma(z) = z$. The  Virasoro constraint 
\begin{equation}
    0 = T_{\rm AdS} + T_{\rm int} = - \beta \der \gamma - \frac{1}{2}\der \phi \der \phi - \frac{Q_{\phi}}{\sqrt{2}} \der^2 \phi  + T_{\rm int} \, , 
\end{equation}
with $Q_\phi = -\frac{1}{\sqrt{k-2}}$, then allows us to solve for $\beta$, giving 
\begin{equation}
\label{beta long string w=1}
    \beta = - \frac{1}{2}\der \phi \der \phi - \frac{Q_{\phi}}{\sqrt{2}} \der^2 \phi + T_{\rm int} \, . 
\end{equation}
Finally, we use the currents $J^+ = \beta$ and $J^3 = (\beta \gamma) - \sqrt{\frac{k-2}{2}} \der \phi$ to rewrite the \textit{spacetime} Virasoro modes. Starting from Eq.~\eqref{Ln Wakimoto}, we get  
\begin{eqnarray}
    \Ll_n^{(1)} &=& \oint dz \left[(n+1)\gamma^n J^3 - n \gamma^{n+1} J^+ \right] \nn \\
    &=& \oint dz \left[\beta z^{n+1} - (n+1) z^{n} \sqrt{\frac{k-2}{2}} \der \phi\right] 
    \label{Tst(1) long string} \\
    &=& \oint dz \, z^{n+1}\left[- \frac{1}{2}\der \phi \der \phi - \frac{Q_\phi - Q_\phi^{-1}}{\sqrt{2}} \der^2 \phi + T_{\rm int}\right] \, , \nn
\end{eqnarray}
where we have inserted \eqref{beta long string w=1} in the last step. 
Hence, it is tempting to identify the spacetime stress tensor $\Tt^{(1)}$ (in the unit winding sector, hence the superscript) with the term in brackets appearing in the final expression. 
%More precisely, for this interpretation to make sense, we should be able to replace $z \to x$, but, at least heuristically, this is exactly what we expect form the fact that (1) the long string itself is, roughly speaking, on top of the boundary, and (2) in the $\w=1$ sector $\gamma$ should behave as the appropriate covering map, i.e.~the main contributions to correlators of excitations on top of the long string comes from vertex operators inserted at $z = \gamma(z) \sim \Gamma(z) = x$. 
Eq.~\eqref{Tst(1) long string} suggests that it contains a factor analogous to the internal CFT on the worldsheet, combined with a non-compact scalar $\phi$. Importantly, the latter comes with a \textit{modified} background charge, namely 
\begin{equation}
\label{Q-HCFT def}
   Q = Q_\phi - Q_\phi^{-1} = \frac{k-3}{\sqrt{k-2}}. 
\end{equation}
In particular, we see that it changes sign as we move from $k>3$ to $k<3$, and vanishes precisely at the tensionless string point. We will come back to this in the next section. For now, we note that,  
as expected, the corresponding central charge gives 
\begin{equation}
    c_{\rm st}^{(1)} = 1 + 6 Q
^2 + c_{\rm int} = 1 + 6 \frac{(k-3)^2}{k-2} + 26 - \frac{3k}{k-2} = 6 k \, .
\end{equation}
Recall that we have already obtained this formula and inferred the presence of the effective charge $Q$ from a different perspective, namely when discussing the long string spectrum in Sec.~\ref{sec: sym orb spectrum 1}, see Eqs.~\eqref{Q alpha Eberhardt} and \eqref{cseed}.

So far we have described the theory on a single long string. The maximum number of long strings available is set by $n_1$, and given that all such long strings are indistinguishable, one is, once again, tempted to propose that the holographic CFT should be related to the following symmetric product orbifold: 
\begin{equation}
    {\rm Sym}^{n_1} \left[
    \R_{Q} \times M_{\rm int}
    \right] \, ,
    \label{asymptotic HCFT}
\end{equation}
where $M_{\rm int}$ stands for the non-linear sigma-model with the internal manifold as its target space. As we know, this indeed has $c_{\rm st} = 6 k n_1$.  Of course, this description can only be an approximate one, since our considerations are mostly about symmetries and kinematics, and they hold in a regime where the free field description is accurate. Moreover,  we have also ignored the short string sector.  Nevertheless, it is encouraging that this proposal is in precise agreement with the long string spectrum derived in Sec.~\ref{sec: sym orb spectrum 1}. Interestingly, we see that the non-compact direction of the holographic CFT (more precisely the center-of-mass field $\phi_{\rm CM}$ given by the gauge-invariant sum over all the $\phi_i$ in different copies of the seed theory) should be identified with the bulk radial direction.   

We can make the connection to the twisted sectors of the symmetric orbifold theory even more explicit by including spectral flow. For a long string winding $\w$ times around the AdS$_3$ asymptotic boundary we can take $\gamma(z) = z^\w$. We first discuss the spacetime R-symmetry generators for simplicity, which were derived in Sec.~\ref{sec: bosonic AdS3 spectrum}: 
\begin{equation}
    \Kk^{a \, (\w)}_n = \oint dz \, \gamma^{n} K^a(z) = \oint dz \, z^{\w n} K^a(z) \,.
\end{equation}
For these to be single-valued, the modes of the must take fractional values, i.e.~we must have $n \in \frac{1}{\w} \mathbb{Z}$. This is exactly what one expects in the $\w$-twisted sector of a symmetric orbifold theory. As for the spacetime Virasoro modes, we get
\begin{equation}
\label{Lnw 1}
    \Ll_n^{(\w)} = \oint dz \left[ \frac{z^{\w n+1}}{\w}\left(- \frac{1}{2}\der \phi^{(\w)} \der \phi^{(\w)} - \frac{Q}{\sqrt{2}} \der^2 \phi^{(\w)} + T_{\rm int} - \frac{\delta}{z^2} \right)  - \frac{z^{\w n} (\w-1)}{\sqrt{2}Q_\phi \w} \der \phi^{(\w)}  \right] \, ,
\end{equation}
where we have included a possible normal ordering constant entering when solving the Virasoro condition. This was ignored for $\w=1$, hence $\delta$ should vanish in that particular case. This ordering constant, which induces a  shift in $\Ll^{(\w)}_0$, was computed in \cite{Yu:1998qw}, giving $\delta = -\frac{1}{2}(\w+2)(\w-1)$. We further make use of the transformation of $\phi$ when adding $\w-1$ units of spectral flow, which reads $\phi \to \phi^{(\w)} = \phi - \frac{(w-1)}{\sqrt{2}Q_\phi} \log |z|^2$, see Eq.\eqref{spectral flow wakimoto}. Up to an additional shift of the zero mode, this is precisely what is needed to remove the term linear in $\der \phi$ in \eqref{Lnw 1}, leading to   
\begin{equation}
    \Ll_n^{(\w)} = \frac{1}{\w}\oint dz \left[ z^{\w n+1}\left(- \frac{1}{2}\der \phi \der \phi - \frac{Q}{\sqrt{2}} \der^2 \phi+ T_{\rm int} \right) \right] + \delta_{n,0} \frac{k(\w^2-1)}{4k} \, .
\end{equation}
As reviewed in Appendix \ref{sec: appD SymOrbifolds}, this mirrors the usual procedure for defining the \textit{untwisted} generators in $\hat{\Ll}_n$ in  symmetric orbifolds models (with \textit{seed} central charge $6k$) in terms of the fractionally-moded ones, namely
\begin{equation}
    \Ll_{\frac{n}{\w}} = \frac{1}{\w}\hat{\Ll}_{n} + \frac{k(\w^2-1)}{4\w} \delta_{n,0} \, , 
    \qquad n \in \mathbb{Z} \, . 
\end{equation}

The study of long strings thus  strongly suggests that, at least in the weakly coupled region, the Virasoro and R-symmetry generators of the theory should be well described by the holographic CFT given in \eqref{asymptotic HCFT}. The continuous sector of the worldsheet spectrum computed in Sec.~\ref{sec: sym orb spectrum 1} provides further evidence. In the following section we construct some of the relevant untwisted operators explicitly. We then study the appropriate deformation of the boundary theory, which takes into account the presence of a non-trivial potential  in the interior of AdS$_3$, and leads to the appearance of short strings, identified with bound states from this holographic perspective. 

\subsection{Operators in the untwisted sector of the spacetime theory}

Let us first briefly pause in order to discuss the interpretation of our results from the previous section. We have focused on the radial field $\phi$. From the worldsheet point of view, discussed at length in Sec.~\ref{sec: bosonic AdS3 spectrum}, we know that (near the boundary) this behaves as a free non-compact boson satisfying $
\phi(z,\zb) \phi(0) \sim - \log |z|^2$. The background  charge is $Q_\phi$, while the energy momentum tensor and central charge read 
\begin{equation}
    T(z) = -\frac{1}{2}\der \phi \der\phi - \frac{Q_\phi}{\sqrt{2}} \der^2 \phi \, , \qquad 
    c = 1 + 6 Q_\phi^2. 
\end{equation}
In our conventions, vertex operators $V_{\beta} = e^{\sqrt{2}\beta \phi}$ have weights 
\begin{equation}
    \Delta\left[V_\beta\right] = \beta (Q_\phi-\beta). 
\end{equation}
The (Euclidean) path integral definition for correlation functions is given by \begin{equation}
    \langle \prod_{i=1}^n V_{\beta_i}(x_i)\rangle = \int D\phi \, e^{-S} \prod_{i=1}^n V_{\beta_i}(x_i). 
\end{equation}
By using the action
\begin{equation}
    S= \frac{1}{
    4\pi
    %2\pi
    }\int d^2x
    \sqrt{g}\left[
    2
    \der  \phi \derb \phi + \frac{Q_\phi}{\sqrt{2}} R^{(2)} \phi \right] \, ,
\end{equation}
together with $n$ insertions of the form  
\begin{equation}
V_{\beta_i} = \exp \left[\sqrt{2} \beta_i\phi(x_i,\xb_i)\right] = \exp \left[\int d^2x \sqrt{2} \beta_i \phi(x,\xb) \delta(x-x_i) \right]    \,,
\end{equation}
we see that the contribution from the large (constant) $\phi$ region scales as 
\begin{equation}
    \langle \prod_{i=1}^n V_{\beta_i}(x_i)\rangle \sim  \exp \left[ \sqrt{2}\left(\sum_{i=1}^n\beta_i 
    -
    Q_\phi\right)\phi \right]\, ,
    \label{Louville path int}
\end{equation}
where we have used that for the sphere $\int d^2x \sqrt{g}R = 8\pi$.  

The first type of operator one considers are the analogues of the flat space plane waves, which have 
\begin{equation}
    \beta = \frac{Q_\phi}{2} + i p \, , \qquad p \in \R \, , 
\end{equation}
which form the usual Liouville spectrum.
Applying the above analysis to the two-point functions one finds that, due to the presence of the zero mode, they are delta-function normalizable. In the language of \cite{Seiberg:1990eb,Teschner:1999ug}  these are the macroscopic states.  
On the other hand, the so-called microscopic operators have  $\beta \in \R$. These are local operators when their wave-functions diverge in the weakly coupled regime. From the worldsheet action \eqref{Swzw H3 beta quantum} we identify the dilaton $\Phi$ as $\Phi \sim  Q_\phi \phi$ (up to a positive constant), and since  $Q_\phi < 0$ the coupling $g_s = e^{\Phi}$ vanishes in the limit $\phi \to + \infty$. This can also be seen by introducing a Liouville-type potential to regularize the $\phi \to - \infty$ region, which we identify as the strong coupling region in the bulk; we will come back to this type of deformation below. For now, we note that, as it follows from \eqref{Louville path int}, at large $\phi$ the wave-function associated to $V_\beta$ behaves as $e^{\sqrt{2}\left(\beta - \frac{Q_\phi}{2}\right) \phi}$, hence leading to the condition 
\begin{equation}
\label{normalizability Liouville 1}
    \beta > \frac{Q_\phi}{2}. 
\end{equation}

This should agree with the string theory perspective, where $\phi$ is identified with the radial field in SL(2,$\R$) WZW model, and local operators of the holographic CFT correspond to the non-normalizable modes of the operators \eqref{Vjm Wakimoto}, see Eq.~\eqref{Vjx semiclassical}. Comparing the two expressions we get $\beta = \frac{(j-1)}{\sqrt{k-2}}$ and $Q_\phi = - \frac{1}{\sqrt{k-2}}$. Since we are working at $\w=0$, this accounts for the worldsheet weight  $\Delta = \beta (Q_\phi-\beta) = -\frac{j(j-1)}{k-2}$ because the additional powers of $\gamma$ and $\gammab$ in the SL(2,$\R$) vertex operators \eqref{Vjm Wakimoto} are dimensionless. Moreover,    Eq.~\eqref{normalizability Liouville 1} translates into  $j>\frac{1}{2}$ as expected.  

On the other hand, we have learned that from the point of view of the boundary theory, we should consider a dual non-compact boson, albeit with an effective charge $Q = Q_\phi-Q_\phi^{-1} = \frac{k-3}{\sqrt{k-2}}$. Note that for $k>3$ this has the opposite sign as compared to $Q_\phi$. When this happens, one finds that for the HCFT the weak coupling regime is at $\phi \to - \infty$. The condition analogous to Eq.~\eqref{normalizability Liouville 1} hence reads 
$    \beta < Q/2$.
In terms of the worldsheet SL(2,$\R$) spin, this gives $j < \frac{k-1}{2}$. We thus recover the range \eqref{Djrange} from the holographic perspective. Conversely, for $k<3$ both weak coupling regions lie at the AdS$_3$ boundary. A more explicit discussion of the differences between these two cases is provided in the following section. 

Let us now consider long string $n$-point functions. In the worldsheet language, all insertions have $\re \beta_i = \frac{Q_\phi}{2}$,  hence the path integral is well behaved since at large $\phi$ we get an exponential suppression driven by  
\begin{equation}
    \re \sqrt{2}\left( \sum_{i=1}^n \beta_i - Q_\phi\right) \phi  = \frac{1}{\sqrt{2}}(n-2) Q_\phi\phi = - \frac{n-2}{\sqrt{2(k-2)}}\phi 
    \, . 
\end{equation}
However, from the boundary point of view we would like to think of exponential operators of the form $e^{\sqrt{2}\alpha \phi}$ as living in the untwisted sector of the symmetric orbifold theory. Given that we want to identify the latter with the singly-flowed sector of the worldsheet WZW model, by evaluating the corresponding spacetime weight $h$ derived in Eq.~\eqref{h for long strings} for $\w=1$ and $h_{\rm int}=0$ and comparing with $h = \alpha (Q-\alpha)$, we find that we should take
\begin{equation}
\label{def beta orbifold}
    \alpha = \frac{j+\frac{k}{2}-2}{\sqrt{k-2}}. 
\end{equation}
This makes sense because long strings should correspond to macroscopic operators with 
\begin{equation}
    \alpha = \frac{\frac{1}{2}+is+\frac{k}{2}-2}{\sqrt{k-2}} = \frac{Q}{2} + i p \, , \qquad p \equiv  \frac{s}{\sqrt{k-2}} \, .
\end{equation}
In other words, the holographic dictionary for operators in the untwisted sector of the boundary theory reads  
\begin{equation}
    e^{\sqrt{2}\alpha \phi(x)} \quad  \leftrightarrow \quad \int d^2z \, V_{j=\frac{1}{2}+i s,h=\alpha (Q-\alpha)}^{\w=1} (x,z). 
\end{equation}

Extrapolating these formulas to the short string sector, we find that the relation between the exponents $\alpha$ and $\beta$ defined above is consistent with the series identifications of Eq.~\eqref{series Id SL2 1}. The latter shows that unflowed operators can be identified as operators with $\w=1$ upon implementing the replacement $j \to \frac{k}{2}-j$. In particular, this applies to our worldsheet construction of the spacetime symmetry generators $\Ii$, $\Tt(x)$ and $\Kk^a(x)$ of Sec.~\ref{sec: Exact theory, unflowed sector}, which was based on the unflowed bosonic SL(2,$\R$) primary $V_{j=1}(x,z)$. In short, $\Ii$, $\Tt(x)$ and $\Kk^a(x)$ can also be understood as operators with $\w=1$. For instance, the integrand of the identity operator defined in \eqref{def I(x)} can be rewritten as  
\begin{equation}
\label{Id w=1}
   \Ii = \int d^2z  (J \bar{J} V_1)(x,\xb,z,\zb) \sim \int d^2z \, V_{\tilde{j},h=-\tilde{\jmath}-1 + \frac{k}{2}=0}^{\w=1} (x,\xb,z,\zb)\,, \qquad \tilde{\jmath} = \frac{k}{2}-1 \, ,  
\end{equation}
where we have used that in the frame with $\w=1$ we have $J^-_{-1} = \tilde{J}_0^-$.
For $\Kk^a(x)$ and $\Tt(x)$, we need to work with operators which are not spacetime scalars, but the idea is analogous. Hence, we can indeed see these operators as living in the untwisted sector of the holographic CFT. 
As a check, we note that taking the limit $\alpha \to Q$ should give a representation of the identity operator, and from Eq.~\eqref{def beta orbifold} we see that this indeed corresponds to taking $j \to \tilde{\jmath} = \frac{k}{2}-1$.

We should perhaps provide a small clarification. From the spacetime point of view, the relevant spacetime fields should be interpreted as the (gauge-invariant) sum over the corresponding images over all copies of the seed theory. In other words, in this section, whenever we wrote a boundary field such as $\phi(x,\xb)$, we really meant $\sum_{i=1}^{n_1} \phi_i$. With this notation, the rest of the untwisted spectrum is built by combining exponentials in $\phi$ with currents and primary fields of the sigma model with target space $M_{\rm int}$. The latter will translate almost trivially to the corresponding worldsheet fields belonging to the internal sector, which is why we have ignored them in the above discussion.

\subsection{The difference between $k>3$ and $k<3$}

The boundary path integral for the long-string $n$-point function will have a factor 
\begin{equation}
\label{k<3 vs k>3 path int}
    \re \sqrt{2}\left(\sum_{i=1}^n \alpha_i - Q\right) \phi  = \frac{1}{\sqrt{2}}(n-2) Q\phi . 
\end{equation}
This is well-behaved at the boundary $\phi \to +\infty$ only when $Q<0$. Thus, such correlators can only be computed directly from the path integral formalism in models with $k\leq 3$, i.e.~for stringy or even sub-stringy AdS$_3$ spaces. These are the configurations considered in the dualities proposed in \cite{Eberhardt:2018ouy,Balthazar:2021xeh}. 

The situation is qualitatively different when we move from $k\leq 3$ to   $k>3$. Indeed, in the former case, the range for defining normalizable (short string) operators in the worldsheet SL(2,$\R$) model reduces to $\frac{1}{2} < j < 1$. This means that spacetime vacuum state is no longer part of the string theory spectrum since it is created by the operator $\Ii$ constructed in \eqref{def I(x)} in terms of the $j=1$ bosonic vertex\footnote{This has interesting implications for black holes. Given that the BTZ black holes are obtained as quotients of global AdS$_3$, they also become non-normalizable and drop out of the spectrum for $k<3$ \cite{Giveon:2005mi}. In this sense, $k=3$ was interpreted as the string-black hole correspondence point in the AdS$_3$ context in \cite{Giveon:2005mi}. The absence of black holes is another aspect which makes the holographic duality is more tractable for $k<3$ \cite{Balthazar:2021xeh}. }. The same goes for the states created by $\Tt(x)$ and $\Kk(x)$.  However, $\Ii$, $\Tt(x)$ and $\Kk(x)$ still make sense as local operators. The situation is similar to what happens in Liouville theory, where the spectrum is built solely from the macroscopic states, which, in our context, corresponds to the long strings. 

Moreover, Eq.~\eqref{k<3 vs k>3 path int} shows that for $k>3$ the path integral description of the spacetime correlation functions becomes ill-defined. Nevertheless, given that the correlators computed at $k<3$ are analytic in the momenta $\alpha_i$ and the level $k$, it was argued in \cite{Eberhardt:2021vsx} that one can define correlators at $k>3$ by analytic continuation, which suggests that a very similar duality should hold also in this regime. This leads to the proposal of \cite{Eberhardt:2021vsx}, which is strongly supported by the (perturbative) matching between the highly non-trivial residues of the correlators of the putative holographic CFT and those of the worldsheet correlators at genus zero \cite{Hikida:2023jyc,Knighton:2023mhq,Knighton:2024qxd}. Recently, a detailed study of the so-called long-string partition function has led to the same holographic CFT \cite{Aharony:2024fid,Knighton:2024pqh}.
In the following sections we will be agnostic about this discussion and consider both situations on equal footing. 

%%%%%%%%%%%%%%%%%% new stuff

\subsection{Analytic structure of worldsheet correlators and Coulomb gas formalism}

As we have discussed at length, although the long string spectrum is that of a symmetric orbifold CFT, the presence of the shorts strings and the detailed structure of the spacetime correlation functions as derived from the worldsheet tell us that, away from the tensionless limit, the holographic theory is more complicated. It is believed that it, at least in conformal perturbation theory, one should be able to define the boundary theory by including a specific marginal deformation of \eqref{asymptotic HCFT}. We now combine what we have learned from the exact three-point correlators \cite{Dei:2021xgh,Bufalini:2022toj} with a free-field approach to higher point functions \cite{Knighton:2023mhq,Knighton:2024qxd} to provide some intuition about the precise form of this deformation, first discussed in \cite{Balthazar:2021xeh,Eberhardt:2021vsx}. 

The proposed holographic CFT will be of the form 
\begin{equation}
\label{proposed HCFT with def}
    {\rm Sym}^{n_1} \left[\R_Q \times M_{\rm int}\right] \,\,  + \,\, \Phi_{\rm st}{\rm-deformation} \, ,
\end{equation}
where we ask that the deforming operator $\Phi_{\rm st}$ satisfies, at the very least, the following conditions. 
\begin{itemize}
    \item It should be exactly marginal, and independent of the details of the internal CFT.
    
    \item As we know, short string states have wavefunctions that are confined to the interior of AdS$_3$. We can think of the curvature as generating a potential in the radial direction, such that short strings are understood as the corresponding bound states. Given that we have identified the non-compact field in \eqref{proposed HCFT with def} as capturing the physics along this direction, we expect the deformation to generate a (Liouville-type) potential for $\phi$. 

    \item We have also shown that three-point functions have spectral flow selection rules that, upon identifying $\w$ with the spacetime twist, are not those of a symmetric orbifold. This implies that $\Phi_{\rm st}$ should belong to a twisted sector of the spacetime theory, which will turn out to be $\w=2$.  
\end{itemize}

In order to establish a precise holographic duality, one would of course like to match all correlation functions exactly. We will be less ambitious, and attempt to do so at least in conformal perturbation theory\footnote{A similar, successful check was carried out in \cite{Aharony:2024fid,Knighton:2024pqh} by matching the perturbative expansion of the boundary theory (grand canonical) partition function and the long string path integral.}. This is analogous to using free-field techniques. In our context, it is natural to expect this to work as in Liouville theory \cite{Zamolodchikov:1995aa}. That is, instead of deriving the exact correlators, we will only have access to a crucial aspect of their analytic structure:  the computation of the residues of their poles in complex \textit{momentum} space. More precisely, the location of these poles, which in the language of the SL(2,$\R$) WZW model correspond to poles in the space of unflowed spins $j$, is understood in terms of charge conservation conditions for the $\phi$-exponential insertions, further taking into account the presence of the background charge. We should also allow for the presence of the so-called screening operators \cite{DiFrancesco:1997nk}. 

We now discuss the location of such poles for the case of the exact three-point functions obtained in Sec.~\ref{sec: bosonic correlators}. We focus on the poles occurring at special values of $j_1+j_2+j_3$; the rest of them can then be deduced from the reflection symmetry under $j_i \to 1-j_i$ for each $i=1,2,3$. The divergencies of the unflowed structure constants $C(j_1,j_2,j_3)$ follow from Eqs.~\eqref{C123} and \eqref{C123 Gj poles}. However, our analysis of spectrally flowed correlators indicates that, while $C(j_1,j_2,j_3)$ appears in all even-parity cases, for the odd-parity it is replaced by a factor $C(\frac{k}{2}-j_1,j_2,j_3)$. Moreover, one has to carry out the integration over the $y$-variables, which could lead to additional divergencies, and indeed it does in some cases. Up to spin reflections, one finds that the full set of poles is given by \cite{Eberhardt:2021vsx}
\begin{equation}
\label{3pt poles}
j_1+j_2+j_3-2 =  \frac{(m-1)(k-2)}{2} + r \, , \qquad  m,r \in \mathbb{Z}_{\geq 0} \, ,
\end{equation}
with the parity restriction 
\begin{equation}
\label{parity m}
    m \equiv \sum_{i=1}^3(\w_i-1) \,\, {\rm mod} \,\, 2 \, .
\end{equation}
We now interpret this double series of poles from the boundary and worldsheet points of view. 

From the HCFT perspective, we should work in terms of the momenta $\alpha_i$ and the background charge $Q$ defined in Eqs.~\eqref{def beta orbifold} and \eqref{Q-HCFT def}, respectively. Parametrizing $Q$ as 
\begin{equation}
    \label{HCFT b def}
    Q= \frac{1}{b}-b \, , \qquad b = \frac{1}{\sqrt{k-2}}\, ,  
\end{equation}
we get 
\begin{equation}
\label{j constraint alpha}
\sum_{i=1}^3\alpha_i = Q + \frac{m}{2b} + rb \, ,
\end{equation}
which generalizes to $n$-point functions simply by extending the upper limit of the sum.  
Combined with the condition in \eqref{parity m}, this suggests that the $\Phi_{\rm st}$ should have quantum numbers 
\begin{equation}
\label{Phi def quantum numbers}
\w_\Phi = 2 \, , \qquad \alpha_\Phi = -\frac{1}{2b} = -\frac{\sqrt{k-2}}{2} \, .  
\end{equation}
Indeed, this gives 
\begin{equation}
    h[
    \Phi_{\rm st} 
    ] = \frac{k (\w_\Phi^2-1)}{4\w_\Phi}  + \frac{\alpha_\Phi (Q-\alpha_\Phi)}{\w_\Phi} = 1 \, .
\end{equation}
 We will describe the implications for the computation of the HCFT correlation functions in conformal perturbation theory later on.

On the other hand, from the worldsheet perspective, we consider the Wakimoto description, where flowed $x$-basis vertex operators take the form given in Eq.~\eqref{Vw xbasis Wakimoto}, which we reproduce here for later convenience:  
\begin{equation}
\label{Vw xbasis Wakimoto bis}
    V_{jh}^\w(x,z) \equiv e^{(j-1+\frac{k-2}{2}\w)\sqrt{\frac{2}{k-2}} \phi(z)} \left[\frac{\der^\w(\gamma(z)-x)}{\w!}\right]^{\frac{k}{2}\w+j-h-1}
     \delta_\w(\gamma(z)-x) \, .
\end{equation}
In view of the exponential dependence in $\phi$, which has background charge $Q_\phi= - \frac{1}{\sqrt{k-2}}$, in this case it is more useful to rewrite \eqref{3pt poles} as 
\begin{equation}
\label{j-constraint ws}
    \frac{1}{\sqrt{k-2}}\sum_{i=1}^3 \left(j_i -1 + \frac{k-2}{2}\w_i\right)  = Q_\phi + \frac{1}{b} \left[1+\frac{1}{2}\sum_{i=1}^3(\w_i-1) + \frac{m}{2}\right] + r b
     \, ,
\end{equation}
which again generalizes to higher-point functions by extending the corresponding sums. We should be able to interpret the last two terms on the RHS of \eqref{j-constraint ws} as coming from two different screening operators relevant for the Coulomb gas description of the SL(2,$\R$) WZW model. Indeed, we immediately recognize the term proportional to $r$ by going back to the action \eqref{Swzw H3 beta quantum}: it corresponds to the insertion of $r$ times the (integrated) interaction term, which depends on $\phi$ as needed, namely 
\begin{equation}
\label{Sr screening def}
    \Ss_r = \int d^2z e^{-\sqrt{\frac{2}{k-2}}\phi} \beta \betab \, . 
\end{equation} 
We expect this interaction term to become less and less relevant close to the boundary of AdS$_3$\footnote{For a recent attempt to compute correlators with $r>0$ within the path integral formalism, see \cite{Sriprachyakul:2024gyl}.}. 
For the term proportional to $1/b$, the crucial point is that the factor between square brackets can be understood as the positive integer $N$ computing the number of poles of a (genus zero) holomorphic covering map defined by $3$ insertions with spectral flow charges $\w_i$ combined with $m$ additional $\w=2$ insertions! The appearance of such map is consistent with what we have just derived from the boundary point of view. Hence, we have 
\begin{equation}
\label{j-constraint ws N}
    \frac{1}{\sqrt{k-2}}\sum_{i=1}^3 \left(j_i -1 + \frac{k-2}{2}\w_i\right)  = Q_\phi + \frac{N}{b}  + r b
     \, ,
\end{equation}
with 
\begin{equation}
\label{N genus 0 with m}
    N= 1+\sum_{i=1}^3 \frac{\w_i-1}{2} + \frac{m}{2} \, . 
\end{equation}

What is the nature of this second screening operator?
Let us express it as 
\begin{equation}
\label{Screening SN def}
    \mathcal{S}_N = \int d^2z D(z)\bar{D}(\bar{z}) \, , 
\end{equation}
and summarize what we know about $D(z)$ so far:
\begin{itemize}
    \item It should have unit worldsheet weight $\Delta$. 

    \item Based on the discussion above, we do not want to generate new boundary insertions, hence it should have spacetime weight $h=0$. In other words, it should be invisible to the SL(2,$\R$) currents (up to total derivatives). 
    
    \item At the same time, it should have non-trivial OPEs with the individual Wakimoto fields $\gamma$ and $\phi$ (since $J^+ = \beta$). In particular, Eq.~\eqref{j-constraint ws N} suggests that it should be proportional to an exponential of the form $e^{-\sqrt{2(k-2)} \phi}$.  
\end{itemize}
Moreover, the fact that we expect to insert exactly $N$ screening operators in a given correlation function, with $N$ as in Eq.~\eqref{N genus 0 with m}, actually allows us to address a puzzle that was left unsolved in Sec.~\ref{sec: stringy geometry}. There, we mentioned that $\gamma(z)$ behaved precisely as a covering map inside correlators, see Eq.~\eqref{gamma is Gamma}. This was motivated by the OPEs in Eq.~\eqref{OPE gamma Vwxz}, although it raised the question of how to understand the singularities of $\Gamma(z)$ in terms of the free field description of the worldsheet CFT \cite{Eberhardt:2019ywk}.  
We now see that, provided $D(z)$ has a simple pole in its OPE with $\gamma(z)$, the presence of these screening operators explains why there were $N$ additional poles! Importantly, the location of these poles in $z$ is not fixed: it should be integrated over. 

A candidate screening operator was used early on in \cite{Iguri:2007af}, and also more recently in \cite{Hikida:2023jyc}. In the $m$-basis language, it  has $\w=-1$ and $j=2-\frac{k}{2}$, hence it can be written in terms of the Wakimoto fields as \cite{Dei:2023ivl,Knighton:2023mhq}
\begin{equation}
\label{D screening def}
    D(z) = V_{j=2-\frac{k}{2},m=\frac{k}{2}}^{\w=-1}(z) = e^{-\sqrt{2(k-2)} \phi(z)} \delta(\beta(z)) \left(\oint dw \, \gamma(w)\right)^{1-k} \, .
\end{equation}
Here we have made use of the delta function operators introduced in Sec.~\ref{sec: Wakimoto Vjhw} and, as usual, we have not included the analogous anti-holomorphic factors. As we shall see shortly, the final factor on the RHS of Eq.~\eqref{D screening def} indicates that the correlator will include a factor given by the residue of the relevant covering map at the corresponding insertion point. 

The operator in \eqref{D screening def} has negative spectral flow charge, and can thus be equivalently interpreted as an $x$-basis operator inserted at $x\to \infty$. This makes sense for the following reason. As discussed in \cite{Dei:2023ivl,Knighton:2023mhq}, we should remember that our string path integral approach includes the customary Wick rotation to Euclidean signature. The boundary of Euclidean AdS$_3$ has the topology of a two-sphere. Hence, the planar coordinates $\gamma$ and $\gammab$ we are using, which are associated with the classical limit of the corresponding Wakimoto field, are \textit{not} globally well-defined. Indeed, they diverge at, say, the north pole of the boundary $S^2$, where there is a curvature singularity. One way to deal with this would be to consider two different patches, which we would then need to glue appropriately. However, since the problematic region is given by a single point, one can effectively compactify the boundary by including additional insertions at this point, namely $x\to \infty$. From the worldsheet perspective, these additional insertions should induce divergencies of $\gamma(z)$, as expected. 

 \subsection{Holographic matching of residues for $k>3$}

Here we describe the computation of the momentum-space residues of $n$-point functions for bosonic strings in AdS$_3\times M_{\rm int}$ and their holographic matching.  This program was initiated in \cite{Eberhardt:2021vsx,Dei:2022pkr} for a set of three- and four-point functions, starting from their exact expressions, and indeed a perfect agreement was obtained for all cases under consideration. The details, which will not be discussed in these notes, are quite involved at the technical level. Moreover, the method is hard to extend to higher-point functions. Instead, we will discuss the path integral approach considered recently in \cite{Knighton:2023mhq,Knighton:2024qxd}\footnote{The supersymmetric extension was also considered in \cite{Sriprachyakul:2024gyl,Yu:2024kxr,Yu:2025qnw,Knighton:2026wva}.}. We then show how to rederive the corresponding results from the holographic CFT using conformal perturbation theory. This matching constitutes a highly non-trivial consistency check for the proposal of \cite{Eberhardt:2021vsx}. In terms of correlation functions, the duality reads 
\begin{equation}
\label{String HCFT pert matching}
    \sum_{
   g=0}^\infty
   g_s^{2g-2} \int_{{\cal{M}}_{g,n}} \Braket{\prod_{i=1}^n \Vv_{i}(x_i,z_i)}_{\rm String} = C_{S^2}  \Braket{\prod_{i=1}^n N_i \Oo_{i}(x_i,z_i)}_{\rm HCFT}^{c} \,.  
\end{equation}
Here, we have included several factors accounting for the relative normalizations for vertex operators and the full string path integral, namely the factors $N_i$ and $C_{S}^2$. Also, $g_s$ is the string coupling and ${\cal{M}}_{g,n}$ is the genus $g$ moduli space for $n$-point functions. The string correlator is computed in the worldsheet theory, i.e.~the SL(2,$\R$) WZW model at level $k$, while the HCFT correlator is defined in terms of the deformed symmetric orbifold model \eqref{proposed HCFT with def}.

As anticipated above, we will only be able to access the residues of both sides of \eqref{String HCFT pert matching}. For the sake of simplicity, we focus on the genus zero case and consider vertex operators which are trivial in the internal sector. Moreover, due to technical complications, so far only the infinite series of residues that correspond to setting $r=0$ in \eqref{j-constraint ws} were matched successfully. 

Finally, let us comment on the superscript $c$ included in the boundary correlator on the RHS of Eq.~\eqref{String HCFT pert matching}. This signals that one should only consider the so-called \textit{connected} contributions. Indeed, and as in the tensionless limit, we would like to identify the string worldsheet as the covering space involved in the computation of the spacetime correlator, and the LHS of \eqref{String HCFT pert matching} computes the contributions from a single, connected worldsheet\footnote{See \cite{Eberhardt:2020bgq} for a detailed discussion of these contributions in the context of tensionless strings.}. The power counting arguments in Appendix \ref{sec: appD SymOrbifolds} show that, at large $N$, the connected HCFT correlation functions are dominated by the contributions from  genus zero covering surfaces.

 \subsubsection{Free fields near the boundary of  AdS$_3$ and spectral flow% : a sketch of the of the path integral computation
 }
\label{sec: free fields k>3}
 
Let us consider the integrand of the $g=0$ term on the LHS of \eqref{String HCFT pert matching}. Omitting the anti-holomorphic factors and the string coupling constant as usual, we can express it as  
\begin{equation}
    \Braket{c(z_1)c(z_2)c(z_3)}
    \int \prod_{i=4}^n d^2z_i 
    \Braket{\prod_{i=1}^n V_{j_ih_i}^{\w_i}(x_i,z_i)}
    \Braket{\prod_{i=1}^n V_{{\rm int},i}(z_i)} \, ,
\end{equation}
where we have factorized the contributions coming from the reparametrization ghosts, which simply gives simply $\Braket{c(z_1)c(z_2)c(z_3)} = z_{12}z_{23}z_{13}$, and from the internal sector. Although it is not necessary, we assume that the latter trivializes. One should keep in mind that the value of each $h_i$ is related to the spin $j_i$ and the weight of the corresponding internal operator due to the Virasoro condition.  

We will work in the near-boundary region of AdS$_3$. In other words, we employ free-field techniques and do not include any insertion of the Wakimoto interaction term \eqref{Sr screening def}. Hence, we only take into account the screening operator defined in Eqs.~\eqref{Screening SN def} and \eqref{D screening def}, and set  
\begin{equation}
\label{corr screenings}
    \Braket{\prod_{i=1}^n V_{j_ih_i}^{\w_i}(x_i,z_i)} = 
    \sum_{N=0}^{\infty} \frac{p^N}{N!} 
    \Braket{\prod_{i=1}^n V_{j_ih_i}^{\w_i}(x_i,z_i) \prod_{a=1}^N \int d^2\lambda_a D(\lambda_a)}\, .
\end{equation}
Although this does not capture the exact result, it will compute the relevant residues. In \eqref{corr screenings} we have inserted a constant $p$ accounting for the normalization of $D(z)$. This can be adjusted by shifting $\phi$ with a constant, which ultimately reflects the freedom in choosing the value of the parameter $\nu$ appearing in the SL(2,$\R$) two-point function \eqref{SL2 bosonic 2point}. 

The RHS of \eqref{corr screenings} will be computed as a path integral over all relevant configurations of the Wakimoto fields $\beta, \gamma$ and $\phi$, where the vertex operators take the explicit form \eqref{Vw xbasis Wakimoto bis}. 
The charge conservation condition of the $\phi$-correlator imposes that, out of the whole perturbative series, only the terms satisfying the so-called $j$-constraint in Eq.~\eqref{j-constraint ws N} (with $r=0$) can contribute non-trivially. In other words, if 
% \begin{equation}
% j_1+j_2+j_3-2 =  \frac{(m-1)(k-2)}{2}
% \end{equation} %Answer (8)
\begin{equation}
\sum_{i=1}^n j_i +1 -n=  \frac{k-2}{2}(2-n+m)
\end{equation}
for some non-negative integer $m$, one gets delta-function which fixes  
\begin{equation}
    \label{N genus 0 with m n-point}
    N = 1+\sum_{i=1}^n \frac{\w_i-1}{2} + \frac{m}{2} \, .
\end{equation} 
Including  both the holomorphic and the anti-holomorphic sectors, the relevant $\beta\gamma$ correlator can then be written as 
\begin{equation}
\label{betagamma path integral}
\int\mathcal{D}[\beta\gamma]\,e^{-S_{\beta\gamma}}\prod_{a=1}^{N}\left|\oint_{\lambda_a}\gamma\right|^{-2(k-1)}\delta^{(2)} (\beta(\lambda_a))\prod_{i=1}^{n}\left|\left[\frac{\partial^{\w_i}\gamma(z_i)}{\w_i!}\right]^{\frac{k}{2}\w_i-h_i-j_i}\right|^2\delta_{\w_i}^{(2)}(\gamma(z_i)-x_i)\, ,
\end{equation}
with $S_{\beta\gamma} = (2\pi)^{-1} \int d^2z( \beta\derb \gamma + \betab\der \gammab)$, and where, as usual, in the absolute value squared should be understood as the anti-holomorphic contribution with $\bar{h}_i$ instead if $h_i$. The path integral over $\beta$ can be carried out explicitly by exponentiating the delta-functions $\delta^{(2)}(\beta(\lambda_a))$, giving 
\begin{align}
\int \mathcal{D}[\beta] e^{-S_{\beta\gamma}}\prod_{a=1}^{N}\delta^{(2)}(\beta(\lambda_a)) & =  
\int \mathcal{D}[\beta] e^{-S_{\beta\gamma}} \int \prod_{a=1}^{N} d^2c_a e^{i[c_a \beta(\lambda_a) + 
\bar{c}_a \betab(\bar{\lambda}_a)]} \nn \\
& = \int \prod_{a=1}^{N} d^2c_a \delta^{(2)}\bigg( \bar\partial\gamma-2\pi i\sum_a c_a\delta^{(2)}(z-\lambda_a)\bigg) \, .
\end{align}
Combined with the factors $\delta_{\w_i}^{(2)}(\gamma(z_i)-x_i)$, this tell us that, for the functional integration over $\gamma$, we must consider meromorphic functions with $N$ simple poles -- located at $z=\lambda_a$, and with $c_a$ as the corresponding residues, which are integrated over -- which further behave as $\gamma(z) \approx x_i + a_i (z-z_i)^{\w_i}$, for some constants $a_i$, near each of the original insertions. We know by now that these are precisely the branched covering maps $\Gamma(z)$ we have discussed at length in Secs.~\ref{sec: bosonic correlators} and \ref{sec:tensionless}.    

Let us be slightly more explicit about the integration one must carry out, while referring the interested reader to \cite{Knighton:2024qxd} for more details. The Riemann-Roch theorem states that  (at genus zero) the space of meromorphic functions $\gamma(z)$ with $N$ simple poles at specific locations which further satisfy $\gamma(z \to \infty) = \Gamma_\infty$ for some constant $\Gamma_\infty$ has complex dimension $N+1$. Roughly speaking, this is because one can always write such functions as 
\begin{equation}
\label{Gamma sum over poles}
    \gamma(z) = \Gamma_\infty + \sum_{a=1}^N 
    \frac{c_a}{z-\lambda_a} \, . 
\end{equation}
This means that the path integral over $\gamma$ localizes on a finite-dimensional integral over the constant $\Gamma_{\infty}$ and the residues $c_a$. We also know that the derivative of $\gamma(z)$ must read 
\begin{equation}
\label{der Gamma m}
    \der \gamma(z) = C^\Gamma \frac{\prod_{i=1}^3 (z-z_i)^{\w_i-1}\prod_{l=1}^m (z-\zeta_l)}{\prod_{a=1}^N (z-\lambda_a)^2} \, , 
\end{equation}
for some constant $C^\Gamma$. Now, in the string computation we also have to integrate over the locations of the poles, i.e.~the $\lambda_a$ for $a=1,\dots,N$, and also over the $z_i$ for $i=4,\dots,n$. However, due to the presence of the delta functions $\delta_{\w_i}^{(2)}(\gamma(z_i)-x_i)$, we must impose a considerable number of constraints. The final integration space thus has complex dimension
\begin{equation}
    N+1+N+n-3-\sum_{i=1}^n \w_i = m \, .
\end{equation}
This is precisely the amount of extra twist two insertions we expect from the dual perspective. At these points $\zeta_\ell$, we have $\gamma(z) \approx \xi_\ell + b_\ell (z-\zeta_\ell)^{2}$ for some constants $\xi_\ell$ and $b_\ell$, with $\ell=1,\dots,m$. To facilitate the holographic interpretation, the remaining $m$ complex integrals can be taken over the boundary insertion points $\xi_\ell$. Of course, one must take into account a non-trivial Jacobian ${\cal J}$, which was derived in \cite{Knighton:2024qxd}. 
In the end, the relevant contribution from the $\beta\gamma$ path integral \eqref{betagamma path integral}, further integrated over $z_i$ with $i=1,\dots,4$ and $\lambda_a$ with $a=1,\dots,N$ can be written in terms of the covering map data as  
\begin{equation}
    \int d^2\xi_1 \dots d^2\xi_m \, \Jj^{-1} \prod_{a=1}^N |c_a|^{-2(k-1)}
    \prod_{i=1}^n \left|a_i^{j_i+\frac{k}{2}\w_i-h_i-1}\right|^2
\end{equation}
with 
\begin{equation}\label{eq:jacobian-main-text}
\mathcal{J}=m! |z_{12}z_{13}z_{23}|^2|C^\Gamma|^{-4}\prod_{i=1}^n|\w_i a_i|^{\w_i+1} 
\prod_{\ell=1}^m|2b_\ell|\,.
\end{equation}
More precisely, this is the contribution one gets for each map, and the final formula will further include a discrete sum over all possible covering maps. 

We are thus nearly done. It remains to include the delta-function-stripped $\phi$-correlator 
\begin{equation}
\begin{aligned}
&\Braket{\prod_{a=1}^{N}e^{-\sqrt{2(k-2)} \phi}(\lambda_a)\prod_{i=1}^{n}e^{\left(j_i-1+\frac{k-2}{2}\w_i\right)\sqrt{\frac{2}{k-2}} \phi}(z_i)}'=\prod_{a<b}|\lambda_a-\lambda_b|^{-4(k-2)
}\\
&\qquad \times\prod_{i,a}|z_i-\lambda_a|^{4\left(j_i-1+\frac{k-2}{2}\w_i\right)} \prod_{i<j}|z_i-z_j|^{-\frac{4}{k-2}\left(j_i-1+\frac{k-2}{2}\w_i\right)\left(j_j-1+\frac{k-2}{2}\w_j\right)}\,,
\label{phicorrwick}
\end{aligned}
\end{equation}
which can also be written in terms of the same covering map data. Finally, the residue of the genus zero string $n$-point function gives \cite{Knighton:2023mhq,Knighton:2024qxd}
\begin{align}\label{final residues ws}
\begin{aligned}
\mathop{\mathrm{Res}}_{\sum_{i=1}^n (j_i - 1) = (m+2-n) \frac{k-2}{2}} &  \,\, g_s^{-2} \int_{{\cal{M}}_{0,n}} \Braket{\prod_{i=1}^n \Vv_i(x_i)} =  \frac{p^N}{g_s^{2}m!}\,  \left(2^{\frac{k}{2}-2}\right)^m \prod_{i=1}^{n}\w_i^{2-2j_i-\frac{k(\w_i+1)}{2}} \\
&\hspace{-1cm}\times \sum_{{\rm connected} \,  \Gamma}\int\mathrm{d}^2\xi_1\ldots\mathrm{d}^2\xi_m\,|C^{\Gamma}|^k\prod_{a=1}^{N}|c_a|^{-k}\prod_{i=1}^{n}\left|a_i^{-h_i+\frac{k(\w_i-1)}{4}}\right|^2\prod_{\ell=1}^{m}|b_{\ell}|^{\frac{k}{2}-2}\\
&\hspace{-1cm}\times\prod_{i<j}|z_i-z_j|^{-4\alpha_i\alpha_j}\prod_{i,\ell}|z_i-\zeta_{\ell}|^{-4\alpha_i\alpha_\Phi}\prod_{\ell<k}|\zeta_{\ell}-\zeta_{k}|^{-4\alpha_\Phi^2} \, .
\end{aligned}
\end{align}
In the following section, we show how this is reproduced from the boundary theory.

\subsubsection{Correlators of the Holographic CFT in perturbation theory}
\label{HCFT pert correlators}

We now consider the deformed symmetric orbifold theory \eqref{proposed HCFT with def}. The relevant part of the seed CFT is  described by the action
\begin{equation}
    S_0= \frac{1}{
    2\pi
    %2\pi
    }\int d^2x
    \sqrt{g}\left[
    \der  \phi \derb \phi + \frac{Q}{2\sqrt{2}} R^{(2)} \phi \right] \, ,
\end{equation}
where $Q$ was defined in \eqref{Q-HCFT def}. As discussed above, untwisted exponential vertex operators $V_{\alpha} = e^{\sqrt{2}\alpha \phi} $ have weights $h\left[V_\alpha\right] = \alpha (Q-\alpha)$. 
More generally, we would like to consider similar operators living in the twisted sectors of the theory. These operators, which will be denoted $V_{\alpha}^\w$, are obtained by combining the $V_{\alpha}$ with the so-called bare twist operators $\sigma_\w$, which enforce the corresponding boundary conditions. Hence, we have $V_{\alpha}^\w \sim \sigma_\w e^{\sqrt{2}\alpha \phi}$. More precisely, we use the gauge-invariant version of these operators, see Appendix \ref{sec: appD SymOrbifolds}. Their weights read
\begin{equation}
\label{h def HCFT}
    h\left[V_\alpha^\w\right]
    = \frac{k (\w^2-1)}{4\w} + \frac{\alpha(Q-\alpha)}{\w} \, , 
\end{equation}
where we have used that $c_{\rm seed} = 1+6 Q^2+c_{\rm int} = 6k$. 
Euclidean correlation functions can then be defined in terms of the  path integral, 
\begin{equation}
    \Braket{ \prod_{i=1}^n V_{\alpha_i}^{\w_i}(x_i)}_\mu = \int D\phi \, e^{-S} \prod_{i=1}^n V_{\alpha_i}(x_i),  
\end{equation}
with 
\begin{equation}
\label{S symorb + mu pert}
    S = S_{\rm sym \, orb} + \mu \int d^2x \,  \Phi_{\rm st} \, , 
\end{equation}
where we have included the deforming marginal operator $\Phi_{\rm st} = V_{\alpha_\Phi}^2$, with $\alpha_\Phi$ defined as  in Eq.~\eqref{Phi def quantum numbers}. This comes with a coupling constant $\mu$; we only care about the fact that $\mu$ is non-zero since its value can be adjusted by means of a constant shift in $\phi$. 

The presence of the exponential operators imposes a charge conservation condition, which implies that only a single term in the perturbative expansion
\begin{equation}
\label{pert expansion HCFT}
   \Braket{\prod_{i=1}^n V_{\alpha_i}^{\w_i}(x_i)}_\mu = \sum_{m=0}^\infty \frac{(-\mu)^m}{m!} \int  d^2 \xi_1 \dots 
    d^2 \xi_m
    \Braket{ \prod_{i=1}^n V_{\alpha_i}^{\w_i}(x_i)
    \prod_{\ell=1}^m 
    V_{\alpha_\Phi}^2(\xi_\ell)
    }_{\mu=0}
\end{equation}
contributes non-trivially. At genus zero, the corresponding value of $m$ is determined by the fact that, in the free-field approach, the correlator appearing in the integrand will be proportional to the delta function 
\begin{equation}
\label{delta charge conservation pert HCFT}
    \delta\left(\sum_{i=1}^n\alpha_i +  m \alpha_\Phi 
    -
    Q \right) \, .
\end{equation}
As in Liouville theory \cite{Zamolodchikov:1995aa}, this indicates that, in the exact theory,  the correlation functions on the LHS of Eq.~\eqref{pert expansion HCFT} must have simple poles when the $\alpha_i$ are chosen in such a way that 
the argument of the delta function in \eqref{delta charge conservation pert HCFT} vanishes. This happens when the $j$-constraint \eqref{j constraint alpha} with $r=0$ holds. The residue at this pole is then given by the RHS of  \eqref{pert expansion HCFT},  evaluated at the corresponding value of $m$, and with the delta function stripped off. 

The correlation function on the RHS  of \eqref{pert expansion HCFT} is computed in the undeformed theory. The usual techniques for symmetric orbifold CFTs \cite{Lunin:2000yv,Pakman:2009zz,Dei:2019iym} are briefly reviewed in Appendix \ref{sec: appD SymOrbifolds}. As explained above, we focus on the connected contributions. For the result to be non-vanishing, there must exist at least one covering map $\Gamma(z)$ satisfying 
\begin{align}
\begin{aligned}
    &\Gamma(z \sim z_i) \approx x_i + a_i (z-z_i)^{\w_i} \, , \qquad 
    \forall \,\, i = 1,\dots n \, , \\
    & \Gamma(z \sim \zeta_\ell) \approx \xi_\ell + b_\ell (z-\zeta_\ell)^{2} \, , \qquad 
    \,\,\,\forall \,\, \ell = 1,\dots m \, , 
\end{aligned}
\end{align}
for some coefficients $a_i$ and $b_\ell$, and where $z$ is the coordinate on the covering space. This must be a function of the form \eqref{Gamma sum over poles} with $N$ simple poles, while its derivative is given by Eq.~\eqref{der Gamma m}. 
The final result for the $n$-point function consists of a sum over such covering maps, such that each term is expressed as a factor containing the covering map data times a free-field correlator to be evaluated on the covering space. Explicitly, one finds \cite{Dei:2019iym}
\begin{align}
\label{eq:sym-correlators}
&\Braket{ \prod_{i=1}^n V_{\alpha_i}^{\w_i}(x_i)
    \prod_{\ell=1}^m 
    V_{\alpha_\Phi}^2(\xi_\ell)
    }_{\mu=0}^c =  \sum_{\Gamma} 
n_1^{1-\frac{n+m}{2}}|C^{\Gamma}|^k\prod_{a=1}^{N}|c_a|^{-k} \\
&\hspace{1cm}\times \prod_{i=1}^{n}\left(\w_i^{\frac{1}{2}-\frac{k(\w_i+1)}{2}}|a_i|^{-2h_i+\frac{k(\w_i-1)}{2}}\right) \prod_{\ell=1}^{m}\left(2^{\frac{1}{2}-\frac{3k}{2}}|b_{\ell}|^{\frac{k-4}{2}}\right)\Braket{\prod_{i=1}^{n}V_{\alpha_i}(z_i)\prod_{\ell=1}^{m}V_{\alpha_\Phi}(\zeta_{\ell})}'\, , \nn
\end{align}
where the prime indicates that we remove the delta function from the correlator on the RHS. Finally, performing the free-field contractions leads to the final (genus zero) expression \cite{Eberhardt:2021vsx,Knighton:2024qxd}
\begin{equation}\label{final residues HCFT}
\begin{split}
\mathop{\mathrm{Res}}_{\sum_{i=1}^n\alpha_i +  m \alpha_\Phi 
    =
    Q} &\Braket{\prod_{i=1}^n V_{\alpha_i}^{\w_i}(x_i)}_\mu^c = \frac{1}{m!}\,  \left(-2^{\frac{1}{2}-\frac{3k}{2}}\mu\right)^m n_1^{1-\frac{n+m}{2}}\prod_{i=1}^{n}\w_i^{\frac{1}{2}-\frac{k(\w_i+1)}{2}} \\
&\hspace{-1cm}\times \sum_{{\rm connected}\, \Gamma}\int\mathrm{d}^2\xi_1\ldots\mathrm{d}^2\xi_m\,|C^{\Gamma}|^k\prod_{a=1}^{N}|c_a|^{-k}\prod_{i=1}^{n}|a_i|^{-2h_i+\frac{k(\w_i-1)}{2}}\prod_{\ell=1}^{m}|b_{\ell}|^{\frac{k}{2}-2}\\
&\hspace{-1cm}\times\prod_{i<j}|z_i-z_j|^{-4\alpha_i\alpha_j}\prod_{i,\ell}|z_i-\zeta_{\ell}|^{-4\alpha_i\alpha_\Phi}\prod_{\ell<k}|\zeta_{\ell}-\zeta_{k}|^{-4\alpha_\Phi^2} \, .
\end{split}
\end{equation}
This precisely matches the (small $g_s$) string result \eqref{final residues ws} provided we identify 
\begin{equation}
    g_s = \frac{1}{\sqrt{n_1}} \, , \qquad 
    \mu = -  2^{2k-\frac{5}{2}}\sqrt{n_1 p}\,,
\end{equation}
and fix the normalization constants in \eqref{String HCFT pert matching} to be 
\begin{equation}
    C_{S^2} = p \, , \qquad 
    N_i =  p^{\frac{\w_i-1}{2}}\w_i^{\frac{3}{2}-2j_i} \sqrt{n_1} \, .
\end{equation}
In the exact theory, this should hold up to factors which may only depend on $k$, and large $n_1$ corrections, see \cite{Eberhardt:2021vsx}.  

Needless to say, it would be extremely interesting to extend the matching to the full set of residues \eqref{j-constraint ws N}, i.e.~including those with $r>0$. These correspond to subleading corrections in the language of the proposed holographic CFT \cite{Eberhardt:2021vsx}.

\subsection{Comments on discrete states and grand canonical ensembles}

Let us finish this section by commenting on two aspects of the proposal in Eq.~\eqref{proposed HCFT with def} for the conformal field theory holographically dual to bosonic strings in AdS$_3\times M_{\rm int}$ that are somewhat more subtle to address. 

We have seen above that long string states living near the boundary of AdS$_3$ are in one-to-one correspondence with the so-called macroscopic states (in the language of \cite{Teschner:1999ug,Seiberg:1990eb}) of the boundary symmetric orbifold CFT, for which the seed CFT is a non-compact, Liouville-type theory. The holographic dictionary in this sector of the theory identifies worldsheet long-string vertex operators with unflowed spin $j$, spectral flow charge $\w$ and internal weight $h_{\rm int}$, and boundary operators with twist $\tau=\w$, $\phi$ exponent $\alpha$ as in Eq.~\eqref{def beta orbifold} and a spacetime weight $h$ given by Eq.~\eqref{h for bosonic sym orbifold HCFT}. One might then ask: what about the discrete states? To answer this, let us, for a moment, assume that long string two- and three-point functions have been matched \textit{exactly}. Of course, this exact matching would require a better handling of the marginal deformation included in the definition of the proposed  HCFT. If achieved, it would mean that the corresponding holographic long string propagators contain the factors 
\begin{equation}
    \frac{\gamma(j+\frac{k\w}{2}-h)}{\gamma(1-j+\frac{k\w}{2}-h)} 
\end{equation}
appearing in Eq.~\eqref{SL2 2point bosonic with flow}. This would mean that the boundary CFT knows about the reflection coefficient for spectrally flowed states.  
The interesting point is that this coefficient includes simple poles at the locations of the discrete states, i.e.~at $m=h-\frac{k}{2}\w=j+n$ and $m  =h-\frac{k}{2}\w = -j-n$ with $n\in \mathbb{N}_0$. It was argued in \cite{Eberhardt:2021vsx} that these should be understood as LSZ-type poles, in the sense of  \cite{Aharony:2004xn}. The location of these poles signals the presence external particles  going on-shell. Hence, even though the short-string states are, for the most part, invisible in the undeformed symmetric orbifold theory, they seem to become part of the spectrum when the deformation is included. This makes sense, since we view them as bound states, which live in the bulk of AdS$_3$, and we have identified the boundary Liouville field with the radial direction in the dual gravitational theory.

Finally, we discuss an interesting, albeit more heuristic link between the holographic picture developed over the last few pages and the discussion about the fact that the spacetime "identity" operator $\Ii$ is not exactly the spacetime identity. 
Consider the boundary theory, described by the action \eqref{S symorb + mu pert}. The marginal operator $\Phi_{\rm st}$ comes with the coupling constant $\mu$. This  implies that the integrated insertion of this operator into correlation functions will satisfy 
\begin{equation}
    \int d^2x\langle \Phi_{\rm st}(x) \prod_{i=1}^n V_{\alpha_i}^{\w_i} (x_i)\rangle_\mu = - \frac{\der}{\der \mu}  \langle \prod_{i=1}^n V_{\alpha_i}^{\w_i} (x_i) \rangle_\mu \, .
\end{equation}
This was verified explicitly in the exact theory in \cite{Eberhardt:2021vsx} for $n=2$. 
On the other hand, the worldsheet action written in terms of Wakimoto fields reads  
\begin{equation}
\label{Swzw H3 beta quantum nu}
    S = \frac{1}{2\pi} \int d^2z \left(\frac{1}{2}\der \phi \derb \phi  + \frac{Q_\phi}{4 \sqrt{2}} R^{(2)}\phi + \betab \der \bar{\gamma} + \beta
    \derb \gamma  - 
    \nu^{-1} e^{-\sqrt{\frac{2}{k-2}}\phi} \beta \betab \right) \, .      
\end{equation}
Here, in comparison with Eq.~\eqref{Swzw H3 beta quantum nu}, we have included a conventional coupling constant for the interaction term, which can be scaled away by shifting $\phi$. The latter property implies that $\nu$ is precisely the constant that was included in the worldsheet two-point function, where the term proportional to $\delta(j-j')$ scales as $\nu^{1-2j}$,  see Eqs.~\eqref{SL2 bosonic 2point} and \eqref{def B(j)}. Moreover, it also guarantees that the constant $p$ appearing in Sec.~\ref{sec: free fields k>3} above must scale as $p \sim \nu^{2-k}$. 

Now, consider the operator $\Ii$, previously discussed in Sec.~\ref{sec: Exact theory, unflowed sector}. The leading expression was given in Eq.~\eqref{I def leading}. By inserting the explicit form of the SL(2,$\R$) currents \eqref{Ja Wakimoto} into the exact expression given in \eqref{def I(x)}, the subleading term of $\Ii$ can be obtained from that of $V_1$, namely 
\begin{equation}
    V_1(x) \sim \delta(\gamma-x) + \frac{1}{\pi} e^{-\sqrt{\frac{2}{k-2}}\phi}|\gamma-x|^{-4} \, .
\end{equation}
This gives \cite{Giveon:2001up}
\begin{equation}
    \Ii \sim \int d^2z \left[ \der \gamma \derb \gammab \delta(\gamma-x) + \frac{1}{\pi} e^{-\sqrt{\frac{2}{k-2}}\phi} \beta \betab \right]\, , 
\end{equation}
which shows that $\Ii$ is closely related with to the  Wakimoto interaction term. In particular, we have \cite{Eberhardt:2023lwd}
\begin{equation}
    \langle \Ii \prod_{i=1}^n V_{j_ih_i}^{\w_i} (x_i,z_i)\rangle_\mu = - \frac{\der}{\der \nu} \langle  \prod_{i=1}^n V_{j_ih_i}^{\w_i} (x_i,z_i) \rangle_\mu \, .
\end{equation}
Importantly, the holographic matching derived in the previous section shows that boundary and worldsheet coupling constants are related by 
\begin{equation}
    \mu \sim p^{1/2} \sim \nu^{1-\frac{k}{2}} \, .
\end{equation}
Combined with the above discussion, this tells us that there should exist a relation of the form 
\cite{Eberhardt:2021vsx}
\begin{equation}
\label{Id is Phi}
    \Ii \sim   \int d^2x \, \Phi_{\rm st}(x) \, ,
\end{equation}
which is consistent with the fact that $\Ii$ is $x$-independent. 

What is the interpretation of the identification proposed in Eq.~\eqref{Id is Phi}? At the level of the boundary action, we can, roughly speaking, rewrite the marginal deformation as implementing a modification of the form 
\begin{equation}
    S_0 \to S = S_0 - \mu  \int d^2x \, \Phi_{\rm st} \sim S_0 - \mu \Ii \, . 
\end{equation}
But, as discussed in Sec.~\ref{sec: wakimoto syms}, the expectation value of the operator $\Ii$ counts, roughly speaking, the number of strings in the background, namely $n_1$. Hence, we find that (in the saddle-point approximation to the path integral) the deformation effectively implements a Legendre transform. The result is that the marginal deformation that defines the holographic CFT can be understood as a realization of the fact that, as anticipated in \cite{Kim:2015gak}, one should  \textit{unfix} $n_1$ in favour of working with a given $\mu$, now viewed as a chemical potential. The latter will then be holographically related to the string coupling. This resonates with the interpretation of the boundary theory as a grand canonical ensemble of CFTs \cite{Eberhardt:2020bgq,Eberhardt:2021vsx}, see also   \cite{Kim:2015gak,Aharony:2024fid,Knighton:2024pqh}.

%%%%%%%%%%%%%%%%%%%%%%%%%%%%%
%%%%%%%%%%%%%%%%%%%%%%%%%%%%%
\newpage
%%%%%%%%%%%%%%%%%%%%%%%%%%%%%%%

\section{Three roads to little string theory}
\label{sec: 3 roads to LST}

We now present an interesting application of the worldsheet formalism developed in the previous chapters that goes beyond the description of strings in AdS$_3$. The object of interest remains the concept of holography, but now in a more general sense. More precisely, we focus on the NS5-brane theory. Although the relevant supergravity backgrounds are not asymptotically AdS, a suitable decoupling limit can still be defined. The decoupled theory is non-gravitational but remains stringy nonetheless. It is known as Little String Theory. 

We present three alternative descriptions for the dynamics of these little strings: the first one based on the dual gravitational picture, the second one given by a marginal deformation of the AdS$_3$ worldsheet model, and the third one based on an irrelevant $T\bar{T}$-type deformation of the holographic CFT.   

\subsection{The NS5-brane decoupling limit}

Consider a stack of $n_5$ NS5-branes. The supergravity configuration corresponds to the $n_1 \to 0$ ($f_1(r) \to 1$) limit of  \eqref{NS5F1 metric}, namely \cite{Callan:1991at}
\begin{equation}
\label{NS5 metric}
    ds^2 =  -dt^2 + dy^2 + f_5(r)(dr^2 + r^2 d\Omega^2_3) + dz_i dz^i \,, \quad  
    e^{2\Phi} = g_s^2 f_5(r) \, , \quad 
    f_5(r) = 1+\frac{r_5^2}{r^2} \, .
\end{equation}
Here $r_5^2 = \alpha' n_5$, the five-branes are extended along the $t,y,z^i$ directions with $i=1,\dots,4$, and there is also the magnetic NSNS 3-form $H = 2 n_5 \ep_{S^3}$. We take the $y$-circle to have radius $R_y$. The string coupling asymptotes to $g_s$ at infinity and diverges at the origin. We zoom in on the near-horizon region by rescaling $r \to \vep r$ and taking $\vep \to 0$ while going to small coupling with $g_s \to \vep$, such that $r/g_s$ remains fixed\footnote{We could have kept a fiducial  $g_s$ as in some of the literature, but it would lack a precise meaning. On the other hand, energy scales will be important in our discussion, hence we will find it useful to reintroduce various factors $\alpha'$ which were set to one in previous sections.}. This effectively amounts to dropping the "1+" in $f_5$, leading to 
\begin{equation}
\label{NS5 metric decoupling}
    ds^2 =  -dt^2 + dy^2 + d\phi^2 + r_5^2 d\Omega^2_3 + dz_i dz^i \,, \quad  
    \Phi =  - \frac{1}{r_5} \phi \,  .
\end{equation}
where we have defined the radial coordinate $\phi = - r_5\log \left(\frac{r}{r_5}\right)$. In this limit the dilaton becomes linear in $\phi$, while the radius of the 3-sphere becomes constant and proportional to $\sqrt{n_5}$ in string units, giving a geometry of the form 
\begin{equation}
    {\rm Mink}^{1,5} \times \R_\phi \times S^3, 
\end{equation}
where, as in the AdS$_3 \times S^3 \times T^4$ case, string propagation can be studied exactly. The relation between $\Phi$ and $\phi$ implies that the linear dilaton worldsheet field will have charge $Q_\phi = -\frac{1}{\sqrt{n_5}}$, such that the matter central charge of the corresponding worldsheet CFT is 
\begin{equation}
    c= (6 + 3) + \left(1 + 6 Q_\phi^2 + \frac{1}{2}\right) + \left(\frac{3 (n_5-2)}{n_5} + \frac{3}{2} \right) = 15, 
\end{equation}
where we have included the fermionic contributions and accounted for the fact that the bosonic SU(2) level gets shifted to $n_5-2$, see Sec.~\ref{sec: AdS3xS3xT4}. As usual in critical superstring theory, this is cancelled by the ghost contributions.  

The above near-horizon limit is reminiscent of those used for D-branes in the context of the AdS/CFT correspondence in order to reach the gravitational dual of boundary theory IR. There are however, several important differences. Most notably, for D-branes one usually takes $\alpha' \to 0$, while here we take $g_s \to 0$ and keep $\alpha'$ fixed, which has the effect of decoupling the bulk strings while keeping the stringy modes at finite mass. Nevertheless, one might ask: is there some sort of holography at play for NS5-branes? It was argued in \cite{Itzhaki:1998dd,Boonstra:1998mp,Aharony:1998ub,Kutasov:2001uf} that the answer is yes. We now briefly discuss the qualitative motivations behind this proposal, and the main features of the putative boundary theory. Although the above discussion is valid for both type II string theories, some details  depend on whether we are in type IIA or in type IIB. 

NS5-branes are dynamical objects. In the type IIB case, their fluctuations are described by a supersymmetric six-dimensional theory with $\N = (1,1)$  whose low-energy limit is given by U($n_5$) Super Yang-Mills \cite{Seiberg:1997zk}. One can infer this from the D5-brane perspective by means of S-duality. In the D5 frame the six-dimensional gauge coupling satisfies $g_{\rm D5}^{2} = \tilde{g}_s \tilde{\alpha}'$. This is weakly coupled in the IR, which is why in the dual description $\Phi$ diverges as we take $\phi \to \infty$. By starting with $g_{\rm D5}^2$ and applying the S-duality transformations $\tilde{g}_s \to g_s^{-1}$ and $\tilde{\alpha}' \to g_s \alpha'$ we get the NS5 gauge coupling $g_{\rm NS5}^{2} = \alpha'$. The theory on the NS5-branes thus remains interacting even in the $g_s\to 0$ limit. 

Roughly speaking, for energies below $l_s^{-1}$ one has a SYM description. Its non-renor\-malizability indicates that new degrees of freedom must come into play above this scale. (More precisely, at large $n_5$ one should work in terms of the 't Hooft coupling, which comes with an extra factor of $n_5$, suggesting that the breakdown rather takes place at energies $E^2 \sim 1/n_5\alpha'$.) These UV degrees of freedom are fundamental strings, whose tension reads $(2\pi\alpha')^{-1} \sim g_{\rm NS5}^{-2} $. This shows that they can also be interpreted as non-perturbative instanton excitations of the low-energy gauge fields, which are string-like objects in six dimensions. Though it is not surprising to find dynamical strings since we have kept $\alpha'$ finite, we find that they are pinned to the five-brane worldvolume. Moreover, there are no massless spin-two modes in the spectrum \cite{Kutasov:2001uf}, hence we are dealing with a non-gravitational theory.

The type IIA case is slightly more involved, as the IR fixed point is non-trivial. It corresponds to the six-dimensional (2,0) SCFT. Here S-duality leads to an M-theory description. As $\phi \to \infty$ the dual metric becomes that of $n_5$ coincident M5-branes, for which the near-horizon geometry is AdS$_7 \times S^4$, the supergravity dual of the (2,0) theory. Another stringy feature of LST is that it inherits the T-duality properties from those of type II strings since this commutes with the decoupling limit described above. Hence, the type IIA and type IIB versions of LST are related to each other upon compactification. At energies where the little string dynamics become relevant, the corresponding theories actually become non-local.

If the background \eqref{NS5 metric decoupling} is to provide a gravitational dual for these non-critical, non-gravitational six-dimensional string theories, one should at least be able to match the (low energy) BPS spectra on both sides. This was done in \cite{Aharony:1998ub} (exactly in $\alpha'$) for the full set of BPS of chiral operators using the low-energy field theory description of the NS5-theory and the worldsheet model for strings moving in the linear dilaton geometry. On the gravity side one also finds a continuum of long string states above the gap set by the NS5 't Hooft scale $n_5\alpha'$, as expected. 

It is also possible to study the thermodynamics of the system. However, the fact that the dilaton diverges in the IR makes the holographic description somewhat unreliable. 
Thankfully, we know one way\footnote{Another possibility was discussed in \cite{Giveon:1999px,Giveon:1999tq}.} to fix this! The IR behavior can be regularized by including $n_1$ fundamental string sources. Further compactifying the $z^i$ directions on a four-torus leads to the background of Eqs.~\eqref{NS5F1 metric} and \eqref{NS5F1 dilaton and Bfield}. Upon implementing the five-brane decoupling limit in this context (i.e.~going to the near-horizon of the NS5-branes, but not of the F1 strings), we obtain  
\begin{equation}
%\label{NS5 metric decoupling}
    ds^2 =  f_1^{-1} \left(-dt^2 + dy^2\right) + d\phi^2 + r_5^2 d\Omega^2_3 + dz_i dz^i \,, \quad  
    e^{2\Phi} =  f_1^{-1} r_5^2 e^{-2\phi} \,  .
\end{equation}
with $f_1 = 1+ r_1^2 e^{-2\phi}$, with $r_1^2 = \alpha' n_1 / v_4$. One also has $H = 2 n_5(1 + (1-f_1^{-1}) *_6)\ep_{S^3}$, where the AdS$_3$ part can be rewritten as \begin{equation}
\label{NS5 B-field decoupling}
H|_{{\rm AdS}_3} = 2r_1^2 e^{-2\phi} f_1^{-2} dt \wedge dy \wedge d\phi = \frac{1}{2}d\left[\frac{1}{1+r_1^2 e^{-2\phi}} dt\wedge dy\right] \, .     
\end{equation}
The dilaton is now finite in the $\phi \to - \infty $ limit, and remains small everywhere as long as we have $n_1 \gg n_5$. This geometry  thus  interpolates between  the linear dilaton background in the UV and AdS$_3\times S^3 \times T^4$ in the IR, see Eq.~\eqref{AdS3S3T4 metric}. The latter is obtained in the string decoupling limit where we take $r,\alpha' \to 0$ with $r/\alpha'$ fixed. We also note that upon rescaling $\phi \to \phi - \frac{1}{2} \log \left(\alpha' n_1 n_5 /v_4\right)$ we get 
\begin{equation}
\label{final metric NS5 decoupling}
    f_1 \to 1+n_5^{-1} e^{-2\phi} \,, \qquad 
    e^{2\Phi} \to  \frac{v_4}{n_1}\left(1+n_5^{-1} e^{-2\phi}\right)^{-1} \, ,
\end{equation}
which will be useful later on.

\subsection{Solvable worldsheet deformations}

In the previous section we have described the step by step procedure that takes us from the asymptotically flat NS5-F1 configuration (compactified on $T^{4}$) to the IR AdS$_3\times S^3$ description, including the intermediate linear dilaton regime associated to the NS5-brane decoupling. In short, in the first step we drop the "$1+$" in the NS5-brane harmonic function $f_5$ only, while in the second one we do the same for the $f_1$ harmonic function associated to the fundamental string sources. We now describe how the inverse process can be realized, at least for the second step. This is done by implementing a suitable deformation of the worldsheet model, which effectively \textit{adds back} the "$1+$" in $f_1$ \cite{Forste:1994wp,Giveon:1999zm}. Here we switch to the analogous bosonic configuration in order to make contact with the discussion in Sec.~\ref{sec: bosonic AdS3 spectrum}, hence some factors of $n_5$ are replaced by $k$ or $k-2$. At any rate, we are mostly interested in the large $k$ regime where the difference is negligible.  

Let us recall the AdS$_3$ worldsheet model, whose action is given in terms of the Wakimoto fields by 
\begin{equation}
    S_0 = \frac{1}{2\pi} \int d^2z \left( \frac{1}{2}\der \phi \der \phi  + \frac{Q_\phi}{4\sqrt{2}} R^{(2)}\phi + \betab \der \bar{\gamma} + \beta
    \derb \gamma  - e^{-\sqrt{\frac{2}{k-2}}\phi}
    \beta \betab  \right).      
\end{equation}
The holomorphic null current $J^+ = \beta$ has protected dimension $\Delta=1$. One can thus deform the model by the marginal current-current operator $J^+\bar{J}^+ = \beta \betab$, leading to 
\begin{equation}
    S = \frac{1}{2\pi} \int d^2z \left[\frac{1}{2} \der \phi \der \phi  + \frac{Q_\phi}{4\sqrt{2}} R^{(2)}\phi + \betab \der \bar{\gamma} + \beta
    \derb \gamma  - 
    \beta \betab \left(1+e^{-\sqrt{\frac{2}{k-2}}\phi}  \right)\right].      
\end{equation}
In the IR region of the spacetime geometry $\phi \to -\infty$, that is, the strongly coupled regime of the worldsheet theory, the effect of the deformation is negligible. Conversely, in the large $\phi$ regime the new term becomes important. In particular, it changes the asymptotics of the spacetime configuration. By integrating out the Lagrange multipliers $\beta$ and $\betab$, considering the classical large $k$ limit and redefining $\phi \to \sqrt{2k}\left(\phi + \log \frac{k}{2}\right)$, we obtain    
\begin{equation}
    S = \frac{1}{2\pi} \int d^2z \left\{k \der \phi \der \phi   + \frac{\derb\gamma \der \bar{\gamma}}{\left(1+k^{-1}e^{-2\phi} \right)} + \frac{1}{4}R^{(2)} \log \left[e^{-2\phi}(1+k^{-1}e^{-2\phi})^{-1}  + {\rm const}  \right] \right\}.  
\end{equation}
Here the dilaton factor is necessary for the worldsheet beta-functions to vanish, and fixes $\Phi$ up to a multiplicative constant, such that in particular $\Phi \sim - \phi$ at large $\phi$. The background we have landed on is thus exactly the ($t,y,\phi$) sector of that given in Eqs.~\eqref{NS5 metric decoupling}-\eqref{final metric NS5 decoupling} for $k \approx n_5$. In the supersymmetric setup the analogous procedure leads to the full NS5-F1 background in the five-brane decoupling limit.    

Once this deformation is included, the geometry is not AdS$_3$ anymore, and the spacetime conformal symmetry is broken (except in the IR). Nevertheless,  it turns out that at the worldsheet level one can still treat the theory exactly. This is because, as we now show, the deformed geometry is still sourced purely by NSNS fluxes and takes the form of a coset space\footnote{An alternative, equivalent approach based on the Wakimoto fields with modified boundary conditions induced by the so-called TsT solution generating technique was used in \cite{Apolo:2019zai}.}. 

Coset models are obtained by gauging some of the conserved currents of an underlying WZW model. The starting point is not, strictly speaking, the SL(2,$\R$) model, but rather a model whose target space constitutes a simple 3+2 dimensional extension. More explicitly, we include an auxiliary time direction $\ttt$ together with an auxiliary spatial circle $\yt$, and consider the coset \cite{Giveon:2017myj}
\begin{equation}
    \frac{{\rm SL}(2,\R) \times \R_{\ttt} \times {\rm U}(1)_{\yt}}{\R \times U(1)} \,. 
\end{equation}
We then gauge the null chiral currents
\begin{equation}
{\cal J} =  J^+ + i (\der \yt - \der \ttt) \, , 
\qquad 
{\cal \Jb} =  \Jb^+ + i (\derb \yt + \derb \ttt) \, , 
\end{equation}
which correspond to the isometries 
\begin{equation}
(x^-,\gamma) \to (x^- + \alpha,\gamma + \alpha) \, , \qquad 
(x^+,\gammab) \to (x^++ \bar{\alpha},\gammab+ \bar{\alpha}) \, ,   
\end{equation}
 with $x^\pm = \yt \pm \ttt$. Following \cite{Hull:1989jk,Chung:1992mj}, the gauge invariant Lagrangian reads 
\begin{equation}
    L_{\rm gWZW} = k\left[
    \der \phi \derb \phi + e^{2\phi}(\derb \gamma + \bar{A} ) (\der \gammab + A )
    \right] + (\der x^+ + A)(\derb x^- + \bar{A}) \, .
\end{equation}
In two dimensions the chiral gauge fields $A$ and $\bar{A}$ act as Lagrange multipliers. By integrating them out and gauge-fixing $\yt = \ttt = 0$ we obtain 
\begin{equation}
    L_{\rm gWZW} = k \der\phi \derb \phi + 
     \frac{\derb\gamma \der \bar{\gamma}}{\left(1+k^{-1}e^{-2\phi} \right)} +
     \frac{1}{4}R^{(2)} \log \left[e^{-2\phi}(1+k^{-1}e^{-2\phi})^{-1} + {\rm const} \right]\, . 
\end{equation}
This shows that in the IR $\phi \to - \infty$ region we basically gauge away the auxiliary directions and recover the AdS$_3$ configuration, while in the UV we reobtain the interpolating background discussed above. Of course, we could have gauged-fixed $\gamma = \gammab = 0$ equivalently. Gauging away $\yt$ and $\ttt$ is more natural from the IR point of view, since there we have the undeformed AdS$_3$. On the other hand, gauging away $\gamma$ and $\gammab$ makes more sense in the UV since the geometry asymptotes to a linear dilaton background where the geometry is flat and the spatial circle has a fixed radius. 

We now note that upon reinserting $\alpha'$ in the above formulas by rescaling $k \to \alpha' k$ (this is the dimensionful squared radius of the AdS$_3$ factor in the IR) one effectively gets 
\begin{equation}
    1+k^{-1}e^{-2\phi} \to \frac{1}{\alpha'}
    \left(\alpha'+k^{-1}e^{-2\phi} \right) \,. 
\end{equation}
Hence, it makes sense to think of the gauging parameter, i.e.~the new factor in the metric, B-field and dilaton, as the inverse string tension $\alpha'$ \cite{Giveon:2017myj}. The fact that this has units of length$^2$ is consistent with the discussion carried out in the following section, where the procedure is interpreted in terms of an $(h,\hb) = (2,2)$ irrelevant deformation of the boundary theory. It also makes sense from the holographic perspective, as we have seen that the deformation should be negligible in the (string) near-horizon limit where we take $\alpha' \to 0$. 

As discussed in \cite{Chung:1992mj}, when the currents are chiral and null the model is automatically anomaly free.  The string worldsheet model is then built from the \textit{ungauged} action -- that of the \textit{upstairs} 3+2 dimensional model, together with the internal sector and reparametrization ghosts, which simply go along for the ride -- combined with an additional system of $\tilde{b}\tilde{c}$ ghosts associated with the null gauging procedure. The usual  holomoprhic and anti-holomorphic BRST charges acquire additional contributions of the form $\oint dz \, \tilde{c} {\cal J}$ and $\oint d\zb \, \bar{\tilde{c}} {\cal \Jb}$, respectively, ensuring that the coset operators are gauge-invariant. 

Given that our gaugings involve the SL(2,$\R$) currents $J^+$ and $\bar{J}^+$, it is useful to work directly with their eigenstates. Since $J^+ \sim \der_x$, and $\Jb\sim \der_{\xb}$, this means that we consider momentum-space operators in the AdS$_3$ subsector. These are then combined with exponentials in $\ttt$ and $\yt$ to define 
\begin{equation}
\label{Vops coset}
    {\cal V}_{j,h}^\w(p,\bar{p},z,\zb) = \int d^2x e^{i(px + \bar{p}\xb)} V_{j h}^\w(x,\xb,z,\zb) e^{-i (E \ttt + P_{\yt,L} \yt_{L} + P_{\yt,R} \yt_{R})} V_{\rm int}(z,\zb)\, , 
\end{equation}
where we have included the possibility of having non-trivial winding around on the $\yt$-circle, so that 
\begin{equation}
    P_{\yt,L/R} = \frac{n_{\yt}}{R_{\yt}} \pm \frac{\w_{\yt} R_{\yt}}{\alpha'} \, .
\end{equation}
Gauge invariance then implies 
\begin{equation}
\label{gauge constraints coset}
p + (E-P_{\yt,L}) = 0 \, , \qquad \bar{p} + (E+P_{\yt,R}) = 0  \, .
\end{equation}
On the other hand, the Virasoro conditions for primary operators now include contributions from the exponentials, giving 
\begin{eqnarray}
\label{Virasoro cond TTbar}
    - \frac{j(j-1)}{k-2} - \w h + \frac{k}{4}\w^2- \frac{\alpha'}{4} \left(E^2 - P_{\yt,L}^2\right) + h_{\rm int} -1 &=& 0 \, , \\
    - \frac{j(j-1)}{k-2} - \w \bar{h} + \frac{k}{4}\w^2- \frac{\alpha'}{4} \left(E^2 - P_{\yt,R}^2\right) + \bar{h}_{\rm int} -1 &=& 0 \, .
\end{eqnarray}
We have argued that $\ttt$ and $\yt$ can be thought of as parametrizing the asymptotic circle, hence we interpret $E$  as the energy in the dual theory and $n_{\yt}$ as the corresponding momentum. 

We now consider a state with $\w=0$ and $\w_{\yt} = 1$, and also set $n_{\tilde{y}}=0$ for simplicity, hence $\hb_{\rm int} = h_{\rm int}$. After redefining the energy as that of the fluctuation above the winding contribution by replacing $E \to E + R_{\yt}/\alpha'$, the (sum of the) Virasoro conditions gives
\begin{equation}
   \left(E + \frac{R_{\yt}}{\alpha'}\right)^2 - \left(\frac{R_{\yt}}{\alpha'}\right)^2 = \frac{2}{\alpha'} \left(-\frac{2j(j-1)}{k-2} + h_{\rm int} + \hb_{\rm int} - 2\right)\, . 
\end{equation}
We now recall that for a singly-wound long string in the \textit{undeformed} model we had 
\begin{equation}
    h_0^{(1)} = - \frac{j(j-1)}{k-2} + \frac{k}{4} + h_{\rm int} -1 \,, \qquad \bar{h}_0^{(1)} = - \frac{j(j-1)}{k-2} + \frac{k}{4} + \hb_{\rm int} -1 
\end{equation}
hence we can rewrite the energy in the \textit{deformed} model in terms of undeformed quantities as
\begin{equation}
\label{TTbar energy w=1 from worldsheet}
   \left(E + \frac{R_{\yt}}{\alpha'}\right)^2 - \left(\frac{R_{\yt}}{\alpha'}\right)^2 = \frac{2}{\alpha'} \left(h^{(1)}_0 + \hb^{(1)}_0 - \frac{k}{2}\right)\, . 
\end{equation}
Repeating the exercise for states with arbitrary $\w_{\yt}$ gives 
\begin{equation}
\label{TTbar energy w>1 from worldsheet}
   \left(E + \frac{\w_{\yt} R_{\yt}}{\alpha'}\right)^2 - \left(\frac{\w_{\yt} R_{\yt}}{\alpha'}\right)^2 = \frac{2}{\alpha'} \left(h^{(1)}_0 + \hb^{(1)}_0 - \frac{k}{2}\right)\, . 
\end{equation}
We will recover this result from the holographic perspective in the following section. 

The curious reader might wonder why we are comparing states which wind around the $\tilde{y}$-circle in the coset model with states which wind around the AdS$_3$ circle in the original theory, i.e.~spectrally flowed (long string) states. The point is that although we construct the vertex operators in Eq.~\eqref{Vops coset} by combining the elements of the auxiliary \textit{upstairs model}, which contains both an SL(2,$\R$) factor and the auxiliary circle and time coordinates, in the gauged theory there is only a single physical spatial circle. Indeed, we are free to gauge away either $\ttt$ and $\yt$ or $t$ and $y$. This is clear in the geometrical analysis, where the UV circle becomes the AdS$_3$ boundary circle in the near-horizon limit. 
At the level of defining the winding charges in the quantum theory, this manifests as a residual discrete gauge symmetry. This was discussed in detail for a related model in \cite{Martinec:2018nco,Bufalini:2021ndn,Bufalini:2022wzu}. It implies that operators with windings $(\w,\w_y)$ have equivalent descriptions in terms of operators with windings $(\w+q,\w_y-q)$ with $q \in \mathbb{Z}$, i.e.~one can trade one winding charge for the other. This is easily seen at the level of the spectrum. Indeed, the gauge constraints \eqref{gauge constraints coset} identify the energy $E$ and momentum $P_y$ with the AdS$_3$ ones, suggestsing that we can extend the relation $E_0 (h_0,\bar{h}_0)$ to an analogous relation between the quantities in the deformed theory. For instance, focusing on states with $h_0 = \bar{h}_0$ for simplicity, the Virasoro conditions for a spectrally flowed operator with $\w=1$ and $\w_y=0$ give
\begin{equation}
    E^2 + \frac{2}{\alpha'} \left(2h - \frac{k}{2}\right)  = \frac{2}{\alpha'} \left(2 h^{(1)}_0 - \frac{k}{2}\right)\, .
\end{equation}
This coincides with \eqref{TTbar energy w=1 from worldsheet} upon identifying $E R_{\yt} = 2 h - \frac{k}{2}$. More generally, the identification reads $E R_{\yt} = 2 h - \frac{k}{2}\w$, in agreement with the fact that we are computing the energy \textit{above} the winding contribution. 

The coset description of the NS5-F1 linear dilaton background can also be used to compute correlation functions exactly. In turn, these can be seen as providing a precise definition for the corresponding holographic observables, namely Little String Theory correlators \cite{Asrat:2017tzd,Giribet:2017imm,Cui:2023jrb}. The above construction also shows that these are best described in momentum space, which resonates with the expectation that the holographic theory is now non-local. The characteristic scale of this non-locality is set by the deformation parameter, which we have identified  with $\alpha'$, i.e.~the fundamental length scale of the little strings. 
In short, for long string operators with definite momentum $p$ (and $\bar{p}$) the deformed correlators are obtained from the undeformed ones simply by replacing the original weight $h_0$ by the momentum-dependent expression $h = h_0 + \frac{\alpha'}{\w} p \bar{p}$, and similarly for ${\hb}_0$. These are nothing but the weights derived from the modified Virasoro conditions \eqref{Virasoro cond TTbar}, combined with the gauge constraints which relate $E$ and $P_y$ with $p$, $\bar{p}$ and the deformation parameter.   

\subsection{Single-trace TTbar}

We now provide yet another intepretation of the linear dilaton background  directly in terms of the holographic theory. In the previous section we have  seen that one can obtain the NS5-F1 linear dilaton background by a marginal 
deformation of the worldsheet theory associated to the AdS$_3$ description.  From the spacetime point of view, this represents a flow from the IR regime 
towards the UV, where the geometry is not asymptotically AdS anymore. This inverse
 RG flow can be understood as an irrelevant deformation 
of the original boundary CFT, such that the resulting two-dimensional theory is not conformal 
invariant anymore. The LST picture suggests that it is not even a local quantum field 
theory. However, it remains quasi-local below the scale introduced by the irrelevant deformation, and retains Lorentz invariance.

In general one would expect such a deformation to be 
ill-defined, at least on its own, as it might be necessary to  
specify which new degrees of freedom should be integrated in, and how they couple to 
the low-energy modes. Nevertheless, in the present context the fact that the 
worldsheet description remains exactly solvable suggests that observables such as 
the energy spectrum and the correlation functions should actually be computable in terms of  
their undeformed expressions. This is one of the trademarks of the so-called $T\bar{T}$
deformations, which lead to UV complete theories. 

Let us briefly review the basic facts about these very special deformations, 
following \cite{MonicaCERN,Guica:2025jkq}. For any quasi-local Lorentz-invariant two-dimensional theory one can 
define a conserved (symmetric) energy-momentum tensor. One can then build 
the $T\bar{T}$ operator, given by\footnote{For the generalization to higher 
dimensions it is more useful to think of this as ${\cal O}_{T\bar{T}} =  {\rm det}\, T_{\mu \nu}$. } 
\begin{equation}
{\cal O}_{T\bar{T}} \equiv T \bar{T} - \Theta^{2} \, ,  
\end{equation}
where $T = T_{xx}$, $\bar{T} = T_{\xb\xb}$, and $\Theta = T_{x\xb}$. The latter
vanishes in a CFT, but is non-trivial in more general situations. The 
operator ${\cal O}_{T\bar{T}}$ is well defined (up to total derivatives) since there 
are no short-distance singularities in $T(x+\vep,\xb + \bar{\vep}) \bar{T}(x,\xb) - 
\Theta (x+\vep,\xb + \bar{\vep}) \Theta(x,\xb)$. This follows from the fact that 
the expectation value $\langle T(x,\xb) \bar{T}(y,\bar{y}) - \Theta (x,\xb) \Theta(y,\bar{y}) 
\rangle$ is actually coordinate-independent as a consequence of translational invariance. Another important property of this operator is that 
its expectation value factorizes. 
This holds upon quantizing the theory on a cylinder, 
where the spatial coordinate $\sigma$ is compactified on a circle of radius $R$. 
Indeed, 
consider the expectation value of $T(x,\xb) \bar{T}(y,\bar{y}) - \Theta (x,\xb) \Theta(y,\bar{y})$ 
in an energy-momentum eigenstate $|n\rangle = |E_n,P_n\rangle$. Upon inserting an identity 
built out of a basis of similar eigenstates $|n'\rangle$, we find that the result is 
consistent with coordinate indepencence iff the only contribution comes from the term 
with $n=n'$, implying  that 
\begin{equation}
\label{Ottbar n-vev}
\langle n | {\cal O}_{T\bar{T}} |n \rangle = 
- \frac{1}{4} \left( 
\langle n | T_{\tau \tau} |n \rangle \langle n | T_{\sigma\sigma} |n \rangle -
\langle n | T_{\tau \sigma} |n \rangle^2  \right) = - \frac{1}{16 \pi^2 R^2} \left[ (1+ R \der_R) E_n + 
P_n^2\right]
 \,,
\end{equation}
where $\tau$ denotes (Euclidean) time coordinate. 

The $T\bar{T}$ 
operator has weight (2,2), and defines an irrelevant deformation when added to the action, with a dimensionful parameter $\lambda$ in front. In other words, the action 
of the $T\bar{T}$-deformed theory is defined incrementally by the flow equation  
\begin{equation}
\der_\lambda S = - 2 \int d^2 x {\cal O}_{T\bar{T}}^{(\lambda)},  
\end{equation}   
where the superscript on the RHS indicates that the definition of the deforming operator must be updated at each step.
One of the remarkable properties of $T\bar{T}$-deformed QFTs is that several observables can be computed exactly for \textit{finite} values of the parameter $\lambda$ in terms of the corresponding the undeformed quantities. One of these is the spectrum of excitations on the cylinder. Indeed, a small change in $\lambda$ produces a variation in the energy eigenvalue of an eigenstate $|n\rangle$ of the form   
\begin{equation}
    \der_\lambda E_n = -2  \int d^2x \langle n | {\cal O}_{T\bar{T}}^{(\lambda)} | n \rangle \, .
\end{equation}
Inserting \eqref{Ottbar n-vev} on the RHS, including a factor $R$ from the integration, and further focusing on states with $P_n=0$ for simplicity then leads to a flow equation for the energy: 
\begin{equation}
\label{TTbar E flow eq}
    2 \pi \der_\lambda E_{n} = 
    E_n \der_R E_n \, .
\end{equation}

When the starting point is a two-dimensional CFT, we find it useful to express this in terms of the weights $h$ and $\hb$. We ask that the solution $E_{h,\hb}(\lambda,R)$ to \eqref{TTbar E flow eq} satisfies 
\begin{equation}
    R E_{h,\hb}(\lambda=0,R) = \left(h + \hb - \frac{c}{12}\right) \, ,
\end{equation}
where $c$ is the central charge at the IR fixed point, and $-c/12$ corresponds to the corresponding Casimir energy. Imposing this initial condition, one finds that the solution is specified by the quadratic relation 
\begin{equation}
\label{TTbar energy}
   \left(E_{h,\hb} + \frac{\pi R}{\lambda}\right)^2 = \left(\frac{\pi R}{\lambda}\right)^2 + \frac{2 \pi}{\lambda}\left(h + \hb - \frac{c}{12}\right) 
   \, . 
\end{equation}
Hence, provided we identify 
\begin{equation}
    R = R_{\yt}, \qquad 
    h = h^{(1)}_0 \, \qquad 
    \hb = \hb^{(1)}_0 \, \qquad 
    \lambda = \pi \alpha' \, \qquad 
    c = c^{(1)}_{\rm st} = 6 k \, ,  
\end{equation}
the $T\bar{T}$-deformed energy spectrum  precisely reproduces the worldsheet result given in Eq.~\eqref{TTbar energy w=1 from worldsheet}.

This shows that the long string spectrum in the singly-wound sector of the worldsheet coset theory considered above can be interpreted as that of the $T\bar{T}$-deformation of the \textit{seed} theory in the symmetric orbifold of Eq.~\eqref{proposed HCFT with def}. 
If this interpretation is be taken seriously,  one should establish (1) how it generalizes to higher-twist sectors of the symmetric orbifold CFT, and (2) what is the effect of the deformation in Eq.~\eqref{proposed HCFT with def} which accounts for the fact that the IR holographic CFT is actually \textit{not} a symmetric orbifold. 

For the first question, there are two possible natural answers. One of them is to simply consider the $T\bar{T}$-deformation of the full symmetric orbifold. However, we note that the operator ${\cal O}_{T\bar{T}}$ is a double-trace operator, i.e.~it involves a double sum over all copies of the theory. At zero-th order in $\lambda$ we have $T(x) = \sum_{i=1}^{n_1} T_i(x)$, hence ${\cal O}_{T\bar{T}}(x,\bar{x}) = T(x) \bar{T}(\xb) = \sum_{i,j=1}^{n_1} T_i(x)\bar{T}_j(\xb)$. From the string point of view this is a multi-particle state,  which seems at odds with the fact that we have been able to describe the linear dilaton background in terms of a  marginal deformation of the worldsheet theory by the current-current operator $J^+(z) \bar{J}^+(\zb)$. From the exact AdS$_3$ description we know that the spacetime holomorphic energy-momentum tensor $\Tt(x)$ is constructed on the worldsheet as in Eq.~\eqref{def T(x)}. The double-trace $T\bar{T}$-deformation under consideration should  then correspond to an insertion involving not one but two integrals over the worldsheet coordinates. 

In the context of symmetric orbifolds there is, however, another possibility: one can maintain the orbifold structure and consider the symmetric orbifold model where the seed theory itself is $T\bar{T}$-deformed. For the case at hand, we thus consider the orbifold model 
\begin{equation}
    {\rm Sym}^{n_1} \left[\left(R^Q_\phi \times M_{\rm int}\right)_{T\bar{T}}\right] \, .
\end{equation}
In this case, the energies in the twisted sectors will be related to those in the $T\bar{T}$-deformed seed in exactly the same way as what was described in  previous sections, see for instance \eqref{h for bosonic sym orbifold HCFT}, which is in agreement with the result obtained in Eq.~\eqref{TTbar energy w>1 from worldsheet}.  
This proposal was put forward in \cite{Giveon:2017nie}, and is known as single-trace $T\bar{T}$. In this case the irrelevant deformation is generated by the following (2,2) operator: 
\begin{equation}
    {\cal O}_{{\rm st}\,T \bar{T} } (x,\xb) = \sum_{i=1}^{n_1} T_i(x)\bar{T}_i(\xb) \, .
\end{equation}
Upon adding the integrated version of this operator to the action we have, roughly speaking,  
\begin{equation}
\delta S = \lambda \int d^2x \,  {\cal O}_{{\rm st}\,T \bar{T} } (x,\xb) \sim \lambda \Ll_{-1} \bar{\Ll}_{-1}   \, . 
\end{equation}
This makes sense because these Virasoro spacetime modes are seen from the worldsheet as $\Ll_{-1}= \der_x = J^+_0$ and $\bar{\Ll}_{-1}= \der_{\xb} = \bar{J}^+_0$. From the worldsheet perspective, the natural candidate for the exact deformation is obtained by extending the holomorphic structure in the definition of $\Tt(x)$ given in  Eq.~\eqref{def T(x)} to the anti-holomorphic sector. In other words, in the worldsheet language we have 
\begin{equation}
\label{def TTbar ws single trace}
    {\cal O}_{{\rm st}\,T \bar{T} }^{\rm ws} (x,\xb) = \frac{1}{4}\int d^2z \, \left[
    (\der_x J)\der_x + 2 (\der_x^2 J)
    \right] \left[
    (\der_{\xb} \bar{J})\der_{\xb}+ 2 (\der_{\xb}^2 \bar{J})
    \right] V_1  (x,\xb,z,\zb) 
    \,. 
\end{equation}
to the worldsheet action. This indeed has spacetime weights $h = \hb = 2$. Moreover, integrating by parts  
we get
\begin{eqnarray}
  \int d^2 x  {\cal O}_{{\rm st}\,T \bar{T} }^{\rm ws} (x,\xb) &=& \frac{1}{4}\int d^2x \,  d^2z \, 
  (\der_x^2 J)(\der_{\xb}^2 \bar{J})V_1 (x,\xb,z,\zb) \nn \\
  &=& \int d^2x d^2z J^+(z) \Jb^+(\zb) V_1(x,\xb,z,\zb)\, .
\end{eqnarray}
The only singular term in the $JV_1$ OPE is a total derivative, hence we can treat this as a simple product. Since at leading order $V_1 \sim \delta(x-\gamma)$, the integration over $x$ trivializes, leading to 
\begin{equation}
    \int d^2 x  {\cal O}_{{\rm st}\,T \bar{T} }^{\rm ws} (x,\xb) = 
    \int d^2z J^+(z) \Jb^+(\zb) \, , 
\end{equation}
which is consistent with the current-current deformation considered in the previous section.

Finally, we address the second question raised above, related to the deformation of the symmetric orbifold discussed in Sec.~\ref{sec: holography} and its implications for the  $T \bar{T}$ interpretation of the irrelevant flow of the spacetime theory. 
Clearly, the notion of single-trace $T\bar{T}$-deformation relies heavily on the symmetric orbifold structure. Consequently, it is not clear how one would define it when considering a  generic point in the moduli space of the holographic CFT as the starting point of the deformation. Although a precise answer is presently not known, we note that one can take the worldsheet deformation given in \eqref{def TTbar ws single trace} as a holographic \textit{definition} for the generalization of the single-trace deformation, at least along the NSNS locus. 

As we have discussed in detail, in the supersymmetric setting the symmetric orbifold point corresponds to the background with $n_5=1$. 
In such situations the spectrum only contains long strings, and we expect the coset model to be the exact dual of the symmetric orbifold built out of $n_1$ copies of the  $T\bar{T}$-deformed (4,4) SCFT with target space $T^4$. On the other hand, for $n_5\neq 1$ the precise matching holds only for the long string spectrum, while short strings and correlation functions are not expected to preserve the symmetric orbifold structure. This is the situation considered recently in \cite{Dei:2024sct,Dei:2025ilx}.

\subsection{Little string thermodynamics}

We now discuss the thermodynamics, which provide one of the main tests for the proposed triality between type IIB little string theory, the worldsheet theory for the linear dilaton NS5-F1 background, and the so-called (single-trace) $T\bar{T}$-deformed holographic CFT. 

We start by constructing black holes in the asymptotically flat backgrounds under consideration. Focusing on the zero-momentum case for simplicity, the line element reads 
\begin{equation}
    ds^2 = f_1^{-1} \left( f dt^2 +
    dy^2\right) + f_5 \left(
    f^{-1} dr^2 + r^2 d\Omega^3
    \right) + dz_i dz^i \, ,
\end{equation}
while the dilaton and NSNS 3-form take the form 
\begin{equation}
    H = 2 q_5 \vep_{S^3} 
    + \frac{2 q_1}{r^3 f_1^2} dt \wedge dy \wedge dr \ , \  
    e^{2\Phi } = g_s^2f_5 f_1^{-1}\, . 
\end{equation}
Here the harmonic functions are 
\begin{equation}
    f = 1 - \frac{r_0^2}{r^2} \, , \qquad 
    f_{1,5}= 1 + \frac{r_{1,5}^2}{r^2} \, , \qquad q_{1,5}^2 = r_{1,5}^{2}(r_{1,5}^{2} + r_{0}^{2}) \,,  
\end{equation}
with $q_1 = n_1 g_s^2 \alpha'/v_4$ and $q_5= n_5 \alpha'$, 
such that the  horizon is located at $r= r_0$. In the $r_0 \to 0$ limit, we recover the extremal BTZ black hole, i.e.~the ground state in the (spacetime) R sector considered in the last few pages. We can compute the associated entropy as a function of $r_0$ by evaluating the horizon area (in Einstein frame) and plugging it in the Bekenstein-Hawking formula 
\begin{equation}
    S = \frac{A_{\rm h}}{4G_N} = \frac{2 \pi R_y v_4 r_0^3}{g_s^2l_s^4} \sqrt{f_1 (r_0) f_5(r_0)}\, ,
\end{equation}
where $G_N = 8\pi^6 g_s^2 l_s^4$ is the ten-dimensional gravitational coupling. One can also compute the Arnowitt-Deser-Misner (ADM) mass and obtain the energy above extremality   
\begin{equation}
    E = M_{\rm ADM } - E_{\rm ext}\, , \qquad E_{\rm ext} =   2\pi R_y  n_1 T_{\rm F1} +  (2\pi)^5  R_y v_4 l_s^4  n_5 T_{\rm NS5} \, ,  
\end{equation}
where the tensions are $T_{\rm F1} = (2\pi \alpha')^{-1}$ and $T_{\rm NS5} = [(2\pi)^5 l_s^6 g_s^2]^{-1}$. This gives \cite{Maldacena:1996ky}  
\begin{equation}
    E = \frac{R_y v_4 r_0^2}{2 l_s^4 g_s^2} \left[3+ \frac{2}{r_0^2} (r_1^2 + r_5^2 - q_1-q_5)\right] \, . 
\end{equation}
We find that in the asymptotically flat spacetime we have $E \sim r_0^2$ and $S \sim r_0^3$  at high energies, so that $S \sim E^{3/2}$. 

This changes drastically in the five-brane decoupling limit. As before, we rescale $r \to g_s r$, $r_0 \to g_s r_0$ and take $g_s \to 0$ (with $l_s$ fixed). Since we effectively drop the constant term in $f_5$, we now get $S\sim r_0^2\sim E$ at high energies. More precisely, the relation between $S$ and $E$ reads 
\begin{equation}
\label{LST grav entropy}
    S = 2\pi \sqrt{n_5 l_s^2 E^2 + 2 n_1 n_5 R_y E} \, .
\end{equation} 
Finally, at low energies we can ignore the first term on the RHS, so that 
\begin{equation}
\label{S cardy SUGRA}
    S \approx 2 \pi \sqrt{2 n_1 n_5 ER_y }\, .  
\end{equation}

Eq.~\eqref{S cardy SUGRA} reproduces the Cardy formula for a CFT with central charge $c = 6 n_1 n_5$ on a circle of radius $R_y$, namely $S_{\rm CFT} \sim 2\pi \sqrt{c E R_y/3}$. Here we can use the relation $E R_y= h + \hb - \frac{c}{12}$ and set $h = \hb$ since our black holes have no angular momentum. On the other hand, for very high energies we find 
\begin{equation}
    S \approx \beta_H E \, , \qquad  \beta_H = 2 \pi \sqrt{n_5 \alpha'}. 
\end{equation}
The  is known as a Hagedorn regime \cite{Hagedorn:1965st,Atick:1988si}, and $\beta_H$ defines the (inverse) Hagedorn temperature. It is characteristic of systems where the density of states grows exponentially at high energies, and can be seen as a trademark for stringy degrees of freedom.

Hagedorn behavior dominating the high-energy regime of the theory is a strong indication that the dual LST theory is indeed a theory of strings. There is, however, an important difference with critical string theory. For critical strings in flat space, the Hagedorn temperature is set by the fundamental string tension, i.e.~$\beta_H \sim \sqrt{\alpha'}$. For little strings we have found $\beta_H \sim \sqrt{n_5 \alpha'}$. This suggests that the effective tension in LST is lowered by a factor of $n_5$. We can interpret this as coming from the fact that fundamental strings bound to the NS5-branes can fractionate their winding charge in units of $n_5^{-1}$, hence the name \textit{little strings}.  

The above picture can also be derived  from the $T\bar{T}$ description. We first consider the $T\bar{T}$ deformation of a single copy of the holographic  CFT. The result obtained in Eq.~\eqref{TTbar energy} describes how each of the individual energy eigenvalues evolves upon including the irrelevant deformation, as a function of the deformation parameter $\lambda$. This shows that the different levels of the original CFT do not cross each other as we climb up the RG flow. Hence, we can still use $h$ and $\hb$ to label the different states, and the entropy is given by the Cardy formula $S^{(1)} = S^{(1)}(h,\hb,c)$. Importantly, the expression of $S^{(1)}(E)$ \textit{does} change, as can be derived directly from \eqref{TTbar energy}. Focusing on the sector with $h = \hb$, from Eq.~\eqref{TTbar energy} we get 
\begin{equation}
    S^{(1)} = 2 \pi \sqrt{\frac{c^{(1)}}{3} \left(2h  - \frac{c^{(1)}}{12}\right)} = 2 \pi \sqrt{ n_5 \left(2 E R+ \frac{\lambda}{\pi} E^2\right)} \, .
\end{equation}
At least at the orbifold point, the holographic theory is interpreted as the symmetric orbifold of $T\bar{T}$-deformed CFTs, where $n_1$ denotes the total number of copies of the seed theory. For fixed total energy $E$, the entropy is extremized (and actually maximized) when each copy carries an equal energy fraction $E/n_1$. We conclude that 
\begin{equation}
    S(E,c) = n_1 S^{(1)}(E/n_1,c^{(1)}) = 
    2 \pi \sqrt{n_5 \left(2 n_1 E R+ \frac{\lambda}{\pi} E^2\right)} \, , 
\end{equation}
As before, the single-trace $T\bar{T}$ thermodynamics reproduce the gravitational entropy in Eq.~\eqref{LST grav entropy} upon identifying $R = R_y$ and $\lambda = \pi \alpha'$. 
The fact that we have obtained a precise matching for all values of  $n_1$ and $n_5$ suggests that this entropy is dominated by the physics of long strings, otherwise the deformation of the symmetric orbifold would have had a non-trivial impact. 

Before finishing, we should mention a further highly non-trivial check of the duality under consideration that was presented in \cite{Georgescu:2022iyx,Chakraborty:2023wel}. There, the authors studied the symmetry structure of the $T\bar{T}$-deformed theory both from the symmetric orbifold point of view and from the gravitational, asymptotically linear dilaton background. From both perspectives it was found that  the Virasoro $\times$ Virasoro symmetry of the undeformed theory survives in the deformed one, although the generators become field-dependent.

%%%%%%%%%%%%%%%%%%%%%%%%%%%%%
%%%%%%%%%%%%%%%%%%%%%%%%%%%%%
\newpage
%%%%%%%%%%%%%%%%%%%%%%%%%%%%%%%

% \section{Application II: black holes and their microstates}

% \subsection{Coset models. The Cigar.}

% \subsection{BTZ Black holes in AdS$_3$}

% \subsection{Sub-stringy geometries and correspondence point for AdS$_3$ black holes}

% \subsection{Black hole microstates and Heavy-Light correlators}

% %%%%%%%%%%%%%%%%%%%%%%%%%%%%%
% \newpage
% %%%%%%%%%%%%%%%%%%%%%%%%%%%%%%%

%%%%%%%%%%%%%%%%%%%%%%%%%%
\section{Final comments}

The propagation of strings and superstrings in AdS$_3$ and related backgrounds, and the holographic dualities they embody, constitute an extremely rich and quite explicit playground for studying quantum gravity. Although this incarnation of the AdS/CFT correspondence was first explored more than 25 years ago, it remains a highly exciting and rapidly evolving subject, continuing to yield both qualitative and quantitative new insights into key aspects of gravity and holography, deep in the stringy regime. 

In these lectures we have focused on a particular, but central, corner of the story: the computation of the genus-zero worldsheet correlators, discussing in detail the difficulties coming from the inclusion of spectrally flowed vertex operators. We have also described the dynamics of long strings, and the subsequent implications regarding the nature of the holographic CFT. The evidence suggests that, from the point of view of the AdS$_3$ boundary, the theory takes the form of a marginally deformed symmetric orbifold CFT, or, more precisely, a grand canonical ensemble of such CFTs with different central charges. Moreover, we have shown that, in the tensionless limit, the deformation turns off, as correlators localize on configurations where the spectral flow charges and the worldsheet and boundary insertion points allow for the existence of the appropriate holomorphic covering maps. In this limit, the holographic duality is essentially proven. 

This constitutes a remarkable stepping stone toward a controlled understanding of AdS$_3$ holography at 
$k>3$, and helps build intuition for how holography may operate in more general and higher-dimensional settings.
Needless to say, many aspects of gravity in AdS$_3$ in general and strings in particular are not covered in these lecture notes. It is only fair we end these notes by providing an (inevitably incomplete) list of such topics, accompanied by the corresponding references.

A comprehensive set of important results have been derived in recent years in the context of the tensionless string in AdS$_3\times S^3\times T^4$. Most of them were derived using the hybrid formalism of \cite{Berkovits:1999im,Eberhardt:2018ouy} based on the PSU(1,1|2)$_1$ supersymmetric WZW model. They include the computation of string partition functions \cite{Eberhardt:2020bgq,Eberhardt:2021jvj}, as well as the extension of several of the ideas discussed in the main text to higher genera   \cite{Eberhardt:2020akk,Knighton:2020kuh}, $T\bar{T}$ deformations \cite{Dei:2024sct} and AdS$_3$ orbifold theories \cite{Gaberdiel:2023dxt}.
More generally, D-brane states were studied from the worldsheet perspective in  \cite{Gaberdiel:2021kkp,Rajaraman:2001cr,Lee:2001gh,Ponsot:2001gt,Giveon:2001uq,Bachas:2000fr,Israel:2004jt,Parnachev:2001gw}.
String propagation in the related  AdS$_3\times S^3\times S^3\times S^1$ geometry was considered in \cite{Elitzur:1998mm,Gukov:2004ym,Eberhardt:2017pty,Tong:2014yna,Eberhardt:2017fsi,Eberhardt:2019niq,Gaberdiel:2024dva,Witten:2024yod,Eloy:2023zzh}. Some additional recent results regarding symmetric orbifolds CFTs include \cite{Belin:2020nmp,Apolo:2022fya,Belin:2019rba,Benjamin:2022jin,Hikida:2023jyc,Hikida:2020kil}.
The subject of $J\bar{T}$- and $T\bar{T}$-deformations and non-AdS was also explored, see  \cite{Jiang:2019epa,He:2025ppz,Apolo:2019zai,Guica:2025jkq} for recent reviews.  
Integrability and other exact approaches for strings in AdS$_3$ backgrounds including RR flux were investigated, see  \cite{Demulder:2023bux,Seibold:2024qkh} and references therein.
Finally, the study of tensionless strings in AdS$_5$ (dual to free $\N=4$ SYM) was initiated in   \cite{Gaberdiel:2021qbb,Gaberdiel:2021jrv}.

Applications of worldsheet techniques similar to the AdS$_3$ ones to both two-dimensional black holes \cite{Witten:1991yr,Nappi:1992as}  and three-dimensional black holes \cite{Banados:1992gq,Maldacena:1998bw} were discussed in  \cite{David:2002wn,Giveon:2005mi,Kutasov:2005rr,Detournay:2005fz, Kazakov:2000pm,Ribault:2003ss,Fotopoulos:2004ut,Nakayama:2005pk,Hanany:2002ev,Giveon:2013ica,Massai:2025nci,Natsuume:1996ij,Hemming:2001we,Hemming:2002kd,Rangamani:2007fz,Parsons:2009si,Ashok:2021ffx}.
Further applications to black hole microstates in the context of the fuzzball program \cite{Mathur:2005zp,Bena:2022rna} were put forward in \cite{Bufalini:2021ndn,Bufalini:2022wzu,Bufalini:2022wyp,Martinec:2022okx,Martinec:2017ztd,Martinec:2019wzw,Martinec:2018nco,Martinec:2020gkv,Dei:2024uyx,Martinec:2021vpk,Martinec:2025xoy,Massai:2025nci}.
Other related  models for which the worldsheet theory is exactly solvable include \cite{Collier:2023cyw,Collier:2024kmo}.
More generally,  considerable effort was devoted to studying pure gravity in AdS$_3$, see for instance \cite{Collier:2023fwi,Collier:2024mgv,Chandra:2022bqq,Cotler:2018zff,Afkhami-Jeddi:2020ezh,Cotler:2020ugk,Mertens:2022irh,Maxfield:2020ale,Giombi:2008vd,Harlow:2018tqv,Li:2008dq,Maloney:2007ud,Carlip:2005zn}. Higher-spin theories in AdS$_3$ were also considered in \cite{Vasiliev:2012vf,Gaberdiel:2010pz,Gaberdiel:2012uj,Gaberdiel:2011zw,Kraus:2011ds,Ammon:2012wc}. Finally, AdS$_3$ has been a fruitful playground for testing the relation between entanglement entropy and black holes \cite{Hartman:2013mia,Penington:2019npb,Almheiri:2019hni,Engelhardt:2014gca,Lewkowycz:2013nqa,Calabrese:2009qy,Calabrese:2004eu,Nishioka:2009un,Rangamani:2016dms,Ryu:2006ef,Ryu:2006bv}.    
% Hybrid formalism
% Lorentzian correlators?
% Pure AdS3 gravity/holography
% Integrability AdS3 R flux
% Higher spin?
% Defects/boundaries
% Entanglement entropy
% Wormholes
% Localization
% Flat space limit?
    
%\com{add more refs}

    %\item Other exact WS theories  

%%%%%%%%%%%%%%%%%%%%%%%%%%%%%
\acknowledgments

It is a pleasure to thank Sergio Iguri, Juli\'an H. Toro, Monica G\"uica,  Silvia Georgescu, Davide Bufalini, David Turton,  Soumangsu Chakraborty, Andrea Dei, Bob Knighton and Vit Sriprachyakul for discussions and comments on this manuscript. 
I am also grateful to IPhT for the opportunity to give this course, in particular the organizers Riccardo Guida, Pierfrancesco Urbani and Monica Guica. 
During that period my work was supported by the ERC Consolidator Grant 772408-Stringlandscape. It is now supported by IFLP and CONICET (Argentina). 

\newpage

\appendix

\section{WZW models}
\label{sec: appB WZW models}

In this appendix we review some standard concepts of CFTs in two  dimensions together with the basic aspects of WZW models. We mostly follow \cite{DiFrancesco:1997nk,Ginsparg:1988ui,Blumenhagen:2009zz,Tong:2009np,EberhardtLectures}.

\subsection{Basic concepts of CFT$_2$ and WZW models}

Let us consider a $d$-dimensional manifold $\M$ parametrized by coordinates $X^{\mu}$, with $\mu = 1,\dots,d$. We are interested in describing the motion of a string on $\M$, which we refer to interchangeably as the \textit{target space} or the \textit{spacetime}, by thinking of the coordinates on $\M$ as fields $X^{\mu}(\sigma^{\alpha})$, with $\alpha=0,1$. The timelike and spacelike coordinates on the \textit{worldsheet} $\Sigma$ are usually denoted as $\sigma^0=\tau$ and $\sigma^1=\sigma$, respectively.  
The dynamics of the fields $X^\mu$ is captured by the action 
\begin{equation}
\label{Skinsigma}
    S_{\rm kin} = \frac{1}{4\pi \alpha'} \int_\Sigma d^2 \sigma  \, G_{\mu \nu} (X) 
    \der_\alpha X^\mu \der_\beta X^\nu \eta^{\alpha \beta},
\end{equation}
where $G$ is a local metric on $\M$, and $\alpha'$ is a constant with units of length squared, which defines the so-called string tension. This is known as a non-linear $\sigma$-model. We have taken the worldsheet metric to be $\eta = {\rm diag} (-1,1)$. In the string theory context this can always be achieved by making use of the invariance of the Polyakov action under reparametrizations and Weyl transformations. 

We will mostly work with the Euclidean version of the model, where $\sigma^0$ is taken to be space-like, and we replace $\eta^{\alpha \beta}\to \delta^{\alpha \beta}$. Results for Minkowskian worldsheets can then be obtained by analytic continuation. Moreover, we make use of the complex coordinates 
\begin{equation}
    z = \sigma^0 + i \sigma^1 \, , \quad 
    \bar{z} = \sigma^0 - i \sigma^1 \, , \quad 
    \der \equiv \der_z = \frac{1}{2}(\der_0 - i \der_1) \, , \quad
    \derb\equiv \der_{\zb} = \frac{1}{2}(\der_0 + i \der_1) \, ,
\end{equation}
for which the worldsheet metric becomes  ${\rm diag}(1/2,1/2)$. 

One can also consider additional couplings, related to the presence of other background fields. For instance, given a globally-defined closed $3$-form $H$ on $\M$ one can include a term proportional to the Wess-Zumino action, 
\begin{equation}
\label{SWZsigmamodel}
    S_{\rm WZ} = \frac{i}{6\pi \alpha'} \int_{\Omega} H_{\mu\nu\rho}(X) dX^\mu \wedge dX^\nu \wedge dX^\rho = \frac{i}{4\pi \alpha'} \int_{\Sigma} B_{\mu\nu}(X) \der_\alpha X^\mu \der_\beta X^\nu \vep^{\alpha \beta}, 
\end{equation}
where $\vep^{01}=1$, and $\Omega$ is a 3-dimensional manifold whose boundary is $\Sigma$. The factor of $i$ makes the Euclidean action real. The choice of $\Omega$ is irrelevant since $H$ is closed, while the final expression in \eqref{SWZsigmamodel} is only valid when $H$ is exact, $H = dB$ for some two-form $B$. The inclusion of $S_{\rm WZ}$ implies that our string-like object is (electrically) charged under the potential $B_{\mu\nu}$. For non-trivial worldsheet metrics $\gamma_{\alpha \beta}$ one can also include a coupling to a target-space scalar field $\Phi(X)$, namely the dilaton: 
\begin{equation}
    S_{\rm \Phi} = \frac{1}{4\pi \alpha'} \int_\Sigma d^2 \sigma \sqrt{\gamma} \, \alpha' R^{(2)} \Phi(X),
\end{equation}
where $R^{(2)}$ is the two-dimensional Ricci scalar. The fields $G$, $B$ and $\Phi$ constitute the massless fields of the universal NSNS sector of closed string theories.  

Except when $G$ is flat and both $B$ and $\Phi$ are trivial, the non-linear $\sigma$ model is an interacting theory. By expanding in powers of $X$, the backgrounds fields then define a (possibly infinite) set of interaction terms. Focusing on the kinetic sector, we can estimate the corresponding coupling constants by thinking of a configuration where $G$ has a characteristic length $r_G$. Then, roughly speaking, the effective dimensionless coupling is given by $\sqrt{\alpha'}/r_G$. When this is small, the worldsheet theory is weakly coupled and can be treated perturbatively, and the low-energy physics are captured in terms of a gravitational model in $d$ dimensions. In particular, the $\beta$-functions characterizing the renormalization group flow of the $\sigma$-model are identified with the equations of motion (EOMs) of the effective spacetime action. In the string theory context, and at the two-derivative level, this provides a derivation of the EOMs of general relativity coupled with the relevant matter fields. 

Of course, if we are interested in  highly energetic processes it is necessary to include an ever-growing set of higher-derivative terms in the effective action, known as $\alpha'$-corrections, which  quickly becomes quite cumbersome.  
Moreover, when $\sqrt{\alpha'}/r_G$ is not small the worldsheet model is strongly coupled. In order to address both of these problems it would be very useful to be able to study the worldsheet theory beyond the  perturbative regime.  However, it is not known how to do this in arbitrary backgrounds. 

We now introduce Wess-Zumino-Witten (WZW) models, which constitute a set of non-linear $\sigma$-models which can  be treated exactly -- at least in principle -- and provide the crucial building blocks in the construction of non-perturbative string backgrounds. Let us take the target space $\M$ to be the group manifold associated to a Lie group $\mathfrak{G}$. The WZW action is then 
\begin{equation}
\label{SWZW}
    S_{\rm WZW} = S_{\rm kin} + k \Gamma. 
\end{equation}
Here $S_{\rm kin}$ is the kinetic part of the action defined in Eq.~\eqref{Skinsigma}, which can be written in this context as 
\begin{equation}
\label{SkinWZW}
    S_{\rm kin} = \frac{1}{4a^2} \int d^2\sigma \Tr \left[g^{-1} \der^\alpha g g^{-1} \der_\alpha g\right], 
\end{equation}
where we have included an overall constant which will be fixed shortly. For simplicity, we consider the model as defined on the Riemann sphere, $\Sigma=S^2$. The bosonic field $g (z,\bar{z})$ parametrizes the group elements of $\mathfrak{G}$, such that $g^{-1}\der_\alpha g$ is an element of the corresponding Lie algebra $\mathfrak{g}$. The trace in \eqref{SkinWZW} defines the Killing form, and for a suitable basis of a matrix representation of $\mathfrak{g}$, $t^a$, $a=1,\dots,{\rm dim} \, \mathfrak{g}$ we have 
\begin{equation}
    \Tr \left[t^a t^b\right] = 2g^{ab} \qqquad 
    \left[t^a, t^b\right] = i f^{ab}{}_c t^c\,,
\end{equation}
where $f^{ab}{}_c$ are the structure constants and $g^{ab}$ is the Killing metric. 
The second term in $S_{\rm WZW}$ is the WZ part of the action, which takes the form
\begin{equation}
    \label{def WZ term}
    \Gamma = \frac{-i}{24\pi} \int_\Omega d^3y \vep_{\alpha \beta \gamma} \Tr \left[ 
    \tilde{g}^{-1} \der^\alpha \tilde{g}
    \tilde{g}^{-1} \der^\beta \tilde{g}
    \tilde{g}^{-1} \der^\gamma \tilde{g}
    \right]. 
\end{equation}
One must be careful when extending the field $g$ to $\tilde{g}$, which is defined on a given $\Omega$, since the procedure is not unique. For small variations $\tilde{g} \to \tilde{g} + \delta \tilde{g}$, one finds that the ingrand of the corresponding variation $\delta \Gamma$ is a total derivative, hence $\Gamma$ is invariant as long as $\delta \tilde{g}|_{S^2} = \delta g = 0$. On the other hand, when the variation is topologically non-trivial we can use that $(\Omega \cup \Omega') / S^2$ is topologically equivalent to a three-sphere, and compute the difference between two choices from
\begin{equation}
    \Delta \Gamma = 
    \frac{-i}{24\pi} \int_{S^3} d^3y \vep_{\alpha \beta \gamma} \Tr \left[ 
    \tilde{g}^{-1} \der^\alpha \tilde{g}
    \tilde{g}^{-1} \der^\beta \tilde{g}
    \tilde{g}^{-1} \der^\gamma \tilde{g}
    \right]. 
\end{equation}
When $\mathfrak{G}$ is compact and has an SU(2) subgroup, it turns out to be enough to focus on mappings $g:S^3 \to SU(2)$, hence one can choose \cite{DiFrancesco:1997nk}
\begin{equation}
    g(y) = y^0 - i y^i \sigma^i \qqquad 
    (y^0)^2 + y^iy^i = 1 \, \quad i=1,2,3 \,,  
\end{equation}
where $\sigma^i$ are the Pauli matrices, while $y^0,y^i$ are coordinates on $S^3$.  This leads to 
\begin{equation}
    \Delta \Gamma = \frac{\pi}{6} \vep_{ijk} \Tr \left[
    \sigma^i\sigma^j\sigma^k
    \right] = 2 \pi i. 
\end{equation}
We conclude that the exponential of (minus) the action \eqref{SWZW} is invariant as long $k$ takes integer values. The parameter $k$ is known as the \textit{level} of the WZW model. In more general situations, such as the SL(2,$\R$) WZW model discussed in the main text, $k$ may not be quantized. 

WZW models are classically scale invariant, and also invariant under global $\mathfrak{G}_L\times \mathfrak{G}_R$ transformations 
\begin{equation}
g(z,\bar{z}) \to g_L g(z,\bar{z}) g_R^{-1}.      
\end{equation}
In the $\sigma$-model language, these come from the Killing vectors of the target space metric which also leave the background $H$-flux untouched.  The EOMs derived from $S_{\rm WZW}$ read 
\begin{equation}
    \left(1+\frac{a^2k}{4\pi}\right)
    \der \left(g^{-1} \derb g \right) + 
    \left(1-\frac{a^2k}{4\pi}\right)
    \derb \left(g^{-1} \der g \right) = 0.
\end{equation}
Hence, upon taking $k>0$ and choosing $a^2 = 4\pi/k$ we obtain 
\begin{equation}
    \der \left(g^{-1} \derb g \right) = 0 \,, \qquad \derb \left( \der g g^{-1}\right) = 0\,,
\end{equation}
where the second identity is implied by the first one since $\derb \left( \derb g g^{-1}\right) = g \der \left(g^{-1} \derb g \right) g^{-1}$. Hence, WZW models (by which we mean those with $a^2 = 4\pi/k$ from now on) have holomorphic and anti-homorphic conserved currents given by 
\begin{equation}
    J = - k \der g g^{-1} \qqquad 
    \bar{J} = k g^{-1} \derb g. 
\end{equation}
Indeed, the global $\mathfrak{G}_L\times \mathfrak{G}_R$ symmetry  is enhanced to a local one:
\begin{equation}
\label{gLggR-WZW}
g(z,\bar{z}) \to g_L(z) g(z,\bar{z}) g_R^{-1}(\bar{z}) \, .   
\end{equation}
For instance, for $g_L(z) = 1+ \omega(z)$ and $g_R(\zb) = 1$ we have
\begin{equation}
\label{delSWZW}
    \delta_{\w} S_{\rm WZW } 
    = - \frac{1}{2\pi} \int d^2 \sigma  
    \, \der_{\zb} \Tr\left[ 
    \w(z) J(z)\right] = 0 \,.    
\end{equation}
From now on we focus on the holomorphic sector, although analogous formulas can be obtained for the anti-holomorphic one. 
%seguir

Let us now move to the quantum theory. Eq.~\eqref{delSWZW} translates into the (local) Ward identity for the correlation functions of the theory, namely 
\begin{equation}
\label{WardJ-WZW}
    \delta_{\w} \langle X \rangle = - \oint dz \w^a(z) \langle J^a(z) X \rangle, \qquad 
    J(z) \equiv J^a(z) t^a, 
\end{equation}
where we have integrated by parts\footnote{Here and from now on we will omit the factors $(2\pi i)^{-1}$ when dealing with contour integrals. 
}.
On the other hand, the variation $\delta g = \w g$ implies $\delta_\w J = [\w,J] - k \der \w$. By combining this with Eq.~\eqref{WardJ-WZW} we find the operator product expansion 
\begin{equation}
\label{JJOPE-WZW}
    J^a(z)J^b(w) \sim \frac{k g^{ab}}{(z-w)^2} + \frac{i f^{ab}{}_c J^c(w)}{z-w},
\end{equation}
which holds inside correlation functions, and defines the current algebra of the model. In terms of the corresponding modes 
\begin{equation}
J^a(z) = \sum_{n \in \mathbb{Z}} z^{-n-1} J^a_n    \,, 
\end{equation} this gives 
\begin{equation}
\label{JnJmalgebra-WZW}
    \left[J^a_n,J^b_m\right] = \oint_0 dw 
    \oint_w dz \, w^n z^m J^a(z) J^b(\w) =  k n g^{ab} \delta_{n+m,0} + 
    i f^{ab}{}_c J^c_{n+m},
\end{equation}
which defines an affine Lie algebra $\hat{\mathfrak{g}}$ at level $k$. The modes $J^a_n$ and $\bar{J}^b_m$ commute with each other, hence the current algebra factorizes. 

The presence of the two conserved currents $J(z)$ and $\bar{J}(\zb)$ is very powerful. In particular, it implies that WZW models are conformally invariant at the quantum level. In order to see this, we need to show that there is a conserved holomorphic tensor $T(z)$ of weight $h=2$ whose modes generate the Virasoro algebra, captured by the OPE 
\begin{equation}
\label{TTOPE-WZW}
    T(z)T(w) \sim \frac{c/2}{(z-w)^4} + \frac{2 T(w)}{(z-w)^2} + \frac{\der T(w)}{z-w},
\end{equation}
where $c$ is the central charge. This follows from the well-known Sugawara construction. We can start from the classical expression for the energy-momentum tensor and propose 
\begin{equation}
    T(z) \equiv \gamma g_{ab} (J^a J^b)(z) = \gamma \oint_z \frac{dw}{w-z} g_{ab} J^a(w)J^b(z), 
\end{equation}
for some constant $\gamma$, and where the brackets indicate normal ordering. Note that, classically, one has $\gamma = (2k)^{-1}$. However, requiring that the currents are primary fields of weight $h=1$ with respect to $T(z)$, namely 
\begin{equation}
\label{TJOPE-WZW}
    T(z)J^a(w) \sim \frac{J^a(w)}{(z-w)^2} + \frac{\der J^a(w)}{z-w},
\end{equation}
imposes  
\begin{equation}
    \gamma = \frac{1}{2(k+\mathfrak{c})} \,, 
\end{equation}
where $\mathfrak{c}$ is the so-called dual Coxeter number, defined by 
    $f^{ab}{}_c f^{cd}{}_b = 2 \mathfrak{c} g^{ab}$. 
Hence, the value of $\gamma$ receives a quantum correction. Indeed, $k \to \infty$ defines the classical limit, where all modes $J_{n\neq 0}^a$ decouple. As the fields $g(z,\zb)$ are dimensionless, $1/\sqrt{k}$ plays the role of the effective coupling. Once the quantum shift is taken into account, it can be seen that the modes 
\begin{equation}
\label{Lnmodesdef-WZW}
    T(z) = \sum_{n \in \mathbb{Z}} z^{-n-2} L_n \qqquad 
    L_n = \frac{g_{ab} }{2(k + \mathfrak{c})} \left[
    \sum_{m < 0} J^a_m J^b_{n-m} +
    \sum_{m \geq 0} J^a_{n-m}J^b_m 
    \right]
\end{equation}
satisfy the Virasoro algebra
\begin{equation}
    \left[L_n,L_m\right] =   \frac{c}{12} n (n^2-1) \delta_{n+m,0} + 
    (n-m) L_{n+m} \qqquad 
    c = \frac{k \, {\rm dim} \mathfrak{g}}{k + \mathfrak{c}}\,, 
\end{equation}
while the commutation relations with the current modes implied read
\begin{equation}
    \left[L_n,J^a_m\right] =   -m J^a_{n+m}.
\end{equation}
We conclude that WZW models constitute two-dimensional conformal field theories. 

Let us now discuss the primary fields of the theory, which will be denoted as $V_{j,\bar{\jmath}}(z,\zb)$ (and in general do \textit{not} factorize). Here we must distinguish between the usual Virasoro primaries and the so-called affine primary fields. Affine primaries appear in CFTs with symmetry algebras of the type \eqref{JnJmalgebra-WZW}. They transform as in Eq.~\eqref{gLggR-WZW}, which translates into an OPE with the currents of the form 
\begin{equation}
\label{JVOPE-WZW}
    J^a(z)V_{j,\bar{\jmath}}(w,\bar{\w}) \sim \frac{-t^a_j V_{j,\bar{\jmath}}(w,\bar{w})}{z-w}
\,, \qquad 
\bar{J}^a(\zb)V_{j,\bar{\jmath}}(w,\bar{\w}) \sim \frac{V_{j,\bar{\jmath}}(w,\bar{w})t^a_{\bar{j} }}{\bar{z}-\bar{w}}\,,
\end{equation}
where $j$ and $\bar{\jmath}$ label the corresponding   zero-mode representations $R_j \times R_{\bar{\jmath}}$. Of course, in general there will be additional quantum numbers which specify the particular state we are dealing with, $V_{j,\bar{\jmath},m,\bar{m}}(z,\zb)$ with $m = 1,\dots,\,{\rm dim}\, R_{j}$ and $\bar{m} = 1,\dots,\,{\rm dim} \, R_{\bar{\jmath}}$, but we will omit them for now.  In these notes we mainly focus on \textit{diagonal} operators with $\bar{\jmath} = j$, which represent scalar fields. For the state created by a scalar operator $V_{j}(z,\zb)$  acting on the vacuum, 
\begin{equation}
    V_{j}(0) |0\rangle = |j\rangle\,
\end{equation}
the simple pole in the OPE captures the action of the current zero-modes, while the absence of higher-order poles shows that positive modes act as annihilation operators, i.e.~ 
\begin{equation}
\label{AffinePrimaryStates-WZW}
    J_0^a |j\rangle = - t_j^a |j\rangle \qqquad 
    J_n^a |j\rangle = 0 \quad \forall \quad n>0. 
\end{equation}
Note the minus sign in the action of the zero modes, which leads to 
\begin{equation}
    [J_0^a,J_0^b]|j\rangle = 
    (J_0^b t_j^a - 
    J_0^a t_j^b)|j\rangle = 
    (t_j^a J_0^b  - 
    t_j^b J_0^a )|j\rangle = 
    - [t_j^a,t_j^b]|j\rangle = 
    - i f^{ab}{}_c t_j^c |j\rangle =
    i f^{ab}{}_c J_0^c |j\rangle \, ,
    \nn
\end{equation}
as expected.

Importantly, Eq.~\eqref{AffinePrimaryStates-WZW}  automatically implies that  
\begin{equation}
    L_0 |j\rangle =  \frac{g_{ab}t_j^at_j^b}{2(k + \mathfrak{c})} |j\rangle \qqquad 
    L_n |j\rangle = 0 \quad \forall \quad n>0, 
\end{equation}
which shows that the conformal weight $h = h(j)$ is proportional to the quadratic Casimir of the representation. In other words,
\begin{equation}
\label{TVOPE-WZW}
    T(z)V_j(w,\bar{w}) \sim \frac{h V_j(w,\bar{w})}{(z-w)^2} + \frac{\der V_j(w,\bar{w})}{z-w} \qqquad 
    h = \frac{g_{ab}t_j^at_j^b}{2(k + \mathfrak{c})}. 
\end{equation}
The rest of the states in the affine module are of the form $J_{n_1}^{a_1}J_{n_2}^{a_2}\dots |jm\rangle$ with $n_i < 0$. 
In most examples one is interested in the highest-weight representations of the zero-mode algebra $\mathfrak{g}$. These contain a so-called highest-weight state, from which all the other states can be obtained by acting with $J^a_{n \leq 0}$. 

We have shown that affine primary fields are also Virasoro primaries. In most cases it is enough to focus on the physics of such operators. This holds for instance for models with (Euclidean) compact target spaces, such as the SU(2) WZW model.  However, the converse statement is not true. As discussed in the main text, this introduces a number of important additional features when discussing strings in AdS$_3$, described by the WZW model based on the universal cover of SL(2,$\R$), which is both Lorentzian and non-compact. As a consequence, an important part of its spectrum is characterized by operators which are Virasoro primaries, but \textit{not} affine primaries.  

We now discuss correlation functions in WZW models. They must be invariant under the action of both gauge and conformal transformations. For correlators of primary fields, this is captured by the (global) Ward identities, giving 
\begin{equation}
 0=    \sum_{i=1}^n t^a_{j_i} \langle V_{j_1}(z_1,\bar{z}_1) \dots V_{j_n}(z_n,\bar{z}_n)  \rangle,
\end{equation}
 and 
\begin{subequations}
\label{TWard-WZW}
    \begin{eqnarray}
     0 &=&   \sum_{i=1}^n \der_{z_i} \langle V_{j_1}(z_1,\bar{z}_1) \dots V_{j_n}(z_n,\bar{z}_n)  \rangle, \\
     0 &=&   \sum_{i=1}^n (z_i \der_{z_i} + h_i) \langle V_{j_1}(z_1,\bar{z}_1) \dots V_{j_n}(z_n,\bar{z}_n)  \rangle, \\
     0 &=&   \sum_{i=1}^n (z_i^2 \der_{z_i} + 2 z_i h_i) \langle V_{j_1}(z_1,\bar{z}_1) \dots V_{j_n}(z_n,\bar{z}_n)  \rangle \,, 
    \end{eqnarray}
\end{subequations}
respectively.
As is well known, Eqs.~\eqref{TWard-WZW} completely fix the coordinate dependence of two- and three-point functions. Moreover, there are additional constraints coming from the Sugawara construction itself. The most notable one is known as the Knizhnik-Zamolodchikov equation. The definition \eqref{Lnmodesdef-WZW} for the mode $L_{-1}$ in terms of the currents implies that for any affine primary state $|j\rangle$, 
\begin{equation}
    \left( L_{-1} - \frac{g_{ab}}{k+\mathfrak{c}} 
    J^a_{-1}J^b_{0}\right) |j\rangle = 0. 
\end{equation}
This leads to a differential equation for correlation functions of WZW models, namely 
\begin{equation}
\label{KZequation-WZW}
    \left[ \der_{z_i} + 
    \frac{g_{ab}}{k+\mathfrak{c}}
    \sum_{j \neq i}
    \frac{t^{a}_i \otimes t^b_j}{z_i - z_j}\right] 
    \langle V_{j_1}(z_1,\bar{z}_1) \dots V_{j_n}(z_n,\bar{z}_n)  \rangle, 
\end{equation}
where we have used that $L_{-1}\sim \der_z$, the OPE \eqref{JVOPE-WZW}, and 
\begin{eqnarray}
    \langle V_{j_1}(z_1,\bar{z}_1) \dots 
    (J_{-1}^a V_{j_i})(z_i,\zb_i)
    \dots V_{j_n}(z_n,\bar{z}_n)  \rangle & = & \oint_{z_i} \frac{dz}{z-z_i} 
    \langle J^a(z) V_{j_1}(z_1,\bar{z}_1)  \dots V_{j_n}(z_n,\bar{z}_n)  \rangle \nn \\
    &=& \sum_{j \neq i} \frac{t^a_j}{z_i-z_j} 
    \langle J^a(z) V_{j_1}(z_1,\bar{z}_1)   \dots V_{j_n}(z_n,\bar{z}_n)  \rangle 
    \nn . 
\end{eqnarray}
In the last step we have inverted the contour of integration and used that $J^a(z \to \infty) \sim z^{-2}$, so that there is no pole at infinity. The tensor product in Eq.~\eqref{KZequation-WZW} is used to keep track of the action 
\begin{equation}
    \left(t^a_i \otimes t^a_j \right) V_{j_i}(z_i,\zb_i) V_{j_j}(z_j,\zb_j) \equiv 
    (t^a_iV_{j_i})(z_i,\zb_i) (t^a_jV_{j_j})(z_j,\zb_j).
\end{equation}
The existence of this crucial constraint distinguishes WZW models from generic two-dimensional CFTs. Additional constraints arise when other null vectors are present. 

For some WZW models, the KZ equation leads to exact solutions for four-point functions of primary fields. This is the case, for instance, for SU($N$)$_{k}$-WZW models, with $N,k \in \mathbb{N}$. For this, one uses the conformal Ward identities \eqref{TWard-WZW} to fix the dependence on the insertions $z_i$ and $\zb_i$, with $i=1,\dots,4$, up to an arbitrary function of the cross-ratios, defined as  
\begin{equation}
z \equiv \frac{z_{23}z_{14}}{z_{12}z_{34}} \qqquad z_{ij} = z_i - z_j,     
\end{equation}  
and similarly for $\zb$. Thus, one has 
\begin{equation}
    \langle \prod_{i=1}^4 V_{j_i}(z_i,\zb_i) \rangle = |
    z_{12}^{h_3-h_1-h_2-h_4}
    z_{13}^{h_2+h_4-h_1-h_3}
    z_{23}^{h_1+h_4-h_2-h_3}
    z_{34}^{-2h_4}
    |^2 F_{14}^{23}(z,\zb).  
\end{equation}
The KZ equations can then be written as partial differential equations of the hypergeometric type for the functions $F_{14}^{23}(z,\zb)$, known as (combinations of) the conformal blocks. The exact solution can then be obtained by using the representation theory of SU($N$), and further imposing that the correlator must be invariant under \textit{crossing symmetry}, i.e.~the exchange symmetry between the different fields, which reduces to the identity
\begin{equation}
    F_{14}^{23}(z,\zb) = F_{24}^{13}(1-z,1-\zb).
\end{equation}
The resulting expressions as well as the details of the computation can be found in  \cite{DiFrancesco:1997nk}. 
%%%%%%%%%%%%%%%%%%%%%%%%%%%%%%%

\subsection{The SU(2) model}
\label{sec: SU(2) WZW model}

We now discuss one of the prototypical examples: the WZW model based on SU(2) \cite{Zamolodchikov:1986bd}, for which the target space is $S^3$. We can parametrize the group elements in terms of the Euler angles, 
\begin{equation}
    g = e^{\frac{i}{2}(\psi - \varphi)\sigma^3} 
    e^{i \theta \sigma^1} 
    e^{\frac{i}{2}(\psi + \varphi)\sigma^3} = 
    \left( \begin{array}{cc}
        e^{i\psi}\cos \theta  & i e^{-i\varphi}\sin \theta \\
        i e^{i\varphi}\sin \theta & e^{-i\psi}\cos \theta
    \end{array} \right) \, ,  
\end{equation}
where $\sigma^a$ are the Pauli matrices, $a=1,2,3$.
By inserting this expression for $g$ into the WZW action \eqref{SWZW} one can read off the target space line element and $H$-field, namely \begin{equation}
ds^2 = k \left(d\theta^2 + \sin^2\theta d\varphi^2 + 
\cos^2\theta d\psi^2\right) \, , \qquad
H = k \sin 2\theta d\theta \wedge d\psi \wedge d\varphi 
%=    d \left(
%k \sin^2 \theta d\psi \wedge d\varphi
%\right)
\, . 
\end{equation}
The generators of the Lie algebra are $t^a = \sigma^a/2$. Generic solutions of the classical EOMs  $\bar{\der} (\der g g^{-1}) = 0$ take the factorized form $g (z,\zb)= g_L(z)g_R(\zb)$. From the holomoprhic currents, which we will denote as $k^a(z)$, one can define the ladder operators $k^{\pm}(z) = k^1(z) \pm i k^2(z)$. The currents  satisfy the OPE \eqref{JJOPE-WZW} with $g^{ab} = \delta^{ab}/2$ and $f^{ab}{}_c = \vep^{ab}{}_c$, where $\vep^{012} = 1$. In terms of the corresponding modes this gives 
\begin{equation}
    [k^3_n,k^3_m] = \frac{k}{2} n \delta_{n+m,0}, \qquad 
    [k^3_n,k^\pm_m] = \pm k^\pm_{n+m}, \qquad
    [k^+_n,k^-_m] = 2 k^3_{n+m} + k n \delta_{n+m,0}. 
\end{equation}
The energy-momentum tensor and central charge are 
\begin{equation}
\label{T-c-SU2}
    T(z) = \frac{1}{k+2} (k_a k^a)(z) \qqquad 
    c = \frac{3k}{k+2}. 
\end{equation}
The relevant highest-weight representations of the zero-mode algebra are given by states of the form  
\begin{equation}
    \left\{ |l\, n\rangle, \, l \in \frac{1}{2}\mathbb{N}_0, \, 
    n = -l, -l+1, \dots, l-1, l\right\},  
\end{equation}
where $l$ is the SU(2) spin and $n$ the corresponding projection along the $k^3$ direction. 
The associated vertex operators $W_{ln\bar{n}}(z,\zb)$ transform in such highest-weight representations of the holomorphic and anti-holomoprhic zero-mode algebras, and with $\bar{l} = l$.   
We work with conventions where the currents act as
\begin{equation}
    k^3(z) W_{ln\bar{n}}(w,\bar{w}) \sim \frac{n W_{ln\bar{n}}(w,\bar{w})}{z-w},
    \quad 
    k^\pm(z) W_{ln\bar{n}}(w,\bar{w}) \sim \frac{(l+1\pm n)) W_{l,n\pm 1,\bar{n}}(w,\bar{w})}{z-w}, 
\end{equation}
such that the weights are 
\begin{equation}
\label{Delta-l-SU2}
    \Delta_l = \bar{\Delta}_l = \frac{l(l+1)}{k+2}. 
\end{equation}
By restricting to diagonal states with $l = \bar{l}$ we have chosen to work with the scalar vertex operators. 

The SU(2) Lie algebra can be described in terms of the following first order differential operators: 
\begin{equation}
    -t_0^+ \sim P^+_l = \der_u, \qquad -t^3_0 \sim P^3_l = u \der_u - l, \qquad 
    -t^+_0 \sim P^-_l = - u^2\der_u + 2 l u,   
\end{equation}
where $u$ is known as the isospin coordinate.  Relatedly, we have a one-parameter family of automorphisms of the  affine algebra, which can be understood as translating the currents from $u=0$ to a different point in $u$-space, namely 
\begin{equation}
    k^a(z) \to k^a(u,z) = e^{u k_0^+} 
    k^a(z) e^{-u k_0^+},  
\end{equation}
giving  $k^+(u,z)=k^+(z)$ and 
\begin{equation}
    k^3(u,z) = k^3(z)-u k^+(z), \qquad
    k^-(u,z) = k^-(z)+2u k^3(z)-u^2k^+(z). 
\end{equation}
Using the isospin variable we can encode all states in a given zero-mode representation by defining  
\begin{equation}
    W_{l}(u,\bar{u},z,\zb) = 
    e^{\bar{u} \bar{k}_0^+} e^{u k_0^+}  
    W_{l,-l,-\bar{l}}(z,\zb) e^{-u k_0^+} e^{-\bar{u} \bar{k}_0^+}
    = \sum_{n,\bar{n}=-l}^l
    u^{l-n}\bar{u}^{l-\bar{n}} W_{ln\bar{n}}(z,\zb)\,.
\end{equation}
The operator at the origin $u=0$ is identified as $W_{l}(0,0,z,\zb) =  W_{l,-l,-\bar{l}}(z,\zb)$. Then, we have for these $u$-basis vertex operators 
\begin{equation}
    k^a(z)W_{l}(u,\bar{u},w,\bar{w}) \sim \frac{P_l^a W_{l}(u,\bar{u},w,\bar{w})}{z-w}. 
\end{equation}
This allows us to express the Ward identities \eqref{WardJ-WZW} stemming from the action of the zero-mode algebra as partial differential equations in $u$, leading to  
\begin{subequations}
\label{JWard-SU2}
    \begin{eqnarray}
     0 &=&   \sum_{i=1}^n \der_{u_i} \langle W_{l_1}(u_1,\bar{u}_1,z_1,\bar{z}_1) \dots W_{l_n}(u_n,\bar{u}_n,z_n,\bar{z}_n)  \rangle, \\
     0 &=&   \sum_{i=1}^n (u_i \der_{u_i} - l_i) \langle W_{l_1}(u_1,\bar{u}_1,z_1,\bar{z}_1) \dots W_{l_n}(u_n,\bar{u}_n,z_n,\bar{z}_n)  \rangle, \\
     0 &=&   \sum_{i=1}^n (u_i^2 \der_{u_i} - 2 u_i l_i) \langle W_{l_1}(u_1,\bar{u}_1,z_1,\bar{z}_1) \dots W_{l_n}(u_n,\bar{u}_n,z_n,\bar{z}_n)  \rangle.
    \end{eqnarray}
\end{subequations}
These equations are  analogous (up to some signs) to the conformal Ward identities of Eq.~\eqref{TWard-WZW}, with the spin $l$ playing the role of the holomorphic weight $\Delta$. Hence, we find that the $z$ and $u$ dependence of two- and three-point functions is fully fixed by the symmetries of the system.  A suitable choice for the normalizations leads to 
\begin{equation}
    \langle W_{l_1}(u_1,\bar{u}_1,z_1,\bar{z}_1) W_{l_2}(u_2,\bar{u}_2,z_2,\bar{z}_2)  \rangle = \delta_{l_1,l_2}\frac{|u_{12}|^{4l_1}}{|z_{12}|^{4\Delta_1}}
\end{equation}
and 
\begin{equation}
\label{3pt-SU2-zudep}
    \langle \prod_{i=1}^3W_{l_i}(u_i,\bar{u}_i,z_i,\bar{z}_i) \rangle = C(l_1,l_2,l_3)\Bigg|\frac{u_{12}^{l_1+l_2-l_3}u_{23}^{l_2+l_3-l_1}u_{13}^{l_1+l_3-l_2}}{
    z_{12}^{\Delta_1+\Delta_2-\Delta_3}
    z_{23}^{\Delta_2+\Delta_3-\Delta_1}
    z_{13}^{\Delta_1+\Delta_3-\Delta_2}}\Bigg|^2, 
\end{equation}
where $C(l_1,l_2,l_3)$ are the structure constants of the model. Their explicit expressions were derived in \cite{Zamolodchikov:1986bd}. Once they are fixed, correlators of the original operators $W_{ln\bar{n}}$ can be deduced from the $u$-basis ones by expanding in powers of $u_i$ and isolating the corresponding coefficients. For three-point functions, this imposes the charge conservation condition $n_1+n_2+n_3=0$ and leads to  
the appearance of the well-known Clebsch-Gordan coefficients \cite{Fateev:1985mm}. 

\subsection{Null states in the SU(2) model. Partition function}

We now discuss an important aspect of the SU(2) model at level $k$: the presence of null states in the affine modules. From the current algebra it follows that  
\begin{equation}
\label{Nullstate-SU2}
    (k_{-1}^-)^{k-2l+1}|l,-l\rangle = 0 \qqquad l \leq \frac{k}{2}. 
\end{equation}
This can be seen for instance by showing that all $J^a_{n>0}$ annihilate the state on the LHS of \eqref{Nullstate-SU2}. Since this state is also a descendant, it must be  null.  Alternatively, one can derive it by showing that \cite{EberhardtLectures} 
\begin{equation}
\label{norm proto null states}
    || (k_{-1}^-)^{N}|l,-l\rangle ||^2 = \langle l,l|
    (k_{1}^+)^{N}(k_{-1}^-)^{N}|l,-l\rangle = \prod_{n=1}^N n (k-2l+1-n)\,,
\end{equation}
using induction. Hence, the norm vanishes for $N = k-2l+1$. We will provide an alternative proof in the following section using spectral flow. For now, we describe the consequences of Eq.~\eqref{Nullstate-SU2}.
As in the derivation of the KZ equation, we insert the null state 
\begin{equation}
\label{Nullvertex-SU2}
    k_{-1}^-(u)^{k-2l+1} W_{l}(u,\bar{u},z,\bar{z}) = 0 
\end{equation}
in a given correlation function, which leads to the constraint 
\begin{equation}
     0 =   \left[\sum_{i=1}^n \frac{(u-u_i)^2 \der_{u_i} + 2 l_i (u- u_i)}{z-z_i}\right]^{k-2l+1} \langle W_{l}(u,\bar{u},z,\bar{z})  \prod_{i=1}^n W_{l_i}(u_i,\bar{u}_i,z_i,\bar{z}_i)  \rangle
\end{equation}
by means of the Ward identities \eqref{JWard-SU2}. Setting $n=2$ and using \eqref{3pt-SU2-zudep} then gives 
\begin{equation}
    (l_1+l_2-l)(l_1+l_2-l-1)\dots (l_1+l_2+l-k) C(l,l_1,l_2)=0. 
\end{equation}
From this, we see that for any $l_1,l_2 \leq k/2$,  $C(l_1,l_2,l_3)$ vanishes for all $l_3 > k/2$. Hence, the theory defined by including only the fields with $l \leq k/2$ is self-consistent. Moreover, we can deduce the fusion rules of the model
\begin{equation}
    C(l_1,l_2,l_3) \neq 0 \quad \Leftrightarrow \quad l_1+l_2+l_3 \leq k \quad \text{and} \quad l_i \leq l_j+l_k \quad \forall \, i,j,k=1,2,3.  
\end{equation}
In other words, we recover the well-known result that the product $W_{l_1} \otimes W_{l_2}$ only contains states $W_l$ with $|l_1-l_2|\leq l \leq \, \text{min} \, (l_1+l_2,k-l_1-l_2)$. 

In order to check that this is correct, one may consider the partition function of the SU(2) model and establish modular invariance. The distribution of weights and charges in the theory is captured by the partition function of the theory defined on a torus with modular parameter $\tau$, namely 
\begin{equation}
Z(\tau,\bar{\tau}) = \Tr \left[q^{L_0 - \frac{c}{24}}\bar{q}^{\bar{L}_0 - \frac{c}{24}}\right]\qqquad q = e^{2\pi i \tau}.    
\end{equation}
Conformal symmetry requires $Z(\tau,\bar{\tau})$ to be modular invariant, i.e.~invariant under the action of the $S$ and $T$ transformations, 
\begin{equation}
    T: \, \tau \to \tau+1 \qqquad S: \, \tau \to -\frac{1}{\tau}.
\end{equation}
The partition function decomposes into a sum of contributions from each highest-weight representation of the left- and right-handed current algebras, 
\begin{equation}
    Z(\tau,\bar{\tau}) = \sum_{l,\bar{l}} M_{l \bar{l}} 
    \chi_l(\tau)\chi_{\bar{l}}(\bar{\tau}).
\end{equation}
Although here we have focused on the diagonal case $M_{l\bar{l}} = \delta_{l,\bar{l}}$, there are actually other modular invariant constructions, which fall into the so-called ADE classification \cite{DiFrancesco:1997nk,EberhardtLectures}. 

For the holomorphic sector of the SU(2) model, the corresponding characters read
\begin{equation}
\label{Character-l-SU2}
    \chi_l (\tau,z) = \Tr_l\left[
    q^{L_0 - \frac{c}{24}} y^{k_0^3}
    \right] = \frac{
\Theta_{2l+1}^{(k+2)}(\tau,z)-\Theta_{-2l-1}^{(k+2)}(\tau,z)
    }{\Theta_1^{(2)}(\tau,z)-\Theta_{-1}^{(2)}(\tau,z)} \qqquad y = e^{2\pi i z},  
\end{equation}
where we have restricted to the spin-$l$ sector, introduced a chemical potential in order to keep track of the U(1) charge measured by $k_0^3$, and used the level-$k$ theta functions
\begin{equation}
\label{ExpThetaFunction}
    \Theta_{l}^{(k)}(\tau,z) = \sum_{n \in \mathbb{Z} + \frac{l}{2k}} q^{kn^2} y^{kn}.
\end{equation}
It is instructive to understand how Eq.~\eqref{Character-l-SU2} comes about. The first ingredient that should appear is the global character, 
\begin{equation}
    \label{GlobalCharacter-l-SU2}
    \chi_l(z) = \sum_{n=-l}^l y^{n} = \frac{
    y^{l+\frac{1}{2}} - y^{-l-\frac{1}{2}}}{
    y^{\frac{1}{2}}-y^{-\frac{1}{2}}}. 
\end{equation}
We should also account for the affine descendants. Those coming from the modes $k^3_n$ for a given negative $n$ give 
\begin{equation}
    \sum_{j=0}^\infty q^{j n} = (1-q^n)^{-1}, 
\end{equation}
while the action of $k^\pm_{n<0}$ induces an extra unit shift in the charges, giving analogous contributions with $q^n\to y q^n$ and $q^n\to y^{-1} q^n$. 
Combining these factors and taking into account the central charge \eqref{T-c-SU2} as well as the primary weights \eqref{Delta-l-SU2} would lead to 
\begin{equation}
\label{WrongCharacter-l-SU2}
    \frac{q^{\frac{l(l+1)}{k+2}-\frac{k}{8(k+2)}} \chi_l(z)}
    {
    \prod_{n=1}^{\infty} (1-q^n) (1-y q^n) (1- y^{-1}q^n)
    }.
\end{equation}
The Jacobi's triple product identity shows that the denominator coincides with that of Eq.~\eqref{Character-l-SU2}. However, \eqref{WrongCharacter-l-SU2} only captures the leading term in an expansion in powers of $q$ of the numerator in \eqref{Character-l-SU2}. The correct expression takes the form of an alternating series,
\begin{equation}
\label{ExpCharacter-l-SU2}
   \chi_l (\tau,z) = 
   \frac{q^{\frac{l(l+1)}{k+2}-\frac{k}{8(k+2)}} 
   \left[\chi_l(z) - q^{k+1-2l}\chi_{k+1-l}(z) +  
   q^{k+3+2l}\chi_{k+2+l}(z) - \cdots\right] }
    {
    \prod_{n=1}^{\infty} (1-q^n) (1-y q^n) (1- y^{-1}q^n)
    }.
\end{equation}
Indeed, the expression given in Eq.~\eqref{WrongCharacter-l-SU2} does not take into account the presence of null states in the affine module, which must not be counted. The subtraction of the null state \eqref{Nullstate-SU2} together with its descendants gives precisely the shift $\chi_l(z) \to \chi_l(z) - q^{k+1-2l}\chi_{k+1-l}(z)$. The following term has a similar origin: there is an additional null state among the descendants of \eqref{Nullstate-SU2}, namely 
\begin{equation}
    (k^-_{-2})^{2l+1}(k^-_{-1})^{k-2l+1}|l,-l\rangle.    
\end{equation}
This means that we have to add back this state and its descendants in order to avoid over-counting. The corresponding weight  and spin with respect to the zero-mode algebra imply that this is done by including the contribution $q^{k+3+2l}\chi_{k+2+l}(z)$ in \eqref{ExpCharacter-l-SU2}. This pattern repeats \textit{ad infinitum}, generating the full series expansion of the Theta functions appearing in the numerator of Eq.~\eqref{Character-l-SU2}.  

Finally, one can use the modular properties of the Theta functions to obtain those of the characters, and show that the full partition function is modular invariant. 
We will not reproduce the explicit calculation, which can be found for example in \cite{EberhardtLectures}.  

\subsection{Spectral flow in the SU(2) case}

Another property of the affine SU(2) at level $k$ algebra is the existence of the spectral flow automorphisms 
\begin{equation}
\label{spectral flow - SU2}
    k^\pm_n \to \tilde{k}^\pm_n =  k^\pm_{n\pm \w}, \quad 
    k^3_n \to \tilde{k}^3_n = k^3_n + \frac{k\w}{2} \delta_{n,0} \qqquad \w \in \mathbb{Z}.    
\end{equation}
At the level of the Sugawara construction, this induces 
\begin{equation}
    L_n \to \tilde{L}_n = L_n + \w k^3_n + \frac{k}{4}\w^2 \delta_{n,0}. 
\end{equation}
All relevant SU(2) states are contained in the affine highest-weight modules. The spectral flow automorphisms \eqref{spectral flow - SU2} merely reshuffle primary states and their descendants, not necessarily with the same value of $j$. Conversely, all states in highest-weight representations of the zero-mode algebra in the spectrally flowed frame, i.e.~the one generated by $\tilde{k}_0^\pm = k^\pm_{\pm w}$ and $\tilde{k}^3_0 = k^3_0 + \frac{k\w}{2}$, must correspond to combinations of primaries and descendants in terms of the original states. Indeed, consider a state with spin $l$ and projection $m$ in a flowed frame with $\w>0$, which we denote $|l\, n\, \w\rangle$. This satisfies 
\begin{equation}
    k^3_0|l\, n\, \w\rangle = \left(n - \frac{k\w}{2}\right)|l\, n\, \w\rangle, \quad
    L_0|l\, n\, \w\rangle = \left(\Delta_l  - w n + \frac{k\w^2}{4}\right)|l\, n\, \w\rangle,
\end{equation}
as well as 
\begin{equation}
    k^+_{n>\w}|l\, n\, \w\rangle = k^-_{n>-\w}|l\, n\, \w\rangle = 
    k^3_{n>0}|l\, n\, \w\rangle = L_{n>0}|l\, n\, \w\rangle = 0.
\end{equation}
Thus, although $|l\, n\, \w\rangle $ is clearly not an affine primary, we see that it is a  Virasoro primary. We also note that $k_0^-$ annihilates the state, hence we can read off the spin from (minus) the eigenvalue of $k_0^3$. This shows that the associated vertex operator, which we denote as $W_{ln}^\w (z,\zb)$ and refer to as a  \textit{spectrally flowed primary}, transforms in a spin-$l_\w$ representation with $l_\w = -n+\frac{\w}{2}$. However, these operators  are not always different form each other. By matching the quantum numbers and choosing appropriate normalizations, one obtains the following identifications 
\begin{equation}
\label{SeriesId-SU2}
    W_{l,-l}^\w (z,\zb) = W_{\tilde{l},\tilde{l}}^{\w+1} (z,\zb) \qqquad \tilde{l} = \frac{k}{2}-l.
\end{equation}

Let us pause and consider in more detail the identity \eqref{SeriesId-SU2} for the case $\w=0$, which shows that we can identify the unflowed, lowest-weight state $|l,-l\rangle$ with the highest-weight state of spin $\tilde{l}$ in the spectrally flowed frame with $\w=1$, namely 
\begin{equation}
|l,-l\rangle = |\tilde{l} \, \tilde{l} \, \w = 1\rangle   \,. 
\end{equation}
The latter is annihilated by applying the corresponding lowering operator (at least)  $2\tilde{l}+1 = k - 2l+1$ times. Since in this frame $\tilde{k}_0^- = k_{-1}^-$, this provides an alternative proof for Eq.~\eqref{Nullstate-SU2}.   

As stated above, spectrally flowed primaries are nothing but specific descendants of the original ones. We can use \eqref{SeriesId-SU2} to describe exactly which descendants we are dealing with: 
\begin{eqnarray}
    |l\, n\, \w\rangle &\sim & (k^{-,\w}_0)^{l-n} |l\, l\, \w\rangle \nn \\
    &=& (k^{-}_{-\w})^{l-n} |\tilde{l}, -\tilde{l}, \w-1\rangle \nn \\
    &\sim & (k^{-}_{-\w})^{l-n} (k^{-,\w-1}_0)^{2\tilde{l}} |\tilde{l}, \tilde{l}, \w-1\rangle \nn \\
    &=& (k^{-}_{-\w})^{l-n} (k^{-}_{-\w+1})^{2\tilde{l}} |l, -l, \w-2\rangle 
    \, = \, \dots \, , \nn
\end{eqnarray}
where $k^{a,\w}_n$ are the currents $\tilde{k}^a_n$ in a given frame with spectral flow $\w$. Hence, recalling that $\tilde{l} = \frac{k}{2}-l$, we find 
\begin{equation}
    |l\, n\, \w\rangle \sim  
    \begin{cases}
        (k^{-}_{-\w})^{l-n} (k^{-}_{-\w+1})^{k-2l} (k^{-}_{-\w+2})^{2l} \dots (k^{-}_{-1})^{k-2l} |l, -l\rangle & \w \, \in  \, 2 \mathbb{Z}, \\
        (k^{-}_{-\w})^{l-n} (k^{-}_{-\w+1})^{k-2l} (k^{-}_{-\w+2})^{2l} \dots (k^{-}_{-1})^{2l} |\frac{k}{2}-l, l-\frac{k}{2}\rangle  & \w \, \in \, 2  \mathbb{Z}+1, 
    \end{cases}
\end{equation}
where we have ignored all proportionality constants. This confirms that there are no new states in the spectrally flowed sectors of the theory, which  is in stark contrast with the SL(2,$\R$) case studied in the main text. 
Moreover, we immediately see that there are many other null states in highest-weight affine modules, which generalize Eq.~\eqref{Nullstate-SU2}: 
\begin{equation}
   0 =  \begin{cases}
        (k^{-}_{-\w+1})^{k-2l+1} (k^{-}_{-\w+2})^{2l} \dots (k^{-}_{-1})^{k-2l} |l, -l\rangle & \w \, \in  \, 2 \mathbb{Z}, \\
        (k^{-}_{-\w+1})^{k-2l+1} (k^{-}_{-\w+2})^{2l} \dots (k^{-}_{-1})^{2l} |\frac{k}{2}-l, l-\frac{k}{2}\rangle  & \w \, \in \, 2  \mathbb{Z}+1. 
    \end{cases}
\end{equation}
These are precisely the states we removed when constructing the partition function above. 
We can also see this directly at the level of the characters. Given the spectral flow transformations, the character of a flowed highest-weight representation reads 
\begin{equation}
\label{ytoyqw-SU2}
    \chi_l^\w (q,y) = \Tr_l\left[q^{\tilde{L}_0-\frac{c}{24}}y^{\tilde{k^3_0}}\right] =
    \Tr_l\left[q^{L_0+\w k_0^3 + \frac{k }{4} \w^2-\frac{c}{24}}y^{k^3_0 + \frac{k}{2}\w}\right]
    = q^{\frac{k}{4}\w^2} y^{\frac{k}{2}\w} \chi_l (q,yq^\w). 
\end{equation}
Using \eqref{ExpThetaFunction} we find that for the denominator of the character \eqref{Character-l-SU2} one has 
\begin{equation}
    \Theta_{1}^{(2)} (q,y q^\w) - \Theta_{-1}^{(2)} (q,y q^\w) = (-1)^\w y^{-\w} q^{-\frac{\w^2}{2}} \left [\Theta_{1}^{(2)} (q,y) - \Theta_{-1}^{(2)} (q,y) \right]. 
\end{equation}
For the numerator the result depends on the parity of $\w$. For even $\w$, and up to an overall factor, the effect of the replacement $y \to y q^\w$ can be reabsorbed as a shift in the summation index of \eqref{ExpThetaFunction}, $n \to n - \frac{\w}{2}$. On the other hand, for odd $\w$ we would need a half-integer shift. This can be reinterpreted as a change in the spin, since 
\begin{equation}
    l \, \to \, \frac{k}{2} - l \qquad \Rightarrow 
    \qquad 
    \frac{2l+1}{2(k+2)} \, \to \, 
    \frac{1}{2} - \frac{2l+1}{2(k+2)}. 
\end{equation}
The additional overall factors then combine to cancel those in the final expression of Eq.~\eqref{ytoyqw-SU2}, such that  
\begin{equation}
    \chi_l^\w (q,y) =  \begin{cases}
        \chi_l(q,y) & \w \, \in \, 2\mathbb{Z}, \\
         \chi_{\frac{k}{2}-l}(q,y) & \w \, \in \, 2\mathbb{Z}+1.
    \end{cases}
\end{equation}
As a result, we see that spectral flow leaves the full partition function invariant.

%\subsection{Gauged WZW models}

%%%%%%%%%%%%%%%%%%%%%%%%%%%%%%%%%%%%%%%%%%

% \section{Geometry of AdS$_3$ and BTZ black holes}
% \label{sec: appA BTZ}

% \begin{itemize}
%     \item Global AdS$_3$ vs Poincare patch (HCFT NSNS  vs RR vacuum)
%     \item Lorentzian vs Euclidean
%     \item BTZ black holes as quotients
% \end{itemize}

%\section{Liouville Theory and its relation to the $H_3^+$ model}
%\label{sec: appC Liouville}

% \begin{tcolorbox}[title=Exercise,colback=yellow!10!white,colframe=red!75!black,lowerbox=invisible,
%   savelowerto=\jobname_ex.tex]
%   What to do
%   \tcblower
%   Solution
% \end{tcolorbox}

\section{Symmetric product orbifolds CFTs}
\label{sec: appD SymOrbifolds}

Given a conformal sigma-model with target space $\M$, which we will refer to as the \textit{seed} theory, one can define a family of CFTs as follows: 
\begin{equation}
    {\rm Sym}^N\left( \M\right) \equiv \M^N/S_N \, .
    \label{SymOrbDef}
\end{equation}
Here $N$ is a natural number, $\M^N$ stands for the direct product of $N$ copies of the original CFT on $\M$, say $\M^{(i)}$ with $i=1,\dots,N$, and $S_N$ denotes the permutation group, whose elements  act as 
\begin{equation}
    s: \, X^{(i)} \to  X^{s(i)}
    \,, \qquad s \in S_N \,,
\end{equation}
where $X^{(i)}$ is a generic field in the $i$-th copy of  the theory. 
In Eq.~\eqref{SymOrbDef} the action of the permutation group is gauged. Physical operators of the symmetric orbifold CFT must be gauge-invariant. The simplest example corresponds to operators in the \textit{untwisted} sector of the theory, such as 
\begin{equation}
    X \equiv \sum_{i=1}^N X^{(i)} \, ,
\end{equation}
where we sum over all copies of a given operator in the seed. This sector contains for instance the energy-momentum tensor of the theory, namely\footnote{In this appendix we use $x$ to denote the complex coordinate for the symmetric orbifold CFT in order to be consistent with the notation employed in the main text. Indeed, the symmetric orbifold CFT provides an approximate description of the holographic CFT living on the boundary of AdS$_3$.}
\begin{equation}
    \Tt(x) = \sum_{i=1}^N \Tt^{(i)}(x) \, ,
\end{equation}
from which one derives the central charge 
\begin{equation}
    c = N c_{\rm seed} \, .
\end{equation}
If the seed theory contains other conserved bosonic or fermionic currents, they similarly extend to the full symmetric orbifold CFT.  

One can also construct physical operators belonging to the twisted sectors of the theory. This is achieved by making use of the so-called (bare) twist operators $\sigma_s$. 
Their insertion modifies the boundary conditions for the different fields, giving 
\begin{equation}
    X^{(i)} (e^{2\pi i }x) =  X^{s(i)} (x) \, .
\end{equation}
For instance, for a twist field involving the first $\w$ copies by means of the permutation $(12\dots \w)$ we have  
\begin{equation}
    \sigma_{(12\dots \w)} : 
    X^{(1)} \to 
    X^{(2)} \to 
    \cdots \to X^{(\w)} 
    \to X^{(1)} \, .  
\end{equation}
The insertion of this twist operator at the origin thus imposes the modified twisted boundary conditions  
\begin{equation}
    X^{(i)} (e^{2\pi i} x) = X^{(i+1)} (x) \, ,
\end{equation}
where the indices are understood mod $\w$. This type of operators are called single-cycle twist fields. They will be our main focus since in the holographic applications they are identified with single-string states. Of course, one can also have multi-cycle fields, but these correspond to multi-string states in the bulk picture. Their properties can be studied by fusing single-cycle operators.

We will use the compact notation $\sigma_{\w} \equiv \sigma_{(12\dots \w)}$. Now, consider for instance the operator $\sigma_2$ inserted at the origin $x=0$. This implements $X^{(1)} (e^{2\pi i} x) = X^{(2)}$ and $X^{(2)} (e^{2\pi i} x) = X^{(1)}$. Hence, although we have two functions $X^{(1)}$ and $X^{(2)}$, there is no global distinction between them. We can effectively trivialize the effect of the twist field insertions by going to the covering space, which, at least locally, corresponds to the Riemann surface of the function   $ z = x^{1/2}$. We can also think of this in terms of the inverse transformation, given by the holomorphic map $x = \Gamma(z) = z^2$. In the $z$-plane, we then have a single function $X(z)$, which we identify with, say, $X^{(1)}$ in the upper-half plane, and with $X^{(2)}$ in the lower-half plane. This implies that $X(z)$ is single-valued. 

We can also include another insertion of $\sigma_2$ at infinity, and evaluate the expectation value of the energy momentum tensor $\langle \sigma_2| \Tt(z)| \sigma_2 \rangle$ \cite{Dixon:1986qv}. The latter is dictated by the conformal anomaly generated by the transformation to the covering space. This allows one to derive the (holomorphic) weight of the twist operator, $
    h[\sigma_2]  = \frac{c_{\rm seed}}{16}$. 
More generally, for a (single-cycle) twist operator $\sigma_\w$ involving $\w$ copies of the theory one uses $x = \Gamma(z) = z^{\w}$, which leads to \cite{Lunin:2000yv}  
\begin{equation}
\label{weight sigma w}
    h [\sigma_\w] = \frac{c_{\rm seed} (\w^2-1)}{24\w} \, .  
\end{equation}
This defines the weight of the ground state in the $\w$-twisted sector. 

The spectrum of the theory is then obtained by dressing these bare twist operators with excitations\footnote{To be precise, in order to complete the construction we should take gauge-invariant linear combinations of these operators. This will be discussed shortly.}. The latter can be constructed directly from those of the seed theory. In the $\w$-twisted sector we can do this by using any operator with weights $h_{\rm seed}$ and $\hb_{\rm seed}$ in the seed theory which satisfy $h_{\rm seed} - \hb_{\rm seed} \in \w \mathbb{Z}$. For scalar operators with $h_{\rm seed} = \hb_{\rm seed}$, this leads to weights of the form \cite{Eberhardt:2021vsx}
\begin{equation}
\label{SymOrb spectrum generic}
    h = \frac{c_{\rm seed} (\w^2-1)}{24\w} + \frac{h_{\rm seed}}{\w} \, .
\end{equation}
The factor of $\w$ dividing $h_{\rm seed}$ in the last term  makes sense because (for the two-point function) the relation between the covering space coordinate and the physical one is of the form $z \sim x^{1/\w}$.

In the twisted sector where the insertion of $\sigma_{\w}$ at the origin defines the ground state, one can trade ${X^{(1)},\dots,X^{(\w)}}$ for a basis of operators which diagonalize the twisted boundary conditions. These are defined (up to a phase) as 
\begin{equation}
    X^r (x) \equiv \sum_{j=0}^{\w-1}
    e^{-\frac{2\pi i j r}{\w}} X^{(j)} (x)\,, \quad r = 0, \dots, \w-1 \qquad \Rightarrow 
    \qquad X^r (e^{2\pi i}x) = e^{\frac{2 \pi i r}{\w}} X^r(x) \, .
\end{equation}
For the case of $X = \Tt$, this shows that in the $\w$-twisted sector one can define the fractional Virasoro modes 
\begin{equation}
\label{def Vir fractional modes}
    \Ll_{\frac{n}{\w}} = \oint dx \, x^{\frac{n}{\w}+1} \,  \Tt^r(x) \, \qquad n \in \w \mathbb{Z}-r \, .
\end{equation}
These fractional modes satisfy the Virasoro algebra 
\begin{equation}
    \left[ \Ll_{\frac{n}{\w}} ,\Ll_{\frac{m}{\w}} \right] = 
    \left(\frac{n}{\w}-\frac{m}{\w}\right) \Ll_{\frac{n+m}{\w}} + \frac{c_{\rm seed} \w}{12} \frac{n}{\w} \left(\frac{n^2}{\w^2}-1\right) \delta_{n+m,0}\, ,
\end{equation}
hence the central charge in this sector is $c^{(\w)} = c_{\rm seed} \w$. The fractional generators can be written in terms of \textit{untwisted} ones $\hat{\Ll}_n$ with central charge $c_{\rm seed}$ by means of 
\begin{equation}
    \Ll_{\frac{n}{\w}} = \frac{1}{\w}\hat{\Ll}_n + \frac{c_{\rm seed} (\w^2-1)}{24 \w} \delta_{n,0} \, .
\end{equation}

\subsection{Correlators and large $N$ counting} 

We now move to correlators. We first focus on bare twist fields for simplicity. Consider a sample $n$-point function of the form 
\begin{equation}
    \langle \sigma_{\w_1} (x_1) \cdots  \sigma_{\w_n} (x_n) \rangle \, .
\end{equation}
This can only be non-zero if the product of the corresponding permutations (in a given order) amounts to the identity operator. For instance, for $n=3$ a non-trivial example with $\w_1 = 2$, $\w_2=2$ and $\w_3=3$ is given by  
\begin{equation}
    \langle \sigma_{(12)} (x_1)
     \sigma_{(13)} (x_2)
     \sigma_{(123)} (x_3)\rangle \, .
\end{equation}
Let us focus on connected correlators, namely those that do not factorize trivially because some of the insertions involve a subset of copies of the theory which are untouched by the others. They can be computed via the covering space method \cite{Lunin:2000yv}. For this one needs to construct a holomorphic map $\Gamma: \Sigma \to S^2$ satisfying 
\begin{equation}
\label{Gamma map exp}
    \Gamma(z \sim z_i) = x_i + a_i (z-z_i)^{\w_i} + \cdots  
\end{equation}
for all $i = 1,\dots,n$. Here $x$ is the complex coordinate on the physical sphere, while $z$ is the coordinate on the covering surface $\Sigma$. Hence, $\Sigma$ has ramifications of order $r_i = \w_i-1$ at each of the insertion points. Let us further denote the total number of \textit{active} copies of the theory by $m$. For instance, $m = 3$ in the example above because all insertions involve only the copies $1$, $2$ and $3$. The value of $m$ sets the total number of sheets of the covering surface, i.e.~the number of preimages $\Gamma^{-1}(x)$ at a generic point $x$ (which was denoted as $N$ in the main text). This is related to the genus $g$ of $\Sigma$ by the Riemann-Hurwitz formula \cite{cavalieri_miles_2016}
\begin{equation}
\label{Riemann-Hurwitz genus g}
    g = 1-m + \frac{1}{2}\sum_{i=1}^{n} (\w_i-1) \, .
\end{equation}
In the example above we have $g=0$, while for instance for $\langle \sigma_{(123)}\sigma_{(123)}\sigma_{(123)}\rangle $ we get $g=1$.

The bare twist operators defined above are clearly not gauge-invariant. One can construct gauge-invariant twist operators $\sigma_{[\w]}$ by summing over the corresponding $S_N$ orbit, 
\begin{equation}
    \sigma_{[\w]} = N_\w \sum_{s\in S_N} \sigma_{s (12\dots\w)s^{-1}} \, ,  \qquad 
    N_{\w} = \frac{1}{\sqrt{\w (N-\w)!N!}} \, .
\end{equation}
Here $N_\w$ is a normalization factor ensuring that $\sigma_{[\w]}$ has unit two-point functions (assuming this was true for the original twist fields $\sigma_\w$). Hence, one can expand a given gauge-invariant $n$-point function in terms correlators of bare twist operators, namely 
\begin{equation}
\label{gauge-invariant corr sym orb}
    \left\langle \prod_{i=1}^n \sigma_{[\w_i]}\right\rangle =  \left( \prod_{i=1}^n N_{\w_i} \right)\sum_{g=0}^{\infty} \sum_{s_1,\dots,s_n \in S_N}
    \left\langle \prod_{j=1}^n \sigma_{s_j(12\dots \w_j)s_j^{-1}}\right\rangle_g \, ,
\end{equation}
where we have included a sum over the genera of the covering maps employed in each case. Note that generically there will be several contributions to a given correlation function coming from different covering maps. 

However, not all of the contributions on the RHS of \eqref{gauge-invariant corr sym orb} are independent. 
%Indeed, the twist operators also have non-trivial monodromies around each other, which take the form \cite{Dei:2019iym}
% \begin{equation}
%     \sigma_{s_2} (e^{2\pi i}z) \sigma_{s_1}(0) = \sigma_{(s_2 s_1)s_2 (s_2 s_1)^{-1}} (z) \sigma_{(s_2 s_1)s_1(s_2 s_1){-1}}(0) \, .  
% \end{equation}
Indeed, for each $\sigma_{\w_j}$ there are exactly $(N -\w_j)! w_j$ choices of $s_j$ that leave $(12\dots \w_j)$ invariant. Finally, by means of the Riemann-Hurwitz formula \eqref{Riemann-Hurwitz genus g}, for each value of  $g$ we have a fixed total number of active copies, denoted by $m = m(g,\w_i)$. We are then left with a sum over contributions generating the orbit of the $S_{m}$ subgroup  associated with permutations of these specific copies of the theory, which is thus independent of $N$. 

We are interested in taking the large $N$ limit of the model. The arguments above show that for correlators of gauge-invariant twist operators, the genus $g$ contribution scales with $N$ as follows: 
\begin{equation}
\label{counting power N}
    \left\langle \prod_{i=1}^n \sigma_{[\w_i]}\right\rangle_g \sim  {{N}\choose{m}} \prod_{i=1}^n \sqrt{\frac{(N-\w_i)! \w_i}{N!}} \sim N^{1-g-\frac{n}{2}}\, .
\end{equation}
where $2m = 2-2g+\sum_{i=1}^n (\w_i-1)$. In the last step we have used Stirling's approximation. In the final expression we see that the exponent of $1/N$ is precisely (half of) the Euler characteristic  of a genus-$g$ Riemann surface with exactly $n$ punctures. At least at first sight, this looks analogous to  the string-theoretical topolgical expansion we would expect from the holographic duality, provided we identify the dual string coupling as 
\begin{equation}
    g_s^2 \sim N^{-1} \,. 
\end{equation}
To be precise, this holds in the strict large $N$ limit, hence the leading contributions come from covering spaces which have the  topology of a two-dimensional sphere. These are precisely the type of correlators  considered in the main text. For a  discussion on how to extend this picture to make sense of the subleading corrections, see \cite{Eberhardt:2021jvj,Aharony:2024fid}.

Once the combinatorial factors are taken into account, all that remains is to actually compute the bare twist correlators. This is done by lifting to the appropriate covering space, where the  twisted boundary conditions induced by these operators are trivialized. The computation can be carried out quite explicitly by following the methods of \cite{Dixon:1986qv,Lunin:2000yv,Dei:2019iym}. Here we simply quote the result, and refer the interested reader to the original publications for the details. At genus zero one obtains    
\begin{equation}
    \left\langle \prod_{i=1}^n \sigma_{\w_i} (x_i)\right\rangle = \left|\prod_{i=1}^n 
    \w_i^{-\frac{c_{\rm seed}(\w_i+1)}{24}}
    a_i^{\frac{c_{\rm seed}(\w_i-1)}{24}}
    \Pi^{-\frac{c_{\rm seed}}{12}} \right|^2 \, .
\end{equation}
Here the $a_i$ stand for the  coefficients involved in the expansion of the relevant covering map around each of the $z_i$, see Eq.~\eqref{Gamma map exp}. These coefficients contain the dependence on the insertion points $x_i$. The $g=0$ covering maps $\Gamma(z)$ associated with three-point functions  are  derived explicitly in Sec.~\ref{sec: bosonic correlators}. Finally, $\Pi$ represents the product of the residues of the covering map at each of its simple poles.  More explicitly, for covering surfaces with $g=0$ the number of simple poles is precisely that of active copies. There are thus $m$ points $\lambda_a$ (with $a = 1,\dots,m$) for which $\Gamma(z\sim \lambda_a) = \frac{c_a}{z-\lambda_a} + \cdots$, hence $\Pi = \prod_{a=1}^m c_a$. For three-point functions this was computed in \cite{Lunin:2000yv}. 

More generally, one can also compute correlators involving excitations built form the operators of the seed theory. Denoting by $\sigma_{\w_i,h_i}$ the operator consisting of a bare twist field dressed with an excitation corresponding to a seed theory operator ${\cal{O}}_i$ one finds the following result \cite{Dei:2019iym}: 
\begin{equation}
    \left\langle \prod_{i=1}^n \sigma_{\w_i,h_i} (x_i)\right\rangle = \left|\prod_{i=1}^n 
    \w_i^{-\frac{c_{\rm seed}(\w_i+1)}{24}}
    a_i^{\frac{c_{\rm seed}(\w_i-1)}{24}-h_i}
    \Pi^{-\frac{c_{\rm seed}}{12}} \right|^2  
    \left\langle \prod_{i=1}^n {\cal{O}}_i (z_i)\right\rangle \, .
\end{equation}
% Answer 7
To be precise, here we have considered the contribution from a given connected genus-zero covering surface. The full large $N$ result should include a sum over all such covering surfaces. For the exact correlator, one should also include the disconnected ones, for which the counting in \eqref{counting power N} works slightly differently, as well as covering surfaces of higher genera\footnote{The disconnected contributions can be treated analogously to disconnected Feynman diagrams in gauge theories. We thank the referee for their comments on this point.} 
.  

\subsection{The chiral ring for the $T^4$ case}

We now discuss the protected sector of the supersymmetric theory for the case where the seed CFT is the $\N = (4,4)$ sigma-model on $T^4$. The Virasoro, supercurrent and R-symmetry generators, denoted as $\Ll_n$, $\Gg_n^{\alpha A}$ and $J^a$ respectively, satisfy the following algebra:  
\begin{eqnarray}
    \left[\Ll_m, \Ll_n\right] &=& 
    (m-n)\Ll_{m+n} + \frac{c}{12} m(m^2-1) \delta_{m+n,0} \,, \\
    \left[\Kk_m^a, \Kk_n^b\right] &=& 
    i \vep^{ab}_{\phantom{ab}c}\Kk_{m+n}^c + \frac{c}{12} m \delta^{ab} \delta_{m+n,0} \, , \\
    \left\{\Gg_m^{\alpha A}, \Gg_n^{\beta B}\right\} &=& 
    (m-n)\vep^{AB} \vep^{\beta\gamma} \left(\sigma^{*a}\right)^\alpha_\gamma \Kk_{m+n}^a-\vep^{AB}\vep^{ \alpha \beta} \Ll_{m+n}  \nn \\
    && - \frac{c}{6} (m^2 - \frac{1}{4}) \vep^{AB} \vep^{\alpha \beta} \delta_{m+n,0} \, ,\\
    \left[\Ll_m, \Kk^a_n\right] &=& \phantom{\frac{1}{2}} - n \Kk^a_{m+n} \, ,\\
    \left[\Ll_m, \Gg^{\alpha A}_n\right] &=& \left(\frac{m}{2}-n\right) \Gg^{\alpha A}_{m+n} \, , \\
    \left[\Kk^a_m, \Gg^{\alpha A}_n\right] &=&  
    \frac{1}{2} \left(\sigma^{*a}\right)^\alpha_\beta 
    \Gg_{m+n}^{\beta A}\, .
\end{eqnarray}
Let us clarify the notation \cite{Avery:2010qw}. Here $a,b,c$ are the SU(2)$_L$ R-symmetry Lie algebra indices, $m,n$ are mode indices (which are integer for bosonic currents and half-integer for the fermionic ones), and $\alpha,\beta,\gamma$ are spinor indices for doublets of SU(2)$_L$. On the other hand, $A,B$ are indices for doublets under the action of the SU(2)$_1$ inside the  outer SO(4) $\sim$ SU(2)$_{1}\times$SU(2)$_{2}$ that generates the $T^4$ rotations. Similar formulas hold in the anti-holomorphic sector, where we use dotted indices $\dot{\alpha}, \dot{\beta}$ and $\dot{A},\dot{B}$ instead. The basic fields of the $T^4$ model are given by the compact scalars $X^i$ and their fermionic partners $\psi^{\alpha \dot{A}}$ and $\bar{\psi}^{\dot{\alpha}\dot{A}}$. One can also use spinor indices for the scalars by means of 
\begin{equation}
    X^{\dot{A}A} \equiv \frac{1}{\sqrt{2}} X^i \left(\sigma^i \right)^{\dot{A} A} \, ,
\end{equation}
where $\sigma^4 \equiv i \text{1}_2$.
In terms of these free fields, normalized such that their OPEs read 
\begin{equation}
    X^i(z) X^j(0) \sim - \delta^{ij} \log |z|^2 \, , \qquad 
    \psi^{\alpha \dot{A}}(z)\psi^{\beta \dot{B}}(0) \sim - \frac{\vep^{\alpha \beta} \vep^{\dot{A}\dot{B}}} {z} \, ,
\end{equation}
the currents are realized as 
\begin{eqnarray}
    \Gg^{\alpha A } &=& \phantom{\frac{1}{4}} \psi^{\alpha \dot{A}} (\der X)^{\dot{B}A} \vep_{\dot{A}\dot{B}} \, , \\
    \Kk^a &=& \frac{1}{4} \vep_{\dot{A}\dot{B}} \vep_{\alpha \beta} \psi^{\alpha \dot{A}} (\sigma^{*a})^\beta_\gamma \psi^{\gamma \dot{B}} \, , \\
    T &=& \frac{1}{2} \vep_{\dot{A}\dot{B}}\vep_{AB}  
(\der X)^{\dot{A}A} (\der X)^{\dot{B}B} + \frac{1}{2}  
\vep_{\dot{A}\dot{B}}\vep_{\alpha\beta} \psi^{\alpha \dot{A}}\psi^{\beta \dot{B}} \, .
\end{eqnarray}

The chiral operators are those that create primary states $|\phi\rangle$ satisfying the usual conditions 
\begin{equation}
    \Ll_1 |\phi\rangle = \Gg_{\frac{1}{2}}^{\alpha A} |\phi \rangle = \Kk_{1}^{a} |\phi \rangle = 0\, , \qquad  
    \Ll_0 |\phi\rangle = h |\phi\rangle  \,, \qquad   \Kk_{0}^{3} |\phi \rangle = q|\phi \rangle \, ,
\end{equation}
where $h$ is the holomorphic weight and $q$ the R-symmetry charge, together with the additional extremality condition
\begin{equation}
    h = q \, .
\end{equation}
(Anti-chiral states have $h = -q$.) These are extremal states in the sense that the anti-commutator
\begin{equation}
    \left\{\Gg_{\frac{1}{2}}^{- A},\Gg_{-\frac{1}{2}}^{+ B}\right\} = \vep^{AB} \left(\Kk_0^3 - \Ll_0\right) \, ,
\end{equation} 
combined with the condition that all norms must be greater or equal to zero imply that all states must satisfy $h \geq |q| $. Hence, chiral states satisfy 
\begin{equation}
    \Gg_{-\frac{1}{2}}^{+A}| \phi \rangle =0\, , \qquad A = 1,2, 
\end{equation}
so that they are invariant under half of the supersymmetry transformations. 

In the untwisted sector of the theory there are four chiral states. They take the form 
\begin{equation}
    |0\rangle \, , \qquad \psi_{-\frac{1}{2}}^{+ \dot{A}} |0\rangle
    \, , \qquad \Kk_{-1}^+|0\rangle \sim \psi_{-\frac{1}{2}}^{+ \dot{1}}\psi_{-\frac{1}{2}}^{+ \dot{2}} |0\rangle\,,
\end{equation}
such that their weights are $h = 0,\frac{1}{2},1$, respectively. A simple way to see that there are no chiral operators with higher weights in this sector is to note that all creation modes have $h>q$, except for $\psi_{-\frac{1}{2}}^{\alpha \dot{A}}$ and $\Kk^+_{-1}$ for which $h=q$, and one cannot act with each of the fermionic modes more than once. More explicitly, this is derived from the relation 
\begin{equation}
    \left\{\Gg_{\frac{3}{2}}^{- A},\Gg_{-\frac{3}{2}}^{+ B}\right\} = \vep^{AB} \left(3 \Kk_0^3 - \Ll_0 - 2\right) \, ,
\end{equation}
where we have used that for a single $T^4$ one has $c=6$. This leads to $0 \leq h\leq 1$ for modes with $h = q$, which extends to gauge-invariant operators in the untwisted sector of the full symmetric orbifold.  

On the other hand,  we must dress the bare twist operators (which have charge zero) appropriately in order to obtain the chiral operators belonging to the twisted sectors. Let us start with the case where we are in the $\w$-twisted sector, with $\w$ odd. In order to add charge at minimal cost in terms of the weight we can act with the fractional modes $\Kk^+_{-\frac{n}{\w}}$ -- defined in analogy to the fractional Virasoro modes of Eq.~\eqref{def Vir fractional modes} -- which have charge $1$ and weight $\frac{1}{\w}$. The first chiral state is obtained by acting with the first $\frac{\w-1}{2}$ fractional modes of $\Kk^+$ as  
\begin{equation}
    |\sigma^-_\w  \rangle= \Kk^+_{-\frac{\w-2}{\w}} \cdots \Kk^+_{-\frac{3}{\w}}\Kk^+_{-\frac{1}{\w}} \sigma_\w |0\rangle \, ,
\end{equation}
where we have omitted the anti-holomorphic modes. Indeed, this has  
\begin{equation}
    q[\sigma^-_\w] = \frac{\w-1}{2}  
\end{equation}
and 
\begin{equation}
    h[\sigma^-_\w] = \frac{\w^2-1}{4\w} + \sum_{p=1}^{(\w-1)/2} \frac{2p-1}{\w} = \frac{\w-1}{2} \, .
\end{equation}
For $\w$ even one must be slightly more careful because the boundary conditions for the fermions change upon going from the physical space to the covering space, hence one must include an additional spin field $S^+_\w$ in order to go to the Ramond vacuum of the corresponding twisted sector \cite{Lunin:2001pw}. The operators $S^+_\w$ have $h= \frac{1}{4\w}$ and $q = \frac{1}{2}$. The lowest-lying chiral primary in this sector is then 
\begin{equation}
    |\sigma^-_\w  \rangle= \Kk^+_{-\frac{\w-2}{\w}} \cdots \Kk^+_{-\frac{4}{\w}}\Kk^+_{-\frac{2}{\w}} S_\w^+\sigma_\w |0\rangle \, ,
\end{equation}
which again has 
\begin{equation}
    h[\sigma^-_\w] = q [\sigma^-_\w] = \frac{\w-1}{2}\, .
\end{equation}
As in the untwisted sector, one can obtain further chiral states by acting with the center-of-mass $\psi_{-\frac{1}{2}}^{+\dot{A}}$ and $\Kk_{-1}^+$. We denote the corresponding operators as $\sigma^{\dot{A}}_\w$ and $\sigma^+_\w$. Their weights are 
\begin{equation}
    h[\sigma^{\dot{A}}_\w] = \frac{\w}{2} \, , \qquad 
    h[\sigma^{+}_\w] = \frac{\w+1}{2} \, .
\end{equation}
It can be shown that this completes the list of chiral operators in the theory given by the symmetric orbifold of $T^4$. Indeed, by using the anticommutator of the fractional modes of $\Gg_{\pm n}^{\alpha A}$ as before one can show that, in the twisted sectors, the bound on the chiral primary weights takes the form 
\begin{equation}
    \frac{\w-1}{2} \leq h \leq \frac{\w+1}{2} \, . 
\end{equation}

Chiral(-chiral) primary operators are the top components in the short representations of the $\N=(4,4)$ algebra. The spectrum and three-point functions of operators belonging to such representations are protected by supersymmetry \cite{deBoer:2008ss,Baggio:2012rr}. They take the same form at any point in the moduli space of the symmetric orbifold SCFT, namely the superconformal manifold generated by all possible supersymmetry-preserving marginal deformations. In particular, for the top components the corresponding structure constants define the so-called chiral ring, given by the regular contributions to the chiral primary OPE in the coincidence limit, namely
\begin{equation}
    (\phi_a \cdot \phi_b)(0) \equiv \lim_{z\to 0} \phi_{a}(z) \phi_b(0)  = \sum_c C_{ab}^c \phi_c (0) \, . 
\end{equation}
Indeed, if $\phi_a$ and $\phi_b$ are chiral primaries there are no singular terms in the OPE 
\begin{equation}
    \phi_a(z) \phi_b(0) \sim \sum_{c}\sum_{n\geq 0} C_{ab}^c \frac{\der^n \phi_c}{z^{h_a + h_b  - h_c - n}} \, ,
\end{equation}
since 
\begin{equation}
    h_a + h_b - h_c -n= q_a + q_b - h_c -n\leq q_c - h_c - n \leq 0\, , 
\end{equation}
where we have used charge conservation together with  the unitarity condition $h_c \geq q_c$. The first regular term thus corresponds to chiral operators with $h_c = q_c$ \cite{Blumenhagen:2009zz}. 
The chiral primary structure constants were computed in \cite{Jevicki:1998bm} for the D1D5 CFT. In the large $N$($= n_1 n_5$) limit they take the following form \cite{Dabholkar:2007ey}: 
\begin{subequations}
\bea
&\braket{\sigma^{--}_{\w_1}\sigma^{--}_{\w_2}\sigma^{--}_{\w_3}}&= \frac{1}{\sqrt{N}}\off{\frac{\of{h_1 + h_2 + h_3-2}^4}{(2h_1-1)(2h_2-1)(2h_3-1)}}^{1/2},\\
&\braket{\sigma^{++}_{\w_1}\sigma^{--}_{\w_2}\sigma^{--}_{\w_3}}&=  \frac{1}{\sqrt{N}}\off{\frac{\of{1+ h_1 - h_2 - h_3}^4}{(2h_1-1)(2h_2-1)(2h_3-1)}}^{1/2},\\
&\braket{\sigma^{++}_{\w_1}\sigma^{++}_{\w_2}\sigma^{--}_{\w_3}}&=  \frac{1}{\sqrt{N}}\off{\frac{\of{h_1+h_2-h_3}^4}{(2h_1-1)(2h_2-1)(2h_3-1)}}^{1/2},\\
&\braket{\sigma^{++}_{\w_1}\sigma^{++}_{\w_2}\sigma^{++}_{\w_3}}&=  \frac{1}{\sqrt{N}}\off{\frac{\of{h_1+h_2+h_3-1}^4}{(2h_1-1)(2h_2-1)(2h_3-1)}}^{1/2},
\eea
\end{subequations}
and
\begin{subequations}
\bea
&\braket{\sigma^{\dot{A}_1\dot{B}_1}_{\w_1}\sigma^{\dot{A}_2\dot{B}_2}_{\w_2}\sigma^{--}_{\w_3}}&=  \frac{1}{\sqrt{N}}\off{\frac{\of{2h_1-1}\of{2h_2-1}}{(2h_3-1)}}^{1/2}\delta^{\dot{A}_1 \dot{A}_2}\delta^{\dot{B}_1\dot{B}_2},\\
&\braket{\sigma^{\dot{A}_1\dot{B}_1}_{\w_1}\sigma^{\dot{A}_2\dot{B}_2}_{\w_2}\sigma^{++}_{\w_3}}&=  \frac{1}{\sqrt{N}}\off{\frac{\of{2h_1-1}\of{2h_2-1}}{(2h_3-1)}}^{1/2}\xi^{\dot{A}_1 \dot{A}_2}\xi^{\dot{B}_1\dot{B}_2},
\eea
\end{subequations}
where $\xi = \sigma^1$ and $\delta = \text{1}_2$.

\bibliographystyle{JHEP}
\bibliography{refs}

\end{document}